\documentclass[a4paper,oneside,openany]{book}

\usepackage{amsmath,amsfonts,amssymb,amsthm,latexsym,xcolor,hyperref,comment,tikz}

\usepackage{graphicx,color,epsfig,rotating}
\usepackage{latexsym}
\usepackage{mismatch_fnt_shortcuts}
\usepackage{booktabs}
\usepackage{datetime}
\usepackage{subcaption}
\usepackage{dsfont}
\usepackage{multicol}
\usepackage{tocloft}

\usepackage[left=1in,right=1in,top=1in,bottom=1in,hmarginratio=1:1]{geometry}

 \newtheorem{theorem}{Theorem}
 \newtheorem{definition}{Definition}
 \newtheorem{proposition}{Proposition}
 \newtheorem{lemma}[theorem]{Lemma}
 \newtheorem{corollary}[theorem]{Corollary}
\newtheorem{remark}{Remark}

\numberwithin{theorem}{chapter}
\numberwithin{definition}{chapter}
\numberwithin{proposition}{chapter}
\numberwithin{remark}{chapter}

\newtheorem{openproblem}{Open Problem}

\graphicspath{{./figures/}}

\newcommand{\overbar}[1]{\mkern 1.25mu\overline{\mkern-1.25mu#1\mkern-0.25mu}\mkern 0.25mu}
\newcommand{\overbarr}[1]{\mkern 3.5mu\overline{\mkern-3.5mu#1\mkern-0.75mu}\mkern 0.75mu}
\newcommand{\mytilde}[1]{\mkern 0.5mu\widetilde{\mkern-0.5mu#1\mkern-0mu}\mkern 0mu}

\newcommand{\mywidehat}[1]{\mkern 1.5mu\widehat{\mkern-1.5mu#1\mkern-0mu}\mkern 0mu}

\DeclareMathOperator*{\argmax}{arg\,max}
\DeclareMathOperator*{\argmin}{arg\,min}
\DeclareMathOperator*{\maximize}{maximize}

\mathchardef\mhyphen="2D

\newcommand{\sgz}{s \ge 0}
\newcommand{\Tclass}{\Tc^n}
\newcommand{\pem}{p_{\mathrm{e},m}}
\newcommand{\CM}{C_{\textsc{m}}}
\newcommand{\CGMI}{C_{\textsc{gmi}}}
\newcommand{\CLM}{C_{\textsc{lm}}}
\newcommand{\CAWGN}{C_{\textsc{awgn}}}
\newcommand{\RMAC}{ \mathcal{R}_{\textsc{m}}^* }
\newcommand{\RMatchedMAC}{ \mathcal{R}_{\mathrm{Matched}}^* }
\newcommand{\Hvhat}{\mywidehat{\boldsymbol{H}}}

\newcommand{\Rbar}{\overbar{R}}
\newcommand{\Rbariid}{\overbar{R}_{\rm iid}}
\newcommand{\Rbarcc}{\overbar{R}_{\rm cc}}
\newcommand{\Hhat}{\mywidehat{H}}
\newcommand{\hhat}{\mywidehat{h}}

\newcommand{\Htilde}{\widetilde{H}}

\newcommand{\Xvhat}{\mywidehat{\boldsymbol{X}}}
\newcommand{\xvhat}{\mywidehat{\boldsymbol{x}}}
\newcommand{\Xhat}{\mywidehat{X}}
\newcommand{\xhat}{\mywidehat{x}}
\newcommand{\Xchat}{\mywidehat{\mathcal{X}}}

\newcommand{\Dibar}{\overbar{D}_1}
\newcommand{\DibarM}{\overbar{D}_{1,\mathrm{Matched}}}

\newcommand{\dimax}{d_{1,\mathrm{max}}}
\newcommand{\Rmatched}{R^*_{\mathrm{Matched}}}
\newcommand{\Dmatched}{D^*_{\mathrm{Matched}}}
\newcommand{\deltatilde}{\widetilde{\delta}}
\newcommand{\CSC}{C_{\textsc{sc}}}

\newcommand{\qtilde}{\widetilde{q}}
\newcommand{\Ybar}{\overbar{Y}}

\newcommand{\LMn}{I_{\textsc{lm},n}}
\newcommand{\Pctilde}{\widetilde{\mathcal{P}}}
\newcommand{\Gammamin}{\Gamma_{\mathrm{min}}}

\newcommand{\RBSS}{R^*_{\mathrm{symm}}}
\newcommand{\What}{\mywidehat{W}}
\newcommand{\RegSC}{\mathcal{R}_{\textsc{sc}}}
\newcommand{\RegRSC}{\mathcal{R}_{\textsc{rsc}}}
\newcommand{\RegMAC}{\mathcal{R}_{\textsc{mac}}}

\newcommand{\Hv}{\boldsymbol{H}}

\newcommand{\btilde}{\widetilde{b}}
\newcommand{\Phat}{\mywidehat{P}}
\newcommand{\Qunifk}{Q_{X^k}^{\mathrm{U}}}
\newcommand{\rsum}{r_{\mathrm{sum}}}

\newcommand{\Ndist}{\mathrm{N}}
\newcommand{\CNdist}{\mathrm{CN}}

\newcommand{\pebar}{\overbar{p}_{\mathrm{e}}}

\newcommand{\pe}{p_{\mathrm{e}}}

\newcommand{\pemax}{p_{\mathrm{e},\mathrm{max}}}

\newcommand{\Eziid}{E_{0}^{\mathrm{iid}}}
\newcommand{\Ezcc}{E_{0}^{\mathrm{cc}}}

\newcommand{\Ezcost}{E_{0}^{\mathrm{cost}}}

\newcommand{\Eriid}{E_{\mathrm{r}}^{\mathrm{iid}}}
\newcommand{\Ercc}{E_{\mathrm{r}}^{\mathrm{cc}}}

\newcommand{\Ercost}{E_{\mathrm{r}}^{\mathrm{cost}}}

\newcommand{\LM}{I_{\textsc{lm}}}
\newcommand{\GMI}{I_{\textsc{gmi}}}
\newcommand{\GCR}{R_{\textsc{gcr}}}

\newcommand{\SetSncc}{\mathcal{S}_{n}}

\newcommand{\rbar}{\overbar{r}}

\newcommand{\Ptilde}{\widetilde{P}}

\newcommand{\ubar}{\overbar{u}}
\newcommand{\Ubar}{\overbar{U}}

\newcommand{\xbar}{\overbar{x}}
\newcommand{\Xbar}{\overbarr{X}}
\newcommand{\xtilde}{\widetilde{x}}
\newcommand{\Xtilde}{\mytilde{X}}
\newcommand{\ybar}{\overbar{y}}

\newcommand{\Qv}{\boldsymbol{Q}}
\newcommand{\uvbar}{\overbar{\boldsymbol{u}}}
\newcommand{\Uvbar}{\overbar{\boldsymbol{U}}}
\newcommand{\Uv}{\boldsymbol{U}}

\newcommand{\xvbar}{\overbar{\boldsymbol{x}}}
\newcommand{\yvbar}{\overbar{\boldsymbol{y}}}
\newcommand{\xvtilde}{\widetilde{\boldsymbol{x}}}

\newcommand{\Xvbar}{\overbarr{\boldsymbol{X}}}
\newcommand{\Xvtilde}{\mytilde{\boldsymbol{X}}}

\newcommand{\Xv}{\boldsymbol{X}}
\newcommand{\Yv}{\boldsymbol{Y}}
\newcommand{\Zv}{\boldsymbol{Z}}

\newcommand{\Yvi}{\boldsymbol{Y}_{\hspace*{-0.5ex}1}}
\newcommand{\Yvii}{\boldsymbol{Y}_{\hspace*{-0.5ex}2}}

\newcommand{\PP}{\mathbb{P}}

\newcommand{\defeq}{\triangleq}

\newcommand{\Eexiid}{E_{\mathrm{ex}}^{\mathrm{iid}}}
\newcommand{\Eexcc}{E_{\mathrm{ex}}^{\mathrm{cc}}}

\newcommand{\Excc}{E_{\mathrm{x}}^{\mathrm{cc}}}

\newcommand{\Av}{\boldsymbol{A}}

\newcommand{\mbar}{\overbar{m}}

\newcommand{\peobar}{\overbar{p}_{\mathrm{e},0}}
\newcommand{\peibar}{\overbar{p}_{\mathrm{e},1}}
\newcommand{\peiibar}{\overbar{p}_{\mathrm{e},2}}
\newcommand{\peiiibar}{\overbar{p}_{\mathrm{e},12}}

\newcommand{\peiii}{p_{\mathrm{e},12}}

\newcommand{\Eiiircc}{E_{\mathrm{r},12}^{\mathrm{cc}}}

\newcommand{\RegLM}{\mathcal{R}_{\mathrm{LM}}}

\newcommand{\SetTiiincc}{\mathcal{S}'_{12,n}}

\newcommand{\Wv}{\boldsymbol{W}}

\newcommand{\Rosc}{R_{0}^{\mathrm{sc}}}
\newcommand{\Risc}{R_{1}^{\mathrm{sc}}}
\newcommand{\Rimac}{R_{1}^{\mathrm{ex}}}
\newcommand{\Riimac}{R_{2}^{\mathrm{ex}}}
\newcommand{\Usc}{U^{\mathrm{sc}}}

\newcommand{\Ximac}{X_{1}^{\mathrm{ex}}}

\newcommand{\Xiimac}{X_{2}^{\mathrm{ex}}}

\newcommand{\Iv}{\boldsymbol{I}}

\newcommand{\bzero}{\boldsymbol{0}}
\newcommand{\bone}{\boldsymbol{1}}

\newcommand{\bell}{\boldsymbol{\ell}}

\makeatletter
\renewcommand{\@chapapp}{Section}
\makeatother

\begin{document}

\title{Information-Theoretic Foundations of Mismatched Decoding}
\author{Jonathan Scarlett$^{1,2}$, Albert Guill\'en i F\`abregas$^{3,4,5}$, \\ Anelia Somekh-Baruch$^{6}$, and Alfonso Martinez$^{4}$ \\ [4mm] 
\small $^{1}$Department of Computer Science, National University of Singapore \\
\small $^{2}$Department of Mathematics, National University of Singapore \\
\small $^{3}$Instituci\'{o} Catalana de Recerca i Estudis Avan\c{c}ats (ICREA) \\
\small $^{4}$Department of Information and Communication Technologies, Universitat Pompeu Fabra \\
\small $^{5}$Department of Engineering, University of Cambridge \\
\small $^{6}$Faculty of Engineering, Bar-Ilan University \\ [4mm] 
\small e-mail: \url{scarlett@comp.nus.edu.sg}, \url{guillen@ieee.org}, \\ \small \url{somekha@biu.ac.il}, \url{alfonso.martinez@ieee.org}}

\date{}

\begin{minipage}{0.95\textwidth}
    \maketitle

    \vspace*{0.6cm}
    
    {\centering {\bf Abstract} \par} 
    
    \medskip
 

Shannon's channel coding theorem characterizes the maximal rate of information that can be reliably transmitted over a communication channel when optimal encoding and decoding strategies are used.  In many scenarios, however, practical considerations such as channel uncertainty and implementation constraints rule out the use of an optimal decoder.  The mismatched decoding problem addresses such scenarios by considering the case that the decoder cannot be optimized, but is instead fixed as part of the problem statement.  This problem is not only of direct interest in its own right, but also has close connections with other long-standing theoretical problems in information theory.
\bigskip 

In this monograph, we survey both classical literature and recent developments on the mismatched decoding problem, with an emphasis on achievable random-coding rates for memoryless channels.  We present two widely-considered achievable rates known as the generalized mutual information (GMI) and the LM rate, and overview their derivations and properties. In addition, we survey several improved rates via multi-user coding techniques, as well as recent developments and challenges in establishing upper bounds on the mismatch capacity, and an analogous mismatched encoding problem in rate-distortion theory.  Throughout the monograph, we highlight a variety of applications and connections with other prominent information theory problems. 

%
%

\end{minipage}

%


\newpage
\setlength{\cftbeforetoctitleskip}{5ex}
\setlength{\cftaftertoctitleskip}{5ex}

\tableofcontents

\chapter*{Notation}
\markboth{\sffamily\slshape Notation}{\sffamily\slshape Notation}
\addcontentsline{toc}{chapter}{Notation}

\begin{center}
    \begin{tabular}{cp{9cm}}
        %
        %
        \multicolumn{2}{l}{{\bf (Introduced in Section \ref{ch:intro})} } \\ [2mm]
        $\Xc$, $\Yc$ & Input and output alphabets\\
        $m$, $\hat{m}$ & Message and its estimate \\
        $M$ & Number of codewords \\
        $n$ & Block length \\
        $R$ & Coding rate \\
        $W$, $W^n$ & Channel transition law and its $n$-letter extension \\
        $q$, $q^n$ & Decoding metric and its $n$-letter extension \\
        $\Cc$ & Codebook \\
        $\xv^{(m)}$ & $m$-th codeword \\
        $\pe$ & Average error probability \\
        $\pemax$ & Maximal error probability \\
        $\CM$ & Mismatch capacity \\
        $C$ & Matched capacity \\ [4mm]
        %
        %
        \multicolumn{2}{l}{{\bf (Introduced in Section \ref{ch:single_user})} } \\ [2mm]
        $H_2$ & Binary entropy function \\
        $\Xv$, $\Yv$ & Transmitted codeword and received sequence \\
        $\Xvbar$ & Non-transmitted codeword \\
        $P_{\Xv}$ & Codeword distribution for random coding \\
        $\Phat_{\xv}$ & Empirical distribution of $\xv$  \\
        $\GMI$, $\LM$ & Generalized mutual information (GMI) and LM rate \\
        $\CGMI$, $\CLM$ & Rates with optimized input distributions \\
        $Q_X$ & Input distribution \\
        $\Pc$ & Set of all distributions on a given alphabet \\
        $\Pc_n$ & Set of all empirical distributions on a given alphabet \\
        $\Ptilde_{XY}$ & Auxiliary distribution in primal rates \\
        $s$, $a$ & Auxiliary parameters in dual rates \\
        $b$ & Auxiliary function on the output alphabet \\
        $\Tc^n$ & Type class \\
        $Q_{X,n}$ & Type approximating $Q_X$ \\
        $\pebar$ & Random-coding error probability \\
        $i_s^n$ & Information density quantity used in proofs \\
        $\CGMI^{(k)}$, $\CLM^{(k)}$ & Multi-letter achievable rates \\
        $\CGMI^{(\infty)}$, $\CLM^{(\infty)}$ & Limiting multi-letter achievable rates \\ [4mm]
        %
        %
        \multicolumn{2}{l}{{\bf (Introduced in Section \ref{ch:single_user_cont})} } \\ [2mm]
        $c$, $\Gamma$ & System cost function and threshold \\
        $a$, $a_{l}$ & Auxiliary cost functions \\
        $c^n$, $a^n$, $a_{l}^n$ & Additive multi-letter extensions of cost functions \\
        $\phi_c$, $\phi_a$, $\phi_{l}$ & Means of cost functions \\
        $\Omega_n$ & Normalizing constant for cost-constrained ensemble \\
        $\Dc_n$ & Constraint set for cost-constrained ensemble \\
        $\LM'$ & Fixed-cost LM rate \\
        $r_l$, $\rbar_l$ & Auxiliary parameters for cost-constrained ensemble \\
        $L$ & Number of auxiliary costs \\
        $\mu$, $\sigma^2$ & Noise mean and variance in additive channels \\ [4mm]
    \end{tabular}

    \begin{tabular}{cp{9cm}}
        %
        %
        \multicolumn{2}{l}{{\bf (Introduced in Section \ref{ch:rate_dist})} } \\ [2mm]
        $\Pi_X$, $\Pi_X^n$ & Source distribution and $n$-letter extension \\
        $\xvhat^{(m)}$ & Codeword in rate-distortion theory \\
        $d_0$, $d_1$ & Encoding metric and true distortion measure \\
        $d_0^n$, $d_1^n$ & Additive $n$-letter extensions of distortion functions \\
        $D_1$, $D_1^*$ & Distortion threshold and distortion-rate function \\
        $Q_{\Xhat}$ & Auxiliary distribution in rate-distortion theory \\
        $\Dibar$, $\Dibar^{(k)}$ & Achievable distortion and multi-letter version \\
        $\Pctilde$ & Constraint set in achievable distortion expression \\
        $d$, $D$ & Distortion measure and distortion level when $d_0 = d_1$ \\
        $\Rmatched$, $\Dmatched$ & Matched rate-distortion and distortion-rate functions \\
    	$D_{\min}$, $D_{\rm prod} $ & Extreme values of random coding distortion level \\
        $\Rbariid$, $\Rbarcc$ & Random coding rate-distortion functions \\
        $\sigma^2$ & Source variance in continuous rate-distortion setting  \\ [4mm]
        %
        %
        \multicolumn{2}{l}{{\bf (Introduced in Section \ref{ch:mac})} } \\ [2mm]
        $M_1$, $M_2$ & Multiple-access channel codebook sizes \\
        $R_1$, $R_2$ & Multiple-access channel rates \\
        $Q_{1}$, $Q_2$ & Multiple-access channel input distributions \\
        $P_{\Xv_{1}}$, $P_{\Xv_{2}}$ & Multiple-access channel codeword distributions \\ [4mm]
        \multicolumn{2}{l}{{\bf (Introduced in Section \ref{ch:multiuser})} } \\ [2mm]
        $M_0$, $M_1$, $\{M_{1u}\}$ & Superposition coding codebook sizes \\
        $R_0$, $R_1$, $\{R_{1u}\}$ & Superposition coding rates \\
        $\{n_u\}$ & Refined superposition coding sub-block lengths \\
        $Q_{UX}$ & Superposition coding input distribution \\
        $P_{\Uv\Xv}$ & Superposition coding codeword distribution \\
        $\Uc$ & Auxiliary alphabet for superposition coding \\
        $\Uv$ & Auxiliary codeword for superposition coding \\
        $p_{{\rm e},\nu}$, $\overbar{p}_{{\rm e},\nu}$ & Multi-user coding error probabilities \\ [4mm]
        %
        %
        \multicolumn{2}{l}{{\bf (Introduced in Section \ref{ch:exponents})} } \\ [2mm]
        $\Eriid$, $\Eziid$ & i.i.d.~exponent and Gallager function \\
        $\Ercc$, $\Ezcc$ & Constant-composition exponent and Gallager function \\
        $\Ercost$, $\Ezcost$ & Cost-constrained exponent and Gallager function \\
        $\Eexcc$, $\Excc$ & Constant-composition expurgated exponent and Gallager function \\
        $\rho$ & Dual error exponent parameter \\
        $E_{\textsc{ck}}$ & Csisz\'ar-K\"orner exponent \\
        $E_{\textsc{rgv}}$ & Random Gilbert-Varshamov exponent \\ 
		$d$, $\Delta$ & Distance function and parameter for RGV exponent \\ [4mm]
        %
        %
        \multicolumn{2}{l}{{\bf (Introduced in Section \ref{ch:converse})} } \\ [2mm]
        $\epsilon_k$ & Error probability for codebook having block length $k$ \\
        $\eta_k$ & Minimal difference of $k$-letter log-metric values \\
        $q_{\max}$ & Maximal metric value \\
        $q_{\max}^k(y^k)$ & Maximal metric value for fixed output sequence \\
        $B$ & Upper bound on $|\log q(x,y)|$ \\
        $\Ac_k$ & Output vectors with a unique metric maximizer \\
        $\Phi_k$ & Conditional probability of random codeword exceeding metric value \\
        $\Xc_q^*$ & Set of inputs with maximal decoding metric difference \\
        $\Mc_{\max}$, $\Mc_{\max,n}$ & Set of maximal conditional joint distributions and type-based variant \\
        $\overline{R}$ & Single-letter upper bound on the mismatch capacity \\
        $\Gc_{\xv}$ & Bipartite graph associated with two channels \\
        $\Ec_{\xv}$ & Edge set associated with two channels 
    \end{tabular}
\end{center}

 

\chapter{Introduction} \label{ch:intro}

Shannon's channel coding theorem \cite{Sha48} characterizes the conditions under which reliable communication is possible over a noisy channel.  This groundbreaking result spawned decades of research on the theory and practice of communication, and still continues to shape the development of practical communication systems.

One of the key assumptions of the channel coding theorem is that the encoder and decoder can be optimized for the specific channel model under consideration.  In particular, the achievability part is usually proved using decoding methods such as maximum-likelihood decoding or joint typicality decoding, both of which directly depend on the channel transition law.

In many practical scenarios, however, one does not have accurate knowledge of the channel.  In addition, even if the channel model is assumed to be known, the implementation of its corresponding optimal decoding rule may be hindered by practical issues such as computational limitations.  As a result, there is substantial motivation to develop a theory of communication that explicitly accounts for channel uncertainty and decoding constraints.

This monograph surveys the topic of {\em mismatched decoding}, in which the decoder is fixed and possibly suboptimal, and only the codebook can be optimized.  As well as addressing pertinent practical issues such as channel uncertainty and complexity limitations, this problem has strong connections with other fundamental problems in information theory, including zero-error communication.  

In this introductory section, we first formally introduce the problem of reliable communication under mismatched decoding in Section \ref{sec:setup_general}.  In Section \ref{sec:applications}, we outline several applications in which mismatched decoding is encountered, as well as formalizing the connections to other theoretical problems in information theory.
In Section \ref{sec:other_models}, we briefly overview other prominent approaches to addressing model uncertainty in information theory, including universal decoding, channels with a state, and adversarial channel models, and we discuss their connections and differences to the mismatched decoding perspective.  Section \ref{sec:outline} outlines the remainder of the monograph.

\section{Problem Setup} \label{sec:setup_general}

\begin{figure}
    \begin{centering}
        \includegraphics[width=0.8\columnwidth]{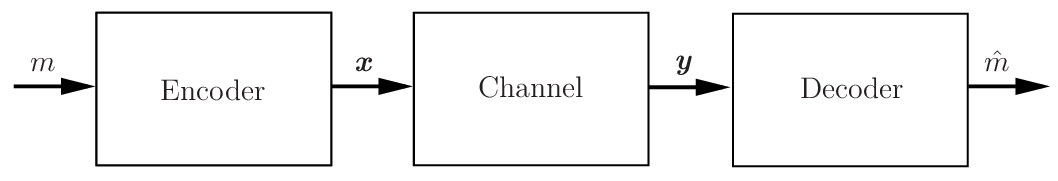}
        \par
    \end{centering}
    
    \caption{Illustration of channel coding with message $m$, channel input $\xv \in \Xc^n$, channel output $\yv \in \Yc^n$, and estimate $\hat{m}$.} \label{fig:CommChannel}
\end{figure}

In this section, we formally introduce the problem of point-to-point channel coding (see Figure \ref{fig:CommChannel}) with mismatched decoding, which will be studied throughout the monograph.

We consider a communication channel defined on an input alphabet $\Xc$ and an output alphabet $\Yc$.  For a given use of the channel, if the input is given by $x\in\Xc$, then the output $y \in \Yc$ is randomly generated according to some channel transition law $W(y|x)$.  It is useful to initially think of $\Xc$ and $\Yc$ being finite, in which case $W(y|x)$ is a conditional probability mass function.  However, when we turn to continuous-alphabet channels, we will use the same notation to represent a probability density function.

We focus our attention on {\em memoryless channels} used without feedback, meaning that different uses of the channel are independent: if we input $\xv=(x_1,\dotsc,x_n)$ in $n$ channel uses, then the output sequence $\yv=(y_1,\dotsc,y_n)$ is generated according to $W^n(\yv|\xv) \triangleq \prod_{i=1}^n W(y_i|x_i)$.  

The goal is to reliably transmit a message $m \in \{1,\dotsc,M\}$ over the channel in $n$ uses.  We assume that $m$ is uniformly random over the $M$ possibilities.  The encoding scheme is represented by a {\em codebook} $\Cc = \{\xv^{(1)},\dotsc,\xv^{(M)}\}$; when the message is $m$, the encoder transmits the corresponding codeword $\xv^{(m)} \in \Xc^n$.

Naturally, the defining feature of the mismatched decoding problem is at the decoder.  In the regular (i.e., matched) channel coding problem, one may consider an optimized decoding rule for producing the estimate $\hat{m}$ as a function of $\yv$ and $\Cc$.  In the {\em mismatched decoding} problem, however, the decoder is fixed as follows:
\begin{equation}
    \hat{m} = \argmax_{j=1,\dotsc,M} q^n(\xv^{(j)},\yv), \quad\text{with}\quad q^n(\xv,\yv) = \prod_{i=1}^n q(x_i, y_i), \label{eq:decoder}
\end{equation}
where $q(x,y)$ is a non-negative function called the {\em decoding metric}.  As a result, our only freedom is in designing the codebook $\Cc$.  We will shortly discuss the issue of tie-breaking (i.e., the possibility of a non-unique maximum in \eqref{eq:decoder}); see Remark \ref{rem:ties} below.

Given the message estimate as in \eqref{eq:decoder}, the {\em error probability} associated with the codebook $\Cc$ is given by
\begin{equation}
    \pe(\Cc) = \PP[\hat{m} \ne m], \label{eq:pe}
\end{equation}
where the probability is with respect to the randomness of the message and the channel.  We seek to characterize the trade-off between the error probability $\pe$, the coding rate $R = \frac{1}{n} \log M$, and the block length $n$.

We briefly mention that \eqref{eq:decoder} can be generalized to the maximization of a metric $q_n(\xv^{(j)},\yv)$ that need not factorize according to a product over the symbols.  The consideration of \eqref{eq:decoder} is analogous to the {\em memorylessness} assumption of the channel, namely, $W^n(\yv|\xv) = \prod_{i=1}^n W(y_i|x_i)$.  The consideration of memoryless channels and product-wise decoding metrics permits analytical tractability, while already coming with many interesting challenges and having several applications (see Section \ref{sec:applications}).  Nevertheless, the study of mismatched decoding is by no means limited to this setting, e.g., see \cite{Lap02,Alv17,Hul17}.

For any given codebook, the decoding rule minimizing $\pe$ is {\em maximum-likelihood} (ML) decoding, corresponding to \eqref{eq:decoder} with $q(x,y) = W(y|x)$.  Accordingly, we will refer to this case as {\em matched decoding}.  The standard Shannon capacity (henceforth called the {\em matched capacity}), denoted by $C(W)$, is defined to be the supremum of all rates such that one can achieve arbitrarily small error probability under matched decoding, and the following definition generalizes this notion to the mismatched case.

\begin{definition} \label{def:CM}
    {\em (Mismatch capacity)}
    For a mismatched memoryless channel described by $(W,q)$, we say that $R$ is an {\em achievable rate} if, for all $\delta > 0$, there exists a sequence of codebooks $\Cc_n$ with $M \ge e^{n(R-\delta)}$ codewords such that $\pe(\Cc_n)\to0$ under the decoding rule \eqref{eq:decoder}.  The {\em mismatch capacity} $\CM = \CM(W,q)$ is defined to be the supremum of all achievable rates.
\end{definition}

We consider $M \ge e^{n(R-\delta)}$ instead of $M \ge e^{nR}$ in this definition in order to ensure that the limit of any achievable rate is also achievable.  This will allow us to safely use the compact terminology ``achievable rate'' in place of ``lower bound on the mismatch capacity''.

\begin{remark} \label{rem:ties}
    {\em (Tie-breaking)}
    In general, it is possible for multiple messages to simultaneously achieve the maximum in \eqref{eq:decoder}; in such cases, we adopt a pessimistic approach and assume that an error always occurs when there is a tie.  One may consider improving the error probability by instead breaking ties uniformly at random, but the improvement amounts to at most a factor of two \cite{Eli55} and thus has no impact on the mismatch capacity.\footnote{A subtle point here, however, is that artificially-constructed tie-breaking rules can increase the mismatch capacity.  For example, if $q(x,y)=0$ for all $(x,y)$ then all codewords will always be tied, leading to zero mismatch capacity under random or pessimistic tie-breaking.  However, in this example, if ties were broken by choosing the tied codeword with the highest likelihood, then the resulting capacity would trivially become the regular Shannon capacity.}
\end{remark}

\begin{remark} \label{rem:max_avg}
    {\em (Maximal vs.~average error probability)}
    While we consider the average error probability in \eqref{eq:pe}, the mismatch capacity is also unchanged when we consider the maximal error probability, defined as
    \begin{equation}
        \pemax(\Cc) = \max_{m=1,\dotsc,M}\PP[\hat{m} \ne m \,|\, \xv^{(m)}~{\rm sent}]. \label{eq:pe_max}
    \end{equation}
    This claim follows via a standard expurgation argument \cite[Sec.~5.6]{Gal68}: For any codebook $\Cc$ of size $M$ satisfying $\pe(\Cc) \le \epsilon$ for some $\epsilon > 0$, we can consider a sub-codebook $\Cc'$ of size $\frac{M}{2}$ whose codewords have the smallest conditional error probability.  Each remaining codeword must satisfy $\PP[\hat{m} \ne m \,|\, \xv^{(m)}~{\rm sent}] \le 2\epsilon$ when the mismatched decoding rule \eqref{eq:decoder} is applied using $\Cc$.  Then, assuming pessimistic tie-breaking, the same follows for the mismatched decoder that uses $\Cc'$, since for any $\xv \in \Cc'$, there are only fewer incorrect codewords $\xvbar \in \Cc'$ (compared to $\xvbar \in \Cc$) that could lead to the error event $q^n(\xvbar,\yv) \ge q^n(\xv,\yv)$.  Hence, $\pemax(\Cc') \le 2\epsilon$, and since the rate of $\Cc'$ is asymptotically the same as $\Cc$, the desired claim follows.
\end{remark}

In the following subsection, we overview several applications of the mismatched decoding, highlighting some cases in which the seemingly innocuous definition of the mismatch capacity $\CM$ should be interpreted with care.

%

\section{Applications of Mismatched Decoding} \label{sec:applications}


\subsection{Channel Uncertainty}

An immediate application of mismatched decoding is that in which the decoder has an incorrect channel estimate $\What(y|x)$ that is used as if it were correct, corresponding to \eqref{eq:decoder} with $q(x,y) = \What(y|x)$.  Such channel uncertainty can arise in many different forms, including incorrect parameter estimates \cite{Csi95}, amplitude mismatch in real channels \cite{Mer95}, phase offsets in complex channels \cite{Mer95}, and incorrect models of additive noise distributions \cite{Lap96a}.  

While this application of mismatched decoding is seemingly natural, the notion of {\em mismatch capacity} often needs to be interpreted with care.  In particular, Definition \ref{def:CM} assumes that the codebook can be optimized, but performing such an optimization may implicitly require knowledge of both $W$ and $q$.  Hence, if $W$ is unknown, then a potentially more meaningful quantity is the rate achieved when a codebook designed for $\What$ is applied to $W$, and used with the above-mentioned decoding metric $q(x,y) = \What(y|x)$.

Fortunately, the achievable rates presented in this monograph will account for both scenarios.  In particular, we will generally provide performance bounds for a {\em given random-coding ensemble}.  We can optimize over the given class of random-coding ensembles to get the best possible lower bound on the mismatch capacity, or in accordance with the above discussion, we can consider the rate for a fixed ensemble (e.g., one which would be capacity-achieving if the true channel were $\What$) as the appropriate performance measure.

\subsection{Fading Channels} \label{sec:app_fading}

In wireless communication scenarios, the channel typically exhibits unknown fluctuations resulting from the signal arriving at the receiver via multiple paths in a dynamic environment.  A widely-adopted complex-valued channel model in this context is the following:
\begin{equation}
    Y_i = H_i X_i + Z_i,
\end{equation}
where $H_i \in \CC$ is a {\em fading coefficient}, and $Z_i \in \CC$ is additive complex Gaussian noise.  The use of a complex-valued model arises naturally when transmitting signals modulated by a high-frequency carrier; see for example \cite{Tse05}.

Since the coefficients $H_i$ are determined by a complex dynamic environment, one cannot expect to know their values precisely.  As a result, one typically performs {\em channel estimation} to obtain estimates $\{\Hhat_i\}_{i=1}^n$ of the coefficients $\{H_i\}_{i=1}^n$, and then uses the estimates as if they were the true coefficients.  In particular, the mismatched version of the maximum-likelihood rule, which would be optimal if the estimates were correct, is given by
\begin{equation}
    \hat{m} = \argmin_{j=1,\dotsc,M} \sum_{i=1}^n |Y_i - \Hhat_i X_i^{(j)} |^2, \label{eq:INTR_weighted_NN}
\end{equation}
where $(X_1^{(j)},\dots,X_n^{(j)})$ is the $j$-th codeword.
As a result, we have a {\em mismatched decoding rule} as in \eqref{eq:decoder}, with decoding metric $q(x,(\hhat,y)) = e^{-|y - \hhat x|^2}$.  Note that $\hhat$ is treated as part of the channel output here, since it is known at the decoder.  In this application, one typically considers random coding techniques with a Gaussian input distribution, rather than (unrealistically) adopting the best codebook as per Definition \ref{def:CM}.

It should be noted that the ``channel'' dictating the conditional distribution of $(\Hvhat,\Yv)$ given $\Xv$ is only memoryless under strong assumptions.  However, even in the presence of memory, the mismatched decoding perspective has led to a variety of powerful theoretical results, including both achievable rates and provable limitations of random coding imposed by the mismatch \cite{Lap02,Wei04}.  

\subsection{Reduced Decoding Complexity} \label{sec:decoding_comp}

Even in cases that the channel $W$ is known, its associated optimal decoding rule may be too complex to implement or even approximate.  In such scenarios, it may be preferable or even essential to adopt a decoding rule that is simpler but suboptimal.  

For instance, while the maximum-likelihood decoder under additive Gaussian noise is the nearest-neighbor decoder, a non-Gaussian noise distribution may lead to a much more complex maximum-likelihood rule.  To circumvent this difficulty, one may simply revert to the nearest-neighbor decoder (or a similar rule) despite its suboptimality, and this becomes a problem of mismatched decoding.  By characterizing the resulting performance, we can understand the extent to which simple decoding rules remain useful beyond the specific channels that they were designed for.

A more detailed example of reducing complexity at the expense of suboptimal decoding is discussed in the following subsection.

\subsection{Bit-Interleaved Coded Modulation} \label{sec:bicm}

The technique of bit-interleaved coded modulation (BICM) \cite{Gui08} is motivated by the question of how binary codes can be employed even when the communication channel induced by a modulation scheme is non-binary.  BICM has been widely adopted in practical settings, including a number of standards.

If the transmitted symbol $x$ belongs to a signal constellation set $\Xc$ with $2^\bicmbits$ elements, we may represent $x$ using a label of $\bicmbits$ bits, $\bv = (b_1,\dotsc,b_\bicmbits)$. We make the dependence explicit by writing the $i$-th bit in the binary mapping of $x$ as $b_i(x)$.
The key element of BICM is the observation that the overall non-binary code $\Cc$ may also be decoded at bit level.  More formally, in place of a potentially complicated symbol-wise decoding metric, one can decode according to a product of bit-wise metrics:
\begin{equation}
    q(x,y) = \prod_{i=1}^\bicmbits q_i\bigl(b_i(x),y\bigr),
    \label{eq:bicm_metric}
\end{equation}
where 
the bit decoding metrics $q_i(b,y)$ (for $i = 1,\dots,\bicmbits$) are given as
\begin{equation}
    q_i(b,y) = \sum_{x\in\Xc} P_{X|B_i}(x|b)W(y|x), \label{eq:bit_metric}
\end{equation}
and where $P_{X|B_i}$ is the conditional distribution induced by the channel input distribution.   The decoding metric \eqref{eq:bicm_metric} corresponds to treating the $\bicmbits$ bits labeling a symbol as if they were independent, and the metric \eqref{eq:bit_metric} for a given bit corresponds to treating all other bits as noise.


A particularly relevant case in practice is that in which the input distribution is uniform (i.e., all constellation symbols occur with equal probability), in which case the conditional probability $P_{X|B_i}(x|b)$ is zero if $b_i(x) \neq b$, and $2^{-(\bicmbits-1)}$ if $b_i(x) = b$. In this case, decoding can proceed using standard binary decoding techniques, e.g.,~forming $\bicmbits$ parallel bit decoding metrics for every channel output.

BICM was first explicitly cast as a mismatched decoding problem in \cite{Mar09}, permitting the application of general mismatched decoding achievable rates  specialized to the decoding metric given in \eqref{eq:bicm_metric}.

\subsection{Finite-Precision Arithmetic}

Another interesting application of mismatched decoding is that of finite-precision arithmetic \cite{Bin99,Sal95}.  We illustrate the general idea through the specific example of an additive white Gaussian noise (AWGN) channel $Y = X + Z$, with $Z \sim \Ndist(0,\sigma^2)$.  

The optimal decoding metric, corresponding to maximum-likelihood decoding, is $\log q(x,y) = -(y-x)^2$, corresponding to nearest-neighbor decoding.  However, such a decoder may be ruled out when hardware constraints limit the precision of the arithmetic.  In such cases, a more feasible decoding rule is given by
\begin{equation}
    \log q(x,y) = - \big( \Phi_Y(y) - \Phi_X(x) \big)^2, \label{eq:q_finite_precision}
\end{equation}
where $\Phi_X$ and $\Phi_Y$ are scalar quantization functions.  This is again a mismatched decoding problem.

We note that this setup is not equivalent to applying the scalar quantizer $\Phi_X$ prior to transmission, as we still allow the transmitted codeword to take arbitrary values; it is only the decoder that is required to work with the quantized values.  In this application, it is often reasonable to assume that the codebook designer knows both $W$ and $q$, which justifies the optimization over the codebook in Definition \ref{def:CM}.

\subsection{Optical Communication}

Optical communication is of considerable practical importance, but the development of precise information-theoretic rates is a notoriously hard problem.  While optical fiber channels with low signal powers are typically well-approximated by an additive white Gaussian noise (AWGN) channel, severe non-linearities hinder the performance at higher powers \cite{Alv17,Sec17}.  In addition, the underlying channel can be far from memoryless, due to phenomena such as inter-symbol interference.


Despite the complexity of the channel, simpler achievable rates can be attained by considering suboptimal decoding rules (see \cite{Sec17} for a recent survey), such as the AWGN decoding metric that ignores channel memory.  In this sense, the corresponding optical communication problem can be viewed through the lens of mismatched decoding \cite{ghozlan2017models}.  

The technique of BICM, outlined in Section \ref{sec:bicm}, is also widespread in the field of optical communication, and a mismatched decoding perspective can again be adopted \cite{Alv17}.  A distinction here is that the resulting achievable rates are typically difficult to compute exactly due to the presence of high-dimensional integrals. Nevertheless, one can accurately approximate the rates using techniques such as Gauss-Hermite quadrature or Monte Carlo integration, as is often done even for standard Gaussian channels when discrete input constellations are employed.

\subsection{Zero Undetected Error Capacity} \label{sec:app_zuec}

In studies of channel coding, it is often useful to let the decoder explicitly output an error when it is not confident enough to make a decision on the transmitted message.  When this is done, one can distinguish between a {\em detected error}, where the decoder chooses to declare an error, and an {\em undetected error}, where the decoder outputs an incorrect estimate of the message.  It is then of interest to understand the trade-off between the two, particularly in cases where undetected errors are considered much more costly.

In the special case of DMCs, an interesting extreme case of this trade-off is the {\em zero-undetected error capacity}, which is the highest achievable rate when the probability of an undetected error must be exactly zero (i.e., the error probability equals the detected error probability) \cite{Csi95,Ahl96}.  For this to occur, the decoder must only output a message estimate if there is a unique codeword for which the output sequence has positive probability; if multiple codewords are  feasible (i.e., consistent with the output), an error must be declared.  For this setting to be meaningful, some of the channel transition probabilities $W(y|x)$ must be zero.

The zero-undetected error capacity is a special case of the mismatch capacity, corresponding to the decoding metric \cite{Csi95}
\begin{equation}
    q(x,y) = \openone\{ W(y|x) > 0 \}, \label{eq:zuec_metric}
\end{equation}
often referred to as the {\em erasures-only} metric.  To see this, note that $q^n(\xv,\yv) = 1$ whenever $\xv$ is feasible, and $q^n(\xv,\yv) = 0$ otherwise.  Hence, assuming ties are broken as errors, the mismatched decoding rule leads to an error if and only if multiple codewords are feasible, as desired. 

\subsection{Zero-Error Capacity} \label{sec:app_zero}

Another interesting special case of the mismatch capacity is the {\em zero-error capacity} \cite{Kor98,Sha56} of a DMC, which is the highest communication rate such that the probability of {\em any} error is exactly zero.  The zero-error capacity is known to depend on the channel $W(y|x)$ only through the $|\Xc| \times |\Xc|$ matrix $\Av_W$ with $(x,x')$-th entry
\begin{equation}
    A_W(x,x') = \openone\{ W(y|x)W(y|x') > 0 \text{ for some } y \in \Yc \},
\end{equation}
where we assume without loss of generality that $\Xc = \{1,\dotsc,|\Xc|\}$.  The matrix $\Av_W$ is the {\em adjacency matrix of the channel graph}, formed by connecting two inputs if and only if they share a common output according to $W$.

With this definition, the zero-error capacity of $W$ is equal to the mismatch capacity of the following channel-metric pair, whose input and output alphabets are both $\Xc$ \cite{Csi95}: 
\begin{equation}
    W_0(x'|x) = \openone\{ x = x' \}, \quad q_0(x,x') = A_W(x,x'). \label{eq:ze_choices}
\end{equation} 
That is, the channel is noiseless, and the metric is the indicator function of $(x,x')$ sharing a common output in $W$, sometimes referred to as being {\em adjacent} or {\em confusable}.

To see that $\CM(W_0,q_0)$ equals the zero-error capacity of $W$, we first recall from Remark \ref{rem:max_avg} that the mismatch capacity is unchanged when the requirement of $\pe$ being arbitrarily small is replaced by the requirement that $\pemax$ (see \eqref{eq:pe_max}) is arbitrarily small.  Hence, it suffices to show that any given sequence of codebooks $\{\Cc_n\}_{n \ge 1}$ attains zero error probability on $W$ (i.e., any transmitted codeword can be uniquely decoded with probability one) if and only if $\pemax(\Cc_n) \to 0$ for the mismatched DMC $(W_0,q_0)$. 


The ``only if'' statement follows from the fact that any zero-error codebook $\Cc_n$ for $W$ yields $q_0^n(\xv,\xv) = 1$ for all $\xv \in \Cc$, and $q_0^n(\xv,\xvbar) = 0$ for all $\xv, \xvbar \in \Cc_n$ with $\xv \ne \xvbar$, due to the choice of $q_0$ in \eqref{eq:ze_choices}.  Thus, $\Cc_n$ satisfies $\pemax(\Cc) = 0$ under the mismatched DMC $(W_0,q_0)$, in which the channel output deterministically equals the input.
To establish the other direction, we note that since both the channel $W_0$ and the decoder (with ties broken as errors) are deterministic, the quantities $\PP[\hat{m} \ne m \,|\, \xv^{(m)}~{\rm sent}]$ defining $\pemax$ in \eqref{eq:pe_max} can only be zero or one.  Thus, the only way to attain $\pemax(\Cc_n) \to 0$ is to have $\pemax(\Cc_n) = 0$ for all sufficiently large $n$.  By the choice of $q_0$, one can only have $\pemax(\Cc_n) = 0$ when all codewords have no common outputs, which implies that $\Cc_n$ is a zero-error codebook for $W$.


A related example can be found in \cite{Csi95}, showing that a graph-theoretic quantity called the {\em Sperner capacity} is also a special case of the mismatch capacity.


%

\section{Other Channel Uncertainty Models} \label{sec:other_models}

The mismatched decoding problem provides one of several perspectives on the role of channel uncertainty in information theory.  An excellent survey comparing the various perspectives can be found in \cite{Lap98}, so we provide only a brief discussion here.

It is well-known that {\em universal decoding rules} exist that are able to achieve the capacity of any DMC when the codebook is optimized.  A prominent example is the maximum empirical mutual information decoder \cite[Ex.~6.20]{Csi11}.  Adopting such a decoder under channel uncertainty would typically lead to strictly better achievable rates compared to a mismatched decoding metric.  However, such decoding rules are {\em multi-letter} in nature, and computationally efficient techniques for implementing or approximating them remain elusive.  In contrast, the decoding rule in \eqref{eq:decoder} is {\em single-letter}, in the sense that it is written as a product over individual symbols, and many practical decoding methods can be viewed as approximating this rule (despite not necessarily implementing it exactly).  

Channels with a state, in which the transition law takes the form $W(y|x,s)$ for some unknown state $s$, also provide a powerful means to address channel uncertainty \cite[Ch.~7]{Elg11}.  In particular, the capacity varies depending on whether the state is known at the encoder and/or decoder, and whether it is known causally or non-causally.  A prominent example is fading channels, where $s$ represents a fading coefficient.  In fact, as evidenced in the previous subsection, the study of channels with a state is by no means disjoint from the mismatched decoding problem.  Rather, channels with a state are simply another class of channels where the decoding rule may be either matched or mismatched; e.g., see \cite{Lap02} for mismatched fading channels, and \cite{Fel16} for mismatched DMCs with a state.

Another widely-adopted approach is to consider {\em adversarial} channel models, taking the form $W_n(\yv|\xv,\sv)$ for some sequence $\sv = (s_1,\dotsc,s_n)$ that can be viewed as being chosen by an adversary (possibly subject to some constraints) in order to hinder the communication.  Hence, one requires small error probability simultaneously for all possible sequences $\sv$.  Under the constraint $s_1 = \dotsc = s_n$ this is referred to as a {\em compound channel} \cite{Bla59}, whereas the case of arbitrary length-$n$ sequences is referred to as the {\em arbitrarily varying channel} (AVC) \cite{Bla60}.   Adversarial channel models are of particular interest for communication in the presence of jamming.


To our knowledge, the study of AVCs has had minimal interaction with the mismatched decoding perspective in the existing literature.  This is because bounds on the capacity of the AVC typically adopt rather complicated multi-letter decoding rules, e.g., not simply maximizing a metric on $(\xv,\yv)$, but instead considering various combinations of triplets $(\xv,\xv',\yv)$ \cite{Csi88}.  In addition, the study of AVCs often leads to very different insights and challenges compared to the mismatched decoding problem, such as the fact that the capacity may differ for the average error probability vs.~maximum error probability criteria, and the consideration of randomized encoding.  See \cite{Lap98} for a detailed overview.

Despite these differences, we briefly mention two examples of connections between adversarial channel models and mismatched decoding: (i) For certain compound channels, the capacity can be achieved by a single-letter mismatched decoding rule \cite{Sti66}; we present this example in Section \ref{sec:ex_compound}.  (ii) It is shown in \cite{Csi95} that the erasures-only decoding metric (see \eqref{eq:zuec_metric}) can achieve the capacity of certain AVCs under deterministic coding and the average-error criterion.

\section{Outline of the Monograph} \label{sec:outline}

Through the monograph, we focus primarily on achievable rates (i.e., lower bounds on the mismatch capacity) via random coding.  This has been the primary focus of the existing literature, with upper bounds on the mismatch capacity generally remaining elusive.  The study of achievable rates in itself comes with many interesting challenges and differences compared to the matched setting.  To name just one example, we will see that both constant-composition codes and multi-user coding techniques can lead to better achievable rates than standard i.i.d.~random coding, in stark contrast with the matched setting in which i.i.d.~random codes are capacity-achieving.

In Section \ref{ch:single_user}, we present achievable rates for discrete memoryless channels, introducing the {\em generalized mutual information} (GMI) for i.i.d.~coding, and the {\em LM rate} for constant-composition random coding.  We present several key properties of both rates, including conditions for positivity, conditions under which the matched capacity is achieved, and ensemble tightness results.  In addition, we present a variety of representative examples from the literature.

In Section \ref{ch:single_user_cont}, we present a generalization of the GMI and LM rate to continuous-alphabet channels, using the idea of random coding with auxiliary costs.  We again provide several representative examples focusing on additive channels and variants of mismatched nearest-neighbor decoding. In particular, these examples include non-Gaussian noise channels and fading channels, both of which have been among the most widespread applications of mismatched decoding in the literature.

In Section \ref{ch:rate_dist}, we momentarily depart from the channel coding problem and survey the role of mismatch in rate-distortion theory.  In this case, the problem becomes one of {\em mismatched encoding}, and we again present achievable rates that are proved via random coding, as well as a multi-letter converse result.  In addition, we overview a different type of mismatch in which the optimal encoder is adopted but a suboptimal random coding distribution is used, highlighting the important special case of Gaussian compression techniques for non-Gaussian sources.

In Section \ref{ch:mac}, we study the mismatched multiple-access channel (MAC), presenting an ensemble-tight achievable rate region for constant-composition random coding, as well as an extension to continuous alphabets.  In Section \ref{ch:multiuser}, we survey further results demonstrating that multi-user coding techniques, including both MAC coding methods and superposition coding, can lead to improved rates for single-user channels.  In particular, we present a refined version of superposition coding that provides the best known achievable rate at the time of writing.

In Section \ref{ch:exponents}, we move beyond achievable rates and present an overview of results on error exponents, which provide refined characterizations of the asymptotic behavior of the error probability at fixed rates.  In particular, we discuss the role of cost-constrained random coding with multiple auxiliary costs for attaining the best known achievable error exponents, highlighting the differences compared to the study of achievable rates alone.

In Section \ref{ch:converse}, we overview recent developments and challenges in obtaining upper bounds on the mismatch capacity.  In accordance with the literature on this topic, we focus primarily on multi-letter bounds that are non-computable but can still provide interesting insight on the problem.  In addition, we present a recent single-letter upper bound.

In Section \ref{ch:other}, we briefly outline several topics in mismatched decoding that were omitted from the main sections, including refined asymptotics, modified channel coding settings, further results on continuous and fading channels, and mismatch in practical coding methods.  For each of these, we give several pointers to the relevant literature.  Conclusions are drawn in Section \ref{ch:conclusion}.

Unless it is explicitly stated otherwise, it should be understood that the results surveyed in this monograph come from the existing literature.  We highlight the main references at the beginning of each section.

\noindent 
\paragraph{Notes on the presentation and assumed background.} We make an effort to provide self-contained proofs of most results, but for certain parts that are highly technical and/or less central to the survey, we instead refer the reader to the relevant literature.

We assume that the reader is familiar with the classical theorems of information theory (e.g., source coding, channel coding, and rate-distortion theory), as well as the associated information measures (e.g., entropy, mutual information, and KL divergence) and their properties (e.g., chain rule, data processing inequality).  As for slightly less standard prior knowledge, the reader is encouraged to learn the basics of the method of types \cite[Ch.~2]{Csi11} and Gallager's proof of the channel coding theorem \cite[Ch.~5]{Gal68} if either of these is unfamiliar.

\noindent 
\paragraph{Notation.} We will introduce our notation throughout the monograph, but for reference, we also summarize some recurring notation here.  Definitions of the most commonly-used mathematical symbols and abbreviations are listed prior to Section \ref{ch:intro}.

The set of all probability distributions on an alphabet $\Xc$ is denoted by $\Pc(\Xc)$, and the set of all conditional distributions on $\Yc$ given $\Xc$ is denoted by $\Pc(\Yc|\Xc)$. More precisely, in the case that $\Xc$ is finite, we write  $P_X \in \Pc(\Xc)$ when $P_X$ is a probability mass function (PMF) on $\Xc$, and similarly for conditional PMFs.   When we consider continuous alphabets, we use the same notation for probability density functions (PDFs).
The set of all empirical distributions on a vector in $\Xc^{n}$ (i.e.~types \cite[Sec. 2]{Csi11}, \cite{GallagerCC}) is denoted by $\Pc_{n}(\Xc)$, and the set of all conditional
empirical distributions (i.e.~conditional types) on $\Yc^n$ given $\Xc^n$ is denoted by $\Pc_n(\Yc|\Xc)$. 
Given $P_X \in\Pc_{n}(\Xc)$, the type class $\Tc^{n}(P_X)$ is defined to be the set of all sequences in $\Xc^{n}$ with type $P_X$.  

We use bold symbols for vectors (e.g.~$\xv$), and we denote the corresponding $i$-th entry using a subscript (e.g.~$x_{i}$).  In the case of random vectors, we use capital letters (e.g., $\Xv$ and $X_i$).  The probability of an event is denoted by $\PP[\cdot]$, and the symbol $\sim$ means ``distributed as''. The marginals of a joint distribution $P_{XY}(x,y)$ are denoted by $P_{X}(x)$ and $P_{Y}(y)$. Similarly, $P_{Y|X}(y|x)$ denotes the conditional distribution induced by $P_{XY}(x,y)$.  Expectation with respect to a joint distribution $P_{XY}(x,y)$ is denoted by $\EE_{P}[\cdot]$. When the probability distribution is understood from the context, we simply write $\EE[\cdot]$.  Given a distribution $Q_X(x)$ and a conditional distribution $W(y|x)$, we write $Q_X\times W$ to denote the joint distribution $Q_X(x)W(y|x)$, and similarly when there are more than two distributions.  

We use standard notation for the entropy $H(X)$, conditional entropy $H(X|Y)$, mutual information $I(X;Y)$, conditional mutual information $I(X;Y|Z)$, KL divergence $D(P\|Q)$ and conditional KL divergence $D(P_{Y|X} \| Q_{Y|X} | P_X)$ \cite{Cov06}.  We often explicitly state the underlying distribution of the random
variables with a subscript (e.g.~$I_{P}(X;Y)$, $I_{\Ptilde}(X;Y)$).
The set of real numbers is denoted by $\RR$, and the set of complex numbers is denoted by $\CC$.  We define $[\alpha]^{+}=\max\{0,\alpha\}$, and we denote the indicator function of an event by $\openone\{\cdot\}$.  The floor function is denoted by $\lfloor\,\cdot\,\rfloor$, and the ceiling function by $\lceil\,\cdot\,\rceil$.  All logarithms have base $e$, and all rates are in units of nats except in the examples, where bits are used. 

 

\chapter{Discrete Memoryless Channels} \label{ch:single_user}

%
%

\section{Introduction} \label{sec:su_intro}

Throughout the monograph, it will be useful to distinguish between channels with finite alphabets and continuous alphabets.  In this section, we focus on the finite-alphabet setting, which will allow us to provide a more elementary introduction using widely-known techniques from the literature.  In Section \ref{ch:single_user_cont}, we will consider the continuous-alphabet setting.

We consider the problem setup as described in Section \ref{sec:setup_general}, in the special case that $\Xc$ and $\Yc$ are finite, and hence $W(y|x)$ represents a conditional probability mass function.  Recall that a non-negative decoding metric $q(x,y)$ dictates the decision rule in \eqref{eq:decoder}.  Collectively, we refer to the pair $(W,q)$ as a {\em mismatched DMC}.

The achievable rates presented in this section are based on {\em random coding}, a ubiquitous tool in information theory in which one proves the existence of good codebooks by bounding the average error probability of a randomly-generated codebook \cite{Sha48}.  A given random construction, characterized by the joint distribution of the codewords, is called a {\em random-coding ensemble}.  In this section, we assume that the codewords are drawn independently from a common distribution $P_{\Xv}$:
\begin{equation}
    \{ \Xv^{(m)} \}_{m=1}^M \sim \prod_{m=1}^M P_{\Xv}(\xv^{(m)}). \label{eq:random_coding}
\end{equation}
In Section \ref{ch:multiuser}, we will turn to alternative ensembles in which some form of dependence is introduced among the codewords.

This section is predominantly based on the works of Hui \cite{Hui83}, Csisz\'ar and K\"orner \cite{Csi81}, Csisz\'ar and Narayan \cite{Csi95}, and Merhav {\em et al.} \cite{Mer95}.

\section{Properties of the Mismatch Capacity} \label{sec:su_properties}

In this subsection, we provide some straightforward properties of the error probability and mismatch capacity that will be useful for motivating the random coding ensembles and interpreting their achievable rates.  

We begin with the following proposition regarding the equivalence of certain decoding metrics, which in fact applies not only to mismatched DMCs, but also general mismatched memoryless channels.

\begin{proposition} \label{prop:equiv_general}
    {\em (Equivalence of metrics)}
    For any memoryless channel $W(y|x)$ and codebook $\Cc = (\xv^{(1)},\dotsc,\xv^{(M)})$, if two decoding metrics $q,\qtilde$ are related according to
    \begin{equation}
    \log \qtilde(x,y) = b(y) + s \log q(x,y) \label{eq:equiv}
    \end{equation}
    for some $b(y)$ and $s > 0$, then we have for all $\yv \in \Yc^n$ that
    \begin{equation}
    \argmax_{j=1,\dotsc,M} \qtilde^n(\xv^{(j)},\yv) = \argmax_{j=1,\dotsc,M} q^n(\xv^{(j)},\yv), \label{eq:equiv_max}
    \end{equation}
    i.e., the two decoding metrics are equivalent.
\end{proposition}

Recalling that $q^n(\xv,\yv) = \prod_{i=1}^n q(x_i,y_i)$, this claim follows by simply writing $\log \qtilde^n(\xv,\yv) = b^n(\yv) + s \log q^n(\xv,\yv)$, where $b^n(\yv) = \sum_{i=1}^n b(y_i)$.  Since $s > 0$, this implies that the two maximization problems in \eqref{eq:equiv_max} are equivalent.

We can strengthen Proposition \ref{prop:equiv_general} for a useful class of codebooks called {\em constant-composition codebooks}.  The {\em composition} of a sequence $\xv$ (also referred to as its {\em type} or {\em empirical distribution}) is defined as
\begin{equation}
    \Phat_{\xv}(x) = \frac{1}{n} \sum_{i=1}^n \openone\{ x_i = x\}, \label{eq:type}
\end{equation}
and a constant-composition codebook is one in which all codewords have the same composition.  Thus, every sequence has the same number of occurrences of any given symbol, and the codewords are permutations of one another.

\begin{proposition} \label{prop:equiv_cc}
    {\em (Equivalence of metrics for constant-composition codebooks)}
    For any memoryless channel $W(y|x)$ and constant-composition codebook $\Cc = (\xv^{(1)},\dotsc,\xv^{(M)})$, if two decoding metrics $q,\qtilde$ are related according to
    \begin{equation}
    \log \qtilde(x,y) = a(x) + b(y) + s \log q(x,y) \label{eq:equiv_a}
    \end{equation}
    for some $a(x)$, $b(y)$, and $s > 0$, then we have for all $\yv \in \Yc^n$ that
    \begin{equation}
    \argmax_{j=1,\dotsc,M} \qtilde^n(\xv^{(j)},\yv) = \argmax_{j=1,\dotsc,M} q^n(\xv^{(j)},\yv),
    \end{equation}
    i.e., the two decoding metrics are equivalent for constant-composition codebooks.
\end{proposition}

The proof is similar to that of Proposition \ref{prop:equiv_general}: We have $\log \qtilde^n(\xv,\yv) = a^n(\xv) + b^n(\yv) + s \log q^n(\xv,\yv)$, and for any constant-composition codebook, the value $a^n(\xv) = \sum_{i=1}^n a(x_i)$ is the same for every codeword, as the codewords are all permutations of one another.

The constant-composition assumption may seem restrictive, but in fact, {\em any} codebook can be reduced to a constant-composition codebook with a negligible loss in the rate.  To see this, note that $\Phat_{\xv}$ in \eqref{eq:type} takes the form $\big(\frac{n_1}{n}, \dotsc, \frac{n_{|\Xc|}}{n} \big)$ with $n_x \in \{0,1,\dotsc,n\}$ for all $x \in \Xc$, and hence, the total number of different compositions is at most $(n+1)^{|\Xc|}$.  As a result, given a codebook with $M = e^{nR}$ codewords, among all possible sub-codebooks with codewords having the same composition, there must exist one whose size is at least
\begin{equation}
    M' \ge \frac{M}{(n+1)^{|\Xc|}}.
\end{equation}
If $\Xc$ is finite, as is the case for a DMC, then it holds for any fixed rate $R$ that $\frac{1}{n}\log M' \to R$ as $n \to \infty$.  In addition, the maximal error probability in the reduced codebook is no higher than that of the original codebook, since for any given transmitted codeword $\xv$, there are only fewer incorrect codewords $\xvbar$ remaining that could lead to the error event $q^n(\xvbar,\yv) \ge q^n(\xv,\yv)$.  Since the mismatch capacity is always the same under the average and maximal error criteria (see Remark \ref{rem:max_avg}), we have the following corollary.

\begin{corollary} \label{cor:equiv_cm}
    {\em (Metrics with equal mismatch capacities)}
    For any memoryless channel $W$ and decoding metrics $q$ and $\qtilde$, we have the following:
    \begin{enumerate}
        \item[(i)] The mismatch capacities of $(W,q)$ and $(W,\qtilde)$ are identical provided that $q$ and $\qtilde$ are related according to \eqref{eq:equiv} for some $b(y)$ and $s > 0$.
        \item[(ii)] If the input alphabet $\Xc$ is finite, then the mismatch capacities of $(W,q)$ and $(W,\qtilde)$ are identical provided that $q,\qtilde$ are related according to \eqref{eq:equiv_a} for some $a(x)$, $b(y)$, and $s > 0$.
    \end{enumerate}
\end{corollary}

\section{Achievable Rates: GMI and LM Rate} \label{sec:su_rates}

In this subsection, we introduce the two most well-known achievable rates in the mismatched decoding literature: The {\em generalized mutual information} (GMI) and the {\em LM rate}.\footnote{We adopt these names because they have become widespread in the literature, but in our opinion, neither of them is ideal.  While the GMI does ``generalize'' mutual information in the sense of recovering it in the special case $q(x,y)=W(y|x)$, the same could be said of any achievable rate that is tight when specialized to the matched case.  The name {\em LM rate} appears to have arisen following \cite{Mer95}, with the acronym implicitly standing for ``lower [bound on the] mismatch [capacity]''.  This would again be equally suitable for {\em any} achievable rate.} These rates are derived using two common random coding techniques known as {\em i.i.d.~random coding} and {\em constant-composition random coding}, which we introduce below.  While the achievable rate for the latter of these was proved first in the literature \cite{Hui83,Csi81}, we find it instructive to start with the former.

Both the GMI and LM rate will be written in two equivalent forms which, at first glance, might appear to be completely unrelated.  The equivalence of the two (stated in Lemma \ref{lem:primal_dual} below) is proved using the method of {\em Lagrange duality} from the theory of convex optimization \cite[Ch.~5]{Boy04}.  Accordingly, we will refer to the two as {\em primal expressions} and {\em dual expressions}, and we will adopt similar terminology throughout the monograph whenever a similar equivalence holds.

We proceed by introducing the random coding ensembles (i.e., choices of the codeword distribution $P_{\Xv}$ in \eqref{eq:random_coding}) and their achievable rates, and then turn to a common discussion of the two in which we interpret both the primal and dual forms.

\subsection{i.i.d.~Random Coding and the GMI}

By far the most widely-studied random-coding ensemble in the information theory literature is {\em i.i.d.~random coding}, defined as follows.  Here and subsequently, $\Pc(\Xc)$ denotes the set of all probability distributions on the alphabet $\Xc$.

\begin{definition}  \label{def:iid}
    {\em (i.i.d.~random coding)}
    Under the {\em i.i.d.~random coding ensemble} with {\em input distribution} $Q_X \in \Pc(\Xc)$,\footnote{Not to be confused with the decoding metric $q(x,y)$.} the codewords are independently drawn from the codeword distribution
    \begin{equation}
        P_{\Xv}(\xv) = \prod_{i=1}^n Q_X(x_i). 
    \end{equation}
    That is, every symbol of every codeword is independently drawn from $Q_X$.
\end{definition}

We will use i.i.d.~random coding to prove the achievability of the {\em generalized mutual information} (GMI), which can equivalently be written in the primal form
\begin{equation}
    \GMI(Q_X)=\min_{\substack{\Ptilde_{XY} \in \Pc(\Xc \times \Yc) \,:\, \Ptilde_Y=P_Y \\ \EE_{\Ptilde}[\log q(X,Y)] \ge \EE_{P}[\log q(X,Y)]}} D(\Ptilde_{XY} \| Q_X \times P_Y) \label{eq:INTR_PrimalGMI}
\end{equation}
with $P_{XY} = Q_X \times W$, or in the dual form,\footnote{A supremum is used for $s \ge 0$ instead of a maximum since it may not be attained by a finite value of $s$; see \eqref{eq:BEC} for an example. }
\begin{equation}
    \GMI(Q_X) = \sup_{\sgz} \sum_{x,y} Q_X(x)W(y|x) \log\frac{q(x,y)^{s}}{\sum_{\xbar} Q_X(\xbar) q(\xbar,y)^{s}}. \label{eq:INTR_RateGMI}
\end{equation}
These expressions are discussed in Sections \ref{sec:interpretations}--\ref{sec:advantages} below.  The following theorem provides a formal statement of achievability, which will be proved in Section \ref{sec:su_proofs}.

\begin{theorem} \label{thm:GMI}
    {\em (Achievability of the GMI)} 
    For any mismatched DMC $(W,q)$ and a given input distribution $Q_X \in \Pc(\Xc)$, the rate $\GMI(Q_X)$ is achievable via i.i.d.~random coding.  Consequently,
    \begin{equation}
        \CM \ge \CGMI \triangleq \max_{Q_X \in \Pc(\Xc)} \GMI(Q_X). \label{eq:CGMI}
    \end{equation}
\end{theorem}

Of course, $\GMI(Q_X)$ depends not only on $Q_X$, but also on $(W,q)$.  For the most part, we leave this dependence implicit, as we will be considering a fixed channel and decoding metric.  However, in some cases, we will make the dependence explicit by writing $\GMI(Q_X,W)$ or $\GMI(Q_X,W,q)$.

\subsection{Constant-Composition Random Coding and the LM Rate} \label{sec:cc_LM}

Another well-known random coding method for channel coding is {\em constant-composition random coding}, which introduces some dependence between the symbols of a given codeword.  In contrast to the matched setting \cite{Csi11,GallagerCC}, we will see that this ensemble can attain higher rates than the i.i.d.~ensemble in the presence of mismatch (see Section \ref{sec:su_examples}).

Before defining the ensemble, we introduce some terminology.  Recalling the notion of a {\em type} in \eqref{eq:type}, we let $\Pc_n(\Xc) \subset \Pc(\Xc)$ be the subset of probability distributions that correspond to the type of some length-$n$ sequence, i.e., each probability equals an integer multiple of $\frac{1}{n}$.  For a given type $P_{X} \in \Pc_n(\Xc)$, we define the {\em type class}
\begin{equation}
    \Tclass(P_{X}) = \big\{ \xv \in \Xc^n \,:\, \Phat_{\xv} = P_{X} \big\}, \label{eq:type_class}
\end{equation}
where $\Phat_{\xv}$ is defined in \eqref{eq:type}.  In words, $\Tclass(P_{X})$ contains all length-$n$ sequences having empirical distribution $P_{X}$.

As with i.i.d.~random coding, the constant-composition codeword distribution is specified by an input distribution $Q_X \in \Pc(\Xc)$.  Given $Q_X$, we define $Q_{X,n} \in \Pc_n(\Xc)$ to be an arbitrary type having the same support as $Q_X$, and satisfying the following:
\begin{equation}
    \|Q_X - Q_{X,n}\|_{\infty} \triangleq \max_{x \in \Xc} |Q_X(x) - Q_{X,n}(x)| \le \frac{1}{n}. \label{eq:closest_type}
\end{equation}
This essentially amounts to rounding each value of $Q_X(x)$ up or down so that it is an integer multiple of $\frac{1}{n}$.

\begin{definition}  \label{def:cc}
    {\em (Constant-composition random coding)}
    Under the {\em constant-composition random coding ensemble} with input distribution $Q_X \in \Pc(\Xc)$, the codewords are independently drawn from the codeword distribution
    \begin{equation}
        P_{\Xv}(\xv) = \frac{1}{ |\Tclass(Q_{X,n})| } \openone\big\{ \xv \in \Tclass(Q_{X,n}) \}, 
    \end{equation}
    where $Q_{X,n} \in \Pc_n(\Xc)$ satisfies \eqref{eq:closest_type}.  That is, each codeword is equiprobable on the set of sequences with empirical distribution $Q_{X,n}$.
\end{definition}

We will use constant-composition random coding to prove the achievability of the {\em LM rate}, which can equivalently be written in the primal form
\begin{equation}
    \LM(Q_X)=\min_{\substack{\Ptilde_{XY} \in \Pc(\Xc \times \Yc) \,:\, \Ptilde_{X}=Q_X, \Ptilde_{Y}=P_Y \\ \EE_{\Ptilde}[\log q(X,Y)] \ge \EE_{P}[\log q(X,Y)]}}I_{\Ptilde}(X;Y) \label{eq:INTR_PrimalLM}
\end{equation}
with $P_{XY} = Q_X \times W$, or in the dual form,
\begin{equation}
    \LM(Q_X)=\sup_{\sgz,a(\cdot)} \sum_{x,y} Q_X(x)W(y|x) \log\frac{q(x,y)^s e^{a(x)}}{\sum_{\xbar} Q_X(\xbar) q(\xbar,y)^{s} e^{a(\xbar)}}. \label{eq:INTR_RateLM}
\end{equation}
These expressions are discussed in Sections \ref{sec:interpretations}--\ref{sec:advantages}, and the following formal statement of achievability is proved in Section \ref{sec:su_proofs}.

\begin{theorem} \label{thm:LM}
    {\em (Achievability of the LM rate)}
    For any mismatched DMC $(W,q)$ and a given input distribution $Q_X \in \Pc(\Xc)$, the rate $\LM(Q_X)$ is achievable via constant-composition random coding.  Consequently, 
    \begin{equation}    
        \CM \ge \CLM \triangleq \max_{Q_X \in \Pc(\Xc)} \LM(Q_X). \label{eq:CLM}
    \end{equation}
\end{theorem}

As with the GMI, we will sometimes write $\LM(Q_X,W)$ or $\LM(Q_X,W,q)$ to make the dependence on the channel and/or decoding metric explicit.

\subsection{Interpretations of the Rates} \label{sec:interpretations}

We proceed by giving some interpretations of the achievable rates, and some intuition as to how they arise from the random-coding analysis.  At this point, it is worth noting that the LM rate is always at least as high as the GMI:
\begin{equation}
    \GMI(Q_X) \le \LM(Q_X). \label{eq:GMI_vs_LM}
\end{equation}
This is easily seen from the fact that the minimization in the primal expression \eqref{eq:INTR_PrimalLM} is more constrained than that in \eqref{eq:INTR_PrimalGMI}, or from the fact that we can lower bound the dual expression \eqref{eq:INTR_RateLM} via the specific choice $a(x) = 0$ in order to obtain \eqref{eq:INTR_RateGMI}.

Although the GMI is weaker than the LM rate, it is still interesting to understand it separately, and to contrast it with the LM rate in order to better understand the differences between the corresponding random-coding ensembles.  Moreover, in several applications of mismatched decoding, the GMI admits a simple analytical expression but the LM rate does not.

\paragraph{Primal expressions.} The idea behind the proof of Shannon's (matched) channel coding theorem is that if $R < I(X;Y)$, then with high probability under random coding, no incorrect codeword $\xvbar$ will be such that the pair $(\xvbar,\yv)$ ``looks like'' it was generated according to $Q_X \times W$ (e.g., in terms of joint typicality, or having a sufficiently high likelihood).  In the mismatched setting, such codewords are not the only problematic ones; errors also occur when $(\xvbar,\yv)$ ``looks like'' it was drawn from {\em any} $\Ptilde_{XY}$ with a higher decoding metric than $Q_X \times W$.  In both \eqref{eq:INTR_PrimalGMI} and  \eqref{eq:INTR_PrimalLM}, the achievable rate is a minimum over all such  $\Ptilde_{XY}$. The $Y$-marginal constraint arises since $\Phat_{\yv} \to P_Y$ by the law of large numbers, and the $X$-marginal constraint in the LM rate arises from the constant-composition construction.

As noted in \cite{Mer95}, the primal expressions also closely resemble {\em rate-distortion functions} in lossy source coding; for instance, in the absence of the constraint $\Ptilde_Y= P_Y$ in \eqref{eq:INTR_PrimalLM}, we would precisely recover the rate distortion function of the ``source'' $X$ with ``reconstruction'' $Y$, distortion measure $d(x,y) = -\log q(x,y)$, and distortion threshold $-\EE_{P}[\log q(X,Y)]$.  See Section \ref{sec:rd_mm_random} for further discussion.

\paragraph{Dual expressions.} Gallager's proof of the channel coding theorem \cite{Gal65} is based on maximum-likelihood decoding, and is particularly amenable to a generalization to maximum-metric decoding.  If this is done naively, one attains a rate of the form \eqref{eq:INTR_RateGMI} with $s = 1$, which is rather limited (e.g., such a rate may be negative) \cite{Fis71}.  However, the achievability of the GMI immediately follows from the equivalence of the decoding metrics $q(x,y)$ and $q(x,y)^s$ for $s > 0$, stated in Proposition \ref{prop:equiv_general} (the case $s=0$ can also trivially be included).  Similarly, the LM rate can be deduced using the equivalence of $q(x,y)$ and $q(x,y)^s e^{a(x)}$ for constant-composition codes, stated in Proposition \ref{prop:equiv_cc}.  Essentially, the improvement of constant-composition random coding over i.i.d.~random coding comes from the additional codebook structure that permits the introduction of $a(\cdot)$ into the bound.

To provide another interpretation, we note that the LM rate can be written as follows (see Lemma \ref{lem:divergence_forms} in Section \ref{sec:properties}):
\begin{equation}
    \LM(Q_X,W) = I(Q_X,W) - \inf_{V_{Y|X} \in \Vc_q } D(W \| V_{Y|X} | Q_X), \label{eq:div_form}
\end{equation}
where $I(Q_X,W) = I(X;Y)$ with $(X,Y) \sim Q_X \times W$, and
\begin{align}
    &\Vc_q = \big\{ V_{Y|X} \in \Pc(\Yc|\Xc) \,:\, V_{Y|X}(y|x) = q(x,y)^s e^{a(x)} e^{b(y)} \nonumber
    \\ &\qquad\qquad\qquad\qquad\qquad\qquad \text{ for some } \sgz, a(x), b(y) \big\},
\end{align}
where we recall that $\Pc(\Yc|\Xc)$ is the set of all conditional distributions on $\Yc$ given $\Xc$.

To understand the form given in \eqref{eq:div_form}, consider the case that $q(x,y) = \What(y|x)$ is a conditional distribution.  In analogy with source coding methods (e.g., Huffman coding) in which one pays a KL divergence penalty for assuming the wrong distribution \cite[Ch.~5]{Cov06}, one might expect to pay a penalty of $D(W\|\What|Q_X)$ here compared to optimal decoding.  The above formulation reveals that this is an {\em upper bound} on the loss, and that it can always be reduced to the infimum of $D(W \| V_{Y|X} | Q_X)$ over all conditional distributions $V_{Y|X}$ yielding an equivalent decoding metric in the sense of Proposition \ref{prop:equiv_cc}.

\subsection{Advantages of the Primal and Dual Forms} \label{sec:advantages}

At this point, it may be unclear why it is beneficial to consider two equivalent formulations of each achievable rate.  The advantages will become increasingly clear throughout the section and the monograph, but we briefly outline some important examples here:
\begin{itemize}
    \item As we already saw above, the two forms can admit interesting alternative interpretations.
    \item The most important properties of the rates (see Section \ref{sec:properties}) are often easier to prove using one particular form, though some can be proved using either of the two.  
    \item As a notable example, when it comes to establishing {\em ensemble tightness} (roughly, showing that the rates cannot be improved under the random coding ensembles considered), the most elementary proofs for DMCs are based on the primal expression.
    \item The dual forms have the notable advantage that substituting {\em any} choices of $\sgz$ and/or $a(\cdot)$ yields a valid achievable rate.  In contrast, the primal expressions are only guaranteed to provide a valid achievable rate under the minimizing joint distribution $\Ptilde_{XY}$.
    \item Another advantage of the dual forms is that they come with direct derivations that readily extend to continuous-alphabet channels (see Section \ref{ch:single_user_cont}).
    \item When it comes to computing the achievable rates numerically, it can be useful to compute both forms and verify that they coincide, as a check for robustness against numerical precision issues.
\end{itemize}
For reasonable-sized alphabets $\Xc$ and $\Yc$, neither the primal nor dual forms pose significant computational challenges {\em for a given input distribution $Q_X$}: The primal expressions are convex minimization problems, and the dual expressions are concave maximization problems, and hence both can be evaluated using off-the-shelf solvers.  In contrast, Lemma \ref{lem:non_concave} below suggests that the optimization over $Q_X$ can be difficult due to non-concavity.

\section{Examples} \label{sec:su_examples}

In this subsection, we provide analytical and numerical evaluations of the GMI and LM rate for a variety of specific DMCs.

\subsection{Binary Channels} \label{sec:binary}

Consider the case that both the channel inputs and outputs are binary, i.e., $|\Xc| = |\Yc| = 2$.  We assume without loss of generality that $\Xc = \Yc = \{0,1\}$.  We will first show that the LM rate is either equal to the matched capacity or zero \cite{Csi95}, and then discuss how the GMI does not exhibit such a dichotomy.

\paragraph{Evaluating the LM rate.} Recall that the LM rate is based on constant-composition codes, and that for any such code, two decoding metrics $q,\qtilde$ are equivalent if
\begin{equation}
    \log \qtilde(x,y) = a(x) + b(y) + s \log q(x,y) \label{eq:equiv_metrics}
\end{equation}
for some constant $s > 0$ and functions $a(x),b(y)$ (see Proposition \ref{prop:equiv_cc}).  To study such equivalences, we introduce the notation
\begin{equation}
    \bell_q = \left[\begin{array}{cc}
    \log q(0,0) & \log q(0,1) \\
    \log q(1,0) & \log q(1,1)
    \end{array}\right] \triangleq \left[\begin{array}{cc}
    \ell_{00} & \ell_{01} \\
    \ell_{10} & \ell_{11}
    \end{array}\right],
\end{equation}
and we write $\bell_{q} \equiv \bell_{\qtilde}$ if \eqref{eq:equiv_metrics} holds.

Notice that choosing $a(x)$ (respectively, $b(y)$) in \eqref{eq:equiv_metrics} amounts to adding or subtracting a constant from each row (respectively, column) of $\bell_q$.  By choosing these constants to equate certain entries to zero, we obtain
\begin{equation}
    \bell_{q}
    \equiv \left[\begin{array}{cc}
    0 & 0 \\
    \ell_{10} - \ell_{00} & \ell_{11} - \ell_{01}
    \end{array}\right]
     \equiv \left[\begin{array}{cc}
    0 & 0 \\
    0 & \ell_{11} - \ell_{01} - \ell_{10} + \ell_{00}
    \end{array}\right].
\end{equation}
By a symmetric argument, we can obtain the same final equivalence with the top-left and bottom-right entries swapped, and in addition, we can sum the two matrices together to obtain
\begin{equation}
    \bell_{q} 
    \equiv \left[\begin{array}{cc}
    \ell_{00} + \ell_{11} - \ell_{01} - \ell_{10} & 0 \\
    0 & \ell_{00} + \ell_{11} - \ell_{01} - \ell_{10}
    \end{array}\right].
\end{equation}
Since multiplying by $s > 0$ in \eqref{eq:equiv_metrics} amounts to scaling the entire matrix, we deduce that as long as $\ell_{11} - \ell_{01} - \ell_{10} + \ell_{00} \ne 0$, one of the following must hold:
\begin{equation}
    \bell_{q} 
    \equiv \left[\begin{array}{cc}
    1 & 0 \\
    0 & 1
    \end{array}\right], 
    \quad \text{or} \quad 
    \bell_{q} 
    \equiv \left[\begin{array}{cc}
    -1 & 0 \\
    0 & -1
    \end{array}\right],
\end{equation}
with the non-zero entries equaling $\mathrm{sign}( \ell_{00} + \ell_{11} - \ell_{01} - \ell_{10} )$.

Now, letting $\delta_1$ and $\delta_2$ denote the transition probabilities $W(1|0)$ and $W(0|1)$ respectively, we observe that upon substituting $\ell_{xy}^* = \log W(y|x)$, it holds that $\ell_{00}^* + \ell_{11}^* - \ell_{01}^* - \ell_{10}^* > 0$ if and only if $\frac{(1-\delta_1)(1-\delta_2)}{\delta_1\delta_2} > 1$, which in turn is equivalent to $\delta_1 + \delta_2 < 1$.  Hence, if $\delta_1 + \delta_2 < 1$ then the ML rule minimizes the Hamming distance, and if $\delta_1 + \delta_2 > 1$, it maximizes the Hamming distance.

If the sign of $\ell_{00} + \ell_{11} - \ell_{01} - \ell_{10} = \log\frac{q(0,0)q(1,1)}{q(0,1)q(1,0)}$ for the decoding metric matches that of the ML rule, then clearly we achieve the matched capacity, as the corresponding decoding rules are equivalent.  If not, then the decoder is minimizing a quantity that is meant to be maximized (or vice versa), and hence we attain a rate of zero.\footnote{This claim can be verified, for example, via Lemma \ref{lem:pos_conds} below.}

\paragraph{Evaluating the GMI.} The GMI is based on i.i.d.~random coding, which can produce general codebooks for which two decoding rules $q,\qtilde$ are only equivalent if 
\begin{equation}
\log \qtilde(x,y) = b(y) + s \log q(x,y) \label{eq:equiv_metrics_iid}
\end{equation}
for some $s > 0$ and $b(y)$ (see Proposition \ref{prop:equiv_general}).
That is, we no longer have the freedom to choose $a(x)$ as in \eqref{eq:equiv_metrics}.  Re-using the notation $\bell_{q} \equiv \bell_{\qtilde}$ to mean equivalence according to \eqref{eq:equiv_metrics_iid}, we have the following whenever $\ell_{00} \ne \ell_{10}$:
\begin{equation}
    \bell_{q}
    \equiv \left[\begin{array}{cc}
    \ell_{00} - \ell_{10} & 0 \\
    0 & \ell_{11} - \ell_{01}
    \end{array}\right]
    \equiv \left[\begin{array}{cc}
    \pm 1 & 0 \\
    0 & \lambda
    \end{array}\right], \label{eq:gmi_equivalences}
\end{equation}
where the $\pm 1$ entry equals the sign of $\ell_{00} - \ell_{10}$, and $\lambda = \pm\big(\frac{\ell_{11} - \ell_{01}}{\ell_{00} - \ell_{10}}\big)$ may be positive or negative.  In this form, the decoder can be interpreted as considering a {\em weighted} Hamming distance.

\begin{figure}
    \begin{centering}
        \includegraphics[width=0.6\columnwidth]{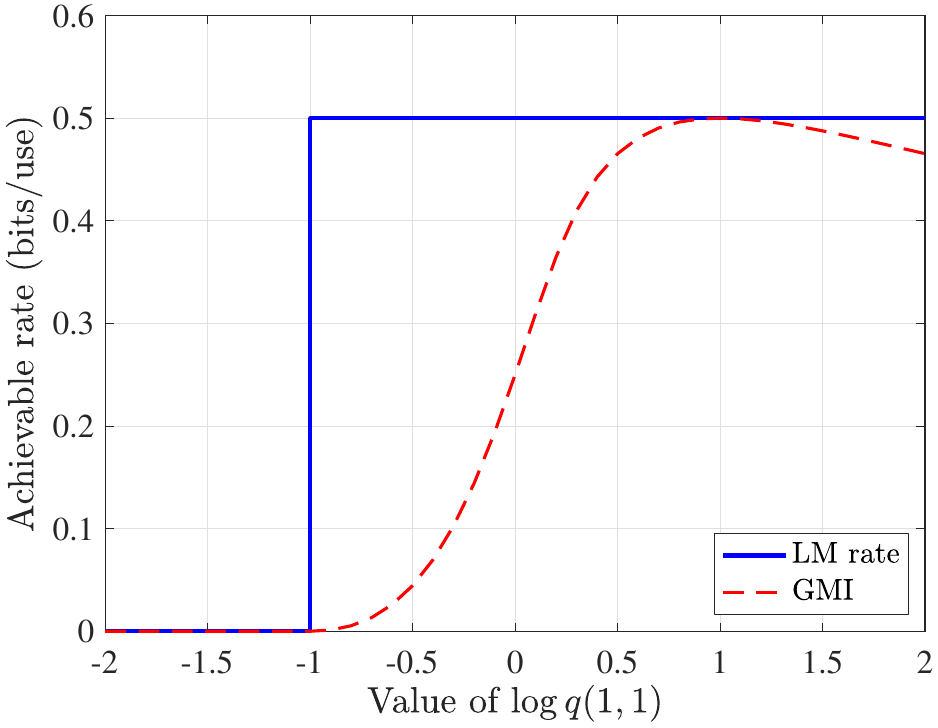}
        \par
    \end{centering}
    
    \caption{Binary example: GMI and LM rate as a function of $\lambda = \log q(1,1)$ with $\delta = 0.11$, under equiprobable inputs and a log-decoding metric of the form \eqref{eq:gmi_equivalences}.} \label{fig:BinaryExample}
\end{figure}

If we reduce both $\log W(y|x)$ and $\log q(x,y)$ to the form in \eqref{eq:gmi_equivalences}, then even if the signs match, the achievable rate can degrade due to the mismatch.  We demonstrate this in Figure \ref{fig:BinaryExample}, where we plot the GMI (as well as the LM rate) in the following setting:
\begin{itemize}
    \item The channel is a BSC with crossover probability $\delta = 0.11$, hence having a matched capacity of roughly $0.5$ bits/use;
    \item The (log-)decoding metric takes the form on the right-hand side of \eqref{eq:gmi_equivalences} with first entry $+1$, and with $\lambda = \log q(1,1)$ taking some value in the range $[-2,2]$.  The choice $\lambda = 1$ recovers optimal minimum Hamming distance decoding.
    \item The input distribution is $Q_X = \big(\frac{1}{2},\frac{1}{2}\big)$, which achieves the matched capacity under matched decoding.
\end{itemize}
In accordance with the above findings, the LM rate is zero for $\lambda < - 1$ (and in fact also for $\lambda = -1$, which can be checked separately) and equals the matched capacity for $\lambda > -1$. On the other hand, the GMI only achieves the matched capacity for $\lambda = 1$.

\subsection{Symmetric Channels and Metrics} 
\label{sec:symmetric}

In the previous example, we saw that the gap between the GMI and LM rate grows when more asymmetry is {\em incorrectly} introduced into the decoding metric.  In this example, we complement this finding via the following claim:  Under matching output-symmetry and an equiprobable input distribution $Q_X$, the GMI and LM rate are identical.

The notion of matching output-symmetry needs some clarification; we adopt the following definition of Gallager \cite[Sec.~4.5]{Gal68} regarding $W$, and extend it to the metric $q$ in the natural way.

\begin{definition} \label{def:output_symm}
    {\em (Output-symmetric channel and metric)}
    A DMC $W(y|x)$ is said to be output-symmetric if there exists a partition $\Yc_1,\dotsc,\Yc_k$ of $\Yc$ such that the following holds: Given the $|\Xc| \times |\Yc|$ transition matrix corresponding to $W$, the $|\Xc| \times |\Yc_j|$ sub-matrix corresponding to each $\Yc_j$ ($j=1,\dotsc,k$) is such that all rows are permutations of each other, and all columns are permutations of each other.  The notion of output-symmetry for a decoding metric $q(x,y)$ is defined similarly.
    
    A mismatched DMC $(W,q)$ is said to exhibit matching output-symmetry if both $W$ and $q$ are output-symmetric with the same partition $\Yc_1,\dotsc,\Yc_k$.
\end{definition}

We briefly give two concrete examples of output-symmetry: For the binary symmetric channel one adopts the trivial partition $\Yc_1 = \Yc = \{0,1\}$, whereas for the binary erasure channel, the two partitions are $\Yc_1 = \{0,1\}$ and $\Yc_2 = \{e\}$ (the erasure symbol).

We can now formally state the following result, which, to the best of our knowledge, has not appeared previously.

\begin{lemma} \label{lem:gmi_lm_symm}
    {\em (GMI and LM rate under output symmetry)}
    For any pair $(W,q)$ exhibiting matching output symmetry, if $Q_X$ is the equiprobable distribution on $\Xc$, then
    \begin{equation}
        \GMI(Q_X) = \LM(Q_X).
    \end{equation}
\end{lemma}

The proof amounts to showing that the choice $a(x) = 0$ in \eqref{eq:INTR_RateLM} satisfies necessary and sufficient optimality conditions, implying that the rate coincides with \eqref{eq:INTR_RateGMI}.  The details are given in Appendix \ref{sec:omitted_proofs}.

In the following example, we show that even when the pair $(W,q)$ satisfies a stronger symmetry assumption (namely, Definition \ref{def:output_symm} restricted to $k = 1$), the achievable rates $\GMI(Q_X)$ and $\LM(Q_X)$ may differ when $Q_X$ is non-equiprobable, and the LM rate can be strictly higher when $Q_X$ is optimized.

\subsection{Zero-Undetected Error Capacity} \label{sec:su_zuec}

As we saw in Section \ref{sec:app_zuec}, this choice $q(x,y) = \openone\{W(y|x) > 0\}$ corresponds to the {\em zero-undetected error capacity}.  Here we consider a specific example from \cite{Ahl96}, in which $\Xc = \Yc = \{0,1,2\}$, and the channel and decoding metric are given by
\begin{align} 
    \Wv & = \left[\begin{array}{ccc}
    0.75 & 0.25 & 0 \\
    0 & 0.75 & 0.25 \\
    0.25 & 0 & 0.75 \\
    \end{array}\right], \qquad
    \qv = \left[\begin{array}{ccc}
    ~~1~\, & ~~1\,~ & ~~0~~ \\
    ~~0~\, & ~~1\,~ & ~~1~~ \\
    ~~1~\, & ~~0\,~ & ~~1~~ \\
    \end{array}\right], \label{eq:Example_q1}
\end{align}
where $x$ indexes the rows and $y$ indexes the columns. 

\begin{figure}
        \begin{subfigure}{.49\columnwidth}
            \centering
            \includegraphics[width=\columnwidth]{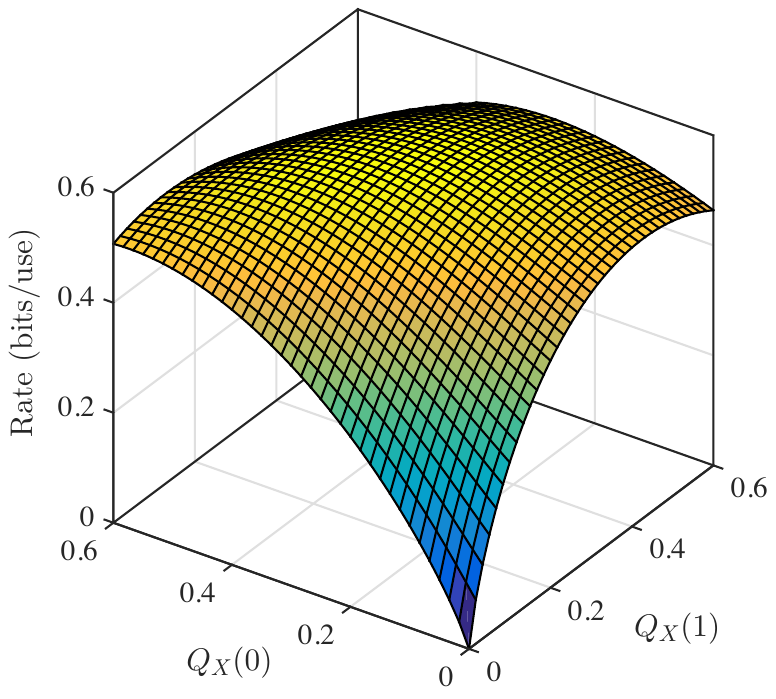} 
            \caption{GMI}
        \end{subfigure}
        \begin{subfigure}{.49\columnwidth}
            \centering
            \includegraphics[width=\columnwidth]{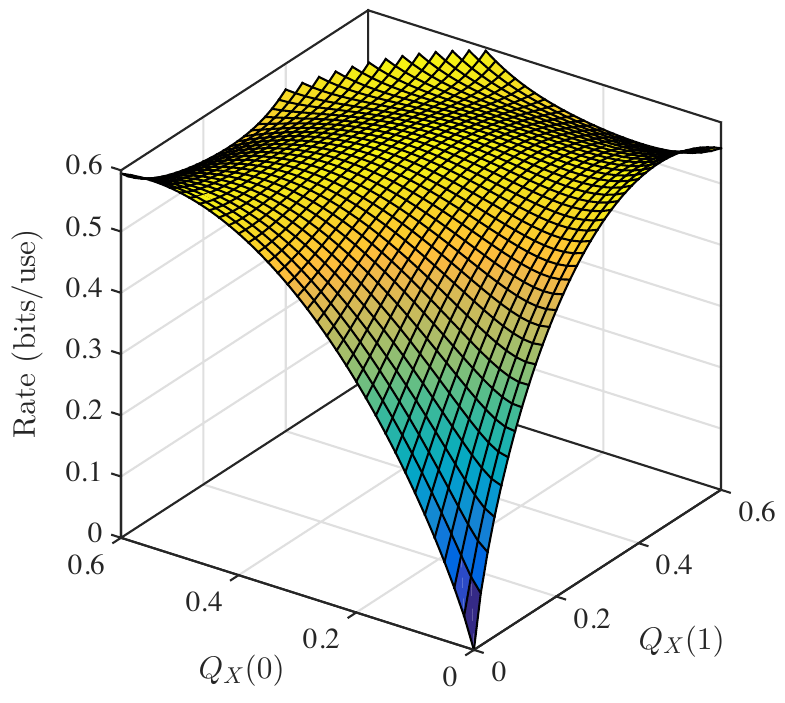} 
            \caption{LM rate}
        \end{subfigure}
    
    \caption{Zero-undetected error capacity example: GMI and LM rate for the channel and decoding metric in \eqref{eq:Example_q1}, as a function of $Q_X(0)$ and $Q_X(1)$.} \label{fig:ZUEC_Rates}
\end{figure}

In Figure \ref{fig:ZUEC_Rates}, we plot the LM rate as a function of $(Q_X(0),Q_X(1))$, with the remaining entry $Q_X(2)$ determined so that the three sum to one.  The optimal choices of $Q_X$ are given as follows:
\begin{itemize}
    \item For the GMI, $Q_X = \big(\frac{1}{3},\frac{1}{3},\frac{1}{3}\big)$ is optimal.  Under this choice, $\GMI(Q_X) = \LM(Q_X) = 0.585$ bits/use; the two rates are identical in accordance with Lemma \ref{lem:gmi_lm_symm}.
    \item For the LM rate, the choice $Q_X = (0.449,0.551,0)$ (or any cyclic shift thereof) is optimal.  Under this choice, $\GMI(Q_X) = 0.502$ bits/use and $\LM(Q_X) = 0.599$ bits/use.
\end{itemize}
We see that the LM rate exceeds the GMI even after the optimization of $Q_X$.  In Section \ref{sec:sc}, we will present an achievable rate that further improves on the LM rate in this example, implying that the LM rate is strictly smaller than the mismatch capacity, i.e.,$\CLM < \CM$ .  Of course, this implies the same for the GMI, which is no higher than the LM rate.


\subsection{Compound Channels} \label{sec:ex_compound}

In this example (taken from \cite{Csi95}), 
we consider the case that the channel is only known to lie in some set $\Wc$, but the exact channel within the set is unknown.  This is commonly referred to as the {\em compound channel} model, and was discussed in Section \ref{sec:other_models}.  We initially focus on the case that $\Wc$ is closed and convex, and then comment on the more general case.

Using i.i.d.~random coding with an input distribution $Q_X$, along with a fixed decoding metric $q$, we can achieve any rate up to 
\begin{equation}
    R_{\Wc}(Q_X,q) \triangleq \min_{W \in \Wc} \GMI(Q_X,W,q), \label{eq:compound0}
\end{equation}
where $\GMI(Q_X,W,q)$ denotes the GMI with an explicit dependence on the channel and metric.  We will show that upon optimizing $Q_X$ and $q$, this recovers the best possible rate (i.e., the compound channel capacity), given as follows in the case that $\Wc$ is closed and convex \cite{Lap98,Csi95}:
\begin{equation}
    C^* = \max_{Q_X} \min_{W \in \Wc} I(Q_X, W) = \min_{W \in \Wc} \max_{Q_X} I(Q_X,W), \label{eq:compound_C}
\end{equation}
where $I(Q_X,W) \triangleq I_{Q_X \times W}(X;Y)$.  

Let $(Q_X^*,W^*)$ denote the pair achieving the saddle-point in \eqref{eq:compound_C}. We choose $q(x,y) = W^*(y|x)$ and $Q_X(x) = Q_X^*(x)$, and proceed by lower bounding the achievable rate in \eqref{eq:compound0}.  For any $W \in \Wc$, setting $s=1$ in \eqref{eq:INTR_RateGMI} yields
\begin{equation}
    \GMI(Q^*_X,W, W^*) \ge \sum_{x,y} Q_X^*(x) W(y|x) \log \frac{ W^*(y|x) }{ P_Y^*(y) }, \label{eq:compound1}
\end{equation}
where $P_Y^*$ is the $Y$-marginal of the joint distribution $Q_X^* \times W^*$.  It now only remains to prove that
\begin{equation}
    \sum_{x,y} Q_X^*(x) W(y|x) \log \frac{ W^*(y|x) }{ P_Y^*(y) } \ge I(Q_X^*, W^*),  \label{eq:compound2}
\end{equation}
since the right-hand side equals $C^*$ in \eqref{eq:compound_C} by definition.  To establish \eqref{eq:compound2}, we first note the following for any $\lambda \in [0,1]$:
\begin{equation}
    I(Q_X^*, \lambda W + (1-\lambda)W^*) \ge I(Q_X^*,W^*), \label{eq:compound3}
\end{equation}
which follows from the convexity of $\Wc$ and the fact that $W^*$ minimizes $I(Q_X^*, W')$ among $W' \in \Wc$.  By taking $\lambda \to 0$, it follows that the derivative of the left-hand side of \eqref{eq:compound3} with respect to $\lambda$ must be non-negative at $\lambda = 0$, and it is a simple differentiation exercise to show that this fact is equivalent to \eqref{eq:compound2}.  It follows that $R_{\Wc}(Q_X^*,W^*) \ge C^*$, as desired.

If we remove the assumption that $\Wc$ is convex, then the best possible single-letter decoding metric $q(x,y)$ may still yield a highly suboptimal rate.  For instance, from the example in Section \ref{sec:binary}, if $\Xc = \Yc = \{0,1\}$ and $\Wc$ contains the deterministic channels with $Y = X$ and $Y = 1-X$, then any maximum-metric decoder gives zero rate.  However, the compound channel capacity is $1$ bit/use, and can be achieved by a standard universal decoder \cite[Ex.~6.20]{Csi11}.

\subsection{Parallel Channels} \label{sec:parallel}

In this example (taken from \cite{Lap96}), we present another channel (in addition to that of Section \ref{sec:su_zuec}) for which the LM rate is known to be strictly smaller than the mismatch capacity.

\begin{figure}
    \begin{centering}
        \includegraphics[width=0.5\columnwidth]{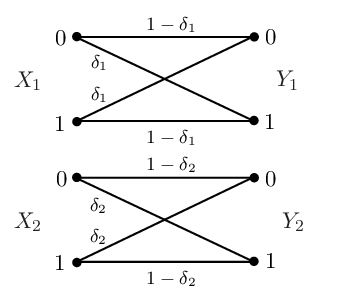}
        \par
    \end{centering}
    
    \caption{Parallel binary symmetric channels with crossover probabilities $(\delta_1,\delta_2)$. } \label{fig:ParallelBSC}
\end{figure}

We let $\Xc = \Yc = \{0,1\}^2$, and write the input and output as binary pairs, $X= (X_1,X_2)$ and $Y = (Y_1,Y_2)$.  The output $Y_1$ is formed by passing $X_1$ through a binary symmetric channel (BSC), and similarly for $Y_2$ and $X_2$, with the two BSCs being independent of each other; see Figure \ref{fig:ParallelBSC}.  The corresponding crossover probabilities are denoted by $\delta_1,\delta_2 \in \big(0,\frac{1}{2}\big)$, and the mismatched decoder incorrectly assumes that both crossover probabilities are equal.
Hence, the channel is given by $W((y_1,y_2)|(x_1,x_2)) = W_1(y_1|x_1)W_2(y_2|x_2)$, and the decoding metric can be taken as
\begin{equation}
    \log q((x_1,x_2),(y_1,y_2)) = \frac{1}{2}\big( \openone\{ x_1 = y_1 \} + \openone\{ x_2 = y_2 \}  \big), \label{eq:parallel_metric}
\end{equation}
meaning that both $x_1 \ne y_1$ and $x_2 \ne y_2$ are penalized equally.

We let $Q_X$ be the equiprobable distribution on $\{0,1\}^2$, and evaluate the LM rate using the primal expression.  For any joint distribution $\Ptilde_{XY}$ satisfying the constraints in \eqref{eq:INTR_PrimalLM}, we have
\begin{align}
        I_{\Ptilde}( X; Y )
        &= I_{\Ptilde}( X_1, X_2; Y_1, Y_2 ) \\
        &= I_{\Ptilde}( X_1; Y_1, Y_2 ) + I_{\Ptilde}( X_2; Y_1, Y_2 | X_1 ) \label{eq:parallel3} \\
        &\ge I_{\Ptilde}( X_1; Y_1 ) + I_{\Ptilde}( X_2; Y_2 ), \label{eq:parallel4}
\end{align}
where \eqref{eq:parallel3} follows from the chain rule, and \eqref{eq:parallel4} follows by trivially lower bounding the first term, and bounding the second term as follows:
\begin{align}
    I_{\Ptilde}( X_2; Y_1, Y_2 | X_1 )
        &= H_{\Ptilde}( X_2 | X_1 ) - H_{\Ptilde}( X_2 | Y_1, Y_2, X_1 ) \\
        &\ge H_{\Ptilde}( X_2 ) - H_{\Ptilde}( X_2 | Y_2 ) \label{eq:parallel4b} \\
        &= I_{\Ptilde}(X_2;Y_2),
\end{align}
where \eqref{eq:parallel4b} uses the independence of $(X_1,X_2)$ following from $\Ptilde_{X}= Q_X$ (with $Q_X$ chosen to be equiprobable on $\{0,1\}^2$), and the fact that conditioning does not increase entropy.

The constraints $\Ptilde_X = Q_X$ and $\Ptilde_Y = P_Y$ in \eqref{eq:INTR_PrimalLM} ensure that $X_1$, $X_2$, $Y_1$, and $Y_2$, are each equiprobable on $\{0,1\}$ under $\Ptilde_{XY}$.  This is only possible if the conditional marginals $\Ptilde_{Y_1|X_1}$ and $\Ptilde_{Y_2|X_2}$ are BSCs. Letting $H_2(\alpha) = \alpha\log\frac{1}{\alpha} + (1-\alpha)\log\frac{1}{1-\alpha}$ denote the binary entropy function, we deduce from \eqref{eq:parallel4} that the following holds for some auxiliary crossover probabilities $\deltatilde_1,\deltatilde_2 \in (0,1)$:
\begin{align}
    I_{\Ptilde}( X; Y )
        &\ge \big(1 - H_2(\deltatilde_1)\big) + \big(1 - H_2(\deltatilde_2)\big) \\
        &\ge 2\bigg(1 - H_2\Big( \frac{\deltatilde_1 + \deltatilde_2}{2} \Big)\bigg) \label{eq:parallel6} \\
        &\ge 2\bigg(1 - H_2\Big( \frac{\delta_1 + \delta_2}{2} \Big)\bigg), \label{eq:parallel7}
\end{align}
where \eqref{eq:parallel6} follows from the concavity of entropy, and \eqref{eq:parallel7} follows since the constraint $\EE_{\Ptilde}[\log q(X,Y)] \ge \EE_{P}[\log q(X,Y)]$ reduces to $\frac{\deltatilde_1 + \deltatilde_2}{2} \le \frac{\delta_1 + \delta_2}{2}$ under the metric in \eqref{eq:parallel_metric}.

\begin{figure}
    \begin{centering}
        \includegraphics[width=0.6\columnwidth]{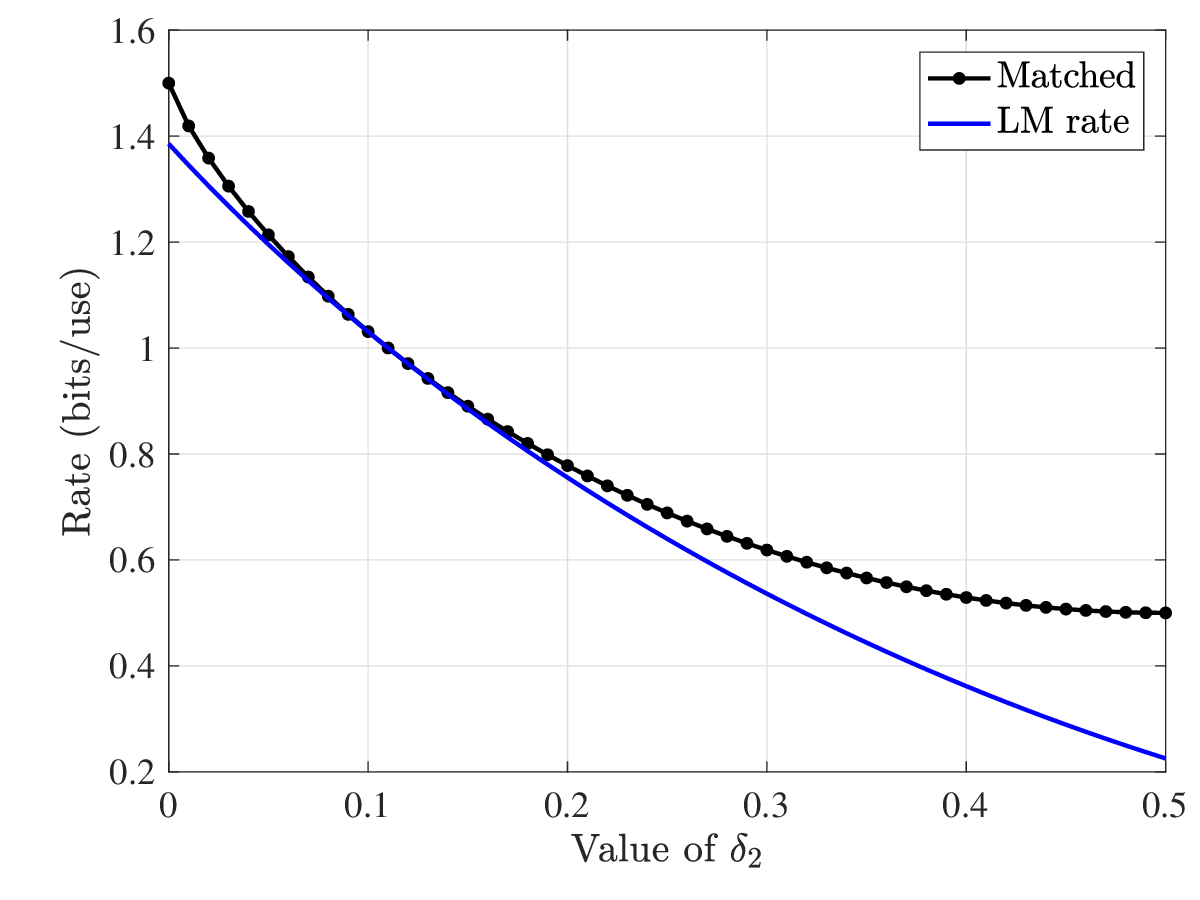}
        \par
    \end{centering}
    
    \caption{Parallel BSC example: Matched and mismatched achievable rates, with $\delta_1 = 0.11$ and various $\delta_2$. } \label{fig:ParallelBSC_Rates}
\end{figure}

We now observe that all the inequalities up to and including \eqref{eq:parallel7} hold with equality when $\deltatilde_1 = \deltatilde_2 = \frac{\delta_1+\delta_2}{2}$, meaning the corresponding joint distribution $\Ptilde_{XY}$ must achieve the minimum in \eqref{eq:INTR_PrimalLM}.  Hence, we have shown that 
\begin{equation}
    \LM(Q_X) =  2\bigg(1 - H_2\Big( \frac{\delta_1 + \delta_2}{2} \Big)\bigg).
\end{equation}
For $\delta_1 \ne \delta_2$, this is strictly smaller than the matched capacity $C = \big(1 - H_2(\delta_1)\big) + \big(1 - H_2(\delta_2)\big)$, which is achieved by equiprobable $Q_X$.  See Figure \ref{fig:ParallelBSC_Rates} for an illustration, where we fix $\delta_1 = 0.11$ and vary $\delta_2$.

We will revisit this example in Section \ref{sec:mu_parallel}, and see that in fact $\CM = C$, and that the LM rate is strictly smaller than the mismatch capacity.

\section{Properties of the GMI and LM Rate} \label{sec:properties}

In this subsection, we overview some of the most important known properties of the GMI and LM rate.

\subsection{Basic Properties}

We start by stating the important property of continuity of the GMI and LM rate as a function of the input distribution and channel.  This property will be used in the derivations of the primal expressions \eqref{eq:INTR_PrimalGMI} and \eqref{eq:INTR_PrimalLM} based on the method of types.

In the following, recall that $\GMI(Q_X,W)$ and $\LM(Q_X,W)$ denote the GMI and LM rate with an explicit dependence on the channel, for a fixed metric $q$.  The following was proved in \cite{Csi95}.

\begin{lemma}  \label{lem:continuity}
    {\em (Continuity)} 
    In the discrete memoryless setting with a given decoding metric $q$, the achievable rates $\GMI(Q_X,W)$ and $\LM(Q_X,W)$ are continuous in the pair $(Q_X,W)$ within the space of pairs satisfying
    \begin{equation}
        \big( Q_X(x) > 0 \text{\rm~and~} q(x,y) = 0 \big) \implies W(y|x) = 0. \label{eq:q_zero_cond}
    \end{equation}
\end{lemma}

The proof is somewhat technical, and is deferred to Appendix \ref{sec:continuity_app}.  We note that the condition \eqref{eq:q_zero_cond} is mild, since both achievable rates are zero when it fails \cite{Csi95}.  Indeed, if \eqref{eq:q_zero_cond} fails then we have $\EE_{Q_X \times W}[\log q(x,y)] = -\infty$, which implies that the choice $\Ptilde_{XY} = Q_X \times P_Y$ in \eqref{eq:INTR_PrimalLM} is feasible, and gives $I_{\Ptilde}(X;Y) = 0$.  At a more intuitive level, if $q(x,y) = 0$ and $Q_X(x)W(y|x) > 0$ then the metric associated with the transmitted codeword will be zero with high probability, precluding correct decoding.

Next, we present necessary and sufficient conditions for the GMI, LM rate, and mismatch capacity to be positive \cite{Csi95}.  Notably, as long as we optimize the input distribution, both i.i.d.~and constant-composition random coding achieve a positive rate whenever there exists {\em any} coding scheme achieving a positive rate.

\begin{lemma} \label{lem:pos_conds}
    {\em (Conditions for positivity)} 
    For any mismatched DMC $(W,q)$ and input distribution $Q_X$, we have the following:
    
    (i) $\GMI(Q_X) > 0 \iff \LM(Q_X) > 0$, with both being positive if and only if
    \begin{equation}
        \EE_{Q_X \times W}[\log q(X,Y)] > \EE_{Q_X \times P_Y}[\log q(X,Y)], \label{eq:pos_cond}
    \end{equation}
    where $P_Y$ is the $Y$-marginal of $P_{XY} = Q_X \times W$.  Moreover, in the case that this condition holds, the primal expressions \eqref{eq:INTR_PrimalGMI} and \eqref{eq:INTR_PrimalLM} remain unchanged when the inequality constraint is replaced by an equality constraint.
    
    (ii) The mismatch capacity $\CM$ satisfies 
    \begin{equation}
        \CM > 0 \iff \CLM > 0 \iff \CGMI>0,
    \end{equation}
    where $\CGMI$ and $\CLM$ are defined in \eqref{eq:CGMI} and \eqref{eq:CLM}, respectively.
\end{lemma}
\begin{proof}
        In the primal expressions of both the GMI and LM rate (see \eqref{eq:INTR_PrimalGMI} and \eqref{eq:INTR_PrimalLM}), the objective function is zero if and only if $\Ptilde_{XY} = Q_X \times P_Y$ is feasible.  For the GMI, this follows since the KL divergence is zero if and only if the two distributions are equal, and for the LM rate, it follows from the marginal constraints and the fact that mutual information is zero if and only if $X$ and $Y$ are independent.  The first claim follows since the choice $\Ptilde_{XY} = Q_X \times P_Y$ is feasible in the minimization problems if and only if \eqref{eq:pos_cond} fails.  
        
        Next, recall that both \eqref{eq:INTR_PrimalGMI} and \eqref{eq:INTR_PrimalLM} are convex minimization problems with linear constraints.  If the inequality constraint were to be removed in either case, the product distribution would be feasible, yielding an objective of zero.  This means that whenever the rate is positive, the inequality constraint is {\em active} (i.e., removing it changes the optimal objective value), and it is then a standard result in convex optimization theory that it must hold with equality \cite[Sec.~5.5]{Boy04}. 
        
        The proof of the second part is rather technical, so it is given in Appendix \ref{sec:omitted_proofs}.
\end{proof}

Lemma \ref{lem:pos_conds} is concerned with the distinction between a zero and non-zero rate, but does not give any indication of how close the achievable rates might be to the mismatch capacity.  The following lemma addresses the other extreme, giving necessary and sufficient conditions for the achievable rate to equal the mutual information $I(Q_X,W) = I(X;Y)$ under $(X,Y) \sim Q_X \times W$, which is the rate that would be achieved under optimal decoding.  Of course, if $Q_X$ is a (matched) capacity-achieving input distribution, then achieving $I(Q_X,W)$ amounts to achieving the matched capacity, and hence also the mismatch capacity.  

\begin{lemma}  \label{lem:opt_conds}
    {\em (Conditions for achieving the matched performance)} 
    For any mismatched DMC $(W,q)$ and input distribution $Q_X \in \Pc(\Xc)$, letting $P_Y$ denote the $Y$-marginal of $P_{XY} = Q_X \times W$, we have:
    \begin{enumerate}
        \item[(i)] $\GMI(Q_X,W) \le I(Q_X,W)$, with equality if and only if
        \begin{equation}
            \inf_{\sgz, b(\cdot)} \max_{\substack{(x,y) \in \Xc \times \Yc \,:\, \\ Q_X(x) > 0, P_Y(y) > 0}} | W(y|x) - q(x,y)^s e^{b(y)} | = 0. \label{eq:GMI_eq_cond_new}
        \end{equation}
        \item[(ii)] $\LM(Q_X,W) \le I(Q_X,W)$, with equality if and only if
        \begin{equation}
            \inf_{\sgz, a(\cdot), b(\cdot)} \max_{\substack{(x,y) \in \Xc \times \Yc \,:\, \\ Q_X(x) > 0, P_Y(y) > 0}} | W(y|x) - q(x,y)^s e^{a(x)} e^{b(y)} | = 0. \label{eq:LM_eq_cond_new}
        \end{equation}
    \end{enumerate}
\end{lemma}

This result will be proved via an alternative formulation of the dual expressions after Lemma \ref{lem:divergence_forms} below.   We note that conditions \eqref{eq:GMI_eq_cond_new} and \eqref{eq:LM_eq_cond_new} are slightly more general than the conditions
\begin{align}
    \log W(y|x) &= b(y) + s \log q(x,y), \label{eq:GMI_eq_cond} \\
    \log W(y|x) &= a(x) + b(y) + s \log q(x,y), \label{eq:LM_eq_cond}
\end{align}
which are easily seen to be sufficient for $\GMI(Q_X) = I(Q_X, W)$ and $\LM(Q_X) = I(Q_X, W)$ respectively: By Propositions \ref{prop:equiv_general}--\ref{prop:equiv_cc}, these conditions imply that mismatched decoding and maximum-likelihood decoding are equivalent when $s > 0$ (whereas if $s = 0$, it is straightforward to show that either of \eqref{eq:GMI_eq_cond}--\eqref{eq:LM_eq_cond} imply $I(Q_X, W) = 0$).  

To give an example of a case where the more general conditions \eqref{eq:GMI_eq_cond_new}--\eqref{eq:LM_eq_cond_new} are required, consider the binary erasure channel (BEC): $\Xc = \{0,1\}$, $\Yc = \{0,e,1\}$, and the channel and metric are described by the matrices \begin{align} 
    \Wv & = \left[\begin{array}{ccc}
    1-\epsilon & \epsilon & 0 \\
    0 & \epsilon & 1-\epsilon \\
    \end{array}\right], \quad
    \qv = \left[\begin{array}{ccc}
    ~~1~\, & ~~1\,~ & ~~0.5~~ \\
    ~~0.5~\, & ~~1\,~ & ~~1~~ \\
    \end{array}\right] \label{eq:BEC}
\end{align}
for some $\epsilon \in (0,1)$.  Due to the entries with $W(y|x) = 0$, there do not exist fixed choices of $s$, $a(\cdot)$, and $b(\cdot)$ such that $W(y|x) = q(x,y)^s e^{a(x)} e^{b(y)}$ for all $(x,y)$.  However, if we raise each entry of $\qv$ to the power of $s > 0$, then in the limit as $s \to \infty$, we recover the standard erasures-only metric $\qtilde(x,y) = \openone\{ W(y|x) > 0 \}$.  Multiplying this metric by $e^{\btilde(y)}$ recovers the maximum-likelihood metric when we let $e^{\btilde(0)} = e^{\btilde(1)} = 1-\epsilon$ and $e^{\btilde(e)} = \epsilon$.  Hence, the condition \eqref{eq:GMI_eq_cond_new} is satisfied (and consequently, so is the more general condition \eqref{eq:LM_eq_cond_new}).


It is also insightful to provide a direct proof that \eqref{eq:GMI_eq_cond}--\eqref{eq:LM_eq_cond} are sufficient; we do this for the GMI, but a similar argument applies to the LM rate.  First note that the upper bound $\GMI(Q_X) \le I(Q_X,W)$ follows immediately from the fact that the choice $\Ptilde_{XY} = P_{XY}$ (with $P_{XY} = Q_X \times W$) is feasible in \eqref{eq:INTR_PrimalGMI}.  Moreover, for any feasible $\Ptilde_{XY}$, if \eqref{eq:GMI_eq_cond} holds then we have
\begin{align}
    &D(\Ptilde_{XY} \| Q_X \times P_Y) \nonumber \\
    &\quad= I_{\Ptilde}(X;Y) + D(\Ptilde_X \| Q_X) \label{eq:primal_matched0} \\
    &\quad\ge I_{\Ptilde}(X;Y) \label{eq:primal_matched0a} \\
    &\quad = H_{\Ptilde}(Y) - H_{\Ptilde}(Y|X) \label{eq:primal_matched1}  \\
    &\quad = H_{P}(Y) + \EE_{\Ptilde}\bigg[ \log\bigg(\frac{\Ptilde_{Y|X}(Y|X)}{W(Y|X)} W(Y|X)\bigg) \bigg] \label{eq:primal_matched2} \\
    &\quad = H_{P}(Y) +  \EE_{\Ptilde}[ \log W(Y|X) ] + D(\Ptilde_{Y|X}\|W|\Ptilde_X) \\
    &\quad \ge H_{P}(Y) +  \EE_{\Ptilde}[ \log W(Y|X) ] \label{eq:primal_matched3} \\
    &\quad \ge H_{P}(Y) +  \EE_{P}[ \log W(Y|X) ] \label{eq:primal_matched4}\\
    &\quad= I_P(X;Y), \label{eq:primal_matched5}
\end{align}
where \eqref{eq:primal_matched0} and \eqref{eq:primal_matched2} use the constraint $\Ptilde_Y = P_Y$, \eqref{eq:primal_matched0a} and \eqref{eq:primal_matched3} follow from the non-negativity of KL divergence, and \eqref{eq:primal_matched4} follows since the metric constraint $\EE_{\Ptilde}[\log q(X,Y)] \ge \EE_P[\log q(X,Y)]$ is equivalent to $\EE_{\Ptilde}[\log W(Y|X)] \ge \EE_P[\log W(Y|X)]$ when \eqref{eq:GMI_eq_cond} holds with $s > 0$ (recall that $\Ptilde_Y = P_Y$).  We conclude that the choice $\Ptilde_{XY} = P_{XY}$ must achieve the minimum in \eqref{eq:INTR_PrimalGMI}.

Next, we provide alternative forms of the dual expressions, which will also allow us to prove Lemma \ref{lem:opt_conds}.  

\begin{lemma} \label{lem:divergence_forms}
    {\em (Alternative formulations of the dual expressions)}
    For any mismatched DMC $(W,q)$, we have the following:
    
    (i) The GMI and LM rate can be written as
    \begin{align}
        \GMI(Q_X,W) &= I(Q_X,W) -   \inf_{\sgz} D(W' \| V'_{s}|P_Y), \label{eq:GMI_div_form} \\
        \LM(Q_X,W) &= I(Q_X,W) -   \inf_{\sgz,a(\cdot)} D(W' \| V'_{s,a}  | P_Y), \label{eq:LM_div_form}
    \end{align}
    where $W'(x|y) = P_{X|Y}(x|y) = \frac{W(y|x)Q_X(x)}{P_Y(y)}$ and $P_Y(y)$ is the marginal of $P_{XY} = Q_X \times W$, and where
    \begin{align}
        V'_{s}(x|y) &= \frac{ Q_X(x) q(x,y)^s }{ \sum_{\xbar} Q_X(\xbar) q(x,y)^s }, \\
        V'_{s,a}(x|y) &= \frac{ Q_X(x) q(x,y)^s e^{a(x)} }{ \sum_{\xbar} Q_X(\xbar) q(\xbar,y)^s e^{a(\xbar)} }.
    \end{align}
    
    (ii) The LM rate can be written as 
    \begin{align}
        \LM(Q_X,W) 
            &= \sup_{\sgz,b(\cdot)}  \sum_{x,y} Q_X(x)W(y|x)\log\frac{ q(x,y)^s e^{b(y)} }{ \sum_{\ybar} P_Y(\ybar) q(x,\ybar)^s e^{b(\ybar)} } \label{eq:LM_b_form} \\
            &= I(Q_X,W) - \inf_{\sgz,b(\cdot)} D(W \| V_{s,b} | Q_X), \label{eq:LM_div_b_form}
    \end{align}
    where
    \begin{equation}
        V_{s,b}(y|x) = \frac{ P_Y(y) q(x,y)^s e^{b(y)} }{ \sum_{\ybar} P_Y(\ybar) q(x,\ybar)^s e^{b(\ybar)} }.
    \end{equation}
\end{lemma}
\begin{proof}
    To prove the first part, we write the LM rate \eqref{eq:INTR_RateLM} as
    \begin{align}
        \LM(Q_X) 
            &= \sup_{\sgz,a(\cdot)} \sum_{x,y} P_Y(y)W'(x|y) \log\bigg(\frac{ V'_{s,a}(x|y) }{ Q_X(x) } \cdot \frac{W'(x|y)}{W'(x|y)}\bigg),
    \end{align}
    where we have used the fact that $P_Y \times W' = Q_X \times W = P_{XY}$, along with the definition of $V'_{s,a}$.  Splitting the logarithm as $\log\big(\frac{ V'_{s,a}(x|y) }{ Q_X(x) } \cdot \frac{W'(x|y)}{W'(x|y)}\big) = \log \frac{ W'(x|y) }{ Q_X(x)} - \log \frac{ W'(x|y) }{ V'_{s,a}(x|y) }$ and re-arranging yields \eqref{eq:LM_div_form}, and the GMI is handled in an identical manner.

    To prove the second part, we define $\phi_a = \EE_Q[a(X)]$ and $\phi_b = \EE_P[b(Y)]$, and proceed as follows:
    \begin{align}
        \LM(Q_X)
            &= \sup_{\sgz,a(\cdot)} \sum_{x,y} Q_X(x)W(y|x)\log q(x,y)^s \nonumber \\
                &\qquad - \sum_{y} P_Y(y) \log \sum_{\xbar} Q_X(\xbar) q(\xbar,y)^s e^{a(\xbar) - \phi_a} \label{eq:LM_rewrite1} \\
            &= \sup_{\sgz,a(\cdot),b(\cdot)} \sum_{x,y} Q_X(x)W(y|x)\log q(x,y)^s \nonumber \\
                &\qquad - \sum_{y} P_Y(y) \log \sum_{\xbar} Q_X(\xbar) q(\xbar,y)^s e^{a(\xbar) - \phi_a} e^{b(y)  - \phi_b} \label{eq:LM_rewrite1a} \\
            &= \sup_{\sgz,a(\cdot),b(\cdot)} \sum_{x,y} Q_X(x)W(y|x)\log q(x,y)^s \nonumber \\
                &\qquad - \log \sum_{\xbar,y} Q_X(\xbar) P_Y(y) q(\xbar,y)^s e^{a(\xbar)  - \phi_a}e^{b(y)  - \phi_b}, \label{eq:LM_rewrite2}
    \end{align}
    where \eqref{eq:LM_rewrite1} follows by re-arranging \eqref{eq:INTR_RateLM}, \eqref{eq:LM_rewrite1a} follows since the factor $e^{b(y)  - \phi_b}$ cancels to zero upon expanding the logarithm, and \eqref{eq:LM_rewrite2} follows by moving the average over $Y$ inside the logarithm using Jensen's inequality, and noting that $b(y)$ can always be chosen to make Jensen's inequality hold with equality (i.e., such that $\sum_{\xbar} Q_X(\xbar) q(\xbar,y)^s e^{a(\xbar) - \phi_a}e^{b(y)-\phi_b}$ does not vary with $y$).
    
    Renaming $(\xbar,y)$ as $(x,\ybar)$ in \eqref{eq:LM_rewrite2} and reversing the steps of \eqref{eq:LM_rewrite1}--\eqref{eq:LM_rewrite2} with the roles of $x$ and $y$ interchanged, we obtain \eqref{eq:LM_b_form}.  Finally, \eqref{eq:LM_div_b_form} follows from \eqref{eq:LM_b_form} by the same argument as the first part.
\end{proof}

The equivalence in \eqref{eq:LM_b_form} can be understood by noting that the primal expression \eqref{eq:INTR_PrimalLM} for the LM rate is symmetric with respect to $X$ and $Y$, so the roles of the two can also be swapped in the dual expression.  In contrast, the GMI does not exhibit such symmetry.

As discussed in Section \ref{sec:interpretations}, the formulation in \eqref{eq:LM_div_b_form} has the natural interpretation that the loss due to the mismatch is the smallest conditional KL divergence $D(W\|V|Q_X)$ among the set of conditional distributions $V(y|x) \in \Pc(\Yc|\Xc)$ that are equivalent to $q(x,y)$ in the sense of Proposition \ref{prop:equiv_cc}.\footnote{Assuming without loss of generality that $P_Y(y) > 0$ for all $y$, we can rename $P_Y(y)e^{b(y)}$ as $e^{\btilde(y)}$ for some $\btilde(y)$.  Moreover, to ensure that $V(y|x)$ is a valid conditional distribution, one must set  $a(x) = -\log\sum_{\ybar} P_Y(\ybar)q(x,\ybar)e^{b(\ybar)}$ when considering \eqref{eq:equiv_a} with $\qtilde(x,y) = V(y|x)$.} For the GMI, the interpretation is slightly less natural, with the loss taking the form $D(W'\|V'_s|P_Y)$ for the ``reverse channel'' $W'(x|y)$.

We conclude by using Lemma \ref{lem:divergence_forms} to prove Lemma \ref{lem:opt_conds}.  We focus on the GMI, but analogous arguments apply for the LM rate.  The idea is to note that $D(W'\|V'_s|P_Y) \ge 0$, with equality if and only if $V'_s(x|y) = W'(x|y)$ for all $y$ such that $P_Y(y) > 0$ \cite[Thm.~2.6.3]{Cov06}.  Suppose first that the infimum in \eqref{eq:GMI_eq_cond_new} is attained by a finite value of $s$.  Writing $V'_s(x|y) = \frac{1}{\Omega(y)} Q_X(x) q(x,y)^s $ with $\Omega(y) = \sum_{\xbar} Q_X(\xbar) q(\xbar,y)^s$, the necessary and sufficient condition $V'_s(x|y) = W'(x|y)$ becomes
\begin{equation}
    W'(x|y) = \frac{1}{\Omega(y)} Q_X(x) q(x,y)^s,
\end{equation}
and substituting $W'(x|y) = \frac{Q_X(x)W(y|x)}{P_Y(y)}$ yields the equivalent condition
\begin{equation}
    W(y|x) = \frac{1}{\Omega(y)} P_Y(y) q(x,y)^s \label{eq:div_cond}
\end{equation}
for all $(x,y)$ with $Q_X(x) > 0$ and $P_Y(y) > 0$.  Defining $b(y)$ to satisfy $e^{b(y)} = \frac{P_Y(y)}{\Omega(y)}$, we find that this condition coincides with \eqref{eq:GMI_eq_cond_new}.  

In the case that the infimum in \eqref{eq:GMI_eq_cond_new} is not attained by a finite value of $s$, the argument is similar but more tedious, so we omit the details.  While the KL divergence is discontinuous in cases where the second argument has a zero transition but the first argument does not, such cases are not relevant here, because we are only considering conditions for vanishing KL divergence.  

\subsection{Ensemble Tightness} \label{sec:ens_tight}

The vast majority of the mismatched decoding literature has focused on achievable rates, whereas upper bounds on $\CM$ have mostly remained elusive (see Section \ref{ch:converse} for an overview).  We have also seen that there are mismatched channels for which the GMI and LM rate are strictly smaller than $\CM$, i.e., the achievable rates are not tight.

Based on these observations, it is natural to ask the following: {\em Is the weakness in the GMI and LM rate due to an inherent weakness in i.i.d.~and constant-composition random coding, or is it due to loose bounds in the mathematical analysis?}  This question is answered by the following {\em ensemble tightness} result \cite{Mer95}, showing that i.i.d.~random coding cannot achieve any rate higher than the GMI, and constant-composition random coding cannot achieve any rate better than the LM rate.  Thus, we cannot hope to obtain improved rates by sharpening the mathematical analysis; we must devise alternative codebook constructions.

In the following, $\pebar = \pebar(n,M)$ denotes the random-coding error probability with block length $n$ and codebook size $M$.

\begin{lemma} \label{lem:ens_tight}
    {\em (Ensemble tightness)} 
     For any mismatched DMC and a given input distribution $Q_X \in \Pc(\Xc)$, we have the following:
     \begin{itemize}
         \item Under i.i.d.~random coding, $\pebar(n,\lfloor e^{nR} \rfloor) \to 1$ as $n \to \infty$ for any $R > \GMI(Q_X)$;
         \item Under constant-composition random coding, $\pebar(n,\lfloor e^{nR} \rfloor) \to 1$ as $n \to \infty$ for any $R > \LM(Q_X)$.
     \end{itemize}
\end{lemma}

The proof is given in Section \ref{sec:su_proofs}.

\subsection{Equivalence of Primal and Dual Expressions}
\label{subsec:primaldual}

Thus far, we have taken for granted that the primal and dual expressions of the GMI \eqref{eq:INTR_PrimalGMI}--\eqref{eq:INTR_RateGMI} are equal, and similarly for the LM rate \eqref{eq:INTR_PrimalLM}--\eqref{eq:INTR_RateLM}.  The following lemma, adapted from \cite{Mer95}, provides the justification for doing so.

\begin{lemma} \label{lem:primal_dual}
    {\em (Primal-dual equivalence)}
    For any mismatched DMC $(W,q)$ and a given input distribution $Q_X \in \Pc(\Xc)$, the right-hand sides of \eqref{eq:INTR_PrimalGMI} and \eqref{eq:INTR_RateGMI} are equal, and similarly for \eqref{eq:INTR_PrimalLM} and \eqref{eq:INTR_RateLM}.
\end{lemma}
\begin{proof}
    We use a direct analysis to show that the primal expression for the LM rate is lower bounded by the dual expression.  The matching upper bound is deduced using Lagrange duality techniques \cite{Boy04}; the details are omitted here, and can be found in \cite[Appendix E]{ScarlettThesis}.  The GMI is handled in a similar (and slightly simpler) manner, so its details are also omitted.
    
    For any $\Ptilde_{XY}$ satisfying the constraints in \eqref{eq:INTR_PrimalLM}, the mutual information can be lower bounded as follows for any $\sgz$ and $a(\cdot)$:
    {\allowdisplaybreaks
    \begin{align}
        &I_{\Ptilde}(X;Y) \nonumber \\
            &~~ = \sum_{x,y} \Ptilde_{XY}(x,y) \log \frac{ \Ptilde_{XY}(x,y) }{ Q_X(x) P_Y(y)} \label{eq:duality1} \\
            &~~ \ge \sum_{x,y} \Ptilde_{XY}(x,y) \log \frac{ \Ptilde_{XY}(x,y) }{ Q_X(x) P_Y(y)} + \sum_{x} a(x)\big( Q_X(x) - \Ptilde_X(x) \big) \nonumber \\
                &\hspace*{3cm} + s \sum_{x,y} \log q(x,y) \big( P_{XY}(x,y) - \Ptilde_{XY}(x,y) \big) \label{eq:duality2} \\
            &~~ = \sum_{x,y} \Ptilde_{XY}(x,y) \log \frac{ \Ptilde_{XY}(x,y) }{ Q_X(x) P_Y(y) q(x,y)^s e^{a(x)} } + \sum_{x} Q_X(x) a(x) \nonumber \\
            &\hspace*{3cm} + s \sum_{x,y} P_{XY}(x,y)\log q(x,y)  \label{eq:duality3} \\
            &~~ \ge \sum_{y} \Ptilde_{Y}(y) \log \frac{ \Ptilde_Y(y) }{ \sum_{\xbar} Q_X(\xbar) P_Y(y) q(\xbar,y)^s e^{a(\xbar)} } + \sum_{x} Q_X(x) a(x) \nonumber \\
            &\hspace*{3cm} + s \sum_{x,y} P_{XY}(x,y)\log q(x,y) \label{eq:duality4} \\
            &= \sum_{x,y} P_{XY}(x,y) \log \frac{ q(x,y)^s e^{a(x)} }{ \sum_{\xbar} Q_X(\xbar) q(\xbar,y)^s e^{a(\xbar)} }, \label{eq:duality5}
    \end{align}}
    where \eqref{eq:duality1} follows from the marginal constraints $\Ptilde_X=Q_X$ and $\Ptilde_Y = P_Y$, \eqref{eq:duality2} follows from the $X$-marginal constraint and the metric constraint $\EE_{\Ptilde}[\log q(X,Y)] \ge \EE_{P}[\log q(X,Y)]$ with $P_{XY} = Q_X \times W$, \eqref{eq:duality3} follows from simple re-arranging, \eqref{eq:duality4} follows from Jensen's inequality in the form of the log-sum inequality $\sum_{i=1}^{n} \alpha_i \log\frac{\alpha_i}{\beta_i} \ge \big(\sum_{i=1}^{n}\alpha_i\big)\log\frac{\sum_{i=1}^{n}\alpha_i}{\sum_{i=1}^{n}\beta_i}$ \cite[Sec.~2.7]{Cov06}, and \eqref{eq:duality5} follows from the marginal constraints along with simple re-arranging.  Taking the supremum over $\sgz$ and $a(\cdot)$ establishes the desired claim.
\end{proof}

\subsection{Additional Properties: Differences to Mutual Information}

Along with the above properties that the GMI and LM rate satisfy, it is useful to highlight certain properties that, despite holding in the matched setting (i.e., for the mutual information), may fail to hold in the mismatched setting.  The following lemmas, adapted from \cite{Csi95,Gan00,Mer95}, provide three such properties.

\begin{lemma} \label{lem:non_tightness}
    {\em (Non-tightness)} 
    There exist mismatched DMCs for which $\CLM < \CM$ (and consequently $\CGMI < \CM$).
\end{lemma}

The claim regarding the GMI follows directly from that of the LM rate, since $\CGMI \le \CLM$ (see \eqref{eq:GMI_vs_LM}).  There are several known examples showing that the LM rate is not tight:
\begin{itemize}
    \item In Section \ref{sec:mac_numerical}, we will return to the parallel BSC example of Section \ref{sec:parallel}, and see that the mismatch capacity equals the matched capacity, which is strictly higher than the LM rate.  
    \item In Section \ref{sec:mu_zuec}, we will return to the zero-undetected error example of Section \ref{sec:su_zuec}, and see that the optimized LM rate can be improved upon;
    \item In Section \ref{sec:binary_input}, we provide an example of a binary-input DMC for which $\CLM$ is strictly smaller than the mismatch capacity, thus providing a counter-example to a reported converse for the binary-input case \cite{Bal95}.
    \item The zero-error capacity ({\em cf.} Section \ref{sec:app_zero}) is a special case of the mismatch capacity, and it is known that i.i.d.~and constant-composition random coding can be strictly suboptimal in this context \cite{Csi95}.
\end{itemize}
The first and last of these examples lead to analytical proofs of Lemma \ref{lem:non_tightness}, whereas the other two are numerical. 

\begin{lemma} \label{lem:non_concave}
    {\em (Non-concavity in $Q_X$)} 
    There exist mismatched DMCs $(W,q)$ such that the functions $\GMI(Q_X)$ and $\LM(Q_X)$ are non-concave in the input distribution $Q_X$.
\end{lemma}

We omit a formal proof, and instead discuss how this is established:
\begin{itemize}
    \item For the LM rate, the non-concavity can be observed in the example in Figure \ref{fig:ZUEC_Rates}.  In the same example, it was shown {\em analytically} in \cite[Ch.~4]{TelatarThesis} that all three permutations of $Q_X = \big(\frac{1}{2},\frac{1}{2},0\big)$ yield the same LM rate, and that this rate is strictly higher than that of  $Q_X = \big(\frac{1}{3},\frac{1}{3},\frac{1}{3}\big)$.  Due to Jensen's inequality, this would not be possible if the LM rate were concave in $Q_X$.
    \item An alternative proof for the LM rate can be found in \cite{Gan00} based on the mismatch capacity per unit cost.
    \item The non-concavity of the GMI can be inferred from a result on the zero-undetected error capacity given in \cite{Bun14}; we provide the details in Appendix \ref{sec:omitted_proofs}.
\end{itemize}

Finally, again recalling the notation $\GMI(Q_X,W)$ and $\LM(Q_X,W)$ with an explicit dependence on the channel, we have the following.

\begin{lemma} \label{lem:dpi}
    {\em (Lack of data processing inequality)} 
     There exist scenarios in which $\GMI(Q_X,W) < \GMI(Q_X,T \circ W)$ and $\LM(Q_X,W) < \LM(Q_X,T \circ W)$ for some transformation $T \,:\, \Yc \to \Yc$, where $\circ$ denotes the composition operation.
\end{lemma}

This result is in contrast to the mutual information $I(Q_X,W)$ with respect to $(X,Y) \sim Q_X \times W$, for which we know that $I(Q_X,W) \ge I(Q_X,T \circ W)$ for any transformation $T$ by the data processing inequality \cite[Sec.~2.8]{Cov06}.

To establish Lemma \ref{lem:dpi}, consider the following example:
Let $\Xc = \Yc = \{0,1\}$, let $Q_X$ be equiprobable on $\{0,1\}$, let $W$ be the noiseless binary channel $X = Y$, and let $q$ be the maximum-likelihood metric of the binary channel with $Y = 1-X$ (i.e., a deterministic flip of the input).  From the example in Section \ref{sec:binary}, $\GMI(Q_X,W) = \LM(Q_X,W) = 0$.  However, if we choose $T$ to represent a deterministic bit flip, we obtain from the same example that $\GMI(Q_X,T \circ W) = \LM(Q_X,T \circ W) = 1$ bit/use.  

In fact, since we know from Lemma \ref{lem:pos_conds} that $\CM > 0 \iff \CLM > 0 \iff \CGMI>0$, this example demonstrates that the mismatch capacity $\CM$ also does not satisfy the data processing inequality.

\section{Proofs of Achievable Rates} \label{sec:su_proofs}

In this subsection, we prove the achievability of the GMI and LM rate, as stated in Theorems \ref{thm:GMI} and \ref{thm:LM}, as well as their ensemble tightness, as stated in Lemma \ref{lem:ens_tight}.  We will provide primal and dual derivations for both the GMI and LM rate.  While there is significant overlap in the proofs of the two rates, we present them separately so that their differences can be properly understood.

Throughout this subsection, we let $\pebar = \pebar(n,M)$ be the error probability averaged over the random codebook.  By the symmetry of the random construction, the random-coding error probability is identical conditioned on any given value of $m$, so we may assume without loss of generality that $m=1$.  Recalling that ties are broken as errors, the starting point of all four achievability proofs will be the following:
\begin{align}
    \pebar 
        &=  \EE\Bigg[ \PP\bigg[ \bigcup_{j=2}^M \Big\{ q^n(\Xv^{(j)},\Yv) \ge q^n(\Xv^{(1)},\Yv) \Big\} \,\bigg|\, \Xv^{(1)},\Yv\bigg] \Bigg] \label{eq:pebar_exact} \\
        &\le \EE\Big[ \min\big\{ 1, (M-1) \PP\big[ q^n(\Xvbar,\Yv) \ge q^n(\Xv,\Yv) \,|\, \Xv,\Yv \big] \big\} \Big], \label{eq:rcu}
\end{align}
with $(\Xv,\Yv,\Xvbar) \sim P_{\Xv}(\xv)W^n(\yv|\xv)P_{\Xv}(\xvbar)$.  In \eqref{eq:pebar_exact}, we have written the probability as an expectation given $(\Xv^{(1)},\Yv)$ for convenience, and in \eqref{eq:rcu} we have applied the {\em truncated union bound}, i.e., the better out of the union bound and the trivial upper bound of one. Equation \eqref{eq:rcu} is a simple extension of the {\em random-coding union bound} \cite{Pol10a} to mismatched decoding.

In our analysis, we will handle the outer expectation over $(\Xv,\Yv)$ using the law of large numbers, which suffices for establishing the achievable rates.  However, both the primal and dual analyses can be strengthened to obtain {\em error exponents} establishing the exponential rate of decay of $\pebar$; see Section \ref{sec:exponents} for details.

\subsection{Preliminaries on the Method of Types} \label{sec:prop_types}

Here we present some useful properties of types that will be used in deriving the primal expressions.  Since these are standard in the literature, we omit the proofs; further details can be found in \cite[Ch.~2]{Csi11} and \cite{GallagerCC}.

Recall from \eqref{eq:type_class}--\eqref{eq:closest_type} that $Q_{X,n}$ is a type (i.e., an empirical distribution) approximating $Q_X$, and that $\Tclass(Q_{X,n})$ is the set of all sequences having type $Q_{X,n}$.  We will also need the notion of a {\em joint type class}, which is simply a type class in the case of a product alphabet, say $\Xc \times \Yc$:
\begin{equation}
    \Tclass(\Ptilde_{XY}) = \big\{ (\xv,\yv) \in \Xc^n \times \Yc^n \,:\, \Phat_{\xv\yv} = \Ptilde_{XY} \big\}, \label{eq:joint_type_class}
\end{equation}
where $\Phat_{\xv\yv}(x,y) = \frac{1}{n}\sum_{i=1}^n \openone\{ (x_i,y_i) = (x,y) \}$.

The properties that we will use are as follows:
\begin{itemize}
    \item There are at most $(n+1)^{|\Xc|}$ types on $\Xc$, and at most $(n+1)^{|\Xc|\cdot|\Yc|}$ joint types on $\Xc \times \Yc$.
    \item The constant-composition codeword distribution can be upper bounded in terms of the corresponding i.i.d.~distribution:
    \begin{equation}
    P_{\Xv}(\xv) \le (n+1)^{|\Xc|} Q_{X,n}^n(\xv), \label{eq:cc_to_iid}
    \end{equation}
    under the choice of $P_{\Xv}$ in Definition \ref{def:cc}.  
    \item For a fixed sequence $\yv \in \Tclass(P_Y)$, an independent codeword $\Xvbar \sim Q_X^n$, and a joint type $\Ptilde_{XY}$ with $\Ptilde_Y = P_Y$, it holds for any $\delta > 0$ and sufficiently large $n$ that
    \begin{align}
        e^{-n (D(\Ptilde_{XY} \| Q_X \times P_Y) + \delta)} &\le \PP\big[ (\Xvbar,\yv) \in \Tclass(\Ptilde_{XY}) \big] \nonumber \\ &\qquad\qquad \le e^{-n D(\Ptilde_{XY} \| Q_X \times P_Y)}. \label{eq:type_prop_iid}
    \end{align} 
    \item For a fixed sequence $\yv \in \Tclass(P_Y)$, a constant-composition codeword $\Xvbar \sim P_{\Xv}$, and a joint type $\Ptilde_{XY}$ with $\Ptilde_X = Q_{X,n}$ and $\Ptilde_Y = P_Y$, it holds for any $\delta > 0$ and sufficiently large $n$ that
    \begin{equation}
        e^{-n (I_{\Ptilde}(X;Y) + \delta)} \le \PP\big[ (\Xvbar,\yv) \in \Tclass(\Ptilde_{XY}) \big] \le e^{-n (I_{\Ptilde}(X;Y)-\delta)}. \label{eq:type_prop_cc}
    \end{equation} 
    \item For $\Xv$ drawn from either the i.i.d.~or constant-composition codeword distribution with input distribution $Q_X$, and $\Yv$ conditionally drawn from $W^n(\,\cdot\,|\xv)$, the joint type $P_{XY}$ of $(\Xv,\Yv)$ satisfies 
    \begin{equation}
        \|P_{XY} - Q_X \times W\|_{\infty} \le \delta \label{eq:LLN}
    \end{equation}
    with probability approaching one, for arbitrarily small $\delta > 0$.
\end{itemize}
The last of these follows directly from the law of large numbers; the rest are non-trivial but still straightforward to derive \cite[Ch.~2]{Csi11}, \cite{GallagerCC}.

\subsection{Primal Derivation of the GMI} \label{sec:pf_GMI_primal}

The key idea of the primal analysis is to observe that if $(\xv,\yv) \in \Tclass(P_{XY})$ and $(\xvbar,\yv) \in \Tclass(\Ptilde_{XY})$, then the condition $q^n(\xvbar,\yv) \ge q^n(\xv,\yv)$ in \eqref{eq:rcu} is equivalent to $\EE_{\Ptilde}[\log q(X,Y)] \ge \EE_{P}[\log q(X,Y)]$.  This follows immediately from the definition of a joint type following \eqref{eq:joint_type_class}.  Hence, for fixed $(\xv,\yv) \in \Tclass(P_{XY})$, we have
\begin{align}
    &\PP\big[ q^n(\Xvbar,\yv) \ge q^n(\xv,\yv) \big] \nonumber \\
    &\qquad = \sum_{\substack{\Ptilde_{XY} \in \Pc_n(\Xc\times\Yc) \,:\, \Ptilde_Y = P_Y, \\ \EE_{\Ptilde}[\log q(X,Y)] \ge \EE_{P}[\log q(X,Y)]}} \PP\big[ (\Xvbar,\yv) \in \Tclass(\Ptilde_{XY}) \big] \label{eq:pf_GMI_primal2} \\
    &\qquad \le \sum_{\substack{\Ptilde_{XY} \in \Pc_n(\Xc\times\Yc) \,:\, \Ptilde_Y = P_Y, \\ \EE_{\Ptilde}[\log q(X,Y)] \ge \EE_{P}[\log q(X,Y)]}} e^{-nD(\Ptilde_{XY} \| Q_X \times P_Y)} \label{eq:pf_GMI_primal3} \\
    &\qquad \le (n+1)^{|\Xc| \cdot |\Yc|} \max_{\substack{\Ptilde_{XY} \in \Pc_n(\Xc\times\Yc) \,:\, \Ptilde_Y = P_Y, \\ \EE_{\Ptilde}[\log q(X,Y)] \ge \EE_{P}[\log q(X,Y)]}} e^{-nD(\Ptilde_{XY} \| Q_X \times P_Y)} \label{eq:pf_GMI_primal4} \\
    &\qquad = (n+1)^{|\Xc| \cdot |\Yc|} \exp\bigg(-n  \min_{\substack{\Ptilde_{XY} \in \Pc_n(\Xc\times\Yc) \,:\, \Ptilde_Y = P_Y, \\ \EE_{\Ptilde}[\log q(X,Y)] \ge \EE_{P}[\log q(X,Y)]}} D(\Ptilde_{XY} \| Q_X \times P_Y) \bigg), \label{eq:pf_GMI_primal5}
\end{align}
where \eqref{eq:pf_GMI_primal3} follows from \eqref{eq:type_prop_iid} and the fact that the joint type of $(\Xvbar,\yv)$ must have $Y$-marginal $P_Y$ (thus constraining $\Ptilde_Y = P_Y$), and \eqref{eq:pf_GMI_primal4} follows since there are at most $(n+1)^{|\Xc| \cdot |\Yc|}$ joint types.

Next, recall from \eqref{eq:LLN} that $\|P_{XY} - Q_X \times W\|_{\infty} \le \delta$ with probability approaching one.  Moreover, the exponent in \eqref{eq:pf_GMI_primal5} equals $\GMI(Q_X)$ in \eqref{eq:INTR_RateGMI} when $P_{XY} = Q_X \times W$.  It can further be shown that this exponent is continuous in the same sense as Lemma \ref{lem:continuity}, using a similar proof as that of the GMI.  Combining these findings, we find that \eqref{eq:pf_GMI_primal5} can be weakened to
\begin{equation}
    \PP\big[ q^n(\Xvbar,\yv) \ge q^n(\xv,\yv) \big] \le e^{-n(\GMI(Q_X \times W) - \delta') } \label{eq:pf_GMI_primal6}
\end{equation}
for arbitrarily small $\delta'$ and sufficiently large $n$.
By upper bounding $\min\{1,\cdot\}$ in \eqref{eq:rcu} by one whenever  $\|P_{XY} - Q_X \times W\|_{\infty} > \delta$, and by \eqref{eq:pf_GMI_primal6} otherwise, we deduce that the error probability vanishes for any rate $R = \frac{1}{n}\log M$ arbitrarily close to the GMI, as desired.

\subsection{Primal Derivation of the LM Rate} \label{sec:pf_LM_primal}

The core part of the argument is very similar to that of the GMI, with the main difference being that we can introduce the constraint $\Ptilde_X = P_X$ by the fact that all codewords have the same composition.  Specifically, for $\Xvbar \sim P_{\Xv}$ following the constant-composition codeword distribution in Definition \ref{def:cc}, and for fixed $(\xv,\yv) \in \Tclass(P_{XY})$, we have
\begin{align}
    &\PP\big[ q^n(\Xvbar,\yv) \ge q^n(\xv,\yv) \big] \nonumber \\
    &\qquad = \sum_{\substack{\Ptilde_{XY} \in \Pc_n(\Xc\times\Yc) \,:\, \\ \EE_{\Ptilde}[\log q(X,Y)] \ge \EE_{P}[\log q(X,Y)]}} \PP\big[ (\Xvbar,\yv) \in \Tclass(\Ptilde_{XY}) \big] \label{eq:pf_LM_primal2} \\
    &\qquad \le \sum_{\substack{\Ptilde_{XY} \in \Pc_n(\Xc\times\Yc) \,:\, \Ptilde_X = P_X, \Ptilde_Y = P_Y, \\ \EE_{\Ptilde}[\log q(X,Y)] \ge \EE_{P}[\log q(X,Y)]}} e^{-n I_{\Ptilde}(X;Y)} \label{eq:pf_LM_primal3} \\
    &\qquad \le (n+1)^{|\Xc| \cdot |\Yc|} \max_{\substack{\Ptilde_{XY} \in \Pc_n(\Xc\times\Yc) \,:\, \Ptilde_X = P_X, \Ptilde_Y = P_Y, \\ \EE_{\Ptilde}[\log q(X,Y)] \ge \EE_{P}[\log q(X,Y)]}} e^{-n I_{\Ptilde}(X;Y)} \label{eq:pf_LM_primal4} \\
    &\qquad\le (n+1)^{|\Xc|\cdot|\Yc|} e^{-n \LM(P_{XY})}, \label{eq:pf_LM_primal5}
\end{align}
where \eqref{eq:pf_LM_primal3} follows from \eqref{eq:type_prop_cc} and the fact that only joint types $\Ptilde_{XY}$ satisfying both $\Ptilde_X = P_X$ and $\Ptilde_Y = P_Y$ have positive probability, \eqref{eq:pf_LM_primal4} follows since there are at most $(n+1)^{|\Xc| \cdot |\Yc|}$ joint types, and \eqref{eq:pf_LM_primal5} follows by upper bounding the maximum over joint types $\Pc_n(\Xc \times \Yc)$ by that over all joint distributions $\Pc(\Xc \times \Yc)$, and defining $\LM(P_{XY})$ to be the LM rate in \eqref{eq:INTR_RateLM} with input distribution $P_X$ and channel $P_{Y|X}$.

The desired result now follows by substituting \eqref{eq:pf_LM_primal5} into \eqref{eq:rcu} and using the continuity of the LM rate ({\em cf.}, Lemma \ref{lem:continuity}).

\subsection{Dual Derivation of the GMI} \label{sec:pf_GMI_dual}

The dual analysis bears some resemblance to that of Gallager \cite[Ch.~5]{Gal68} for maximum-likelihood decoding.  Raising both of the $q^n$ terms to the power of $s > 0$ in \eqref{eq:rcu} clearly leaves the probability unchanged, and upon applying Markov's inequality, we obtain\footnote{We can also consider $s = 0$, since it makes the right-hand side of \eqref{eq:rcu_s} equal one.}
\begin{align}
    \pebar 
    &\le  \EE\Bigg[ \min\bigg\{ 1, M \EE\bigg[ \bigg(\frac{q^n(\Xvbar,\Yv)}{q^n(\Xv,\Yv)}\bigg)^s \,\bigg|\, \Xv,\Yv \bigg] \Bigg\} \bigg] \label{eq:rcu_s} \\
    &\le \PP\Bigg[ \log M + \log \EE\bigg[ \bigg(\frac{q^n(\Xvbar,\Yv)}{q^n(\Xv,\Yv)}\bigg)^s \,\bigg|\, \Xv,\Yv \bigg] \ge \log \delta \Bigg] + \delta, \label{eq:rcu_s_weakened}
\end{align}
where in \eqref{eq:rcu_s} we also used $M-1 \le M$, and \eqref{eq:rcu_s_weakened} holds for any $\delta > 0$ by upper bounding $\min\{1,z\}$ by one when $z > \delta$, and by $\delta$ otherwise.

To simplify the notation, we write \eqref{eq:rcu_s_weakened} as
\begin{equation}
    \pebar \le \PP\big[ i_s^n(\Xv,\Yv) \le \log M - \log \delta \big] + \delta, \label{eq:pe_is}
\end{equation}
where
\begin{equation}
    i_{s}^n(\xv,\yv) = -\log \EE\bigg[ \bigg(\frac{q^n(\Xvbar,\Yv)}{q^n(\Xv,\Yv)}\bigg)^s \,\bigg|\, \Xv,\Yv \bigg].
\end{equation}
We proceed by showing that $i_s^n(\Xv,\Yv)$ can be expressed as a sum of i.i.d.~random variables with mean equaling the objective function in the dual form of the GMI.  We have
\begin{align}
    i_{s}^n(\xv,\yv) 
    & = -\log\sum_{\xvbar} Q_X^n(\xvbar) \bigg(\frac{q^n(\xvbar,\yv)}{q^n(\xv,\yv)}\bigg)^s \label{eq:ChernoffGMI0}\\
    & = -\log\sum_{\xvbar} \prod_{i=1}^n \bigg( Q_X(\xbar_i) \bigg(\frac{q(\xbar_i,y_i)}{q(x_i,y_i)}\bigg)^s \bigg) \\
    & = - \sum_{i=1}^n  \log \sum_{\xbar} Q_X(\xbar) \bigg( \frac{q(\xbar,y_i)}{q(x_i,y_i)} \bigg)^s, \label{eq:ChernoffGMI}
\end{align}
where \eqref{eq:ChernoffGMI} follows by writing $\sum_{\xv} = \sum_{x_1}\dotsc\sum_{x_n}$ and then distributing the sums.  The pair $(\Xv,\Yv)$ is i.i.d.~on $Q_X \times W$, and averaging any summand of \eqref{eq:ChernoffGMI} over $(X_i,Y_i) \sim Q_X \times W$ yields the objective in the dual form of the GMI ({\em cf.}, \eqref{eq:INTR_RateGMI}), as claimed.

By the law of large numbers and the fact that $s$ may be chosen arbitrarily, we conclude from \eqref{eq:pe_is} that if the rate $R = \frac{1}{n} \log M$ is smaller than $\GMI(Q_X) - \delta$, then the upper bound \eqref{eq:rcu_s_weakened} can be upper bounded by $2\delta$ when $n$ is large enough.  Taking $\delta$ arbitrarily small completes the proof.

\subsection{Dual Derivation of the LM Rate} \label{sec:pf_LM_dual}

In the dual derivation of the GMI, the first two steps leading to \eqref{eq:rcu_s_weakened} were not specific to any particular codeword distribution, so they remain valid here with the distribution $(\Xv,\Yv,\Xvbar) \sim P_{\Xv}(\xv)W^n(\yv|\xv)P_{\Xv}(\xvbar)$ adjusted accordingly.  In addition, when $P_{\Xv}$ is the constant-composition codeword distribution, we can further weaken \eqref{eq:rcu_s_weakened} as follows:
\begin{align}
    \pebar 
        &\le \PP\Bigg[ \log M + \log \EE\bigg[ \bigg(\frac{q^n(\Xvbar,\Yv)}{q^n(\Xv,\Yv)}\bigg)^s \frac{e^{a^n(\xvbar)}}{e^{a^n(\xv)}} \,\bigg|\, \Xv,\Yv \bigg] \ge \log\delta \Bigg] + \delta \label{eq:rcu_sa_weakened1} \\
        &\le \PP\Bigg[ \log M + |\Xc|\log (n+1) \nonumber \\
            &\qquad + \log \EE\bigg[ \bigg(\frac{q^n(\Xvbar',\Yv)}{q^n(\Xv,\Yv)}\bigg)^s \frac{e^{a^n(\xvbar)}}{e^{a^n(\xv)}} \,\bigg|\, \Xv,\Yv \bigg] \ge \log\delta \Bigg] + \delta, \label{eq:rcu_sa_weakened2}
\end{align}
where \eqref{eq:rcu_sa_weakened1} holds with $a^n(\xv) = \sum_{i=1}^n a(x_i)$ for any function $a(\cdot)$ since $\Xv$ and $\Xvbar$ have the same composition, and \eqref{eq:rcu_sa_weakened2} holds with $\Xvbar' \sim Q_{X,n}^n$ by upper bounding $P_{\Xv}$ according to \eqref{eq:cc_to_iid}.

Now, similarly to \eqref{eq:pe_is}, we can write \eqref{eq:rcu_sa_weakened2} as
\begin{equation}
    \pebar \le \PP\big[ i_s^n(\Xv,\Yv) \le \log M + |\Xc| \log(n+1) - \log \delta \big] + \delta, \label{eq:pe_isa}
\end{equation}
where
\begin{equation}
    i_{s,a}^n(\xv,\yv) = -\log \EE\bigg[ \bigg(\frac{q^n(\Xvbar',\yv)}{q^n(\xv,\yv)}\bigg)^s \frac{e^{a^n(\Xvbar')}}{e^{a^n(\xv)}}\bigg].
\end{equation}
Since the expectation is taken with respect to an i.i.d.~distribution for $\Xvbar'$, we can expand it in the same way as \eqref{eq:ChernoffGMI0}--\eqref{eq:ChernoffGMI} to obtain
\begin{align}
    i_{s,a}^n(\xv,\yv) = \sum_{i=1}^n  -\log \sum_{\xbar} Q_{X,n}(\xbar) \bigg( \frac{q(\xbar,y_i)}{q(x_i,y_i)} \bigg)^s \frac{e^{a(\xbar)}}{e^{a(x_i)}}. \label{eq:ChernoffLM}
\end{align}
Observe that for any fixed $\xv \in \Tclass(Q_{X,n})$, if we replace $\yv$ by $\Yv \sim W^n(\,\cdot\,|\xv)$ in \eqref{eq:ChernoffLM}, then we get a sum of independent and {\em non-identically distributed} random variables, $nQ_{X,n}(x)$ of which are distributed as $-\log\sum_{\xbar} Q_{X,n} \big( \frac{q(\xbar,Y)}{q(x,Y)} \big)\frac{e^{a(\xbar)}}{e^{a(x)}}$ with $Y \sim W(\cdot|x)$.  The law of large numbers again applies in this case, and the corresponding normalized mean $\frac{1}{n} \EE[i_{s,a}^n(\Xv,\Yv)]$ is given by
\begin{equation}
    \mu_n(s,a) \triangleq -\sum_{x,y} Q_{X,n}(x) W(y|x) \log \sum_{\xbar} Q_{X,n}(\xbar) \bigg( \frac{q(\xbar,y)}{q(x,y)} \bigg)^s \frac{e^{a(\xbar)}}{e^{a(x)}}. \label{eq:mu_sa}
\end{equation}
Using \eqref{eq:pe_isa} and noting that $\frac{|\Xc| \log n - \log \delta}{n} \to 0$, we find that when $R \le \mu_n(s,a) - \delta$, the random-coding error probability vanishes.  The achievability of $\LM(Q_X)$ now follows by recalling that $Q_{X,n}(x) \to Q_X(x)$ by definition, and taking the supremum over $s \ge 0$ and $a(\cdot)$.

\subsection{Ensemble Tightness} \label{sec:pf_ens_tight}

The ensemble tightness of the GMI and LM rate ({\em cf.}, Lemma \ref{lem:ens_tight}) are proved by following the primal achievability proofs, while replacing each upper bounding step by an analogous lower bound.  We give the details for the LM rate (which turns out to be more difficult), and then only briefly comment on the GMI.

Since we have assumed that ties are broken as errors, the {\em exact} random-coding error probability in \eqref{eq:pebar_exact} can be written as
\begin{equation}
    \pebar = 1 - \EE\bigg[ \Big(1 - \PP\big[ q^n(\Xvbar,\Yv) \ge q^n(\Xv,\Yv) \,\big|\, \Xv,\Yv \big] \Big)^{M-1} \bigg] \label{eq:pe_exact}
\end{equation}
with $(\Xv,\Yv,\Xvbar) \sim P_{\Xv}(\xv)W^n(\yv|\xv)P_{\Xv}(\xvbar)$, due to the independence of the codewords.

Since $\big( 1 - \frac{1}{\alpha_n}\big)^{\beta_n} \to 0$ whenever $\alpha_n \ge 1$ and $\lim_{n \to \infty} \frac{\alpha_n}{\beta_n} = 0$, we deduce from \eqref{eq:pe_exact} that $\pebar \to 1$ whenever the following holds for all $(\xv,\yv)$ within some ``typical set'' (i.e., a set with probability tending to one):
\begin{equation}
    \PP\big[ q^n(\Xvbar,\yv) \ge q^n(\xv,\yv) \big] \ge e^{-n(R-\delta)}, \label{eq:ens_tight_cond}
\end{equation}
where $R = \frac{1}{n} \log M$, and $\delta > 0$ is arbitrarily small.

To establish \eqref{eq:ens_tight_cond}, we analyze the left-hand side in terms of joint types, similarly to the primal achievability proof.  Letting $P_{XY} \in \Pc_n(\Xc \times \Yc)$ denote the joint type of $(\xv,\yv)$, we have
\begin{align}
    &\PP\big[ q^n(\Xvbar,\yv) \ge q^n(\xv,\yv) \big] \\
    &\qquad = \sum_{\substack{\Ptilde_{XY} \in \Pc_n(\Xc\times\Yc) \,:\, \\ \EE_{\Ptilde}[\log q(X,Y)] \ge \EE_{P}[\log q(X,Y)]}} \PP\big[ (\Xvbar,\yv) \in \Tclass(\Ptilde_{XY}) \big] \label{eq:pf_LM_tight2} \\
    &\qquad \ge \sum_{\substack{\Ptilde_{XY} \in \Pc_n(\Xc\times\Yc) \,:\, \Ptilde_X = P_X, \Ptilde_Y = P_Y, \\ \EE_{\Ptilde}[\log q(X,Y)] \ge \EE_{P}[\log q(X,Y)]}} e^{-n (I_{\Ptilde}(X;Y) + \delta)} \label{eq:pf_LM_tight3} \\
    &\qquad \ge \max_{\substack{\Ptilde_{XY} \in \Pc_n(\Xc\times\Yc) \,:\, \Ptilde_X = P_X, \Ptilde_Y = P_Y, \\ \EE_{\Ptilde}[\log q(X,Y)] \ge \EE_{P}[\log q(X,Y)]}} e^{-n (I_{\Ptilde}(X;Y) + \delta)}, \label{eq:pf_LM_tight4}
\end{align}
where \eqref{eq:pf_LM_tight3} follows from \eqref{eq:type_prop_cc} and the fact that only joint types $\Ptilde_{XY}$ satisfying both $\Ptilde_X = P_X$ and $\Ptilde_Y = P_Y$ have positive probability, and \eqref{eq:pf_LM_tight4} follows by lower bounding the summation by the maximum.

We would now like to show that we can lower bound \eqref{eq:pf_LM_tight4} in terms of a similar maximization over {\em all} joint distributions, not only joint types.  Intuitively, the two should be essentially equivalent due to the fact that all joint distributions are increasingly close to the nearest joint type as $n \to \infty$.   However, due to the constraints in \eqref{eq:pf_LM_tight4}, the proof is surprisingly tricky in general.  We provide the formal statement as follows, and defer the proof to Appendix \ref{sec:continuity_app}.

\begin{lemma} \label{lem:lb_technical}
    {\em (Passing from types to general distributions)}
    For any decoding metric $q(x,y)$ and joint type $P_{XY} \in \Pc_n(\Xc \times \Yc)$, it holds that
    \begin{align}
        &\min_{\substack{\Ptilde_{XY} \in \Pc_n(\Xc\times\Yc) \,:\, \Ptilde_X = P_X, \Ptilde_Y = P_Y, \\ \EE_{\Ptilde}[\log q(X,Y)] \ge \EE_{P}[\log q(X,Y)]}} I_{\Ptilde}(X;Y) \nonumber \\ & \qquad \le \min_{\substack{\Ptilde_{XY} \in \Pc(\Xc\times\Yc) \,:\, \Ptilde_X = P_X, \Ptilde_Y = P_Y, \\ \EE_{\Ptilde}[\log q(X,Y)] \ge \EE_{P}[\log q(X,Y)]}} I_{\Ptilde}(X;Y) + \delta \label{eq:lb_technical}
    \end{align}
    for arbitrarily small $\delta > 0$ and sufficiently large $n$.
\end{lemma}

By Lemma \ref{lem:lb_technical} and \eqref{eq:pf_LM_tight4}, we have
\begin{equation}
    \PP\big[ q^n(\Xvbar,\yv) \ge q^n(\xv,\yv) \big] \ge e^{-n(\LM(P_{XY}) + 2\delta ) },
\end{equation}
where $\LM(P_{XY})$ denotes the LM rate \eqref{eq:INTR_PrimalLM} with input distribution $P_X$ and channel $P_{Y|X}$.  As a result, recalling the condition \eqref{eq:ens_tight_cond} and the fact that $\|P_{XY} - Q_{X} \times W\|_{\infty} \le \delta$ with probability approaching one, we conclude that $\pebar \to 1$ whenever
\begin{equation}
    R \ge \max_{P_{XY} \,:\, \|P_{XY} - Q_{X} \times W\|_{\infty} \le \delta} \LM(P_{XY}) + 3\delta.
\end{equation}
Since the LM rate is continuous in $P_{XY}$ by Lemma \ref{lem:continuity}, and $\delta$ can be arbitrarily small, this establishes the ensemble tightness of $\LM(Q_X)$.

For the GMI, similar steps are applied, including the lower bound in \eqref{eq:type_prop_iid}.  However, the analog of Lemma \ref{lem:lb_technical} turns out to be simpler to prove; see Appendix \ref{sec:continuity_app} for details.

\section{Multi-Letter Improvements} \label{sec:multi_LM}

Since the GMI and the LM rate do not achieve the mismatch capacity in general ({\em cf.}, Lemma \ref{lem:non_tightness}), it is natural to ask how we might achieve improved rates.  The first proposed method for improvement, due to Csisz\'ar and Narayan \cite{Csi95}, was to apply the GMI and LM rate to the {\em product channel}, described by
\begin{align}
    W^2((y_1,y_2)|(x_1,x_2)) &= W(y_1|x_1)W(y_2|x_2) \\
    q^2((x_1,x_2),(y_1,y_2)) &= q(x_1,y_1)q(x_2,y_2).
\end{align}
Indeed, it is easy to see that if the rate $R^{(2)}$ is achievable for $(W^2,q^2)$, then the rate $R = \frac{1}{2} R^{(2)}$ is achievable for $(W,q)$.    This argument corresponds to coding over {\em pairs} of symbols, and perhaps surprisingly, this can lead to strict improvements in the mismatched setting.

More generally, we can code over $k$-tuples of symbols for any fixed integer $k$, and divide the resulting achievable rate for $(W^k,q^k)$ by $k$ to get the achievable rate for $(W,q)$.   By doing so, we obtain the following:
\begin{gather}
    \CGMI^{(k)} = \frac{1}{k} \max_{ Q_{X^k} } \GMI( Q_{X^k}, W^k, q^k ) \\
    \CLM^{(k)} = \frac{1}{k} \max_{ Q_{X^k} } \LM( Q_{X^k}, W^k, q^k ), \label{eq:CLMk}
\end{gather}
where on the right-hand sides we use the GMI and LM rate with an explicit dependence on the channel and metric.  Formally, we have the following.

\begin{lemma}\label{lem: Multi-letter extensions}
    {\em (Multi-letter extensions)}
    For any mismatched DMC $(W,q)$ and any positive integer $k$, the rates $\CGMI^{(k)}$ and $\CLM^{(k)}$ are achievable, and consequently, so are the rates $\CGMI^{(\infty)} \triangleq \sup_{k \in \ZZ_{>0}} \CGMI^{(k)}$ and $\CLM^{(\infty)} \triangleq \sup_{k \in \ZZ_{>0}} \CLM^{(k)}$.  Moreover, there exist mismatched DMCs for which $\CGMI^{(2)} > \CGMI$ and $\CLM^{(2)} > \CLM$.
\end{lemma}

There are at least two cases where the strict improvement stated in the second part has been established:
\begin{itemize}
    \item We will see that the zero-undetected error capacity example of Section \ref{sec:su_zuec} serves as a numerical example when we return to it in Section \ref{sec:mu_zuec}.
    \item An analytical proof of the second part is given in \cite{Csi95} based on the reduction to zero-error capacity ({\em cf.}, Section \ref{sec:app_zero}).
\end{itemize}
Csisz\'ar and Narayan \cite{Csi95} conjectured that $\CLM^{(k)}$ approaches the mismatch capacity as $k \to \infty$ (i.e., $\CLM^{(\infty)} = \CM$), but this conjecture remains open in general.  We revisit this idea in detail in Section \ref{ch:converse}.

 

\chapter{Continuous-Alphabet Memoryless Channels} \label{ch:single_user_cont}

\section{Introduction} \label{sec:cont_intro}

While discrete memoryless channels (DMCs) are convenient to analyze and capture a variety of interesting communication settings, there is substantial motivation to understand channels defined on continuous alphabets, such as the real or complex numbers. 

In the absence of mismatch, coding theorems for continuous-alphabet channels are often obtained by performing an increasingly fine quantization of the inputs and outputs.  However, such an approach is less suitable in the mismatched setting for at least two reasons.  Firstly, quantizing the output amounts to changing the decoder, and it is non-trivial to study the resulting impact on the original mismatched decoder of interest.  Secondly, it is often of interest to characterize the performance of random coding under a specific continuous input distribution (e.g., Gaussian).

An inspection of the dual i.i.d.~analysis in Section \ref{sec:su_proofs}, yielding the GMI, reveals that it directly extends to continuous-alphabet memoryless channels (see also Section \ref{sec:iid_cont} below).  However, we saw that the constant-composition ensemble, defined on finite alphabets, yields the improved LM rate.  One of the main goals of this section is to present a unified random coding technique, known as {\em cost-constrained random coding}, that generalizes the LM rate to the continuous setting, as well as achieving it for DMCs (see Section \ref{sec:cost}).

In Section \ref{sec:su_cont_examples}, we will apply the GMI and LM rate to three continuous-alphabet mismatched channels of interest: The additive white Gaussian noise (AWGN) channel with a mismatched signal level, non-Gaussian additive channels where an AWGN-type coding scheme is employed anyway, and a channel fading scenario with imperfect knowledge of the fading coefficients.

This section is predominantly based on the works of Ganti {\em et al.} \cite{Gan00} and Scarlett {\em et al.} \cite{Sca14c}, and includes examples based on the works of Merhav {\em et al.} \cite{Mer95}, Lapidoth \cite{Lap96a}, and Lapidoth and Shamai \cite{Lap02}.

\section{Problem Setup} \label{sec:cont_setup}

We continue to consider the setup described in Section \ref{sec:setup_general}, but we drop the assumption that $\Xc$ and $\Yc$ are finite. For concreteness, we focus primarily on real alphabets, i.e., $\Xc = \Yc = \RR$; however, the analysis will extend directly to other alphabets such as $\CC$ or $\RR^d$, as well as finite or countably infinite discrete alphabets.

We let $W(y|x)$ be a conditional probability density function, and consider input distributions $Q_X(x)$ in the form of probability densities.  As a result, the decoding metric $q(x,y)$ is also defined on $\RR\times\RR$.  We initially focus on general choices of $(W,q)$, but we will later specialize to the important class of additive noise channels, taking the form $Y = X + Z$ with $Z$ representing random noise.

For most continuous-alphabet channels, the capacity is infinite unless suitable restrictions are placed on the input.  To address this, we assume throughout this section that the transmitted codeword $\xv = (x_1,\dotsc,x_n)$ must satisfy an {\em input constraint} of the form
\begin{equation}
    \frac{1}{n}\sum_{i=1}^n c(x_i) \le \Gamma \label{eq:input_constr}
\end{equation}
for some cost function $c(x)$ and threshold $\Gamma$.  Perhaps the most widely-adopted special case is a power constraint, in which $c(x) = x^2$, and $\Gamma$ is the permitted per-symbol power averaged over the block.  We denote the minimal cost of a single symbol as
\begin{equation}
    \Gammamin = \inf_{x \in \Xc} c(x), 
\end{equation}
and we assume throughout this section that $\Gamma > \Gamma_{\min}$.

\begin{definition} \label{def:CM_cost}
    {\em (Input-constrained mismatch capacity)}
    The input-constrained mismatch capacity $\CM(\Gamma)$ of the pair $(W,q)$ with input constraint $(c,\Gamma)$ and $\Gamma > \Gammamin$ is defined to be the supremum of all achievable rates under the decoding rule \eqref{eq:decoder} for codebooks $\Cc$ satisfying $\frac{1}{n}\sum_{i=1}^n c(x_i) \le \Gamma$ for all  $\xv \in \Cc$.
\end{definition}

Before proceeding, we present the following useful lemma \cite{Gan00}.

\begin{lemma} \label{lem:concave_gamma}
    {\em (Concavity in $\Gamma$)}
    Under the preceding setup, the function $\CM(\Gamma)$ is a concave non-decreasing function of $\Gamma \in (\Gammamin,\infty)$.
\end{lemma}

Our use of this lemma will be in noting that concavity implies continuity, though the stronger concavity property is also of interest in its own right.  The non-decreasing property is immediate from Definition \ref{def:CM_cost}, and concavity follows from a standard concatenation argument \cite{Gan00}: Choose the first $\lambda n$ symbols from a codebook achieving $\CM(\Gamma_1)$, and the last $1-\lambda n$ symbols from a codebook achieving  $\CM(\Gamma_2)$.  The total cost is $\lambda\Gamma_1 + (1-\lambda) \Gamma_2$, and the rate $\lambda \CM(\Gamma_1) + (1-\lambda)\CM(\Gamma_2)$ is achieved.  Recalling that we are considering decoding metrics of the form $q^n(\xv,\yv) = \prod_{i=1}^n q(x_i,y_i)$, we can factorize $q^n$ as the product of the metrics for the first $\lambda n$ and last $(1-\lambda) n$ symbols.  Hence, the product nature of the codebook implies that the overall mismatched decoder is equivalent to separately decoding the two blocks, and the overall error probability is upper bounded by the sum of the two blocks' error probabilities.

\section{i.i.d.~Random Coding and the GMI} \label{sec:iid_cont}

Similarly to the discrete setting, we fix a density function $Q_X$ and consider i.i.d.~random coding of the form 
\begin{equation}
    P_{\Xv}(\xv) = \prod_{i=1}^n Q_X(x_i).
\end{equation}
In this case, the dual analysis of Section \ref{sec:pf_GMI_dual} applies without change upon replacing the summations by integrals, and we recover a continuous-alphabet counterpart of the GMI \cite{Kap93}, stated in the following.  

\begin{theorem} \label{thm:GMI_cont}
    {\em (GMI for continuous-alphabet channels)}
    For any mismatched memoryless channel $(W,q)$, under i.i.d.~random coding with input distribution $Q_X$, the random coding error probability vanishes for any fixed rate $R <  \GMI(Q_X)$, where
    \begin{equation}
        \GMI(Q_X)=\sup_{\sgz}\EE\Bigg[\log\frac{q(X,Y)^{s}}{\EE[q(\Xbar,Y)^{s}\,|\, Y]}\Bigg], \label{eq:GMI_cont}
    \end{equation}
    with $(X,Y,\Xbar) \sim Q_X(x)W(y|x)Q_X(\xbar)$. 
\end{theorem}

This result can also be inferred from a more general analysis given in Section \ref{sec:su_cont_proofs} below.

Theorem \ref{thm:GMI_cont} does not {\em immediately} provide a lower bound on the input-constrained mismatch capacity in the sense of Definition \ref{def:CM_cost}, as some codewords may violate the input constraint.  However, we can form a feasible codebook using a standard expurgation argument similar to that of Remark \ref{rem:max_avg}: As long as $\EE_{Q}[c(X)] < \Gamma$, any given codeword will indeed be feasible with probability approaching one (e.g., by the law of large numbers).  Hence, for arbitrarily small $\epsilon > 0$, the average number of infeasible codewords is at most $\epsilon M$ for sufficiently large $n$, and Markov's inequality implies that the probability of having $\frac{M}{2}$ or more infeasible codewords is at most $2\epsilon$.  In the high-probability case that this event does not occur, the sub-codebook with only the feasible half (or more) of the codewords achieves vanishing error probability with rate approaching $\GMI(Q_X)$.  Hence, defining the optimized GMI as
\begin{equation}
    \CGMI(\Gamma) \triangleq \sup_{Q_X \,:\, \EE_{Q}[c(X)] \le \Gamma} \GMI(Q_X),  \label{eq:opt_bound_GMI}
\end{equation}
we have for any $\Gamma > \Gammamin$ that
\begin{equation}
    \CGMI(\Gamma) \le \CM(\Gamma). \label{eq:CM_bound_GMI}
\end{equation}
The case $\EE_{Q}[c(X)] = \Gamma$ is justified by the concavity (and hence continuity) of $\CM(\Gamma)$ in $\Gamma$, stated in Lemma \ref{lem:concave_gamma}.

\section{Cost-Constrained Random Coding and the LM Rate} \label{sec:cost}

Based on the findings of Section \ref{ch:single_user}, one would hope to be able to achieve a continuous variation of the LM rate ({\em cf.}, Theorem \ref{thm:LM}) in order to improve on the GMI.  In this subsection, we present a codeword distribution constructed precisely for this purpose \cite{Gan00}.  Specifically, we fix an input distribution $Q_X$ and real-valued function $a(\cdot)$, and consider
\begin{equation}
    P_{\Xv}(\xv) = \frac{1}{\Omega_n} \prod_{i=1}^{n}Q_X(x_{i})\openone\big\{\xv\in\Dc_{n}\big\},\label{eq:PX_cost}
\end{equation}
where
\begin{gather}
    \Dc_{n}\triangleq\bigg\{\xv\,:\,\bigg|\frac{1}{n}\sum_{i=1}^{n}c(x_{i})-\phi_{c}\bigg| \le \delta, ~ \bigg|\frac{1}{n}\sum_{i=1}^{n}a(x_{i})-\phi_{a}\bigg| \le \delta \bigg\},  \label{eq:SU_SetDn} \\
    \phi_c \triangleq \EE_Q[c(X)], \quad \phi_a \triangleq \EE_Q[a(X)],  \label{eq:SU_SetDn_defs}
\end{gather}
and where $\delta > 0$ is a parameter, and $\Omega_n$ is a normalizing constant.  Hence, $P_{\Xv}$ follows an i.i.d.~distribution conditioned on both $c(\cdot)$ and $a(\cdot)$ having an empirical mean close to the true mean.

The intuition behind this choice of $P_{\Xv}$ is as follows:
\begin{itemize}
    \item If $\EE_Q[c(X)] < \Gamma$, where $\Gamma$ is the threshold in \eqref{eq:input_constr}, then the first constraint in \eqref{eq:SU_SetDn} with sufficiently small $\delta$ ensures that all codewords satisfy the input constraint \eqref{eq:input_constr}.
    \item The dual analysis in Section \ref{sec:su_proofs} for constant-composition random coding used the fact that $\sum_{i=1}^{n}a(x_{i})$ is the same for all codewords and any $a(\cdot)$, in order to deduce the improved LM rate rather than the GMI.  For $P_{\Xv}$ in \eqref{eq:PX_cost}, this property is {\em approximately} true for {\em one particular function} $a(\cdot)$.  The deviation by $\delta$ does not impact the achievable rate in the limit $\delta \to 0$, and the fact that this property holds for only one function is also sufficient given that we are free to design that function to our liking.
\end{itemize}
We refer to the ensemble corresponding to the choice of $P_{\Xv}$ in \eqref{eq:PX_cost} as {\em cost-constrained random coding}, but it is important to understand the distinction between $c(\cdot)$ and $a(\cdot)$: The {\em system cost} $c(\cdot)$ corresponds to a constraint imposed by the problem formulation, whereas the {\em auxiliary cost} $a(\cdot)$ is intentionally introduced in order to improve the random coding performance.

While the choice of constraints in \eqref{eq:SU_SetDn} will suffice for our purposes, other variations exist in the literature, such as one-sided system costs \cite[Sec.~7.3]{Gal68} and choices of $\delta$ that decrease with $n$ \cite{Sca14c}.  The choices of $\delta$ could also differ for the two constraints, but this would not affect the results that we present corresponding to $\delta \to 0$.

%

We formally state the achievability of a generalization of the LM rate in the following theorem \cite{Gan00}, which will be proved in Section \ref{sec:su_cont_proofs}.

\begin{theorem} \label{thm:LM_cont}
    {\em (LM rate for continuous-alphabet channels)}
    For any mismatched memoryless channel $(W,q)$ with input constraint $(c,\Gamma)$, and any input distribution $Q_X$ such that $\EE_{Q}[c(X)] < \Gamma$, the following rate is achievable via cost-constrained random coding with suitably-chosen $a(\cdot)$ and $\delta > 0$:
    \begin{equation}
        \LM(Q_X)=\sup_{\sgz,a(\cdot)}\EE\Bigg[\log\frac{q(X,Y)^{s}e^{a(X)}}{\EE[q(\Xbar,Y)^{s}e^{a(\Xbar)}\,|\, Y]}\Bigg], \label{eq:LM_cont}
    \end{equation}
    where $(X,Y,\Xbar) \sim Q_X(x)W(y|x)Q_X(\xbar)$, and the supremum is over all $a(\cdot)$ such that $\EE_{Q}[a(X)]$ is finite.
\end{theorem}

Analogously to \eqref{eq:CM_bound_GMI}, we deduce the following from Theorem \ref{thm:LM_cont}: Defining the optimized LM rate 
\begin{equation}
    \CLM(\Gamma) \triangleq\sup_{Q_X \,:\, \EE_{Q}[c(X)] \le \Gamma} \LM(Q_X)
\end{equation}
we have for any $\Gamma > \Gammamin$ that
\begin{equation}
    \CLM(\Gamma) \le \CM(\Gamma). \label{eq:CM_bound_LM}
\end{equation}
Here we do not need to expurgate any codewords violating the power constraint, since we have ensured that there is zero probability of such an event happening.

Theorem \ref{thm:LM_cont} comes with an important caveat not present in the constant-composition random coding result of Theorem \ref{thm:LM}.  Under constant-composition random coding, we can introduce an {\em arbitrary} function $a(\cdot)$ in the {\em analysis}, and then optimize the final result to deduce the LM rate.  In contrast, in the present setting, we need to choose a {\em specific} function $a(\cdot)$ in the random-coding {\em construction} itself. 
The optimal function $a(\cdot)$ in \eqref{eq:LM_cont} clearly depends on both the channel $W$ and the decoding metric $q$.  As a result, the cost-constrained approach to achieving the LM rate requires knowing both of these.  In fact, even when they are known, explicitly constructing an optimal choice of $a(\cdot)$ may be difficult, since the supremum in \eqref{eq:LM_cont} corresponds to an infinite-dimensional optimization problem.  In the following subsection, we present a more general random coding ensemble and analysis addressing these considerations.

A related fact is that while the decoding metrics $q(x,y)$ and $q(x,y)^s e^{a(x)} e^{b(y)}$ (with $s > 0$) always yield the same mismatch capacity when $\Xc$ is finite ({\em cf.}, Corollary \ref{cor:equiv_cm}), it is unclear whether this is true for infinite alphabets.  The ability to include $a(x)$ in this claim for finite $\Xc$ is based on the fact that any good code also has a good constant-composition sub-code, and this fact has no obvious extension to the continuous-alphabet case.  However, from \eqref{eq:LM_cont}, one can at least deduce that the metrics $q(x,y)$ and $q(x,y)^s e^{a(x)} e^{b(y)}$ yield the same LM rate.

\section{Cost-Constrained Random Coding with Multiple Auxiliary Costs} \label{sec:cost_multi}

We first consider the question of how cost-constrained random coding performs with a suboptimal choice of the auxiliary cost $a(\cdot)$ (again considering the limit $\delta \to 0$).  A natural guess is that we can achieve the rate \eqref{eq:LM_cont} with the supremum over $a(\cdot)$ replaced by the fixed choice.  Such a rate is indeed achievable, but it turns out that a better rate can be derived by introducing an additional auxiliary parameter, allowing us to replace $a(\cdot)$ by $r a(\cdot)$ for any $r \in \RR$ (see Theorem \ref{thm:LM_fixed} below).


We know that in order to achieve the LM rate, having a single optimized auxiliary cost is enough.  However, when such optimization is not possible, it is also of interest to understand the behavior in the presence of {\em multiple} auxiliary costs \cite{Sca13e,Sca14c}.  Generalizing \eqref{eq:PX_cost}, we fix $Q_X \in \Pc(\Xc)$ and let $L$ denote the number of auxiliary costs, denote the $l$-th one by $a_l(\cdot)$ and its mean by $\phi_l = \EE_Q[a_l(X)]$, and consider the codeword distribution
\begin{equation}
    P_{\Xv}(\xv) = \frac{1}{\Omega_n} \prod_{i=1}^{n}Q_X(x_{i})\openone\big\{\xv\in\Dc_{n}\big\},\label{eq:PX_cost_multi}
\end{equation}
where
\begin{align}
    \Dc_{n}\triangleq\bigg\{\xv\,:\,&\bigg|\frac{1}{n}\sum_{i=1}^{n}c(x_{i})-\phi_{c}\bigg| \le \delta, \nonumber \\
    &\bigg|\frac{1}{n}\sum_{i=1}^{n}a_l(x_{i})-\phi_l\bigg| \le \delta, \forall l=1,\dotsc,L \bigg\}. \label{eq:SU_SetDn_multi}
\end{align}

The following theorem states an achievable rate for random coding under this more general ensemble.

\begin{theorem} \label{thm:LM_fixed}
    {\em (Achievable rate with fixed auxiliary costs)}
    For any mismatched memoryless channel $(W,q)$ with input constraint $(c,\Gamma)$, and any input distribution $Q_X$ such that $\EE_Q[c(X)] < \Gamma$, the following rate is achievable via cost-constrained random coding with fixed auxiliary costs $a_1(\cdot),\dotsc,a_L(\cdot)$: 
    \begin{equation}
        \LM'(Q_X,\{a_l\}_{l=1}^L)=\sup_{\sgz,\{r_l\}_{l=1}^L}\EE\Bigg[\log\frac{q(X,Y)^{s}e^{\sum_{l=1}^L r_l a_l(X)}}{\EE[q(\Xbar,Y)^{s}e^{\sum_{l=1}^L r_l a_l(\Xbar)}\,|\, Y]}\Bigg], \label{eq:LM_cont2}
    \end{equation}
    where $(X,Y,\Xbar) \sim Q_X(x)W(y|x)Q_X(\xbar)$. 
\end{theorem}

The intuition behind the presence of the parameters $\{r_l\}_{l=1}^L$ is that since $a_l(\cdot)$ has an empirical mean within $\delta$ of the true mean, $r_l a_l(\cdot)$ has an empirical mean within $|r_l| \delta$ of the true mean.  Hence, by letting $\delta$ be sufficiently small, we can perform the analysis as if the $l$-th auxiliary cost function were scaled by $r_l$.

In fact, we could also include $rc(\cdot)$ (with $r \in \RR$) inside both of the exponential terms in \eqref{eq:LM_cont2} to potentially improve the rate further.  We omit such terms to avoid cumbersome notation, and since the resulting rate could equivalently be recovered by including an additional auxiliary cost with $a_l(\cdot) = c(\cdot)$.

We note the following relation between the achievable rates: 
\begin{equation}
    \LM(Q_X) \ge \LM'(Q_X,\{a_l\}_{l=1}^L) \ge \GMI(Q_X), \label{eq:relations}
\end{equation}
where the right inequality follows by setting all values of $r_l$ to zero in \eqref{eq:LM_cont2}, and the left inequality follows since given any choices of auxiliary costs and parameters in \eqref{eq:LM_cont2}, we recover the objective in \eqref{eq:LM_cont} upon setting $a(x) = \sum_{l=1}^L r_l a_l(x)$.  A generalization of the right inequality in \eqref{eq:relations} is that adding further auxiliary costs can never reduce the achievable rate.

The preceding results and discussions apply to both discrete and continuous channels.  In the discrete case, it is also worth noting that if one chooses $L = |\Xc|$ and $a_l(x) = \openone\{ x = l \}$ (assuming without loss of generality that $\Xc = \{1,\dotsc,|\Xc|\}$), then the distribution in \eqref{eq:PX_cost_multi} produces codewords whose types are within $\delta$ of $Q_X$ in the entry-wise sense.  Hence, this choice recovers a similar ensemble to the constant-composition ensemble.  However, it is not necessary to increase the number of auxiliary costs as a function of the alphabet size to attain the LM rate, as Theorem \ref{thm:LM_cont} shows that a single auxiliary cost is sufficient.

\section{Proofs of Achievable Rates} \label{sec:su_cont_proofs}

Here we present the proof of the fixed-cost LM rate in Theorem \ref{thm:LM_fixed}, which readily implies the GMI (Theorem \ref{thm:GMI_cont}) by setting $L=0$, and the LM rate (Theorem \ref{thm:LM_cont}) by setting $L=1$ and optimizing $a_1(\cdot)$.  The analysis is similar to the dual constant-composition analysis of Section \ref{sec:su_proofs}, but we provide the details for completeness.

We start with the following key property of the cost-constrained ensemble, which follows directly from the law of large numbers.

\begin{lemma} \label{prop:SU_SubExpCost2}
    {\em (Normalizing constant for cost-constrained coding)} 
    For the cost-constrained codeword distribution in \eqref{eq:PX_cost_multi} with $L$ fixed auxiliary costs, if $\phi_c$ and $\{\phi_l\}_{l=1}^L$ are finite, then $\Omega_n \to 1$ as $n \to \infty$ for any fixed $\delta > 0$.
\end{lemma}


We now consider the bound \eqref{eq:rcu_s_weakened}, which holds for a general codeword distribution, and is repeated here for convenience:
\begin{equation}
    \pebar \le \PP\Bigg[ \log M + \log \EE\bigg[ \bigg(\frac{q^n(\Xvbar,\Yv)}{q^n(\Xv,\Yv)}\bigg)^s \,\bigg|\, \Xv,\Yv \bigg] \ge \log\delta \Bigg] + \delta, \label{eq:rcu_s_weakened2}
\end{equation}
where $(\Xv,\Yv,\Xvbar) \sim P_{\Xv}(\xv)W^n(\yv|\xv)P_{\Xv}(\xvbar)$, and $\delta > 0$ is arbitrary.  To form a convenient upper bound, we define $a_l^n(\xv) \triangleq \sum_{i=1}^n a_l(x_i)$, and observe that the constraint set in \eqref{eq:SU_SetDn_multi} implies that
\begin{equation}
    1 \le \frac{ e^{ \sum_{l=1}^L (r_la_l^n(\xvbar) + |r_l| \delta n) } }{ e^{ \sum_{l=1}^L (r_l a_l^n(\xv) - |r_l| \delta n) } } = \frac{ e^{ \sum_{l=1}^L r_l a_l^n(\xvbar) } }{ e^{ \sum_{l=1}^L r_l a_l^n(\xv) } } \cdot e^{2\delta n \sum_{l=1}^L |r_l|}
\end{equation}
for any parameters $\{r_l\}_{l=1}^L$, and any codewords $\xv$ and $\xvbar$ such that $P_{\Xv}(\xv) > 0$ and $P_{\Xv}(\xvbar) > 0$.  Hence, defining $\rsum = \sum_{l=1}^L |r_l|$ for brevity, we obtain
\begin{equation}
    \log \EE\bigg[ \bigg(\frac{q^n(\Xvbar,\yv)}{q^n(\xv,\yv)}\bigg)^s \bigg] \le \log \EE\bigg[ \bigg(\frac{q^n(\Xvbar,\yv)}{q^n(\xv,\yv)}\bigg)^s \frac{ e^{ \sum_{l=1}^L r_l a_l^n(\Xvbar) } }{ e^{ \sum_{l=1}^L r_l a_l^n(\xv) } } e^{2\delta \rsum n} \bigg]. 
\end{equation}
Substituting into \eqref{eq:rcu_s_weakened2} and using the fact that $P_{\Xv}(\xv) \le \frac{1}{\Omega_n} Q_X^n(\xv)$ (see \eqref{eq:PX_cost_multi}), we obtain
\begin{align}
    \pebar &\le \frac{1}{\Omega_n} \PP\Bigg[ -\log \EE\bigg[ \bigg(\frac{q^n(\Xvbar',\Yv)}{q^n(\Xv',\Yv)}\bigg)^s \frac{ e^{ \sum_{l=1}^L r_l a_l^n(\Xvbar') } }{ e^{ \sum_{l=1}^L r_l a_l^n(\Xv') } } \,\bigg|\, \Xv',\Yv \bigg] \nonumber \\ &\qquad\qquad\qquad\qquad\qquad\qquad\le \log\frac{M}{\delta\Omega_n} + 2\delta\rsum n \Bigg] + \delta, \label{eq:rcu_cost}
\end{align}
where $(\Xv',\Yv,\Xvbar') \sim Q_X^n(\xv)W^n(\yv|\xv)Q_X^n(\xvbar)$ is now a triplet of random vectors with the corresponding $(X'_i,Y'_i,\Xbar_i)$ being i.i.d.~with respect to $i=1,\dotsc,n$.   For notational convenience, we rewrite \eqref{eq:rcu_cost} as
\begin{equation}
        \pebar \le \frac{1}{\Omega_n} \PP\Big[ i_{s,\{r_l\}}(\Xv',\Yv) \le \log M - \log(\delta \Omega_n) + 2\delta\rsum n \Big] + \delta, \label{eq:rcu_cost2}
\end{equation}
where
\begin{equation}
    i_{s,\{r_l\}}(\xv,\yv) \triangleq -\log \EE\bigg[ \bigg(\frac{q^n(\Xvbar',\yv)}{q^n(\xv,\yv)}\bigg)^s \frac{ e^{ \sum_{l=1}^L r_l a_l^n(\Xvbar') } }{ e^{ \sum_{l=1}^L r_l a_l^n(\xv) } } \bigg]. \label{eq:i_sr}
\end{equation}
Since the average in \eqref{eq:i_sr} is taken with respect to $\Xvbar' \sim Q_X^n$, we can expand it in the same way as \eqref{eq:ChernoffGMI0}--\eqref{eq:ChernoffGMI} to obtain
\begin{align}
    i_{s,\{r_l\}}(\xv,\yv) = -\sum_{i=1}^n  \log \sum_{\xbar} Q_{X}(\xbar) \bigg( \frac{q(\xbar,y_i)}{q(x_i,y_i)} \bigg)^s \frac{e^{\sum_{l=1}^L r_l a_l(\xbar)}}{e^{\sum_{l=1}^L r_l a_l(x_i)}}. \label{eq:ChernoffCost}
\end{align}
Hence, when we substitute the pair $(\Xv',\Yv)$, we have a sum of i.i.d.~random variables, each having mean
\begin{align}
    &\mu(s,\{r_l\}_{l=1}^L) \nonumber \\ &= -\sum_{x,y} Q_X(x) W(y|x) \log \sum_{\xbar} Q_{X}(\xbar) \bigg( \frac{q(\xbar,y)}{q(x,y)} \bigg)^s \frac{e^{\sum_{l=1}^L r_l a_l(\xbar)}}{e^{\sum_{l=1}^L r_l a_l(x)}}. \label{eq:mu_cost}
\end{align}
By applying the law of large numbers in \eqref{eq:rcu_cost} and noting that $\Omega_n \to 1$ (see Lemma \ref{prop:SU_SubExpCost2}) and $\delta$ is arbitrarily small, we conclude that $\mu(s,\{r_l\}_{l=1}^L)$ is an achievable rate.  Optimizing over $\sgz$ and $\{r_l\}_{l=1}^L$ establishes Theorem \ref{thm:LM_fixed}.

\section{Ensemble Tightness Results} \label{sec:cont_tightness}

We know from Section \ref{ch:single_user} that in the finite-alphabet setting, the GMI and LM rate are ensemble-tight for the i.i.d.~and constant-composition ensembles.  In addition, when the input constraint is absent or trivial (e.g., $c(x) = 0$ for all $x$, and $\Gamma = 0$), or when one of the auxiliary costs $a_l(\cdot)$ is equal to $c(\cdot)$ (which can be assumed without loss of generality), it can also be shown using the techniques of Section \ref{ch:single_user} that $\LM'(Q_X,\{a_l\}_{l=1}^L)$ in Theorem \ref{thm:LM_fixed} is ensemble-tight for the cost-constrained ensemble when specialized to any mismatched DMC \cite{Sca14c}.

In this subsection, we present an ensemble tightness result for the GMI with continuous alphabets, which holds under mild technical assumptions, and follows from the large-deviations analysis of a distinct but related rate-distortion setup \cite{Dem02}.   Related proofs of ensemble tightness for certain additive noise channels can also be found in \cite{Lap96a,Lap02,Wei04}, and an example from \cite{Lap96a} with a simpler direct argument will be illustrated in the proof of Theorem \ref{thm:gaussian} below.

We will also discuss a {\em plausibility argument} for the ensemble tightness of the LM rate under cost-constrained random coding in the continuous-alphabet setting, but we leave open the problem of establishing this rigorously.

The large deviations result that we utilize \cite[Thm.~1]{Dem02} is stated as follows without proof.  We note that the roles of $X$ and $Y$ are reversed here compared to \cite{Dem02}, and the negative log-metric $- \log q(x,y)$ plays the role of the distortion function in \cite{Dem02}.  This result will be re-stated in Section \ref{sec:rd_mm_random} (see Lemma \ref{lem:large_dev_rd} therein) in the context of rate-distortion theory, using notation that more closely resembles that of \cite{Dem02}. 

\begin{lemma} \label{lem:large_dev}
    {\em (Large deviations result for i.i.d.~random coding)}
    Fix an input distribution $Q_X$, an output distribution $P_Y$, and a decoding metric $q$ satisfying $q(x,y) \in (0,1]$ for all $(x,y)$, and define\footnote{The {\em essential infimum} of a function $g(X)$ with respect to $X\sim Q_X$ is defined to be supremum of $t \in \RR$ for which $\PP_{Q}[ g(X) > t ] = 1$.  The essential supremum is defined analogously. \label{foot:ess_inf} }
    \begin{gather}
        \gamma_{\rm min} = \EE_{P_Y}\big[ {\rm ess\,inf}_{Q_X}\,\{-\log q(X,Y)\} \big], \label{eq:gamma_min}\\
        \gamma_{\rm prod} = \EE_{Q_X\times P_Y}\big[ -\log q(X,Y) \big]. \label{eq:gamma_avg}
    \end{gather}
    Then, if $\gamma_{\rm prod} < \infty$, we have for any $\gamma \in ( \gamma_{\rm min}, \gamma_{\rm prod})$ that the following holds with probability one with respect to $\Yv$:
    \begin{align}
        &\lim_{n \to \infty} - \frac{1}{n} \log \PP\big[ -\log q^n(\Xvbar, \Yv) \le n\gamma \,|\, \Yv \big] \nonumber \\
        & \qquad = \sup_{s \ge 0}\Big\{ -s\gamma - \EE\big[ \log \EE[ q(\Xbar,Y)^s \,|\,Y ] \big] \Big\}, \label{eq:large_dev}
    \end{align}
    where $(\Xvbar,\Yv) \sim Q_X^n(\xvbar)P_Y^n(\yv)$ and $(\Xbar,Y) \sim Q_X(\xbar)P_Y(y)$.
\end{lemma}

Using this result, we can establish the ensemble tightness of the GMI for continuous-alphabet memoryless channels, stated in the following theorem.  We adopt some technical assumptions that follow from those regarding $q$, $\gamma_{\rm min}$, and $\gamma_{\rm prod}$ in Lemma \ref{lem:large_dev}, and will be discussed shortly.

\begin{theorem} \label{thm:tightGMI}
    {\em (Ensemble tightness of the GMI)} Consider a mismatched memoryless channel $(W,q)$ and an input distribution $Q_X$, and let $P_Y$ be the marginal distribution of $P_{XY} = Q_X \times W$.  Then, under the assumptions 
    \begin{gather}
        q(x,y) \in (0,q_{\max}], ~ \forall (x,y), ~~\text{for some }q_{\max} < \infty, \label{eq:tight_as0}  \\
        \EE_{Q_X\times P_Y}[\log q(X,Y)] > -\infty, \label{eq:tight_as1} \\
        \EE_{Q_X\times W}[\log q(X,Y)] < \EE_{P_Y}\big[ {\rm ess\,sup}_{Q_X}\,\{\log q(X,Y)\} \big], \label{eq:tight_as2} \\
        \EE_{Q_X\times W}[\log q(X,Y)] > \EE_{Q_X\times P_Y}[\log q(X,Y)], \label{eq:tight_as3}
    \end{gather}
    we have for any $R > \GMI(Q_X)$ that $\pebar(n,\lfloor e^{nR} \rfloor) \to 1$ as $n \to \infty$ under i.i.d.~random coding with input distribution $Q_X$.
\end{theorem}

Before presenting the proof, we briefly discuss the assumptions.  We observe from \eqref{eq:tight_as0} that $q(x,y) = 0$ is disallowed, but we are not aware of any continuous-alphabet examples that consider decoding metrics taking value zero.  Similarly, 
the upper bound $q(x,y) \le q_{\max}$ states that the metric is bounded, and we are not aware of any counter-examples considered previously.  In particular, for standard distance-based decoding metrics such as $q(x,y) = e^{-(y-x)^2}$ (see Section \ref{sec:su_cont_examples}), \eqref{eq:tight_as0} is satisfied with $q_{\max} = 1$.

The assumption \eqref{eq:tight_as1} rules out scenarios in which incorrect codewords yield a log-decoding metric with a very heavy tail in the negative direction; such situations are not commonly encountered.  Similarly, \eqref{eq:tight_as2} rules out pathological scenarios where the transmitted codeword always has the highest metric possible.  This can occur naturally for certain mismatched DMCs (e.g., see Section \ref{sec:app_zuec}), but we are not aware of any such examples in continuous-alphabet settings.


Finally, we note that the condition \eqref{eq:tight_as3} was already considered in Lemma \ref{lem:pos_conds}, where it was shown that, at least for DMCs, the mismatch capacity is zero when this fails to hold.  While we are not aware of an analogous claim for continuous-alphabet channels, it is shown in \cite[Prop.~1]{Gan00} that the LM rate (and therefore, the GMI) is indeed zero when \eqref{eq:tight_as3} fails.

\begin{proof}[Proof of Theorem \ref{thm:tightGMI}]
    We first argue that we can consider the case that $q_{\max} = 1$ in \eqref{eq:tight_as0} without loss of generality.  Indeed, if $q_{\max} > 1$, then an equivalent decoding metric is attained by dividing all values by $q_{\max}$, and the resulting normalized metric is upper bounded by one.

    We start with the exact expression for the error probability in \eqref{eq:pe_exact}:
    \begin{align}
        \pebar &= 1 - \EE\bigg[ \Big(1 - \PP\big[ q^n(\Xvbar,\Yv) \ge q^n(\Xv,\Yv) \,\big|\, \Xv,\Yv \big] \Big)^{M-1} \bigg] \\
            &= 1 - \EE\bigg[ \Big(1 - \PP\big[ - \log q^n(\Xvbar,\Yv) \le - \log q^n(\Xv,\Yv) \,\big|\, \Xv,\Yv \big] \Big)^{M-1} \bigg],
    \end{align}
    where $(\Xv,\Yv,\Xvbar) \sim Q_X^n(\xv)W^n(\yv|\xv)Q_X^n(\xvbar)$, and $M = \lfloor e^{nR} \rfloor$.  Observe that $-\log q^n(\Xv,\Yv) = \sum_{i=1}^n - \log q(X_i,Y_i)$ is an i.i.d.~summation with mean $n\EE[-\log q(X,Y)]$, where $(X,Y) \sim Q_X \times W$.  Hence, by the law of large numbers, we have for any $\delta > 0$ and sufficiently large $n$ that the following holds with $\gamma = \EE[-\log q(X,Y)] - \delta$:
    \begin{align}
        \pebar \ge 1 - \delta - \EE\bigg[ \Big(1 - \PP\big[ - \log q^n(\Xvbar,\Yv) \le n \gamma \,\big|\, \Yv \big] \Big)^{M-1} \bigg], \label{eq:tightGMI_lb}
    \end{align}
    This inner probability matches that in \eqref{eq:large_dev}, and we proceed by verifying the technical conditions of Lemma \ref{lem:large_dev}.

    Multiplying both sides in \eqref{eq:tight_as1}--\eqref{eq:tight_as3} by $-1$ gives $\gamma_{\rm min} < \EE[-\log q(X,Y)] < \gamma_{\rm prod} < \infty$, where $\gamma_{\rm min}$ and $\gamma_{\rm prod}$ are defined in \eqref{eq:gamma_min}--\eqref{eq:gamma_avg}.  Hence, under the choice $\gamma = \EE[-\log q(X,Y)] - \delta$, we have for sufficiently small $\delta$ that $\gamma \in ( \gamma_{\rm min}, \gamma_{\rm prod})$, and Lemma \ref{lem:large_dev} gives
    \begin{align}
        &\lim_{n \to \infty} - \frac{1}{n} \log \PP\big[ -\log q^n(\Xvbar, \Yv) \le n\gamma\,|\, \Yv \big] \nonumber \\
        &\qquad = \sup_{s \ge 0}\Big( -s\gamma - \EE\big[ \log \EE[ q(\Xbar,Y)^s \,|\,Y ] \big] \Big) \label{eq:gamma_star_limit}
    \end{align}
    almost surely with respect to $\Yv$.  

    Observe that when $\delta$ is replaced by zero in the choice of $\gamma$, the right-hand side of \eqref{eq:gamma_star_limit} simplifies as follows:  Defining $(X,Y,\Xbar) \sim Q_X(x)W(y|x)Q_X(\xbar)$, we have
    \begin{align}
        &\sup_{s \ge 0}\Big( -s\EE[-\log q(X,Y)] - \EE\big[ \log \EE[ q(\Xbar,Y)^s \,|\,Y ] \big] \Big)  \nonumber \\
            &\qquad = \sup_{s \ge 0} \EE\bigg[ \log \frac{ q(X,Y)^s }{ \EE[ q(\Xbar,Y)^s \,|\,Y ] } \bigg] \\
            &\qquad = \GMI(Q_X).
    \end{align}
    Thus, assuming momentarily that we can replace $\delta$ by zero in \eqref{eq:tightGMI_lb}, we deduce that $\pebar \to 1$ for $R > \GMI(Q_X)$, using $M = \lfloor e^{nR} \rfloor$ and the fact that $\big( 1 - \frac{1}{\alpha_n}\big)^{\beta_n} \to 0$ whenever $\lim_{n \to \infty} \frac{\alpha_n}{\beta_n} = 0$.

    To justify taking $\delta \to 0$, it suffices to show that the right-hand side of \eqref{eq:large_dev} is continuous with respect to $\gamma \in (\gamma_{\rm min},\gamma_{\rm prod})$.  This follows from the fact that the pointwise supremum of a collection of linear functions is convex \cite[Sec.~3.2.3]{Boy04}, and a convex function is continuous when restricted to the region in which it is finite.  The assumptions of Lemma \ref{lem:large_dev} ensure that \eqref{eq:large_dev} is finite for $\gamma \in (\gamma_{\rm min},\gamma_{\rm prod})$ (see the discussion following \cite[Thm.~1]{Dem02}), which implies the desired claim.
\end{proof}

{\bf Discussion.} The key tool in proving Lemma \ref{lem:large_dev} is the G\"artner-Ellis theorem in large deviations theory; the interested reader is referred to \cite{Dem02} for the details. 
Intuitively, the parameter $s$ in the dual expression for the GMI can be viewed as a Chernoff parameter resulting from applying the Chernoff bound to $\sum_{i=1}^{n}\log q(\Xbar_i,y_i)$.  

Similarly, in the fixed-cost LM rate of Theorem \ref{thm:LM_fixed} each parameter $r_l$ can be viewed as combining two Chernoff parameters: one for $\sum_{i=1}^n a(\Xbar_i) \ge n(\phi_a - \delta)$ and one for $\sum_{i=1}^n a(\Xbar_i) \le n(\phi_a + \delta)$.  By taking the difference of two arbitrary positive parameters, we end up with an arbitrary real-valued parameter.  In light of this interpretation, we expect that counterparts of Lemma \ref{lem:large_dev} and Theorem \ref{thm:tightGMI} should hold for the LM rate and cost-constrained random coding, but currently this remains an open problem.

\section{Examples} \label{sec:su_cont_examples}

In this subsection, we evaluate the preceding achievable rates for three continuous-alphabet channels.  For the sake of analytical tractability, all examples are based on a Gaussian input distribution and a form of nearest-neighbor decoding; however, the three are still fundamentally distinct due to the different forms of the mismatch.

\subsection{Mismatched Signal Level} \label{sec:ex_signal_level}

We consider a variant of the additive white Gaussian noise (AWGN) channel in which the input is scaled by an unknown constant $\alpha > 0$:
\begin{equation}
    Y = \alpha X + Z,
\end{equation}
where $Z \sim \Ndist(0,\sigma^2)$ for some noise power $\sigma^2 > 0$, and $X$ and $Z$ are independent.  We consider a power constraint corresponding to \eqref{eq:input_constr} with $c(x) = x^2$ and some $\Gamma > 0$.  Accordingly, we adopt a Gaussian input distribution $Q_X$ corresponding to $X \sim \Ndist(0,\Gamma)$, which is capacity-achieving in the matched case.

The optimal decoding rule minimizes $\|\yv - \alpha\xv\|_2^2$, and requires knowledge of $\alpha$.  We consider a scenario in which the decoder is unaware of the scaling effect, and therefore acts as if $\alpha = 1$:\footnote{More generally, one could consider a decoder that has an estimate $\hat{\alpha}$ of $\alpha$ and decodes accordingly.  Taking $\hat{\alpha} = 1$ in fact entails no loss of generality, since one can reduce to this scenario by re-scaling (and accordingly modifying the noise level $\sigma^2$ and input power $\Gamma$.) }
\begin{equation}
    \hat{m} = \argmin_{ j=1,\dotsc,M } \| \yv - \xv^{(j)} \|_2^2.  \label{eq:nn_dec}
\end{equation}
This is the {\em nearest-neighbor} decoding rule, and corresponds to the decoding metric $q(x,y) = e^{-(y-x)^2}$. 

In this subsection, we will derive the following expressions for the GMI and LM rate \cite{Mer95}.

\begin{theorem} \label{thm:ex_signal_level}
    {\em (Mismatched signal level)}
    Under the preceding mismatched signal level setting with input power $\Gamma > 0$, noise power $\sigma^2 > 0$, and signal level $\alpha > 0$, we have the following:
    
    (i) The GMI is given by
    \begin{equation}
        \GMI(Q_X) = \sup_{\sgz} \frac{1}{2}\log\big( 1 + 2s\Gamma \big) + \frac{s(\alpha^2 \Gamma + \sigma^2)}{ 1 + 2\Gamma s } - s\big( (\alpha-1)^2 \Gamma + \sigma^2 \big). \label{eq:ex_sig_GMI}
    \end{equation}
    
    (ii) The LM rate is given by
    \begin{equation}
        \LM(Q_X) = \frac{1}{2}\log\bigg( 1 + \frac{\alpha^2 \Gamma}{\sigma^2} \bigg), \label{eq:C_AWGN_alpha}
    \end{equation}
    and equals the matched (and hence mismatched) capacity subject to a power constraint $\Gamma$.
\end{theorem}

The intuition behind the fact that we achieve the matched capacity in the second part is that the mismatched and maximum-likelihood decoding rules are equivalent for codebooks in which all codewords have the same power $\|\xv\|_2^2$.  By using cost-constrained random coding with an auxiliary cost $a(x) = x^2$, we can ensure that the powers are {\em nearly} identical, and that we still achieve the optimal rate.

\begin{figure}
    \begin{centering}
        \includegraphics[width=0.6\columnwidth]{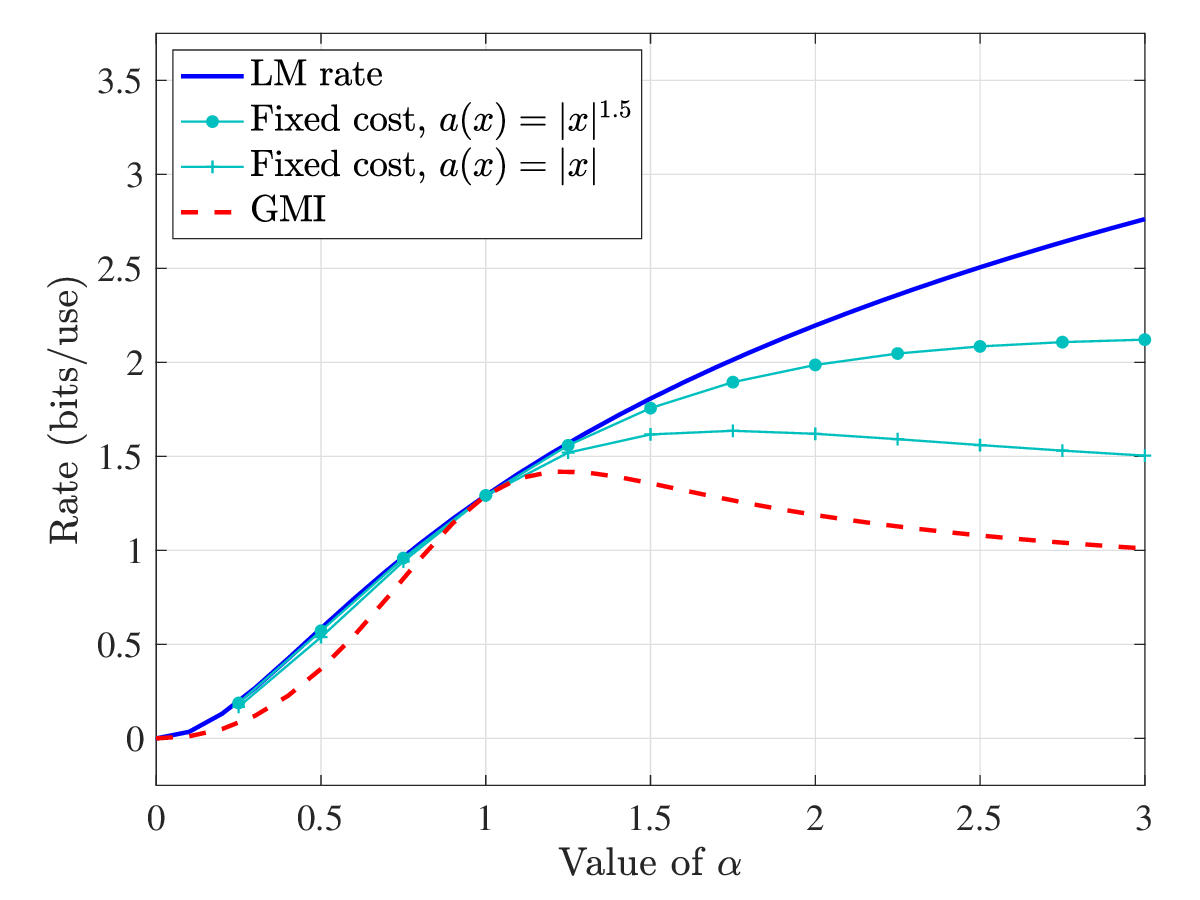}
        \par
    \end{centering}
    
    \caption{Unknown signal level example: GMI, LM rate, and its variation with a suboptimal auxiliary cost, each plotted as a function of $\alpha$. } \label{fig:UnkownSignal}
\end{figure}

We compare the GMI and LM rate numerically in Figure \ref{fig:UnkownSignal}. In addition, we plot the achievable rate of Theorem \ref{thm:LM_fixed} with a single auxiliary cost (i.e., $L=1$) under the suboptimal choices $a_1(x) = |x|$ and $a_1(x) = |x|^{1.5}$.  These rates were evaluated using numerical integration, along with gradient ascent for optimizing $(s,r_1)$.  

We observe that the LM rate exceeds the GMI for all $\alpha > 0$, except for the matched case $\alpha = 1$.  Moreover, the GMI is seen to saturate as $\alpha \to \infty$, while the LM rate grows unbounded.  Moreover, even with a suboptimal auxiliary cost, we get a strict improvement over the GMI, and nearly match the LM rate for $\alpha < 1$.  The choice $a_1(x) = |x|^{1.5}$ provides a better rate than $a_1(x) = |x|$, as should be expected due to the fact that the former more closely resembles the optimal choice $a_1(x) = x^2$.

\subsection*{Proof of Theorem \ref{thm:ex_signal_level}}

We first consider the LM rate, and then the GMI.

\paragraph{Evaluation of the LM rate.} The right-hand side of \eqref{eq:C_AWGN_alpha} is the matched capacity of the AWGN channel with input scaled by $\alpha$.  To see that it is achieved by the LM rate, we write the mismatched metric $q$ and the optimal (maximum-likelihood) metric $q^*$ as
\begin{gather}
    \log q(x,y) = - y^2 - x^2 + 2xy \\
    \log q^*(x,y) = -y^2 - \alpha^2 x^2 + 2\alpha xy,
\end{gather}
which implies that $ \log q^*(x,y) = \alpha \log q(x,y) + (\alpha - \alpha^2) x^2 + (\alpha - 1)y^2$.  By Lemma \ref{lem:opt_conds} (whose proof extends to the continuous setting), we conclude that $\LM(Q_X)$ equals the matched decoding rate (i.e., the mutual information $I(X;Y)$), which in turn is known to be given by \eqref{eq:C_AWGN_alpha}.

\paragraph{Evaluation of the GMI.} We write the GMI in \eqref{eq:GMI_cont} as
\begin{align}
\GMI(Q_X)  
    &=\sup_{\sgz}~ s \EE\big[ \log q(X,Y) \big]  - \EE\big[ \log \EE[q(\Xbar,Y)^{s}\,|\, Y] \big] \\
    &= \sup_{\sgz}~ - s \EE\big[ (Y-X)^2 \big]  - \EE\big[ \log \EE[e^{-s(Y-\Xbar)^2}\,|\, Y] \big], \label{eq:ex_sig_GMI3}
\end{align}
where we have substituted $q(x,y) = e^{-(y-x)^2}$.  We then obtain \eqref{eq:ex_sig_GMI} by evaluating the relevant Gaussian integrals according to $Y = \alpha X + Z$ with $X \sim \Ndist(0,\Gamma)$ and $Z \sim \Ndist(0,\sigma^2)$.  The technical details of these integral evaluations are omitted.  

\subsection{Mismatched Noise Distribution -- Nearest-Neighbor Decoding for Additive Non-Gaussian Channels} \label{sec:ex_noise}

Continuing with the theme of additive noise channels, we turn to a particularly important setting in which the noise distribution is mismatched.  The input-output relationship is given by
\begin{equation}
    Y = X + Z,
\end{equation}
where $X$ and $Z$ are independent.  We again consider an input constraint according to \eqref{eq:input_constr} with $c(x) = x^2$ (i.e., a power constraint of $\Gamma > 0$), and we assume that $Z$ follows an {\em arbitrary distribution} having a finite mean and variance:
\begin{equation}
    \EE[Z] = \mu, \quad \var[Z] = \sigma^2 \label{eq:Z_power}
\end{equation}
for some $\mu \in \RR$ and $\sigma^2 > 0$.

\paragraph{AWGN case.} As we already saw in the previous example, when the noise is Gaussian, the matched capacity is given by
\begin{equation}
    \CAWGN(\Gamma,\sigma^2) = \frac{1}{2}\log\bigg( 1 + \frac{\Gamma}{\sigma^2} \bigg). \label{eq:C_AWGN}
\end{equation}
Moreover, when $\mu = 0$, the optimal decoding rule is nearest-neighbor decoding according to \eqref{eq:nn_dec}, which we repeat here:
\begin{equation}
    \hat{m} = \argmin_{ j=1,\dotsc,M } \| \yv - \xv^{(j)} \|_2^2.  \label{eq:nn_dec2}
\end{equation}
More generally, replacing $\yv$ by $\yv - \mu\bone$ in \eqref{eq:nn_dec2} gives the optimal rule, where $\bone$ is the vector of ones.  The capacity-achieving input distribution under the power constraint $\EE[X^2] \le \Gamma$ is given by $X \sim \Ndist(0,\Gamma)$.

\paragraph{Non-Gaussian noise.} 
Gaussian noise is convenient to analyze mathematically, is often well-motivated in applications, and has been studied extensively from both theoretical and practical viewpoints.   On the other hand, it is certainly not the only type of additive noise of interest.  

In general, under non-Gaussian noise, the {\em matched} channel capacity is given by $C = \sup_{Q_X \,:\,\EE_Q[X^2] \le \Gamma} I(X;Y)$.  However, achieving this may be difficult for several reasons: The distribution $P_Z$ may not be known, and even if it is known, the required codebook structure and decoder may be prohibitively complex.  

Motivated by these considerations, we consider the following question \cite{Lap96a}: {\em If we adopt a Gaussian coding scheme, but apply it to a channel that actually consists of non-Gaussian additive noise, how well will it perform?}  More precisely, we consider random coding with input distribution $Q_X \sim \Ndist(0,\Gamma)$, along with the nearest-neighbor decoding rule \eqref{eq:nn_dec2} corresponding to $q(x,y) = e^{-(y-x)^2}$.  These choices are optimal for the {\em zero-mean} AWGN channel; we will also comment on the alternative choice $q(x,y) = e^{-(y-\mu-x)^2}$ corresponding to optimal decoding for $\Ndist(\mu,\sigma^2)$ noise. 

If we apply such a Gaussian coding scheme to a non-Gaussian channel, we may expect a performance degradation due to the fact that we have designed a code for the wrong channel.  On the other hand, we may expect a higher rate due to the fact that for a given variance $\var[Z] = \sigma^2$, Gaussian noise is the most harmful in the matched scenario (i.e., yields the smallest capacity) \cite[Ex.~9.21]{Cov06}.  Intriguingly, the following result reveals that these effects cancel each other, and all noise distributions lead to the same rate as Gaussian noise.  The first part of this theorem can be found in \cite{Lap96a}, and the second part is a straightforward variant that is new to the best of our knowledge.

\begin{theorem} \label{thm:gaussian}
    {\em (Gaussian coding schemes for non-Gaussian channels)}
    Under the preceding mismatched nearest-neighbor decoding setting with input distribution $Q_X \sim \Ndist(0,\Gamma)$ and an arbitrary additive noise distribution $P_Z$ having mean $\mu$ and variance $\sigma^2$, we have the following:
    
    (i) The GMI is given by
    \begin{equation}
        \GMI(Q_X) = \frac{1}{2}\log\bigg( 1 + \frac{\Gamma}{\mu^2 + \sigma^2} \bigg). \label{eq:ex_noise_GMI}
    \end{equation}
    
    (ii) The fixed-cost LM rate with a single auxiliary cost $a_1(x) = x$ is given by
    \begin{equation}
        \LM'(Q_X,a_1) = \frac{1}{2}\log\bigg( 1 + \frac{\Gamma}{\sigma^2} \bigg). \label{eq:ex_noise_LM}
    \end{equation}
\end{theorem}

In the proof of Theorem \ref{thm:gaussian} (given below), we will additionally present an elementary proof of ensemble tightness for the GMI that is specific to this setting \cite{Lap96a}.  In contrast, we make no claims regarding the ensemble tightness of the fixed-cost LM rate.

Theorem \ref{thm:gaussian} can be viewed both positively and negatively.  On the positive side, we achieve $\CAWGN$ even though we designed the code for the wrong channel, meaning that Gaussian codes form a {\em robust} communication scheme.  On the negative side, since Gaussian noise is the most harmful noise, the matched capacity is strictly higher than $\CAWGN$ in non-Gaussian settings, so we have failed to achieve the best possible rate.  Stated differently,  by designing for the worst case noise, we are guaranteed to achieve the worst-case capacity and no better.

\begin{figure}
    \begin{centering}
        \includegraphics[width=0.6\columnwidth]{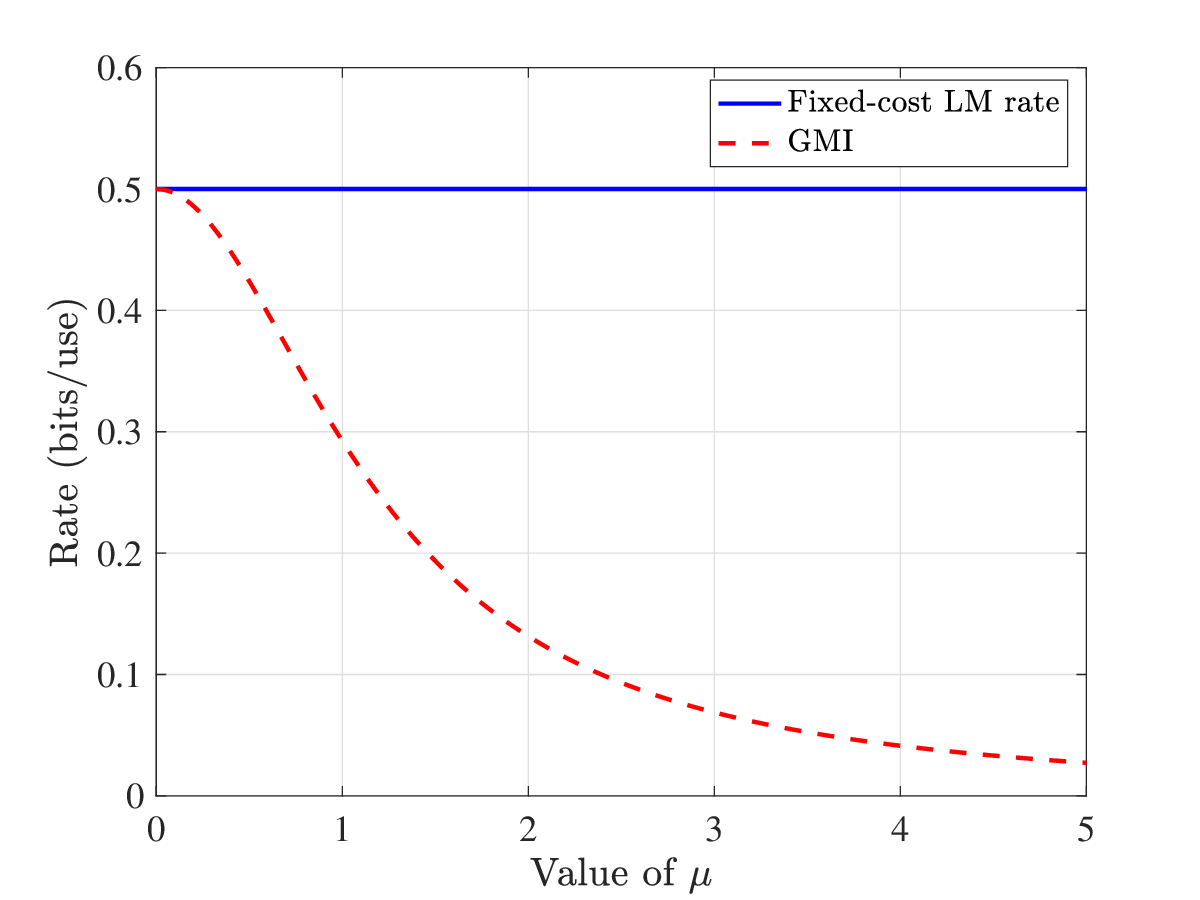}
        \par
    \end{centering}
    
    \caption{Mismatched noise distribution example: GMI and fixed-cost LM rate as a function of $\mu$ with $\Gamma = \sigma^2 = 1$. } \label{fig:NonGaussian}
\end{figure}

From the second part of Theorem \ref{thm:gaussian}, we can conclude that the improvement of the fixed-cost LM rate amounts {\em at least} to replacing $\EE[Z^2] = \mu^2 + \sigma^2$ by $\var[Z] = \sigma^2$.  The gap between the two can be arbitrarily large; we demonstrate this in Figure \ref{fig:NonGaussian}, where we plot the rates as a function of $\mu$ with $\Gamma = \sigma^2 = 1$.

To see why the choice $a(x) = x$ in the second part removes the dependence on the mean $\mu$, we consider the decoding rule $\qtilde(x,y) = e^{-(y - \mu - x)^2}$, which provides a natural counterpart to the above choice when $\mu$ is known.  Expanding the square yields
\begin{equation}
    -\log \qtilde(x,y) = (y-x)^2 - 2\mu y + 2\mu x + \mu^2. \label{eq:ex_noise_qtilde}
\end{equation}
The term $2 \mu y$ has no impact on the decoder, since it depends only on the output.  On the other hand, the term $2\mu x$ does have an impact in general.  However, if we use a codebook such that $\sum_{i=1}^n x_i$ is the same for each codeword $\xv$, then there is no impact.  By using cost-constrained random coding with $a(x) = x$, we ensure that $\sum_{i=1}^n x_i$ is {\em nearly} the same for every codeword, which is enough to ensure that the LM rates corresponding to $q(x,y) = e^{-(y-x)^2}$ and $\qtilde(x,y)$ above are the same.

Under the decoding metric $\qtilde(x,y)$, the GMI also achieves the improved rate $\frac{1}{2}\log\big( 1 + \frac{\Gamma}{\sigma^2} \big)$.  However, such a choice amounts to subtracting the mean $\mu$ at the output, and hence, it requires the mean to be known in the first place.  In contrast, using the fixed-cost LM rate, we attain the improvement without knowledge of $\mu$.

\subsection*{Proof of Theorem \ref{thm:gaussian}}

\paragraph{Evaluation of the GMI.} We rewrite \eqref{eq:GMI_cont} as
\begin{align}
    \GMI(Q_X)  
    &=\sup_{\sgz} s \EE\big[ \log q(X,Y) \big]  - \EE\big[ \log \EE[q(\Xbar,Y)^{s}\,|\, Y] \big] \\
    &= \sup_{\sgz} \frac{1}{2}\log\big( 1 + 2s\Gamma \big) + \frac{s(\Gamma + \mu^2 + \sigma^2)}{ 1 + 2\Gamma s } - s\big( \mu^2 + \sigma^2 \big), \label{eq:ex_noise_GMI_s}
\end{align}
where \eqref{eq:ex_noise_GMI} follows by substituting $q(x,y) = e^{-(y-x)^2}$ and evaluating the expectations explicitly; this requires some technical but elementary integration.  Differentiating the right-hand side of \eqref{eq:ex_noise_GMI_s}, it is a straightforward exercise to verify that the choice $s=\frac{1}{2(\mu^2 + \sigma^2)}$ makes the derivative vanish.  Hence, since the objective function defining the GMI is always concave in $s$, we conclude that this choice of $s$ must be globally optimal.  With some algebraic manipulation, this choice leads to \eqref{eq:ex_noise_GMI}.

\paragraph{An elementary ensemble-tightness proof for the GMI.} 

While the ensemble-tightness of the GMI is already established in Theorem \ref{thm:tightGMI}, it is instructive to provide a direct proof that is not only significantly simpler, but also comes with an interesting alternative proof of the achievability part.  The analysis proceeds in three steps, outlined as follows:
\begin{itemize}
    \item By the circular symmetry of $P_{\Xv} = Q_X^n$, and similar symmetry in the nearest-neighbor decoding rule, it can be verified that the conditional error probability $\PP[\mathrm{error} \,|\, \zv]$ depends on the noise realization $\zv$ only through its power $\|\zv\|^2$ \cite{Lap96a}. 
    \item Moreover, the conditional random-coding error probability $\PP[\mathrm{error} \,|\, \zv]$ is non-decreasing as a function of $\|\zv\|^2$.  To see this, we suppose that $\|\yv - \xvbar\|_2 \le \|\yv - \xv\|_2$ for some incorrect $\xvbar$ and $\yv = \xv + \zv$, and proceed by showing that the same is true with $\yv' = \xv + (1+\delta)\zv$ for some $\delta > 0$:
    \begin{align}
    \|\yv' - \xvbar\|_2 
    &= \|\yv + \delta\zv - \xvbar\|_2  \\
    &\le \|\yv - \xvbar\|_2 + \delta \|\zv\|_2 \label{eq:gaussian_mono2} \\
    &\le \|\zv\|_2 + \delta \|\zv\|_2 \label{eq:gaussian_mono3} \\
    &= \|\yv' - \xv\|_2,
    \end{align}
    where \eqref{eq:gaussian_mono2} follows from the triangle inequality, and \eqref{eq:gaussian_mono3} follows since $\xv$ is closer to $\yv$ than $\xvbar$ is.
    \item As a result, the asymptotic behavior of the error probability is the same for any noise distribution such that $\frac{1}{n}\|\Zv\|^2 \to \EE[Z^2]$ almost surely.  In particular, by the well-known fact that \eqref{eq:ex_noise_GMI} is the (tight) rate achieved under $\Ndist(0,\EE[Z^2])$ additive Gaussian noise, we deduce that it is also the (ensemble-tight) rate achieved for any memoryless channel with the same second moment $\EE[Z^2] = \mu^2 + \sigma^2$.
\end{itemize}
The interested reader is referred to \cite{Lap96a} for the details.

%

\noindent \paragraph{Evaluation of the fixed-cost LM Rate.} 
We rewrite \eqref{eq:LM_cont2} as
\begin{align}
    \LM'(Q_X,a)
    &=\sup_{\sgz, r} s \EE\big[ \log q(X,Y) \big] + r \EE[a(X)]  \nonumber \\ 
        &\qquad\qquad\qquad - \EE\big[ \log \EE[q(\Xbar,Y)^{s} e^{r a(\Xbar)} \,|\, Y] \big] \\
    &= \sup_{\sgz, r} \frac{1}{2}\log\big( 1 + 2s\Gamma \big) - \frac{\Gamma \cdot  r(r+4s\mu)}{2(1+2\Gamma s)} \nonumber \\
         &\qquad\qquad\qquad + \frac{s(\Gamma + \mu^2 + \sigma^2)}{ 1 + 2\Gamma s } - s\big(\mu^2 + \sigma^2 \big), \label{eq:ex_noise_LM_sa}
\end{align}
where \eqref{eq:ex_noise_LM_sa} follows by substituting the choices of $q$ and $a_1(\cdot) = a(\cdot)$ and evaluating the expectations.  Only the second term in \eqref{eq:ex_noise_LM_sa} depends on $r$, and it is a simple differentiation exercise to show that the optimal choice is $r = -2s\mu$.  Substituting this choice and suitably canceling terms, we obtain
\begin{equation}
    \LM'(Q_X,a) = \sup_{\sgz} \frac{1}{2}\log\big( 1 + 2s\Gamma \big) + \frac{s(\Gamma + \sigma^2)}{ 1 + 2\Gamma s } - s\sigma^2 . \label{eq:ex_noise_LM_s}
\end{equation}
This matches \eqref{eq:ex_noise_GMI_s} with $\sigma^2$ in place of $\mu^2 + \sigma^2$.  Hence, since we already know that \eqref{eq:ex_noise_GMI_s} leads to \eqref{eq:ex_noise_GMI}, we deduce that \eqref{eq:ex_noise_LM_s} leads to \eqref{eq:ex_noise_LM}.

\subsection*{Variations and Generalizations}

The first part of Theorem \ref{thm:gaussian} holds not only for i.i.d.~Gaussian codes, but also {\em shell codes}, in which $\Xv$ is drawn uniformly from the sphere of radius $\sqrt{n\Gamma}$ \cite{Lap96a}.  This distribution can be thought of as corresponding to cost-constrained random coding with $a(x) = x^2$ and $\delta = 0$, and indeed, it can be verified analytically that the fixed-cost rate of Theorem \ref{thm:LM_fixed} does not improve on the GMI in this example when $L=1$ and $a_1(x) = x^2$.

The first part of Theorem \ref{thm:gaussian} has also been extended to additive multiple-access channels \cite{Lap96a}, fading channels \cite{Lap96a,Lap02,Wei04} (see also Section \ref{sec:ex_fading}), and parallel additive noise channels \cite{Lap96a}, and refined characterizations of the fixed-error asymptotics have been given \cite{Sca17a}.
In addition, regarding the second part of Theorem \ref{thm:gaussian}, a similar improvement replacing the second moment $\EE[Z^2]$ by the variance $\var[Z]$ is known to hold in the low-SNR regime under binary phase shift keying (BPSK) and constant-composition random coding \cite{Gan00}.

\subsection{Fading Channels} \label{sec:ex_fading}

In our final example, we turn to a class of channels exhibiting {\em fading}, which is fundamental to wireless communication scenarios ({\em cf.}, Section \ref{sec:app_fading}).  Fading channels are usually modeled as being {\em complex-valued}, and we adopt such an approach here.\footnote{The achievability results (i.e., GMI and LM rate) we have given for real-valued channels are directly applicable to the complex-valued setting, and the proofs apply without change.}

The setup that we consider adopts fairly strong memorylessness assumptions to keep the analysis simple, while still conveying some of the main ideas behind mismatch in fading channels.  A more comprehensive treatment (including this special case) can be found in \cite{Lap02,Wei04}. 

We consider a memoryless time-varying model of the form
\begin{equation}
    Y_i = H_i X_i + Z_i, \label{eq:fading}
\end{equation}
where $X_i \in \CC$ is the input, $Z_i \in \CC$ is additive noise, and $H_i \in \CC$ is a {\em fading coefficient}.  We assume that $\{Z_i\}_{i=1}^n$ is i.i.d.~on $\CNdist(0,\sigma^2)$ (i.e., the variance is $\frac{\sigma^2}{2}$ each for the real and imaginary parts), and that $\{H_i\}$ is an i.i.d.~sequence with density function $P_H$.  In addition, we assume that $\Xv=(X_1,\dotsc,X_n)$, $\Hv=(H_1,\dotsc,H_n)$, and $\Zv=(Z_1,\dotsc,Z_n)$ are all mutually independent.

\paragraph{Perfect channel knowledge.} If each random realization $H_i = h_i$ were known perfectly at the decoder, then due to the Gaussianity of the noise, the optimal decoding rule would be the following {\em weighted} version of the nearest-neighbor rule:
\begin{equation}
    \hat{m} = \argmin_{j = 1,\dotsc,M} \sum_{i=1}^n \big| y_i - h_i x_i^{(j)} \big|^2. \label{eq:nn_fading}
\end{equation}
In addition, under a power constraint $\EE[|X|^2] \le \Gamma$ (i.e., $c(x) = |x|^2$), the capacity is achieved by a complex Gaussian input distribution $\CNdist(0,\Gamma)$, and is given by \cite{Tse05,Lap02}
\begin{equation}
    \CAWGN(\Gamma,\sigma^2,P_H) = \EE\bigg[ \log\Big( 1 + \frac{|H|^2 \Gamma}{\sigma^2} \Big) \bigg], \label{eq:C_AWGN_fading}
\end{equation}
where $H \sim P_H$, and $\Gamma$ is the input power.  Note that unlike the previous examples, there is no factor of $\frac{1}{2}$ here, as we are in the complex-valued setting rather than the real setting.

In fact, the achievability of \eqref{eq:C_AWGN_fading} extends to non-Gaussian noise distributions analogously to the first part of Theorem \ref{thm:gaussian} above \cite{Lap96a}, but we focus on Gaussian noise here for simplicity.  

\paragraph{Imperfect channel knowledge.} Our focus in this example is on {\em uncertainty in the fading coefficients $H_i$}.  To address this, we adopt a simple uncertainty model in which
\begin{equation}
    H_i = \Hhat_i + \Htilde_i, \quad \EE[\Htilde_i | \Hhat_i] = 0,
\end{equation}
where $\Hhat_i$ is a possibly-random estimate of $H$ known at the decoder, and $\Htilde_i$ represents an unknown {\em conditionally zero-mean} error term. 
We make the simplifying assumption that the pairs $\{( \Hhat_i, \Htilde_i )\}_{i=1}^n$ are i.i.d.~with respect to $i=1,\dotsc,n$, and independent of the channel input and noise.

In the case that the joint density function of $(\Hhat_i, \Htilde_i)$ is unknown (or even when it is known but difficult to design a corresponding optimal coding scheme), it is natural to apply weighted nearest-neighbor coding similarly to \eqref{eq:nn_fading}, but with each weight given only by the realization $\Hhat_i = \hhat_i$ of the corresponding estimate:
\begin{equation}
    \hat{m} = \argmin_{j = 1,\dotsc,M} \sum_{i=1}^n \big| y_i - \hhat_i x_i^{(j)} \big|^2. \label{eq:nn_fading2}
\end{equation}
This is a mismatched decoding rule, in the sense that it would be optimal under a model of the form $Y = \Hhat X + Z$ with zero-mean Gaussian noise, in contrast with the true model.  The corresponding decoding metric is given by $q(x,(y,\hhat)) = e^{-(y-\hhat x)^2}$.  Note that here and subsequently, since $\Hhat$ is known at the decoder, it is treated as part of the output, so that $(Y,\Hhat)$ plays the role of the usual $Y$.

The following result gives an upper bound for the GMI and a condition under which it holds with equality \cite{Lap02}; to our knowledge, the LM rate is yet to be studied in fading scenarios.

\begin{theorem} \label{thm:ex_fading}
    {\em (Fading channels)}
    Consider the preceding complex-valued channel fading setup with a known estimate $\Hhat$ at the output, and a conditionally zero-mean error term $\Htilde$.  Under i.i.d.~random coding with $X \sim \CNdist(0,\Gamma)$, along with weighted nearest-neighbor decoding according to \eqref{eq:nn_fading2}, the GMI satisfies
    \begin{equation}
        \GMI(Q_X) \le \EE\bigg[ \log \bigg( 1 + \frac{|\Hhat|^2 \Gamma}{\EE\big[|\Htilde|^2 \,|\, \Hhat \big]\Gamma + \sigma^2} \bigg) \bigg], \label{eq:fading_GMI}
    \end{equation}
    and moreover, equality holds when $\EE\big[|\Htilde|^2 \,|\, \Hhat = \hhat \big]$ takes the same value for all possible $\hhat$.
\end{theorem}

We observe that \eqref{eq:fading_GMI} takes a similar form to the matched capacity \eqref{eq:C_AWGN_fading}, except that the signal power in the numerator is multiplied by $|\Hhat|^2$ instead of $|H|^2$, and the denominator contains an extra term $\EE\big[ |\Htilde|^2\,|\, \Hhat \big] \Gamma$.  The intuition is that we are treating the unknown error term $\Htilde$ as noise, and since that term is multiplied by $X$, the overall additional ``noise power'' given $\Hhat$ is $\EE\big[ |\Htilde|^2\,|\, \Hhat \big] \Gamma$.

\begin{figure}
    \begin{centering}
        \includegraphics[width=0.575\columnwidth]{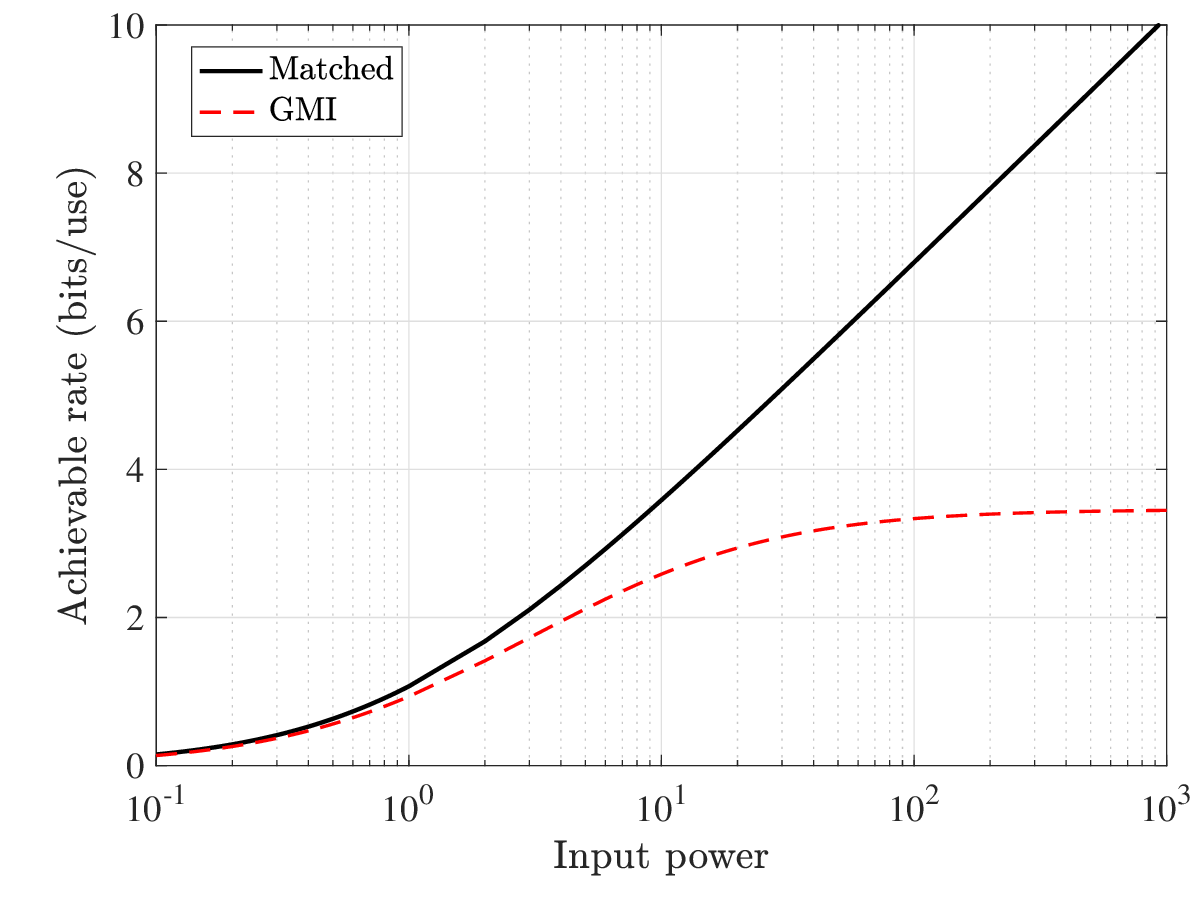}
        \par
    \end{centering}
    
    \caption{Fading channel example: Matched capacity and GMI  as a function of $\Gamma$ with $\sigma^2=1$, $|\Hhat|^2 = 1$ (deterministically) and $\EE[|\Htilde|^2 \,|\,\Hhat = \hhat] = 1$ for all $\hhat$. } \label{fig:FadingExample} \vspace*{-2ex}
\end{figure}

Theorem \ref{thm:ex_fading} reveals that mismatch in the fading coefficient can be much more harmful than mismatch in the noise distribution ({\em cf.}, Section \ref{sec:ex_noise}).  For instance, supposing for simplicity that $\Hhat$ takes some {\em deterministic} non-zero value, we find that $\GMI(Q_X)$ remains {\em bounded in the high-power regime $\Gamma \to \infty$}, converging to $\log\big(1 + \frac{|\Hhat|^2}{\EE[\Htilde^2]}\big)$.  In contrast, the matched capacity \eqref{eq:C_AWGN_fading} grows unbounded.  An illustration is given in Figure \ref{fig:FadingExample}, where we plot the GMI and matched capacity as a function of $\Gamma$ when $\sigma^2=1$, $|\Hhat|^2 = 1$ (deterministically), and $\EE[|\Htilde|^2 \,|\,\Hhat = \hhat] = 1$ for all $\hhat$.

\subsection*{Proof of Theorem \ref{thm:ex_fading}}

Recall that the pair $(Y,\Hhat)$ plays the role of the channel output. We first write the GMI as 
\begin{equation}
    \GMI(Q_X) = \sup_{\sgz} \EE\big[ \GMI(Q_X,s,\Hhat) \big], \label{eq:GMI_init_fading}
\end{equation}
where
\begin{align}
    \GMI(Q_X,s,\hhat) 
        &= s\EE[\log q(X,(Y,\hhat))] - \EE\big[ \log \EE\big[ q(\Xbar,(Y,\hhat))^s) \big] \big] \\
        &= -s\EE[ |Y - \hhat X|^2 ] - \EE\big[ \log \EE\big[ e^{-|Y - \hhat \Xbar|^2}  \,|\,Y \big] \big] 
\end{align}
with implicit conditioning on $\Hhat = \hhat$ throughout.  Notice that this expression is similar to that of \eqref{eq:ex_sig_GMI3} in the mismatched signal level example.  While we are working with complex-valued (rather than real-valued) random variables here, we can still explicitly evaluate the expectations in a similar manner, leading to
\begin{equation}
    \GMI(Q_X,s,\hhat) =  \log\big( 1 + |\hhat|^2 \Gamma s\big) + \frac{s( |\hhat|^2 \Gamma + \Gamma_{XY}(\hhat) ) }{ 1 + |\hhat|^2 \Gamma s }  -s\Gamma_{XY}(\hhat), \label{eq:ex_fading_GMIs}
\end{equation}
where we write $\Gamma_{XY}(\hhat) = \EE[|\Htilde|^2 \,|\, \Hhat = \hhat] \Gamma + \sigma^2$ for brevity. 

The optimization over $s > 0$ in \eqref{eq:GMI_init_fading} is challenging in general.  However, we can upper bound \eqref{eq:GMI_init_fading} by moving the supremum inside the expectation, thus allowing us to choose $s = s(\hhat)$ as a function of $\hhat$, at the expense of only deducing an upper bound on the GMI instead of its exact value.  With this added flexibility, it is a simple exercise to differentiate \eqref{eq:ex_fading_GMIs} and verify that the choice $s(\hhat) = \frac{1}{\Gamma_{XY}(\hhat)}$ yields a derivative of zero.  Since the objective function is concave in $s$, we conclude that such a choice is globally optimal.  Substituting this optimal choice, denoted by $s^*(\hhat)$, we obtain the following after some algebraic manipulation:
\begin{equation}
\GMI(Q_X,s^*(\hhat),\hhat) =  \log\bigg( 1 + \frac{|\hhat|^2 \Gamma}{\Gamma_{XY}(\hhat)}\bigg). \label{eq:ex_fading_GMIs2}
\end{equation}
The proof of \eqref{eq:fading_GMI} is concluded by averaging over $\Hhat$ and substituting $\Gamma_{XY}(\cdot)$.  To establish the claimed condition for equality, we note that if $\EE\big[|\Htilde|^2 \,|\, \Hhat = \hhat \big]$  does not depend on $\hhat$, then neither does $\Gamma_{XY}(\hhat)$, and thus similarly for $s(\hhat)$.  Hence, in this case, the order of supremum and expectation in \eqref{eq:GMI_init_fading} is inconsequential, meaning that the upper bound holds with equality.


%

 

\chapter{Mismatch in Rate Distortion Theory} \label{ch:rate_dist}

\section{Introduction} \label{sec:rd_intro}

Thus far, we have focused on channel coding with a mismatched decoding rule.  In this section, we turn to the distinct problem of source coding subject to a fidelity constraint, i.e., {rate-distortion theory} \cite[Ch.~10]{Cov06}.  As with channel coding, this represents one of the most fundamental coding problems in information theory.

The setup of rate-distortion theory (detailed in Section \ref{sec:rd_setup} below) consists of a memoryless {source}, an {encoder} mapping source sequences to indices, and a {decoder} mapping these indices to a reconstructed sequence.  The quality of the reconstruction is measured by a {distortion function}.  Assuming that the decoder is deterministic, we can view the set of all reconstructed sequences as forming a codebook.  In addition, given such a codebook, the optimal encoding rule is to map the source sequence to the index corresponding to the codeword with the lowest distortion.

If the encoder does not know the true distortion measure under consideration, or if the optimal encoding rule is ruled out due to implementation constraints, then it is natural to consider a variant of this problem with {\em mismatched encoding}, in which the encoder minimizes a given (possibly suboptimal) {\em encoding metric} between the source and its reconstruction.  Some motivating examples are given in Section \ref{sec:rd_discussion}. 

As well as being of interest in its own right, the mismatched rate-distortion problem has interesting analogies with the mismatched channel coding problem.  Among other things, we will see that the multi-letter extension ({\em cf.}, Section \ref{sec:multi_LM}) of a certain random-coding achievability result has a matching converse.  In contrast, for channel coding it remains open as to whether the multi-letter extension of the LM rate coincides with the mismatch capacity in general.

In Sections \ref{sec:rd_mm_random} and \ref{sec:gaussian_rd}, we consider a different type of mismatch, in which a single distortion measure is adopted throughout the system, but a suboptimal random coding distribution is used.  Section \ref{sec:rd_mm_random} considers discrete memoryless sources, while Section \ref{sec:gaussian_rd} focuses on Gaussian coding for non-Gaussian sources.  In the latter setting, in analogy with the mismatched channel coding example of Section \ref{sec:ex_noise} (Gaussian coding with non-Gaussian noise), we find that all real-valued sources with a given second moment yield the same rate-distortion trade-off as the Gaussian source.

This section is predominantly based on the work of Lapidoth \cite{Lap97}, with various aspects also relating to Sakrison \cite{Sak70,Sak69}, Yang and Kieffer \cite{Yan98}, and Dembo and Kontoyiannis \cite{Dem02}.  We note that compared to \cite{Lap97}, we consider a slightly simplified mismatched encoding problem formulation (see below).


\section{Problem Setup} \label{sec:rd_setup}

We consider a discrete memoryless source (DMS), in which each symbol is drawn from a distribution $\Pi_X(x)$ on a finite alphabet $\Xc$.  The resulting $n$-letter distribution is denoted by $\mathrm{\Pi}_X^n(\xv) \triangleq \prod_{i=1}^n \mathrm{\Pi}_X(x_i)$.  The source sequence $\Xv \sim \Pi_X^n$ is received as the input to an {\em encoder} that maps $\Xc^n$ to $\{1,\dotsc,M\}$ for some integer $M$.  The {\em rate} of the code is denoted by $R = \frac{1}{n} \log M$.  

We consider encoding rules that are dictated by a codebook $\Cc = \{ \xvhat^{(1)},\dotsc,\xvhat^{(M)}\}$, where each codeword lies in $\Xchat^n$ for some finite reconstruction alphabet $\Xchat$. In particular, in accordance with the discussion in Section \ref{sec:rd_intro}, we assume that the encoder chooses\footnote{Here we minimize an additive metric, whereas in channel coding we maximized a multiplicative metric.  In both settings, we can readily switch between these notions, e.g., maximizing $\prod_{i=1}^n q(x_i,y_i)$ is equivalent to minimizing $\sum_{i=1}^n (- \log q(x_i,y_i))$.}
\begin{equation}
    m = \argmin_{j = 1,\dotsc,M} d_0^n(\xv,\xvhat^{(j)}), \quad d_0^n(\xv,\xvhat) = \sum_{i=1}^n d_0(x_i,\xhat_i), \label{eq:rd_m}
\end{equation}
where $d_0(x,\xhat)$ is a given {\em encoding metric} on $\Xc \times \Xchat$.
Except where stated otherwise, the results that we present hold under arbitrary tie-breaking methods in \eqref{eq:rd_m}.

We focus on the trivial decoder that maps $m$ to $\xvhat^{(m)}$.  This is in slight contrast to Lapidoth \cite{Lap97}, who allowed the decoder to apply post-processing to the codeword to obtain the final estimate.  The former setup will lead to slightly simpler results, while still conveying the same key ideas and techniques. 

When the source realization is $\xv$ and its reconstruction is $\xvhat$, the distortion incurred is given by $d_1^n(\xv,\xvhat) = \sum_{i=1}^n d_1(x_i,\xhat_i)$ for some {\em true distortion measure} $d_1$ on $\Xc \times \Xchat$.  Our goal is to achieve a normalized distortion $\frac{1}{n} d_1^n(\xv,\xvhat)$ that, with high probability, does not exceed a given threshold $D$.  Throughout this section, we assume that both $d_0$ and $d_1$ are non-negative and finite-valued, and we define $\dimax \triangleq \max_{x,\xhat} d_1(x,\xhat)$.

Formally, we have the following.

\begin{definition}  \label{def:rate_dist}
    {\em (Mismatched rate-distortion and distortion-rate functions)}
    Under the preceding mismatched rate-distortion setup specified by $(\Pi_X,d_0,d_1)$, a rate-distortion pair $(R,D_1)$ is said to be {\em achievable} if, for all $\delta > 0$, there exists a sequence of codebooks (indexed by $n$) with $M \le e^{n(R+\delta)}$ codewords such that
    \begin{equation}
        \PP\bigg[ \frac{1}{n} d_1^n(\Xv,\Xvhat) \ge D_1 + \delta \bigg] \le \delta \label{eq:dist_criterion}
    \end{equation}
    for sufficiently large $n$ (depending on $\delta$), where $\Xv \sim \Pi_X^n$, and $\Xvhat$ is the resulting estimate.  Given $D_1$, the {\em mismatched rate-distortion function} $R^*(D_1)$ is defined to be the smallest $R$ such that $(R,D_1)$ is achievable, and given $R$, the {\em mismatched distortion-rate function} $D_1^*(R)$ is defined to be smallest $D_1$ such that $(R,D_1)$ is achievable.
\end{definition}

\noindent Before proceeding, we state the following useful result, which can be viewed as an analog of the result in Lemma \ref{lem:concave_gamma} concerning the input-constrained mismatch capacity.

\begin{lemma} \label{lem:convex_dist}
        {\em (Convexity of mismatched distortion-rate function)}
        Under the preceding setup, the function $D_1^*(R)$ is convex and non-increasing in $R$.
\end{lemma}

\noindent The monotonicity is immediate from Definition \ref{def:rate_dist}, and the convexity follows by a standard concatenation argument: Use a codebook achieving $D_1^*(R_1)$ for the first $\lambda n$ symbols, and a codebook achieving $D_1^*(R_2)$ for the last $(1-\lambda) n$ symbols, yielding a total distortion of $\lambda D_1^*(R_1) + (1-\lambda) D_1^*(R_2)$ at rate $\lambda R_1 + (1-\lambda) R_2$.

\begin{remark} \label{rem:rd_criteria}
    {\em (Expected vs.~high-probability distortion criteria)} The results that we present under the excess distortion criterion \eqref{eq:dist_criterion} also hold under an average distortion criterion of the form $\EE\big[ \frac{1}{n} d_1^n(\Xv,\Xvhat) \big] \le D_1 + \delta$ (as considered in certain results such as \cite[Thm.~3]{Bro12}).  The achievability result can be deduced from the high-probability result  by noting that \eqref{eq:dist_criterion} implies $\EE\big[ \frac{1}{n} d_1^n(\Xv,\Xvhat) \big] \le D_1 + \delta + \delta \dimax$, and $\delta$ can be scaled down by $1 + \dimax$ since it is arbitrarily small in Definition \ref{def:rate_dist}.  In Section \ref{sec:rd_multi_letter}, we will present a multi-letter converse result for the excess distortion criterion, but its proof will imply the same for the expected distortion criterion.
\end{remark}

\subsection{Discussion and Motivating Examples} \label{sec:rd_discussion}

Here we discuss two motivating examples where the mismatched rate-distortion problem may be of interest.

\paragraph{Unknown distortion measure.} 
The mismatched rate-distortion problem naturally arises in the case that it is not possible to know the true distortion measure $d_1(x,\xhat)$.  For instance, when it comes to compressing an image or audio signal,\footnote{These are mentioned for the sake of intuition, but it should be noted that the memoryless property and the assumption of additive distortion measures should not be expected to hold in such cases.} it is difficult to mathematically quantify what makes a reconstruction ``good''.  Nevertheless, one might still try to perform compression according to some heuristic choice, such as $d_0(x,\xhat) = (x - \xhat)^2$.  

Similarly, after designing a coding scheme for some distortion measure $d_0$, the system designer might change their mind and decide that a different measure $d_1$ is more suitable.  It is then of interest to understand how the performance of the original coding scheme degrades compared to a new scheme based on $d_1$.  If the loss is minimal, one may prefer to simply keep the mismatched design.

In both of these settings, the notion of mismatched rate-distortion region in Definition \ref{def:rate_dist} is questionable, as the optimal codebook depends on both $d_0$ and $d_1$.  Nevertheless, similarly to the channel coding setting, we will state the achievability results for a {\em fixed random coding distribution}, hence allowing us to assess the performance of random codebooks that are designed without knowledge of $d_1$.

\paragraph{Finite-precision arithmetic.} Suppose for concreteness that the true distortion measure is $d_1(x,\xhat) = (x-\xhat)^2$.  If the encoder is constrained to use finite-precision arithmetic, it may consist of first applying a scalar quantizer to $\xv$, and then choosing a similarly-quantized codeword $\xvhat$ from the codebook based on a finite-precision distortion measure.  For instance, letting $\Phi(x)$ be a scalar quantizer, and letting $\Xchat$ be a its range (i.e., a finite subset of $\RR$), one can consider the following distortion measure at the encoder, with $x \in \RR$ and $\xhat \in \Xchat$:
\begin{equation}
    d_0(x,\xhat) = (\Phi(x) - \xhat)^2.
\end{equation}
In this case, assuming the system model is known perfectly, the notion of mismatched rate-distortion region in Definition \ref{def:rate_dist} is justified: One knows that the system is constrained by the finite-precision arithmetic and scalar quantizer, and hence, one can design a codebook specifically targeted at combating this constraint.

\section{Achievability Result} \label{sec:rd_ach}

In this subsection, we provide an achievable distortion level for a given rate, i.e., an upper bound on $D_1^*(R)$.  The proof uses a form of constant-composition random coding, and in this sense, the result can be viewed as a counterpart to the LM rate for channel coding ({\em cf.}, Section \ref{sec:cc_LM}).  A similar analysis could also be performed for i.i.d.~random coding, but this is omitted for brevity.

We fix a reconstruction distribution $Q_{\Xhat} \in \Pc(\Xchat)$, let $Q_{\Xhat,n} \in \Pc_n(\Xchat)$ be a type with the same support as $Q_{\Xhat}$ such that $\|Q_{\Xhat,n} - Q_{\Xhat}\|_{\infty} \le \frac{1}{n}$, and draw each codeword $\Xvhat^{(j)}$ independently from
\begin{equation}
    P_{\Xvhat}(\xvhat) = \frac{1}{|\Tclass(Q_{\Xhat,n})|} \openone\big\{ \xvhat \in \Tclass(Q_{\Xhat,n}) \big\}, \label{eq:P_Xhat}
\end{equation}
where $\Tclass(Q_{\Xhat,n})$ is the type class corresponding to $Q_{\Xhat,n}$.  We have the following \cite{Lap97}.

\begin{theorem} \label{thm:rd_ach}
    {\em (Achievable mismatched distortion-rate function)}
    Under the mismatched rate-distortion setup with a discrete memoryless source $\Pi_X$ and distortion measures $(d_0,d_1)$, the following distortion is achievable at rate $R$ via constant-composition random coding with an auxiliary distribution $Q_{\Xhat} \in \Pc(\Xchat)$:
    \begin{gather}
        \Dibar(Q_{\Xhat},R) = \max_{\Ptilde_{X\Xhat} \in \Pctilde} \EE_{\Ptilde}[ d_1(X,\Xhat) ], \label{eq:D1bar}
    \end{gather}
    where
    \begin{align}\label{eq:setF}
         \Pctilde = \Bigg\{ \Ptilde_{X\Xhat} \,:\, \Ptilde_{X\Xhat} \in \argmin_{ \substack{ \Ptilde_{X\Xhat} \,:\, \Ptilde_X = \Pi_X, \Ptilde_{\Xhat} = Q_{\Xhat}, \\ I_{\Ptilde}(X;\Xhat) \le R } } \EE_{\Ptilde}[d_0(X,\Xhat)]\Bigg\}.
    \end{align}
    Consequently, we have $D_1^*(R) \le \min_{Q_{\Xhat}} \Dibar(Q_{\Xhat},R)$.
\end{theorem}
The proof is given in Section \ref{sec:pf_rd_ach}.  The result can be understood intuitively by interpreting $\Ptilde_{X\Xhat}$ as the joint empirical distribution of $(\xv,\xvhat)$.  The distortion incurred is $d_1^n(\xv,\xvhat) = n\EE_{\Ptilde}[ d_1(X,\Xhat) ]$, and by the nature of the encoding function, the joint type is one that minimizes $\EE_P[d_0(X,\Xhat)]$.  The constraint $\Ptilde_X = \Pi_X$ follows by a basic typicality argument, the constraint $\Ptilde_{\Xhat} = Q_{\Xhat}$ arises from the constant-composition nature of the codebook, and the constraint $I_{\Ptilde}(X;\Xhat) \le R$ arises because the probability of {\em any} random codeword inducing $I_{\Ptilde}(X;\Xhat) > R$ is negligible.  Finally, the maximum in \eqref{eq:D1bar} arises by treating ties in a ``worst-case'' manner, i.e., assuming that when a tie occurs, the selected sequence is the one that maximizes $d_1$.

In the case that $d_0 = d_1$, all elements of $\Pctilde$ provide the same distortion, and the bound of Theorem \ref{thm:rd_ach} simplifies to
\begin{equation}
\DibarM(Q_{\Xhat},R) = \min_{\substack{\Ptilde_{X\Xhat} \,:\, \Ptilde_X = \Pi_X, \Ptilde_{\Xhat} = Q_{\Xhat}, \\  I_{\Ptilde}(X;\Xhat) \le R}} \EE_{\Ptilde}[d_1(X,\Xhat)]. \label{eq:Dbar_matched}
\end{equation}
This corresponds to the usual distortion-rate function \cite[Ch.~10]{Cov06}, but with an additional constraint on the $\Xhat$-marginal.  Moreover, this constraint can be removed by simply taking the minimum over all $Q_{\Xhat}$, since the latter is a free parameter.

We briefly note that Theorem \ref{thm:rd_ach} is only stated and proved for discrete memoryless sources, and handling continuous sources is a potentially interesting open problem.

\section{Examples} \label{sec:rd_examples}

In this subsection, we provide two examples taken from \cite{Lap97}.

\subsection{Parallel Binary Sources} \label{sec:rd_parallel}

We begin with an analytical example showing that the achievable distortion of Theorem \ref{thm:rd_ach} is not tight, i.e., it can be strictly higher than the mismatched distortion-rate function.  This example is closely related to the parallel channel example of Section \ref{sec:parallel}, and leads to analogous observations.  In fact, the analysis is also similar, and we therefore omit some repeated details.

The source and reconstruction alphabets are given by $\Xc = \Xchat = \{0,1\}^2$, and we write $X = (X_1,X_2)$ and $\Xhat = (\Xhat_1,\Xhat_2)$ to denote the corresponding binary pairs.  We let $\Pi_X$ be uniform, so that $\Pi_X(x) = \frac{1}{4}$ for all $x$.  For $x = (x_1,x_2)$ and $\xhat = (\xhat_1,\xhat_2)$, the distortion measures are given as follows for some $\lambda \in [0,1]$:
\begin{align}
    d_0(x,\xhat) &= \lambda \openone\{ x_1 \ne \xhat_1 \} + (1-\lambda) \openone\{ x_2 \ne \xhat_2 \}, \label{eq:d0_parallel} \\
    d_1(x,\xhat) &= \frac{1}{2} \openone\{ x_1 \ne \xhat_1 \} + \frac{1}{2} \openone\{ x_2 \ne \xhat_2 \}. \label{eq:d1_parallel} 
\end{align}
Hence, the mismatched encoder treats the reconstructions of $x_1$ and $x_2$ with different weights, even though the two are treated equally under the true distortion measure.

We set $Q_{\Xhat}$ in Theorem \ref{thm:rd_ach} to be the uniform distribution.  To understand the set $\Pctilde$ in \eqref{eq:setF}, we consider the mutual information constraint $I_{\Ptilde}(X;\Xhat) \le R$.  From the analysis in Section  \ref{sec:parallel}, we know that
\begin{equation}
    I_{\Ptilde}(X;\Xhat) \ge \big( 1 - H_2(\deltatilde_1) \big) + \big( 1 - H_2(\deltatilde_2) \big),
\end{equation}
where $\deltatilde_1$ and $\deltatilde_2$ are transition probabilities associated with the marginals $\Ptilde_{X_1\Xhat_1}$ and $\Ptilde_{X_2\Xhat_2}$.  From the choice of $d_0$ in \eqref{eq:d0_parallel}, the corresponding distortion incurred is $\lambda\deltatilde_1 + (1-\lambda)\deltatilde_2$, and we have
\begin{align}
    &\min_{ \Ptilde_{X\Xhat} \,:\, \Ptilde_X = \Pi_X, \Ptilde_{\Xhat} = Q_{\Xhat}, I_{\Ptilde}(X;\Xhat) \le R } \EE_P[d_0(X,\Xhat)] \nonumber \\
        &\quad \ge \min_{\deltatilde_1,\deltatilde_2\,:\,( 1 - H_2(\deltatilde_1)) + ( 1 - H_2(\deltatilde_2)) \le R}  \lambda\deltatilde_1 + (1-\lambda)\deltatilde_2. \label{eq:rd_equivalence}
\end{align}
In fact, this lower bound holds with equality, since it is attained when $\Ptilde_{\Xhat_1|X_1}$ and $\Ptilde_{\Xhat_2|X_2}$ are independent BSCs with crossover probabilities $(\deltatilde_1^*,\deltatilde_2^*)$ achieving the minimum on the right-hand side.  Moreover, it can be shown that the minimizer is unique on both sides of \eqref{eq:rd_equivalence}.  Under the optimal parameters $(\deltatilde_1^*,\deltatilde_2^*)$ attained in \eqref{eq:rd_equivalence}, the resulting distortion incurred in \eqref{eq:D1bar} is
\begin{equation}
    \Dibar(Q_{\Xhat},R) = \frac{1}{2}\big( \deltatilde_1^* + \deltatilde_2^* \big) \label{eq:parallel_d1}
\end{equation}
due to the choice of $d_1$ in \eqref{eq:d1_parallel}.

Next, since the matched rate-distortion function of a binary symmetric source with Hamming distortion is $\RBSS(D) = 1 - H_2(D)$ bits/symbol \cite[Sec.~10.3.1]{Cov06}, the matched rate-distortion function of the parallel source under consideration is $\Rmatched(D_1) = 2(1 - H_2(D_1))$.  Hence, for a given rate $R$, the distortion $D_1$ incurred is the solution to $R = 2(1 - H_2(D_1))$.  We claim that, in fact, the {\em mismatched} rate-distortion function $D_1^*(R)$ is identical to this matched one.  This can be verified by generating a separate codebook for each source, and letting the overall codebook be the product of the two codebooks generated.  The easiest way to see that this achieves the matched performance is to note that $d_0$ and $d_1$ are equivalent under such a product codebook structure.

\begin{figure}
    \begin{centering}
        \includegraphics[width=0.7\columnwidth]{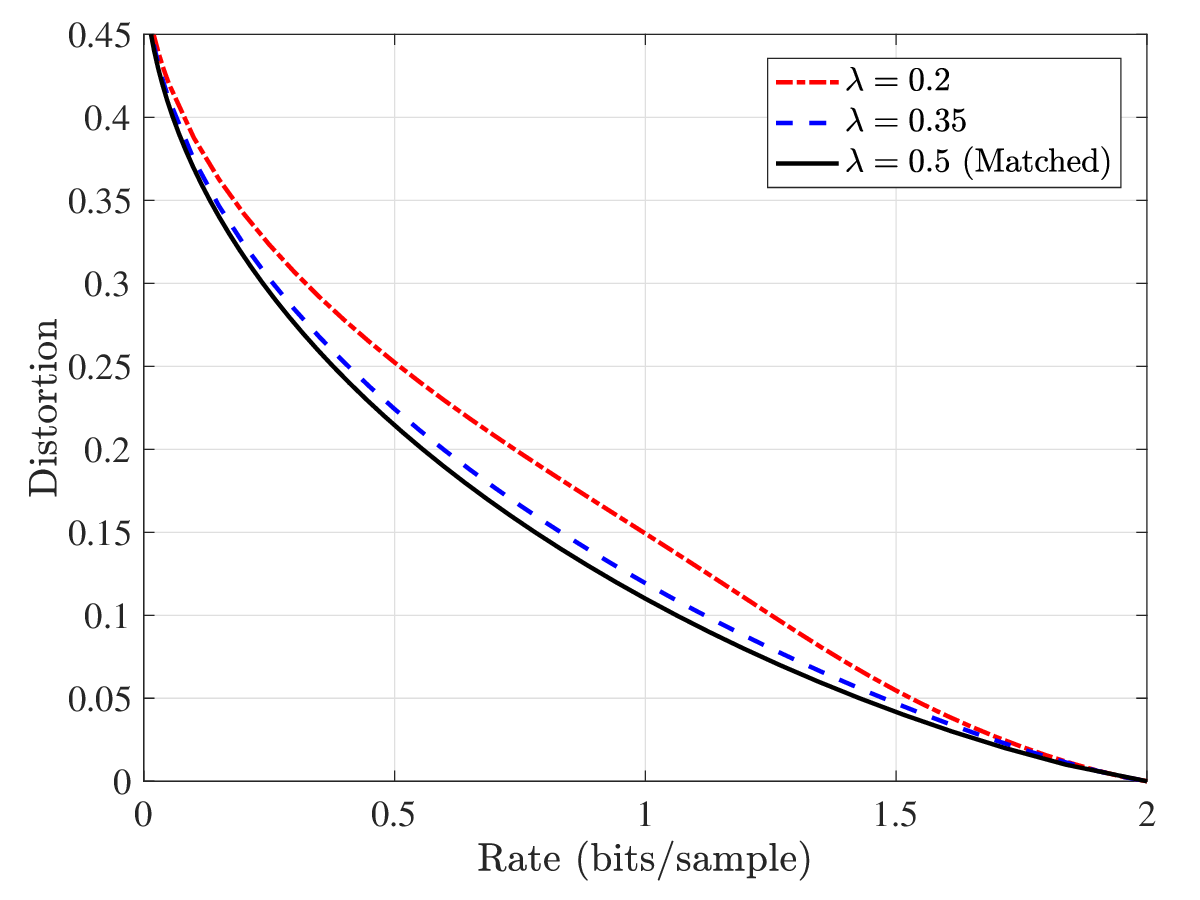}
        \par
    \end{centering}
    
    \caption{Parallel source example: Comparison of matched ($\lambda = 0.5$) and mismatched ($\lambda = 0.2, 0.35$) rate-distortion curves resulting from Theorem \ref{thm:rd_ach}, in the case that the source and the auxiliary distribution are both uniform, i.e., $\Pi_X = Q_{\Xhat} = \big(\frac{1}{4}, \frac{1}{4}, \frac{1}{4}, \frac{1}{4}\big)$. } \label{fig:ParallelRD}
\end{figure}

Summarizing the above, the mismatched distortion rate function yields the optimal value $D_1^*$ satisfying $R = 2(1 - H_2(D_1^*))$, whereas the bound of Theorem \ref{thm:rd_ach} yields the distortion $\frac{1}{2}( \deltatilde_1^* + \deltatilde_2^*)$ under the pair $(\deltatilde_1,\deltatilde_2)$ minimizing \eqref{eq:rd_equivalence}.  Since the latter constrains $(1 - H_2(\deltatilde_1)) + ( 1 - H_2(\deltatilde_2)) \le R$, we can use the strict concavity of $H_2(\cdot)$ to deduce that $\frac{1}{2}( \deltatilde_1^* + \deltatilde_2^*) \ge D_1^*$, with strict inequality when $\delta_1^* \ne \delta_2^*$.  Finally, one can verify that $\delta_1^* \ne \delta_2^*$ whenever $\lambda \ne \frac{1}{2}$ in \eqref{eq:d0_parallel}; see Figure \ref{fig:ParallelRD} for an illustrative numerical example.  Hence, we conclude that Theorem \ref{thm:rd_ach} fails to achieve the mismatched distortion-rate function in such cases.  While we only showed this for uniform $Q_{\Xhat}$, any other choice would be suboptimal even under matched encoding.

This example suggests that Theorem \ref{thm:rd_ach} could be improved via multi-user coding techniques, analogously to the studies of multi-user coding techniques for single-user channel coding surveyed in Section \ref{ch:multiuser}.  To our knowledge, this idea is yet to be investigated beyond the present example.
 
\subsection{Ternary Symmetric Source}

We consider an equiprobable ternary source: $\Xc = \{0,1,2\}$, and $\Pi_X(x) = \frac{1}{3}$ for each $x$.  The distortion measures are given by
\begin{gather}
    d_0(x,\xhat) = \openone\{ x \ne \xhat \}, \\
    d_1(x,\xhat) = \openone\{ x \ne \xhat \cap x \ne 2 \}.
\end{gather}
Hence, the mismatched encoder uses the Hamming distance, whereas under the true distortion measure, no penalty is incurred for $x=2$.

The analysis given in \cite{Lap97} is outlined as follows:
\begin{itemize}
    \item Under the uniform distribution $Q_{\Xhat} = \big(\frac{1}{3},\frac{1}{3},\frac{1}{3}\big)$, the unique minimizer $\Ptilde_{\Xhat|X}$ in \eqref{eq:setF} is a ternary symmetric channel.  Moreover, the transition probability $\delta^*$ between differing symbols is such that $I_{\Ptilde}(X;\Xhat) = R$ (if $R \in [0,\log_2 3]$).  The resulting distortion is $\EE_{\Ptilde}[d_1(X,\Xhat)] = \frac{2}{3} \cdot 2\delta^*$.
    \item Under the distribution  $Q_{\Xhat} = \big(\frac{1}{2},\frac{1}{2},0\big)$, the unique minimizer $\Ptilde_{\Xhat|X}$ in \eqref{eq:setF} maps $x=2$ to $\{0,1\}$ with probability $\frac{1}{2}$ each, and maps $\{0,1\}$ to $\{0,1\}$ according to a BSC .  The crossover probability $\delta^*$ is such that $\frac{2}{3}\big( 1 - H_2(\delta^*) \big) = R$ (if $R \in \big[0,\frac{2}{3}\big]$), and the resulting distortion is $\EE_{\Ptilde}[d_1(X,\Xhat)] = \frac{2}{3} \cdot \delta^*$.
\end{itemize}
The first of these choices of $Q_{\Xhat}$ would be optimal if the true distortion were also $d_0$, whereas the second is optimal under $d_1$.

\begin{figure}
    \begin{centering}
        \includegraphics[width=0.7\columnwidth]{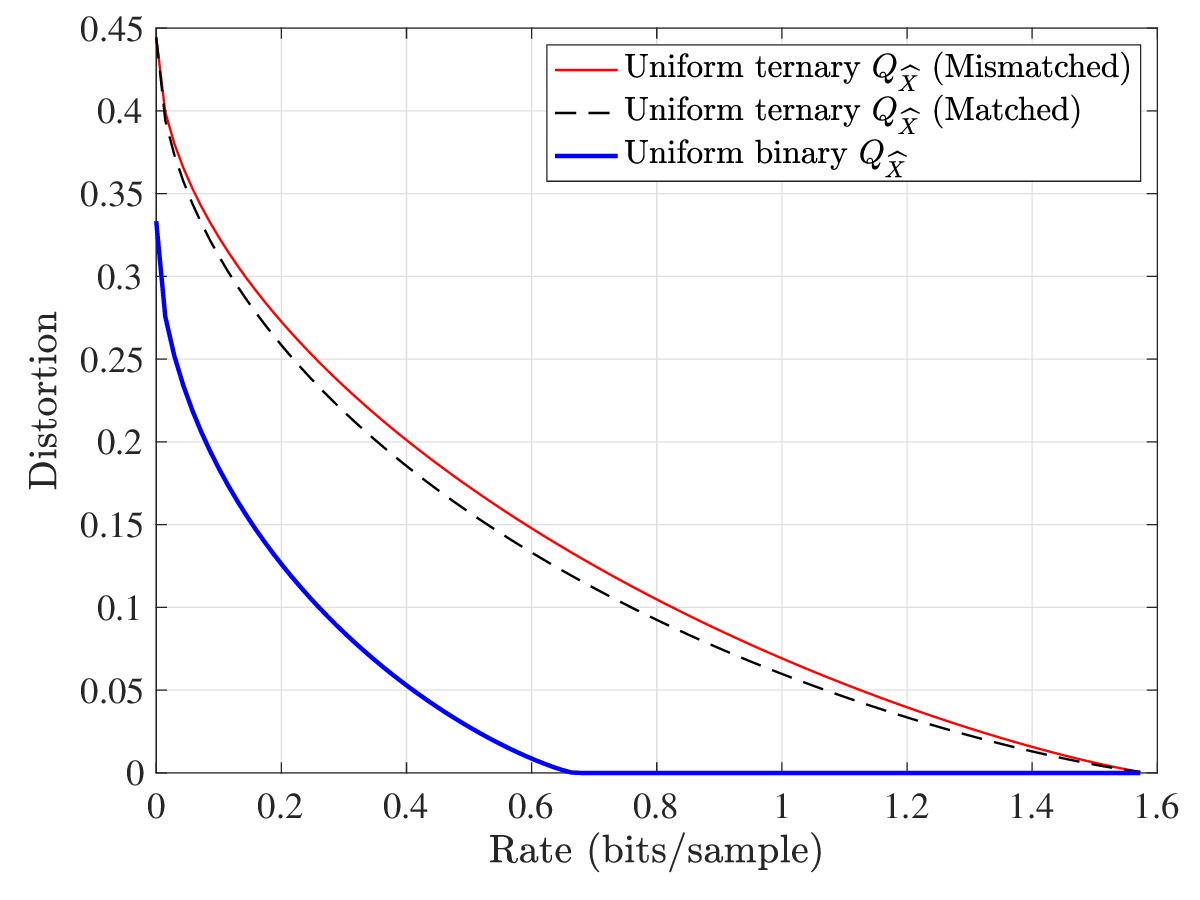}
        \par
    \end{centering}
    
    \caption{Ternary symmetric source example: Upper bounds on the mismatched distortion-rate function under the auxiliary distributions $Q_{\Xhat} = \big(\frac{1}{3}, \frac{1}{3}, \frac{1}{3}\big)$ and $Q_{\Xhat} = \big(\frac{1}{2}, \frac{1}{2}, 0\big)$. } \label{fig:RateDistExample}
\end{figure}

The achievable rate-distortion curves are plotted in Figure \ref{fig:RateDistExample}.  For the case $Q_{\Xhat} = \big(\frac{1}{3},\frac{1}{3},\frac{1}{3}\big)$, we also plot the {\em matched} function given in \eqref{eq:Dbar_matched}, whereas for $Q_{\Xhat} = \big(\frac{1}{2},\frac{1}{2},0\big)$ we found the matched and mismatched curves to coincide.

\section{Achievability Proof} \label{sec:pf_rd_ach}

In this subsection, we prove Theorem \ref{thm:rd_ach}.  We make use of the method of types, including some of the properties stated in Section \ref{sec:prop_types}.

Recall that we generate the codewords $\{\Xvhat^{(m)}\}_{m=1}^M$ independently according to the constant-composition codeword distribution $P_{\Xvhat}$ in \eqref{eq:P_Xhat}.  In accordance with Definition \ref{def:rate_dist}, we would like to show that the random-coding error probability
\begin{equation}
    \pebar \triangleq \PP\bigg[ \frac{1}{n} d_1^n(\Xv,\Xvhat) \ge D_1 + \delta \bigg]
\end{equation}
vanishes as $n\to\infty$, for arbitrarily small $\delta > 0$.  Overloading the notation, we first consider the conditional error probability
\begin{equation}    
    \pebar(\xv) \triangleq \PP\bigg[ \frac{1}{n} d_1^n(\xv,\Xvhat) \ge D_1 + \delta \,\Big|\, \Xv = \xv \bigg].
\end{equation}
In the following, for a given type $P_X \in \Pc_n(\Xc)$ associated with $\xv$, we consider the possible joint types $\Ptilde_{X\Xhat} \in \Pc_n(\Xc \times \Xchat)$ induced between $\xv$ and the codewords $\{\Xvhat^{(j)}\}_{j=1}^M$. 

\begin{lemma} \label{lem:rd_types}
    {\em (Occurrences of joint types)}
    Consider a random codebook $\Cc_n = \{\Xvhat^{(1)},\dotsc,\Xvhat^{(M)}\}$ with codewords of length $n$ drawn independently from $P_{\Xvhat}$ in \eqref{eq:P_Xhat}. For any $\delta' > 0$, conditioned on any $\Xv = \xv$ with $\xv \in \Tclass(P_X)$, the following statements simultaneously hold with probability approaching one as $n \to \infty$:
    \begin{enumerate}
        \item For all $\xvhat \in \Cc_n$, if $(\xv,\xvhat) \in \Tclass(\Ptilde_{X\Xhat})$ for some $\Ptilde_{X\Xhat} \in \Pc_n(\Xc \times \Xchat)$, then it must hold that $\Ptilde_{X} = P_X$ and  $\Ptilde_{\Xhat} = Q_{\Xhat,n}$.
        \item Defining the set
            \begin{align}
                \Sc^{(-)}_n(P_X) = &\Big\{ \Ptilde_{X\Xhat} \in \Pc_n(\Xc \times \Xchat) \,:\, \nonumber \\ &\Ptilde_{X} = P_X, \Ptilde_{\Xhat} = Q_{\Xhat,n}, I_{\Ptilde}(X;\Xhat) \ge R + \delta' \Big\},
            \end{align}
            there does not exist a codeword $\xvhat \in \Cc_n$ such that $(\xv,\xvhat) \in \Tclass(\Ptilde_{X\Xhat})$ with $\Ptilde_{X\Xhat} \in\Sc_n^{(-)}(P_X)$. 
            \item Conversely, for any joint type in the set
            \begin{align}
                \Sc^{(+)}_n(P_X) = &\Big\{ \Ptilde_{X\Xhat} \in \Pc_n(\Xc \times \Xchat) \,:\, \nonumber \\ &\Ptilde_{X} = P_X, \Ptilde_{\Xhat} = Q_{\Xhat,n}, I_{\Ptilde}(X;\Xhat) \le R - \delta' \Big\}, 
            \end{align}
            there exists at least one codeword $\xvhat \in \Cc_n$ for which $(\xv,\xvhat) \in \Tclass(\Ptilde_{X\Xhat})$.
    \end{enumerate}
\end{lemma}
\begin{proof}
    The first claim follows trivially from the assumption that $\xv \in \Tclass(P_X)$ and the fact that we are using constant-composition random coding according to the type $Q_{\Xhat,n}$.
    
    The second and third claims are based on the following expression for the probability that there exists at least one codeword inducing a given joint type $\Ptilde_{X\Xhat}$:
    \begin{equation}
        \PP\bigg[ \bigcup_{j=1}^M \big\{ (\xv,\Xvhat^{(j)}) \in \Tclass(\Ptilde_{X\Xhat}) \big\}\bigg] = 1 - \Big( 1 - \PP\big[ (\xv,\Xvhat) \in \Tclass(\Ptilde_{X\Xhat})  \big] \Big)^M, \label{eq:rd_exact}
    \end{equation}
    where $\Xvhat \sim P_{\Xvhat}$ is independent of $\xv$.  We fix $\epsilon > 0$ and note the following:
    \begin{itemize}
        \item If $\PP\big[ (\xv,\Xvhat) \in \Tclass(\Ptilde_{X\Xhat})  \big] \ge e^{-n(R-\epsilon)}$, then the right-hand side of \eqref{eq:rd_exact} is lower bounded by $1 - (1 - e^{-n(R-\epsilon)})^{e^{nR}} = 1 - \big( (1 - e^{-n(R-\epsilon)})^{e^{n(R-\epsilon)}} \big)^{e^{n\epsilon}} $, which tends to one faster than exponentially since $(1 - e^{-n(R-\epsilon)})^{e^{n(R-\epsilon)}} \to \frac{1}{e}$.
        \item  If $\PP\big[ (\xv,\Xvhat) \in \Tclass(\Ptilde_{X\Xhat})  \big] \le e^{-n(R+\epsilon)}$, then the right-hand side of \eqref{eq:rd_exact} is upper bounded by $1 - (1 - e^{-n(R+\epsilon)})^{e^{nR}} = 1 - \big( (1 - e^{-n(R+\epsilon)})^{e^{n(R+\epsilon)}} \big)^{e^{-n\epsilon}}$, which tends to zero exponentially fast since $(1 - e^{-n(R+\epsilon)})^{e^{n(R+\epsilon)}} \to \frac{1}{e}$, and since for fixed $\alpha > 0 $ it holds that $\big(\frac{1}{\alpha}\big)^z= 1 - z\log\alpha + O(z^2)$ as $z \to 0$.
    \end{itemize}
    The second and third claims of the lemma now follow by combining the above calculations with the fact that $\PP\big[ (\xv,\Xvhat) \in \Tclass(\Ptilde_{X\Xhat}) \big]$ behaves as $e^{-n I_{\Ptilde}(X;\Xhat)}$ times a sub-exponential pre-factor ({\em cf.}, \eqref{eq:type_prop_cc}), taking the union bound over all possible joint types (of which there are only polynomially many), and noting that $\epsilon$ can be arbitrarily small.
\end{proof}

The remainder of the proof of Theorem \ref{thm:rd_ach} contains rather technical continuity arguments; we only present an outline of the arguments here, and refer the interested reader to \cite{Lap97} for further details.

Letting $\Sc_n(P_X)$ be the set of {\em all} joint types of $(\xv,\xvhat)$ induced by some $\xvhat \in \Cc_n$, Lemma \ref{lem:rd_types} states that with high probability, $\Sc^{(+)}_n(P_X) \subseteq \Sc_n(P_X)$ and $\Sc^{(-)}_n(P_X) \cap \Sc_n(P_X) = \emptyset$.  Moreover, by the first part of the lemma, $\Sc^{(-)}$ and $\Sc^{(+)}$ collectively account for all possible joint types except those for which $I_{\Ptilde}(X;\Xhat) \in (R-\delta',R+\delta')$. 

By construction, the encoder maps $\xv$ to an index $m \in \{1,\dotsc,M\}$ such that the corresponding joint type $\Ptilde_{X\Xhat}$ has the smallest possible value of $\EE_{\Ptilde}[ d_0(X,\Xhat) ]$.  In addition, we have $Q_{\Xhat,n} \to Q_{\Xhat}$ by definition, and $P_X \to \Pi_X$ by the law of large numbers.  Hence, and taking $\delta' \to 0$ in Lemma \ref{lem:rd_types}, we can deduce that the joint type $\Ptilde_{X\Xhat}$ of $(\Xv,\Xvhat)$ must be arbitrarily close to a minimizer of $\EE_{\Ptilde}[d_0(X,\Xhat)]$ subject to $\Ptilde_X = \Pi_X$, $\Ptilde_{\Xhat} = Q_{\Xhat}$, and $I_{\Ptilde}(X;\Xhat) \le R$.  Notice that these constraints coincide with those in \eqref{eq:setF}.  In particular, the normalized distortion incurred is $\frac{1}{n} d_1^n(\xv,\Xvhat) = \EE_{\Ptilde}[ d_1(X,\Xhat) ]$, and is therefore upper bounded by $\max_{\Ptilde_{X\Xhat} \in \Pctilde} \EE_{\Ptilde}[ d_1(X,\Xhat) ] + \delta$ for arbitrarily small $\delta > 0$.  

The proof is completed by recalling the definition of an achievable $(R,D)$ pair in Definition \ref{def:rate_dist}, and noting that we have only relied on the high-probability events in Lemma \ref{lem:rd_types} as well as $P_X \to \Pi_X$.

\section{Multi-Letter Improvement and Converse} \label{sec:rd_multi_letter}

We saw in Section \ref{sec:rd_examples} that the upper bound $\Dibar(Q_{\Xhat},R)$ is not tight in general, i.e., it can be strictly higher than the mismatched distortion-rate function $D_1^*(R)$, even after optimizing $Q_{\Xhat}$.   As with the LM rate in channel coding ({\em cf.}, Section \ref{sec:multi_LM}), one can attain improved rates by coding over blocks of $k$ symbols.  Applying Theorem \ref{thm:rd_ach} to the source $\Pi_X^k$, optimizing the input distribution $Q_{\Xhat^{k}}$ on the product alphabet $\Xchat^k$, and normalizing the rate and the distortion incurred by $k$, we obtain an achievability result for the original source.  Specifically, letting $\Dibar(Q_{\Xhat},R,\Pi_X)$ denote \eqref{eq:D1bar} with an explicit dependence on the source, the achievable distortion is given by
\begin{equation}
    \Dibar^{(k)}(R) \triangleq \min_{Q_{\Xhat^k}} \frac{1}{k} \Dibar( Q_{\Xhat^k}, kR, \Pi_X^k ). \label{eq:rd_multi_letter}
\end{equation}
Since the choice of $k$ is arbitrary, it follows that the distortion
\begin{equation}
    \Dibar^{(\infty)}(R) \triangleq \inf_{k \in \ZZ_{>0}} \Dibar^{(k)}(R)
\end{equation}
is also achievable at rate $R$.

While the tightness of the product extension of the LM rate for channel coding (i.e., the Csisz\'ar-Narayan conjecture) remains an open problem, the following theorem reveals that the analogous statement is indeed true for the mismatched rate-distortion problem, at least up to tie-breaking issues \cite{Lap97}.  Specifically, it is true when we assume a {\em pessimistic} view of tie-breaking: If multiple codewords achieve the minimum in \eqref{eq:rd_m}, then the selected codeword is the one with the highest $d_1^n(\xv,\xvhat)$.  If there is still a tie among these, then it makes no difference which one is chosen.

\begin{theorem}
    {\em (Multi-letter converse)}
    Under the mismatched rate-distortion problem specified by $(\Pi,d_0,d_1)$, under pessimistic tie-breaking, the mismatched distortion-rate function satisfies
    \begin{equation}
        D_1^*(R) = \Dibar^{(\infty)}(R).
    \end{equation}
\end{theorem}
\begin{proof}
    We have already established that the distortion $\Dibar^{(\infty)}(R) = \inf_{k \in \ZZ_{>0}} \Dibar^{(k)}(R)$ is achievable, i.e., it is an upper bound on $D_1^*(R)$.  Since the distortion-rate function is continuous in $R$ by Lemma \ref{lem:convex_dist}, it suffices to show that for any $\delta > 0$, we can find an integer $k$ such that $\Dibar^{(k)}(R+\delta) \le D_1^*(R) + c\delta$, where $c > 0$ is a constant not depending on $\delta$.  From \eqref{eq:rd_multi_letter}, such a statement is equivalent to showing that we can find an auxiliary distribution $Q_{\Xhat^k}$ such that
    \begin{equation}
        \frac{1}{k}\Dibar(Q_{\Xhat^k},k(R+\delta),\Pi_X^k) \le D_1^*(R) + c\delta. \label{eq:rd_conv_goal}
    \end{equation}
    By the definition of the mismatched distortion-rate function $D_1^*(R)$ (Definition \ref{def:rate_dist}), for any $\delta > 0$, there exists a sufficiently large block length $k$ and a codebook $\Cc_k = \{ \xvhat^{(1)}, \dotsc, \xvhat^{(M)} \}$ with $M \le e^{k(R+\delta)}$, such that $\frac{1}{k} d_1^{k}(\Xv,\Xvhat) \le D_1^*(R)+\delta$ with probability at least $1-\delta$,\footnote{Here we use $k$ instead of $n$ for the block length to highlight the fact that we consider $\Dibar^{(k)}(R)$ in \eqref{eq:rd_multi_letter} and choose $Q_{\Xhat^k}$ depending on $\Cc_k$.} where $\Xv \sim \Pi_X^k$ and $\Xvhat$ is the resulting estimate.
    Recalling the definition $\dimax = \max_{x,\xhat} d_1(x,\xhat)$ and the non-negativity of $d_1$, we deduce that
    \begin{equation}
        \frac{1}{k} \EE\big[ d_1^{k}(\Xv,\Xvhat) \big] \le D_1^*(R) + \delta(1 + \dimax), \label{eq:d1_avg_ub}
    \end{equation}
    which follows by upper bounding the distortion by $\dimax$ in the case that $\frac{1}{k} d_1^{k}(\Xv,\Xvhat) > D_1^*(R)+\delta$.

    In the following, we denote the mapping from source sequences to compressed sequences under the codebook $\Cc_k$ as $\phi^*(\cdot)$, i.e., $\Xvhat = \phi^*(\Xv)$.  In addition, we let $Q_{\Xhat^{k}}$ be the resulting marginal distribution of $\Xvhat \in \Xchat^k$, corresponding to randomly generating $\Xv \sim \Pi_X^k$ and then setting $\Xvhat = \phi^*(X^k)$.
    
    We now consider the achievable distortion level of Theorem \ref{thm:rd_ach} applied to the multi-letter source $\Pi_X^{k}$:
    \begin{equation}
        \Dibar( Q_{\Xhat^{k}}, k(R + \delta), \Pi_X^{k} ) = \max_{\Ptilde_{X^k\Xhat^k} \in \Pctilde_k} \EE_{\Ptilde}[ d_1^k(X^k,\Xhat^k) ], \label{eq:Dibar_k}
    \end{equation}
    where, in accordance with \eqref{eq:setF}, we have 
    \begin{equation}
        \Pctilde_k = \bigg\{\Ptilde_{X^k\Xhat^k} \,:\, \Ptilde_{X^k\Xhat^k} \in \argmin_{ \substack{ \Ptilde_{X^k\Xhat^k} \,:\, \Ptilde_{X^k} = \Pi_X^k, \Ptilde_{\Xhat^k} = Q_{\Xhat^k}, \\  I_{\Ptilde}(X^k;\Xhat^k) \le k(R+\delta) }} \EE_{\Ptilde}[d_0^k(X^k,\Xhat^k)] \bigg\}.
    \end{equation}
    Note that the $\Xhat$-marginal constraint ensures that $P_{\Xhat^k}$ takes mass only on the elements of $\Cc_k$, of which there are at most $e^{k(R+\delta)}$.  It follows that $I_{\Ptilde}(X^k;\Xhat^k) \le H_{\Ptilde}(\Xhat^k) \le k(R+\delta)$, and hence the rate constraint is redundant (i.e., it is automatically satisfied).  
    
    Since $\phi^*(x^k)$ is a mismatched encoding rule minimizing $d_0$, the joint distribution resulting from $\Ptilde_{\Xhat^k|X^k}(\xhat^k | x^k) = \openone\{ \xhat^k = \phi^*(x^k) \}$ must be in the set $\Pctilde_k$.  Moreover, among {\em all} conditional distributions $\Ptilde_{\Xhat^k|X^k}$ producing the required marginal $Q_{\Xhat^k}$ supported on $\Cc_k$, we claim that the one induced by $\phi^*(x^k)$ achieves the maximum in \eqref{eq:Dibar_k}.  Indeed, this follows directly from our assumption of pessimistic tie-breaking -- if two codewords achieve the same $d_0$, then the one with the highest $d_1$ is selected under $\phi^*$.
    
    As a result, using the bound on the average distortion under $\Cc_k$ in \eqref{eq:d1_avg_ub}, we deduce from \eqref{eq:Dibar_k} that
    \begin{equation}
        \frac{1}{k} \Dibar( Q_{\Xhat^{k}}, k(R + \delta), \Pi_X^{k} ) \le D_1^*(R) + \delta(1 + \dimax).
    \end{equation}
    We have therefore established \eqref{eq:rd_conv_goal} (with $c = 1+\dimax$), as required.
\end{proof}

\section{Mismatched Random Codebooks with Optimal Encoding} \label{sec:rd_mm_random}

The preceding subsections considered the mismatched encoding problem in which the encoder chooses the minimum-distortion codeword with respect to $d_0$, but the actual distortion measure is $d_1$.  In this subsection, we turn to a different form of mismatch in which a common distortion measure $d$ is adopted throughout the system, but a suboptimal random coding distribution is used to generate the codebook.  We are interested in the rate-distortion trade-off for the given random coding ensemble. 
This problem was studied in the case of finite alphabets in \cite{Yan98}, and more general alphabets in \cite{Yan99}.  We refer the reader to \cite{Dem02} for a more detailed survey, and discuss some other notions of mismatch and universality in Section \ref{sec:rd_other}.

We focus primarily on the finite-alphabet setting in this subsection, while occasionally providing references to more general results.  In addition, Section \ref{sec:gaussian_rd} will consider a specific case of significant interest with continuous alphabets, namely, Gaussian codebooks for non-Gaussian sources.

\subsection{Achievable Rate-Distortion Functions}

We consider the setup of Section \ref{sec:rd_setup} with a common distortion measure $d_0 = d_1 = d$, assumed to be non-negative and finite-valued.  In this case, the encoder in \eqref{eq:rd_m} becomes the optimal minimum-distortion rule, which is repeated as follows for convenience: 
\begin{equation}
    m = \argmin_{j = 1,\dotsc,M} d^n(\xv,\xvhat^{(j)}), \quad d^n(\xv,\xvhat) = \sum_{i=1}^n d(x_i,\xhat_i). \label{eq:rd_m2}
\end{equation}
Here the distortion incurred is identical for all tie-breaking strategies.

If the codebook is optimized in accordance with Definition \ref{def:rate_dist}, then the rate-distortion trade-off is characterized by the classical solution of Shannon \cite{Sha59}, stated as follows: 
\begin{equation}
    \Rmatched(D) = \min_{\substack{\Ptilde_{X\Xhat} \,:\, \Ptilde_X = \Pi_X \\\EE_{\Ptilde}[d(X,\Xhat)] \le D }} I_{\Ptilde}(X;\Xhat). \label{eq:rd_matched}
\end{equation}
Here and subsequently, we find it more convenient to consider the rate as a function of the distortion, but the distortion as a function of the rate follows by a simple inversion.  For instance, \eqref{eq:rd_matched} yields the following distortion-rate function:
\begin{equation}
    \Dmatched(R) = \min_{\substack{\Ptilde_{X\Xhat} \,:\, \Ptilde_X = \Pi_X \\ I_{\Ptilde}(X;\Xhat) \le R }} \EE_{\Ptilde}[d(X,\Xhat)]. \label{eq:dr_matched}
\end{equation}

In analogy with the channel coding results surveyed in Section \ref{ch:single_user}, we consider two random-coding ensembles, both of which consist of generating a codebook $\Cc = \{\Xvhat^{(1)},\dotsc,\Xvhat^{(M)}\}$ containing codewords drawn independently from some distribution $P_{\Xvhat}$.  Under the i.i.d.~ensemble, we set $P_{\Xvhat}(\xvhat) = Q_{\Xhat}^n$, and under the constant-composition ensemble, $P_{\Xvhat}$ is as given in \eqref{eq:P_Xhat}.  Similarly to Definition \ref{def:rate_dist}, we are interested in the random-coding error probability
\begin{equation}
    \pebar(n,M,D,Q_{\Xhat},\delta) \triangleq \PP\bigg[ \frac{1}{n} d^n(\Xv,\Xvhat) \ge D + \delta \bigg], \label{eq:rd_pe_bar}
\end{equation}
where the average is with respect to both $\Xv \sim \Pi_X^n$ and the random codebook of size $M$.  We say that a rate-distortion pair $(R,D)$ is achievable if $\pebar(n,\lfloor e^{n(R+\delta)} \rfloor,D,Q_{\Xhat},\delta) \to 0$ for any fixed $\delta > 0$.

The achievable rate-distortion functions for these ensembles are introduced as follows, and will be accompanied by a formal theorem below.  For the i.i.d.~ensemble, the primal form is given by 
\begin{equation}
    \Rbariid(D,Q_{\Xhat}) = \min_{\substack{\Ptilde_{X\Xhat} \,:\, \Ptilde_X = \Pi_X, \\\EE_{\Ptilde}[d(X,\Xhat)] \le D }} D(\Ptilde_{X\Xhat} \| \Pi_X \times Q_{\Xhat}), \label{eq:Rbar_iid}
\end{equation}
and the dual form is given by
\begin{equation}
    \Rbariid(D,Q_{\Xhat}) =\sup_{s \ge 0}~ -\sum_{x} \Pi_X(x) \log \sum_{\xhat} Q_{\Xhat}(\xhat) e^{s(D - d(x,\xhat))}. \label{eq:Rbar_iid_dual}
\end{equation}
Similarly, for the constant-composition ensemble, the primal form is given by
\begin{equation}
    \Rbarcc(D,Q_{\Xhat}) = \min_{\substack{\Ptilde_{X\Xhat} \,:\, \Ptilde_X = \Pi_X, \Ptilde_{\Xhat} = Q_{\Xhat}, \\\EE_{\Ptilde}[d(X,\Xhat)] \le D }} I_{\Ptilde}(X;\Xhat), \label{eq:Rbar_cc}
\end{equation}
and the dual form is given by 
\begin{equation}
    \Rbarcc(D,Q_{\Xhat}) = \sup_{s \ge 0, b(\cdot)}~ -\sum_{x} \Pi_X(x) \log \sum_{\xhat} Q_{\Xhat}(\xhat) e^{s(D - d(x,\xhat))}e^{b(\xhat) - \phi_b} \label{eq:Rbar_cc_dual}
\end{equation}
with $\phi_b = \EE_Q[b(\Xhat)]$.  A proof of the equivalence of \eqref{eq:Rbar_iid}--\eqref{eq:Rbar_iid_dual} can be found in \cite{Dem02}, and that of \eqref{eq:Rbar_cc}--\eqref{eq:Rbar_cc_dual} can be proved in the same way as the LM rate for channel coding.

Interestingly, there is a close connection between these expressions and those of the GMI and LM rate: From \eqref{eq:Rbar_iid} and \eqref{eq:Rbar_cc}, we obtain the primal channel coding rates upon identifying $(X,\Xhat) \leftrightarrow (Y,X)$, $d(x,\xhat) \leftrightarrow -\log q(x,y)$, and $D \leftrightarrow \EE_{Q_X \times W}[ -\log q(X,Y) ]$.  In addition, we have the following analog of the fact that the LM rate is at least as high as the GMI:
\begin{equation}
    \Rbariid(D,Q_{\Xhat}) \le \Rbarcc(D,Q_{\Xhat}). \label{eq:rd_R_cmp}
\end{equation}
It is important, however, to remember that here we would like the rate to be {\em as small as possible} for a given distortion level, in stark contrast with the channel coding setup in which high rates are desired.  Thus, the ordering \eqref{eq:rd_R_cmp} in fact implies that i.i.d.~random coding is {\em preferable} to constant-composition random coding for a given choice of $Q_{\Xhat}$.  While strict inequality is possible in \eqref{eq:rd_R_cmp}, equality holds upon minimizing both sides over $Q_{\Xhat}$ (see the discussion following \eqref{eq:Dbar_matched}).

The intuition as to why i.i.d.~random coding may be preferable is as follows.  Attaining $d^n(\Xv,\Xvhat) < n(D+\delta)$ is equivalent to the existence of a codeword $\Xvhat^{(j)}$ such that $(\Xv,\Xvhat^{(j)})$ induces a joint type $\Ptilde_{X\Xhat}$ satisfying $\EE_{\Ptilde}[d(X,\Xhat)] < D+\delta$.  By a typicality argument, we may assume that the $X$-marginal of $\Ptilde_{X\Xhat}$ is arbitrarily close to $\Pi_X$.  For the $\Xhat$-marginal, we consider two separate cases:
\begin{itemize}
    \item Suppose that the $\Xhat$-marginal of $\Ptilde_{X\Xhat}$ is $Q_{\Xhat}$ (or more generally, the type $Q_{\Xhat,n}$ approximating $Q_{\Xhat}$).  Then, by the properties of types in  Section \ref{sec:prop_types}, the i.i.d.~and constant-composition codeword distributions yield matching behavior (on an exponential scale) in the probability of the event $(\Xv,\Xvhat^{(j)}) \in \Tclass(\Ptilde_{X\Xhat})$.  Note that $I_{\Ptilde}(X;\Xhat) = D(\Ptilde_{X\Xhat} \| \Pi_X \times Q_{\Xhat})$ in this case.
    \item On the other hand, if the $\Xhat$-marginal of $\Ptilde_{X\Xhat}$ differs from $Q_{\Xhat}$, then the event $(\Xv,\Xvhat^{(j)}) \in \Tclass(\Ptilde_{X\Xhat})$ never occurs under the constant-composition ensemble.  However, it may still occur under the i.i.d.~ensemble, with an associated per-codeword probability of roughly $e^{-nD(\Ptilde_{X\Xhat} \| \Pi_X \times Q_{\Xhat})}$.
\end{itemize}
Thus, high-probability success under the constant-composition ensemble essentially implies the same under the i.i.d.~ensemble, but the opposite is not true in general.  In this sense, the i.i.d.~ensemble produces a {\em more diverse} codebook that improves the chances of a low-distortion codeword existing.  Analogous observations were made for the error exponents of Gaussian codebooks in \cite{Zho17}.


In light of this discussion, it is natural to ask why constant-composition coding was used in the proof of Theorem \ref{thm:rd_ach}.  Firstly, we note that the distribution $Q_{\Xhat}$ therein is optimized, rather than being fixed and possibly suboptimal.  In addition, under the setup therein with two distortion measures $d_0$ and $d_1$, an additional subtlety arises: A more diverse codebook may amount to more ways in which the encoder can be led astray due to the mismatch.  In fact, in the parallel source example of Section \ref{sec:rd_parallel} is an example in which additional codebook structure provably eliminates the performance loss due to mismatch.

\subsection{Statement of Results}

In the following, we provide a formal statement of achievability and ensemble tightness for the above rate-distortion functions $\Rbariid$ and $\Rbarcc$.  
To avoid trivial cases, we consider technical assumptions stated in terms of the following definitions:
\begin{gather}
    D_{\min} = \sum_{x} \Pi_X(x) \min_{\xhat \,:\,Q_{\Xhat}(\xhat) > 0} d(x,\xhat), \label{eq:Dmin} \\
    D_{\rm prod} = \sum_{x,\xhat} \Pi_X(x)Q_{\Xhat}(\xhat) d(x,\xhat). \label{eq:Dprod}
\end{gather}
We assume that $D \ge D_{\rm min}$, which is justified by the fact that $D_{\rm min}$ is the average distortion attained when each source symbol $x_i$ is paired with the reconstructed symbol $\xhat_i$ having the lowest possible distortion.  In addition, we can assume without loss of generality that $D < D_{\rm prod}$, since a distortion of at most $D_{\rm prod} + \delta$ is trivially attained with high probability using a ``codebook'' with just a single codeword drawn from $Q_{\Xhat}^n$, due to the law of large numbers.

\begin{theorem} \label{thm:rd_mm_rand}
    {\em (Achievable rate-distortion functions with mismatched random coding)} For any discrete memoryless source $\Pi_X$, distortion function $d$, distortion level $D \in [D_{\min}, D_{\rm prod})$, auxiliary distribution $Q_{\Xhat} \in \Pc(\Xchat)$, and parameter $\delta > 0$, the random-coding error probability \eqref{eq:rd_pe_bar} satisfies the following:
    \begin{itemize}
        \item Under the i.i.d.~ensemble, we have $\pebar(n,\lfloor e^{nR} \rfloor,D,Q_{\Xhat},\delta) \to 0$ for any $R > \Rbariid(D,Q_{\Xhat})$ and $\delta > 0$, and $\pebar(n,\lfloor e^{nR} \rfloor,D,Q_{\Xhat},\delta) \to 1$ for any $R < \Rbariid(D,Q_{\Xhat})$ and sufficiently small $\delta > 0$.
        \item Under the constant-composition ensemble, we have $\pebar(n,\lfloor e^{nR} \rfloor,D,Q_{\Xhat},\delta) \to 0$ for any $R > \Rbarcc(D,Q_{\Xhat})$ and $\delta > 0$, and $\pebar(n,\lfloor e^{nR} \rfloor,D,Q_{\Xhat},\delta) \to 1$ for any $R < \Rbarcc(D,Q_{\Xhat})$ and sufficiently small $\delta > 0$.
    \end{itemize}
\end{theorem}

\begin{remark} \label{rem:rd_cc}
    As we will exemplify in Section \ref{sec:rd_simple_example}, the constraint set in \eqref{eq:Rbar_cc} may be empty, in which case we adopt the convention $\Rbarcc(D,Q_{\Xhat}) = \infty$, meaning that the condition $R > \Rbarcc(D,Q_{\Xhat})$ cannot be satisfied for any rate $R$, and no achievability claim is made for the given pair $(D,Q_{\Xhat})$.  Specifically, this occurs whenever
    \begin{equation}
        D < \min_{\Ptilde_{X\Xhat} \,:\, \Ptilde_X = \Pi_X, \Ptilde_{\Xhat} = Q_{\Xhat}} \EE_{\Ptilde}[d(X,\Xhat)],
    \end{equation}
    and the right-hand side may be strictly larger than $D_{\min}$ in \eqref{eq:Dmin}.
    The reason for this phenomenon is that under a suboptimal choice of $Q_{\Xhat}$, a given typical sequence $\Xv$ may incur significant distortion with {\em every} sequence that has composition $Q_{\Xhat}$.  In contrast, under the i.i.d.~ensemble, although most codewords have composition close to $Q_{\Xhat}$, many codewords also have significantly different compositions.
\end{remark}

The results of Theorem \ref{thm:rd_mm_rand} can be proved in several ways.  For the constant-composition ensemble, the achievability part is in fact a special case of Theorem \ref{thm:rd_ach}, obtained by setting $d_0 = d_1$.  This gives the distortion-rate function in \eqref{eq:Dbar_matched},\footnote{The subscript ``Matched'' in \eqref{eq:Dbar_matched} refers to having $d_0 = d_1$, but the choice of $Q_{\Xhat}$ may still be mismatched in the sense of being fixed and suboptimal.} which can be inverted to obtain the primal rate-distortion expression in \eqref{eq:Rbar_cc}.  The ensemble tightness claim readily follows since the proof in Section \ref{sec:pf_rd_ach} is based on exact error probability expressions and the method of types.  

The i.i.d.~ensemble can be analyzed similarly using types.  Alternatively, an elementary achievability proof that avoids the use of types is possible via the techniques of Gallager \cite[Lemma 9.3.1]{Gal68} and Sakrison \cite{Sak69}.  To further highlight the connections between mismatched channel coding and rate-distortion theory, we instead outline an approach via a large-deviations result of \cite{Dem02}, which we also used in Section \ref{sec:cont_tightness} to prove the ensemble tightness of the GMI for channel coding.

We first form an exact expression for the random-coding error probability under the i.i.d.~ensemble as follows (see also \cite[Thm.~9]{Kos12}): Defining $(\Xv,\Xvhat') \sim \Pi_X^n(\xv) Q_{\Xhat}^n(\xvhat)$, we have
\begin{align}
    \pebar &= \PP\bigg[ \bigcap_{j=1,\dotsc,M} \Big\{ \frac{1}{n} d^n(\Xv,\Xvhat^{(j)}) \ge D + \delta  \Big\} \bigg] \\
    &= \EE\bigg[ \PP\bigg[ \bigcap_{j=1,\dotsc,M} \Big\{ \frac{1}{n} d^n(\Xv,\Xvhat^{(j)}) \ge D + \delta \Big\} \,\Big|\, \Xv \bigg] \bigg] \\
    &= \EE\bigg[ \bigg( \PP\bigg[ \frac{1}{n} d^n(\Xv,\Xvhat') \ge D + \delta \,\Big|\, \Xv \bigg] \bigg)^{M} \bigg] \label{eq:rd_init3} \\
    &= \EE\bigg[ \bigg( 1 - \PP\bigg[ \frac{1}{n} d^n(\Xv,\Xvhat') < D + \delta \,\Big|\, \Xv \bigg] \bigg)^{M} \bigg],\label{eq:rd_init4} 
\end{align}
where \eqref{eq:rd_init3} uses the independence of the random codewords.  Observe that \eqref{eq:rd_init4} bears a strong resemblance to the channel coding counterpart in \eqref{eq:tightGMI_lb}.

One could proceed by characterizing the inner probability in \eqref{eq:rd_init4} using the method of types, via similar steps to Section \ref{sec:pf_rd_ach}.  One may also be tempted to follow the dual analysis of the GMI for channel coding in Section \ref{sec:pf_GMI_dual}; however, the situation is different here, as we require a {\em lower bound} on the tail event in \eqref{eq:rd_init4} to upper bound $\pebar$, whereas the techniques of Section \ref{sec:pf_GMI_dual} (e.g., Markov's inequality) lead to upper bounds on the tail event.\footnote{Note also that $d(x,\xhat)$ here plays the role of $-\log q(x,y)$ in channel coding, so the inner probability in \eqref{eq:rd_init4} considers a lower tail event, instead of an upper tail event.}  Hence, we instead consider the use of the following large deviations result \cite[Thm.~1]{Dem02}, from which both the achievability of $\Rbariid$ and ensemble tightness readily follow.  We state this result in a general form that permits its use even in the continuous-alphabet setting.

\begin{lemma} \label{lem:large_dev_rd}
    {\em (Large deviations result for i.i.d.~random coding)}
    Fix a memoryless source $\Pi_X$, an auxiliary distribution $Q_{\Xhat}$, and a non-negative distortion measure $d(x,\xhat)$, and define\footnote{As stated in Footnote \ref{foot:ess_inf} in Section \ref{sec:cont_tightness}, the {\em essential infimum} of a function $g(\Xhat)$ with respect to $\Xhat\sim Q_{\Xhat}$ is defined to be supremum of $t \in \RR$ for which $\PP_{Q}[ g(\Xhat) > t ] = 1$. }
    \begin{gather}
        D_{\rm min} = \EE_{\Pi_X}\big[ {\rm ess\,inf}_{Q_{\Xhat}}\, d(X,\Xhat) \big], \label{eq:gamma_min_rd}\\
        D_{\rm prod} = \EE_{\Pi_X\times Q_{\Xhat}} \big[ d(X,\Xhat) \big]. \label{eq:gamma_avg_rd}
    \end{gather}
    Then, if $D_{\rm prod} < \infty$, we have for any $D \in ( D_{\rm min}, D_{\rm prod})$ that the following holds with probability one with respect to $\Xv$:
    \begin{equation}
        \lim_{n \to \infty} - \frac{1}{n} \log \PP\big[ d^n(\Xv, \Xvhat') \le nD \,|\, \Xv \big] = \Rbariid(D,Q_{\Xhat}), \label{eq:large_dev_rd}
    \end{equation}
    where $(\Xv,\Xvhat') \sim \Pi_X^n(\xv)Q_{\Xhat}^n(\xvhat)$.
\end{lemma}

With this result in place, the desired claim in Theorem \ref{thm:rd_mm_rand} for $D \in (D_{\rm min}, D_{\rm prod})$ follows from \eqref{eq:rd_init4} (with $M = \lfloor e^{nR} \rfloor$) and the fact that $\big( 1 - \frac{1}{\alpha_n}\big)^{\beta_n} \to 0$ whenever $\alpha_n \ge 1$ and $\lim_{n \to \infty} \frac{\alpha_n}{\beta_n} = 0$ (achievability) and $\big( 1 - \frac{1}{\alpha_n}\big)^{\beta_n} \to 1$ whenever $\alpha_n \ge 1$ and $\lim_{n \to \infty} \frac{\alpha_n}{\beta_n} = \infty$ (ensemble tightness).

The assumption $D \in ( D_{\rm min}, D_{\rm prod})$ in Lemma \ref{lem:large_dev_rd} is motivated in the same way as the discussion following \eqref{eq:Dprod}, except that for general alphabets, strict inequality is assumed in the condition $D > D_{\rm min}$ in order to avoid technical issues when $D = D_{\rm min}$.  In the finite-alphabet setting, the case $D = D_{\rm min}$ can easily be handled separately, and for many distortion functions it amounts to performing lossless compression.  Further discussion can be found in the text following \cite[Thm.~1]{Dem02}.  

\subsection{Example: Bernoulli Source} \label{sec:rd_simple_example}

\begin{figure}
    \begin{centering}
        \includegraphics[width=0.7\columnwidth]{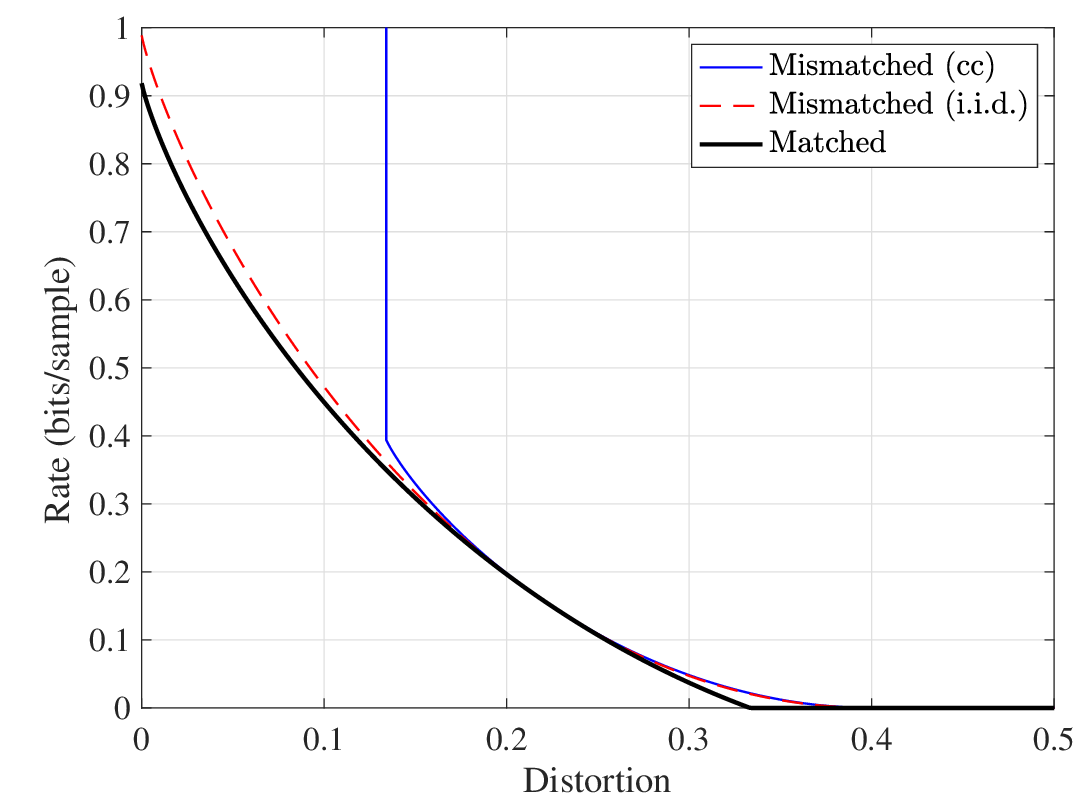}
        \par
    \end{centering}
    
    \caption{Bernoulli source example: Matched and mismatched rate-distortion curves with $X \sim {\rm Bernoulli}\big(\frac{1}{3}\big)$ and the Hamming distortion measure.  In the mismatched case, the auxiliary distribution $Q_{\Xhat}$ is chosen to be ${\rm Bernoulli}\big(\frac{1}{5}\big)$. } \label{fig:rd_simple}
\end{figure}

As a simple example illustrating the preceding achievable rates, we consider the source $X \sim {\rm Bernoulli}(p)$ with $p \le \frac{1}{2}$, and let $d(x,\xhat) = \openone\{ \xhat \ne x \}$ be the Hamming distortion measure.  It is well-known that the matched rate-distortion function is \cite[Sec.~10.3.1]{Cov06}
\begin{equation}
    \Rmatched(D) = [H_2(p) - H_2(D)]_+,
\end{equation}
where $[\alpha]_+ = \max\{0,\alpha\}$.  In addition, the optimal auxiliary distribution $Q_{\Xhat}$ is ${\rm Bernoulli}(q^*)$, with $q^*$ chosen according to one of two cases: (i) if $0 \le D \le p$, then $q^* = \frac{p-D}{1-2D}$; (ii) if $D > p$, then $q^* = 0$.

In Figure \ref{fig:rd_simple}, we compare $\Rmatched$ to $\Rbariid$ and $\Rbarcc$ in the case that $p = \frac{1}{3}$, and $Q_{\Xhat}$ is chosen to be ${\rm Bernoulli}\big(\frac{1}{5}\big)$.  We observe that $\Rbariid$ incurs a small loss compared to $\Rmatched$ across the entire range of distortion levels.  While $\Rbarcc$ behaves similarly to $\Rbariid$ at high distortion levels, it is unable to attain any distortion level below a threshold of roughly $0.13$, regardless of a rate; this is due to the phenomenon discussed in Remark \ref{rem:rd_cc}.  Finally, we note that all three curves coincide when $D = \frac{2}{9} \approx 0.22$; this is due to the fact that the optimal parameter $q^* = \frac{p-D}{1-2D}$ mentioned above equals $\frac{1}{5}$ under this choice, and thus coincides with the choice of $Q_{\Xhat}$.

\subsection{Further Related Work} \label{sec:rd_other}

We conclude this subsection by briefly outlining some other relevant notions of mismatch and universality in the literature on rate-distortion theory.  This outline is far from exhaustive, and further references can be found in \cite{Dem02,Kon06}.

We first briefly mention that the techniques and results surveyed in this subsection are related to (and in some cases directly used in) studies of universal rate-distortion codes \cite{Dem02,Kon06,Ste93,Zha96}.   
Along similar lines, the early works of Sakrison \cite{Sak70,Sak69} introduced the problem of constructing rate-distortion codes that simultaneously achieve low distortion for a class of sources, as a natural analog of the compound channel in channel coding \cite{Bla59}.  This setup is also closely related to the use of Gaussian codes for non-Gaussian sources \cite{Lap97}, which we survey in Section \ref{sec:gaussian_rd}.

In other works such as \cite{Gra75,Gra03}, the goal is also to characterize the degradation in performance when a rate-distortion codebook is designed for some source $\Pi_X$ but then applied to a different source $\Pi'_X$.  This is analogous to the problem considered in this subsection, but is studied in \cite{Gra75,Gra03} beyond the random coding setting, with various results including generalizations of the well-known loss of $D(\Pi_X\|\Pi'_X)$ for Huffman coding \cite[Sec.~5.6]{Cov06}.

\section{Gaussian Compression for Non-Gaussian Sources} \label{sec:gaussian_rd}


In the previous subsection, we focused on discrete memoryless sources with a mismatched random coding distribution.  In this subsection, we turn to a particularly important special case of the continuous-alphabet counterpart of this problem, namely, Gaussian coding for non-Gaussian sources under the squared-error distortion criterion \cite{Lap97}.  This can be viewed as an analogous problem to that of mismatched Gaussian codes for channel coding ({\em cf.}, Section \ref{sec:ex_noise}).  Before formally describing the problem setting, we provide some background on rate-distortion theory with Gaussian sources.

\subsection{Background}

The compression of Gaussian sources subject to a squared-error distortion constraint is one of the most widely-considered rate-distortion settings.  In this setting, the source $\Xv$ is assumed to be i.i.d.~on $\Ndist(0,\sigma^2)$ for some $\sigma^2 > 0$, and the goal is to construct an encoder (mapping $\RR^n$ to $\{1,\dotsc,M\}$) and decoder (mapping $\{1,\dotsc,M\}$ to $\RR^n$) such that $d^n(\xv,\xvhat) = \|\xv - \xvhat\|_2^2$ is small, where $\xvhat$ is the reconstruction.

Under the matched counterpart of Definition \ref{def:rate_dist} with $d_0 = d_1 = d$, the rate-distortion function is given by \cite[Sec.~10.3.2]{Cov06}
\begin{equation}
    \Rmatched(D) = \Big[ \frac{1}{2}\log \frac{\sigma^2}{D} \Big]_+,
\end{equation}
where $[\alpha]_+ = \max\{0,\alpha\}$.  The achievability part can be proved using Gaussian codebooks (to be made precise below), along with an encoder that maps a given source sequence $\xv \in \RR^n$ to the nearest codeword in the codebook:
\begin{equation}
    \xvhat = \argmin_{ j=1,\dotsc,M } \| \xv - \xvhat^{(j)} \|_2^2. \label{eq:rd_nn}
\end{equation}

\subsection{Non-Gaussian Sources}

Suppose that we have an arbitrary memoryless source on $\RR$, i.e., $\Xv \sim \Pi_X^n$ for some $\Pi_X$.  For concreteness, we assume that the underlying distribution is continuous, so that $\Pi_X$ is a density function.  Continuing with the assumption of squared-error distortion $d(x,\xhat) = (x-\xhat)^2$, the matched rate-distortion function in \eqref{eq:rd_matched} specializes as follows:
\begin{equation}
    \Rmatched(D) = \min_{\substack{\Ptilde_{X\Xhat} \,:\, \Ptilde_X = \Pi_X, \\ \EE_{\Ptilde}[ (X-\Xhat)^2 ] \le D}} I(X;\Xhat).
\end{equation}
However, achieving this rate may be difficult -- the source distribution may be unknown, and even if it is known, the required codebook structure may be prohibitively complex.

In this subsection, we address these difficulties by considering the question of how well a Gaussian codebook performs when applied to a non-Gaussian source.  Since we are considering the squared-error distortion $d(x,\xhat) = (x-\xhat)^2$, we still adopt the encoding rule in \eqref{eq:rd_nn}.  We only assume that the source has a fixed second moment:
\begin{equation}
    \EE_{\Pi}[ X^2 ] = \sigma^2.
\end{equation}
One may expect a rate increase due to the suboptimality of the codebook construction, but on the other hand, one may expect a rate decrease due to the fact that Gaussian sources are the hardest to compress for a given second moment under the squared-error distortion criterion \cite[Ex.~10.8]{Cov06}.  Analogously to mismatched channel coding with non-Gaussian noise ({\em cf.}, Section \ref{sec:ex_noise}), we will see that these effects cancel each other, and all sources lead to the same rate-distortion trade-off as the Gaussian source.

While our presentation follows that of Lapidoth \cite{Lap97}, an achievability result was also given in the early work of Sakrison \cite{Sak69} under the additional technical assumption that the $(2+\epsilon)$-th moment of $X$ exists for some $\epsilon > 0$.  


\subsection{Gaussian Codebooks}

We consider two different notions of Gaussian codebooks, which we refer to as the {\em Gaussian i.i.d.~ensemble} and the {\em shell ensemble}.  Fixing $\Gamma \ge 0$, the Gaussian i.i.d.~ensemble considers codewords that are independently distributed according to
\begin{equation}
    \Xvhat \sim \Ndist( \bzero, \Gamma \Iv_n ) \label{eq:G_iid}
\end{equation}
where $\bzero$ is the zero vector and $\Iv_n$ is the $n \times n$ identity matrix.  In contrast, the shell ensemble considers codewords that are independently distributed according to
\begin{equation}
    \Xvhat \sim \mathrm{Uniform}\big( S_n(\Gamma) \big),  \label{eq:G_shell}
\end{equation}
where $S_n(\Gamma) = \{ \xvhat \in \RR^n \,:\, \|\xvhat\|_2^2 = n\Gamma \}$ is the shell of radius $\sqrt{n\Gamma}$ in $n$-dimensional space.   

The parameter $\Gamma$ represents the ``power'' of codewords in the codebook, but in contrast with channel coding, this is not constrained as part of the problem statement.
The optimal choice will turn out to be $\Gamma = \sigma^2 - D$ whenever $D \le \sigma^2$, whereas the case $D > \sigma^2$ trivially yields a rate of zero by always outputting the all-zero sequence (i.e., setting $\Gamma = 0$).

\subsection{Achievability Result}

The above-outlined achievability result is formally stated as follows.

\begin{theorem} \label{thm:rd_gaussian}
    {\em (Gaussian codes for non-Gaussian sources)}
    For any memoryless source $\Pi_X$ with $\EE_{\Pi}[X^2] = \sigma^2$, under the Gaussian i.i.d.~ensemble or shell ensemble with parameter $\Gamma \ge 0$, along with nearest-neighbor encoding according to \eqref{eq:rd_nn}, we have the following for any pair $(R,D)$ and $\delta > 0$:
    \begin{itemize}
        \item If $\Gamma = [\sigma^2 - D]_+$ and $R \ge \big[ \frac{1}{2} \log \frac{\sigma^2}{D} \big]_+ + \delta$, then as $n \to \infty$, we have
        \begin{equation}
            \PP\big[ \|\Xv - \Xvhat\|_2^2 \ge n(D + \delta)  \big] \to 0. \label{eq:Gaussian1}
        \end{equation}
        \item For any choice of $\Gamma$, if $R \le \big[ \frac{1}{2} \log \frac{\sigma^2}{D} \big]_+ - \delta$, then as $n \to \infty$, we have
        \begin{equation}
            \PP\big[ \|\Xv - \Xvhat\|_2^2 \ge n(D - \delta)  \big] \to 1. \label{eq:Gaussian2}
        \end{equation}
    \end{itemize}
\end{theorem}

As with Theorem \ref{thm:gaussian}, this result can be viewed both positively and negatively.  On the positive side, we see that Gaussian codes form a compression scheme that is robust in handling non-Gaussian sources.  On the other hand, Gaussian sources are the hardest to compress for a given second moment \cite[Ex.~10.8]{Cov06}, and Theorem \ref{thm:gaussian} states that we always attain this worst-case performance and no better.

The proof of Theorem \ref{thm:rd_gaussian} is outlined as follows; further details can be found in \cite{Lap97}:
\begin{itemize}
    \item By the circular symmetry of the distributions in \eqref{eq:G_iid}--\eqref{eq:G_shell}, and similar symmetry in the nearest-neighbor encoding rule \eqref{eq:rd_nn}, the conditional error probability $\PP[\mathrm{error} \,|\, \xv]$ depends on the source realization $\xv$ only through $\|\xv\|_2^2$.
    \item For any memoryless source with $\EE[X^2] = \sigma^2$, the law of large numbers implies that $\frac{1}{n}\|\Xv\|_2^2$ converges in probability to $\sigma^2$.
    \item From existing analyses of Gaussian codes for Gaussian sources, one can deduce that under the choice $\Gamma = [\sigma^2 - D]_+$, we have the following characterization of the conditional error probability (in the excess distortion sense) for any $\xv$ satisfying $\frac{1}{n}\|\xv\|_2^2 \to \sigma^2$ as  $n \to \infty$: (i) $\PP[\mathrm{error} \,|\, \xv] \to 0$  when $R > \big[ \frac{1}{2} \log \frac{\sigma^2}{D} \big]_+$; and (ii) $\PP[\mathrm{error} \,|\, \xv] \to 1$ when $R < \big[ \frac{1}{2} \log \frac{\sigma^2}{D} \big]_+$.
    \item This establishes the achievability part, as well as the converse part in the special case that $\Gamma = [\sigma^2 - D]_+$.  The more general converse can be deduced by similar ideas revealing that the preceding choice of $\Gamma$ in fact minimizes the first-order rate threshold.
\end{itemize}
A detailed proof can be found in \cite{Lap97}, and an alternative analysis is given in \cite{Zho17} that also leads to refined asymptotic bounds.  For the i.i.d.~ensemble, it can also be shown that $\Rbariid$ in \eqref{eq:Rbar_iid} evaluates to $\big[ \frac{1}{2} \log \frac{\sigma^2}{D} \big]_+$ when $Q_{\Xhat}$ is chosen as $\Ndist(0,[\sigma^2 - D]_+)$ \cite[Examples~1--2]{Dem02}, meaning that \eqref{eq:Gaussian1}--\eqref{eq:Gaussian2} collectively provide a direct analog of the first part of Theorem \ref{thm:rd_mm_rand}.

 

\chapter{Multiple-Access Channels} \label{ch:mac}

\section{Introduction}

The main focus of this monograph, and of the relevant literature on mismatched decoding, is on the point-to-point channel coding problem.  As outlined in Section \ref{ch:intro}, this is in itself a very challenging problem with many important applications.

In this section, we turn to the problem of coding for the {\em multiple-access channel} (MAC), in which two (or more) users send information to a common receiver.  While this is a harder problem than the already-challenging single-user problem, its study is of interest for a number of reasons:
\begin{itemize}
    \item The analysis of random coding comes with interesting new challenges not present in the single-user problem, both in terms of the achievability and ensemble tightness.
    \item We will find that multiple-access coding techniques can be beneficial even in the single-user setting, leading to rates strictly higher than the LM rate of Section \ref{ch:single_user}.  We will treat this perspective rather briefly in this section, focusing on the parallel BSC example of Section \ref{sec:parallel}, and then explore multi-user coding techniques for single-user channels in detail in Section \ref{ch:multiuser}.
\end{itemize}
The results of this section are mostly due to Lapidoth \cite{Lap96}, but we will generally adopt the more recent proof techniques of Scarlett {\em et al.} \cite{Sca16a}.

\section{Problem Setup}

The problem setting that we consider is in direct analogy with the single-user setting of Section \ref{sec:setup_general}. We focus on the two-user MAC, which is sufficient to convey the key ideas, but the analysis techniques that we present can similarly be applied to any fixed number of users.

The input alphabets are denoted by $\Xc_1$ and $\Xc_2$, and the output alphabet is denoted by $\Yc$.  Except where stated otherwise, these are assumed to be finite, so that we are considering a discrete memoryless MAC (DM-MAC).  The conditional transition law for a single channel use is denoted by $W(y|x_{1},x_{2})$, and the $n$-letter transition law is given by $W^{n}(\yv|\xv_{1},\xv_{2})\defeq\prod_{i=1}^{n}W(y_{i}|x_{1,i},x_{2,i})$.  

Encoder $\nu=1,2$ takes as input a message $m_{\nu}$ uniformly distributed on the set $\{1,\dotsc,M_{\nu}\}$, and transmits the corresponding codeword $\xv_{\nu}^{(m_{\nu})}$ from a codebook $\Cc_{\nu}=\{\xv_{\nu}^{(1)},\dotsc,\xv_{\nu}^{(M_{\nu})}\}$. Given the output sequence $\yv$ generated according to $W^n(\,\cdot\,|\,\xv_1,\xv_2)$, the decoder forms an estimate $(\hat{m}_{1},\hat{m}_{2})$ of the message pair, given by 
\begin{equation}
    (\hat{m}_{1},\hat{m}_{2})=\argmax_{(i,j) \,:\, i\in\{1,\dotsc,M_{1}\}, \, j\in\{1,\dotsc,M_{2}\}}q^{n}(\xv_{1}^{(i)},\xv_{2}^{(j)},\yv),\label{eq:MAC_Metric}
\end{equation}
where $q^{n}(\xv_{1},\xv_{2},\yv)\defeq\prod_{i=1}^{n}q(x_{1,i},x_{2,i},y_{i})$ for some decoding metric $q(x_1,x_2,y)$.  The method of tie-breaking has no impact on any of the results that we present, so we assume that ties are broken as errors.  

The error probability for a given codebook pair $(\Cc_1,\Cc_2)$ is given by
\begin{equation}
    \pe(\Cc_1,\Cc_2) = \PP\big[ (\hat{m}_{1},\hat{m}_{2}) \ne (m_1,m_2) \big].
\end{equation}
For any given codebook pair $(\Cc_1,\Cc_2)$, the decoding rule minimizing $\pe$ is the maximum-likelihood (ML) rule, corresponding to \eqref{eq:MAC_Metric} with $q(x_1,x_2,y) = W(y|x_1,x_2)$.

We consider a random coding scheme in which the codewords for the two users are independently drawn from some distributions $P_{\Xv_1}$ and $P_{\Xv_2}$: 
\begin{equation}
	(\{\Xv_1^{(i)}\}_{i=1}^{M_1},\{\Xv_2^{(j)}\}_{j=1}^{M_2}) \sim \prod_{i=1}^{M_1} P_{\Xv_1}(\xv_1^{(i)}) \prod_{j=1}^{M_2} P_{\Xv_2}(\xv_2^{(j)}).
\end{equation}
The error probability averaged over the random codebooks is denoted by $\pebar(n,M_{1},M_{2})$.

\begin{definition}
    {\em (Mismatch capacity region)}
    For a given mismatched DM-MAC $(W,q)$, a rate pair $(R_1,R_2)$ is said to be achievable if, for all $\delta > 0$, there exist sequences of codebooks $(\Cc_{1,n},\Cc_{2,n})$ of sizes $M_1 \ge e^{n(R_1 - \delta)}$ and $M_2 \ge e^{n(R_2 - \delta)}$ such that $\pe(\Cc_{1,n},\Cc_{2,n}) \to 0$ as $n \to\infty$.  The {\em mismatch capacity region} $\RMAC$ is defined to be the set of all achievable rates.
\end{definition}

\noindent To put our results in context, we recall the {\em matched} capacity region corresponding to $q(x_1,x_2,y)=W(y|x_1,x_2)$ \cite[Ch.~4]{Elg11}:
\begin{equation}
    \RMatchedMAC = \mathrm{conv}\bigg( \bigcup_{Q_1,Q_2} \Rc(Q_1,Q_2) \bigg), \label{eq:MAC_matched_conv}
\end{equation}
where $\mathrm{conv}(\cdot)$ denotes the closure of the convex hull, the union is over all input distributions $Q_1 \in \Pc(\Xc_1)$ and $Q_2 \in \Pc(\Xc_2)$, and the set $\Rc(Q_1,Q_2)$ contains all pairs $(R_1,R_2)$ satisfying
\begin{align} 
    R_1 &\le I(X_1;Y|X_2), \nonumber \\
    R_2 &\le I(X_2;Y|X_1), \nonumber \\
    R_1 + R_2 &\le I(X_1,X_2;Y), \label{eq:MAC_matched}
\end{align}
where $(X_1,X_2,Y) \sim Q_1 \times Q_2 \times W$. These rates are associated with three different types of errors, corresponding to the cases that (i) only the estimate of $m_1$ is incorrect; (ii) only the estimate of $m_2$ is incorrect, and (iii) both are incorrect.  The achievability part is proved by applying random coding with input distributions $Q_1$ and $Q_2$, and using a time-sharing argument \cite[Ch.~4]{Elg11} to achieve the convex hull.

\section{Achievable Rate Region} \label{sec:mac_ach}

In the single-user setting, we compared i.i.d.~random coding and constant-composition random coding for DMCs, and found the latter to provide better achievable rates.  The same is true for the DM-MAC, so to simplify the exposition, we focus exclusively on constant-composition random coding.

For $\nu=1,2$, we fix $Q_{\nu}\in\Pc(\Xc_{\nu})$ and let $P_{\Xv_{\nu}}$ be the uniform distribution on the type class $\Tclass(Q_{\nu,n})$, where $Q_{\nu,n}\in\Pc_{n}(\Xc_{\nu})$ is a type with the same support as $Q_{\nu}$ such that $\|Q_{\nu,n} - Q_{\nu}\|_{\infty} \le \frac{1}{n}$. This yields
\begin{equation}
	P_{\Xv_{\nu}}(\xv_{\nu})=\frac{1}{|\Tclass(Q_{\nu,n})|}\openone\big\{\xv_{\nu}\in \Tclass(Q_{\nu,n})\big\} \label{eq:MAC_Q_X1}
\end{equation}
for $\nu = 1,2$.

We first state and discuss the achievable rate region \cite{Lap96}. We then discuss a number of examples, before proceeding with the proof.


\begin{theorem} \label{thm:MAC_Rate} 
    {\em (MAC achievable rate region)}
    For any mismatched DM-MAC $(W,q)$, and any input distributions $Q_1 \in \Pc(\Xc_1)$ and $Q_2 \in \Pc(\Xc_2)$, we have $\RegLM(Q_1,Q_2) \subseteq \RMAC$, where $\RegLM(Q_1,Q_2)$ is the set of $(R_1,R_2)$ pairs satisfying
    \begin{align}
    R_{1}       & \le\min_{\substack{\Ptilde_{X_{1}X_{2}Y} \,:\, \Ptilde_{X_{1}}=P_{X_{1}}, \, \Ptilde_{X_{2}Y}=P_{X_{2}Y}\\
    \EE_{\Ptilde}[\log q(X_{1},X_{2},Y)]\ge\EE_{P}[\log q(X_{1},X_{2},Y)]}} I_{\Ptilde}(X_{1};X_{2},Y), \label{eq:MAC_R1_LM} \\
     R_{2}       & \le\min_{\substack{\Ptilde_{X_{1}X_{2}Y} \,:\,       \Ptilde_{X_{2}}=P_{X_{2}}, \, \Ptilde_{X_{1}Y}=P_{X_{1}Y}\\
        \EE_{\Ptilde}[\log q(X_{1},X_{2},Y)]\ge\EE_{P}[\log q(X_{1},X_{2},Y)]}}I_{\Ptilde}(X_{2};X_{1},Y), \label{eq:MAC_R2_LM} \\
    R_{1}+R_{2} & \le\min_{\substack{\Ptilde_{X_{1}X_{2}Y} \,:\, \Ptilde_{X_{1}}=P_{X_{1}},\Ptilde_{X_{2}}=P_{X_{2}},\Ptilde_{Y}=P_{Y}\\
            \EE_{\Ptilde}[\log q(X_{1},X_{2},Y)]\ge\EE_{P}[\log q(X_{1},X_{2},Y)], \\ I_{\Ptilde}(X_{1};Y)\le R_{1}, \, I_{\Ptilde}(X_{2};Y)\le R_{2}}} D(\Ptilde_{X_{1}X_{2}Y}\|Q_1 \times Q_2 \times P_{Y}), \label{eq:MAC_R12_LM}
    \end{align}
    where $P_{X_1X_2Y} = Q_1 \times Q_2 \times W$. That is, any rate pair $(R_1,R_2)$ satisfying these conditions is achievable for the mismatched DM-MAC.
\end{theorem}

The proof is given in Section \ref{sec:mac_proof}.  Note that here and subsequently, the minimizations are over all joint distributions $\Ptilde_{X_1X_2Y} \in \Pc(\Xc_1 \times \Xc_2 \times \Yc)$ satisfying the specified constraints, but we let the inclusion in $\Pc(\Xc_1 \times \Xc_2 \times \Yc)$ remain implicit.

Although the rate conditions in \eqref{eq:MAC_R1_LM}--\eqref{eq:MAC_R12_LM} appear complex, they are natural generalizations of the primal expression for the LM rate ({\em cf.}, Section \ref{sec:su_rates}).  The condition in \eqref{eq:MAC_R1_LM} arises by considering a typical realization of $(\Xv_1,\Xv_2,\Yv)$, and studying the probability that some $(\Xvbar_1,\Xv_2,\Yv)$ has a higher decoding metric for some incorrect $\Xvbar_1$.  Thinking of $P_{X_1X_2Y}$ and $\Ptilde_{X_1X_2Y}$ as the corresponding joint types of these triplets, the condition $\EE_{\Ptilde}[\log q(X_{1},X_{2},Y)]\ge\EE_{P}[\log q(X_{1},X_{2},Y)]$ corresponds to the event $q(\Xvbar_1,\Xv_2,\Yv) \ge q(\Xv_1,\Xv_2,\Yv)$.  Moreover, we have $\Ptilde_{X_{2}Y}=P_{X_{2}Y}$ because the two triplets use the same pair $(\Xv_2,\Yv)$, and $\Ptilde_{X_{1}}=P_{X_{1}}$ because we are considering constant-composition random coding.  A similar discussion applies to the condition \eqref{eq:MAC_R2_LM}.

The condition \eqref{eq:MAC_R12_LM} corresponds to some triplet $(\Xvbar_1,\Xvbar_2,\Yv)$ having a higher metric than $(\Xv_1,\Xv_2,\Yv)$, and has some features that are somewhat less standard.  In particular, the minimization problem contains the additional constraints $I_{\Ptilde}(X_{1};Y)\le R_{1}$ and $I_{\Ptilde}(X_{2};Y)\le R_{2}$.  These constraints correspond to the fact that a random codebook $\Cc_{\nu}$ ($\nu = 1,2$) of size $e^{nR_{\nu}}$ is unlikely to produce {\em any} non-transmitted $\Xvbar_{\nu}$ such that the empirical mutual information of $(\Xvbar_{\nu},\Yv)$ satisfies $I_{\Ptilde}(X_{\nu};Y) > R_{\nu}$.  We will see that these constraints are crucial for the ensemble tightness claim (see Section \ref{sec:mac_ens_tight}), in that the weaker achievable rate region with these constraints removed can be strictly smaller.   In more detail, removing the constraints yields a pentagonal region similar to the matched MAC \eqref{eq:MAC_matched}, whereas the region $\RegLM(Q_1,Q_2)$ given above may be non-pentagonal and contain curved segments; see Section \ref{sec:MAC_non_convex} for an example.  Since the right-hand side of \eqref{eq:MAC_R12_LM} is decreasing in $R_1$ and $R_2$, it is not even immediately obvious that $(R_1,R_2) \in \RegLM(Q_1,Q_2) \implies (R_1',R_2') \in\RegLM(Q_1,Q_2)$ when $R'_1 \le R_1$ and $R'_2 \le R_2$, but this property is indeed guaranteed, and can be seen by noting that \eqref{eq:MAC_R12_LM} holds with strict inequality if and only if
\begin{align}
    &\min_{\substack{\Ptilde_{X_{1}X_{2}Y} \,:\, \Ptilde_{X_{1}}=P_{X_{1}},\Ptilde_{X_{2}}=P_{X_{2}},\Ptilde_{Y}=P_{Y} \\ \EE_{\Ptilde}[\log q(X_{1},X_{2},Y)]\ge\EE_{P}[\log q(X_{1},X_{2},Y)]}} 
     \max\Big\{ I_{\Ptilde}(X_1;Y) - R_1, \nonumber \\ &\quad I_{\Ptilde}(X_2;Y) - R_2, D(\Ptilde_{X_{1}X_{2}Y}\|Q_1 \times Q_2 \times P_{Y}) - (R_1 + R_2) \Big\} > 0.
\end{align}
This equivalence will arise in the achievability proof in Section \ref{sec:mac_proof}.

The standard time-sharing argument for multiple-access channels \cite[Ch.~4]{Elg11} applies without change for mismatched maximum-metric decoding, and hence, we deduce from Theorem \ref{thm:MAC_Rate} that
\begin{equation}
    \mathrm{conv}\bigg( \bigcup_{Q_1,Q_2} \RegLM(Q_1,Q_2) \bigg) \subseteq \RMAC,
\end{equation}
where $\mathrm{conv}(\cdot)$ denotes the closure of the convex hull.  For the most part, however, we will focus on random coding under a fixed pair $(Q_1,Q_2)$.  A more detailed investigation of time-sharing for the mismatched MAC can be found in \cite{ScarlettThesis}, where the distinction between {\em explicit time-sharing} and {\em coded time-sharing} is studied.

Since the achievable rate region of Theorem \ref{thm:MAC_Rate} is ensemble-tight (see Section \ref{sec:mac_ens_tight} for a formal statement), it must recover the matched MAC conditions \eqref{eq:MAC_matched} in the case that $q(x_1,x_2,y) = W(y|x_1,x_2)$.  As a sanity check, this can be checked directly via similar steps to \eqref{eq:primal_matched0}--\eqref{eq:primal_matched5} \cite{Lap96}, and by doing so, one finds that the constraints $I_{\Ptilde}(X_{1};Y)\le R_{1}$ and $I_{\Ptilde}(X_{2};Y)\le R_{2}$ in \eqref{eq:MAC_R12_LM} have no impact in the matched case.

\section{Examples} \label{sec:mac_numerical}

\subsection{Parallel Binary Symmetric Channels} \label{sec:mu_parallel}

We return to the parallel binary symmetric channel (BSC) example studied in the single-user setting in Section \ref{sec:parallel} \cite{Lap96}.  We now study the channel as a MAC, with inputs $X_1$ and $X_2$, and output $Y = (Y_1,Y_2)$.  Recall that the channels from $X_{\nu}$ to $Y_{\nu}$ ($\nu = 1,2$) are independent BSCs with crossover probabilities $\delta_1$ and $\delta_2$, and the metric is
\begin{equation}
\log q(x_1,x_2,(y_1,y_2)) = \frac{1}{2}\big( \openone\{ x_1 = y_1 \} + \openone\{ x_2 = y_2 \}  \big). \label{eq:parallel_metric_mac}
\end{equation}
We let the input distributions $Q_1,Q_2$ be equiprobable on $\{0,1\}$.

\paragraph{Analysis of the achievable rate region.} The conditions \eqref{eq:MAC_R1_LM}--\eqref{eq:MAC_R2_LM} are straightforward to analyze, reducing to $R_{\nu} \le 1 - H_2(\delta_{\nu})$ bits/use for $\nu = 1,2$.  We therefore focus our attention on \eqref{eq:MAC_R12_LM}, which is the most interesting of the three.  The main step is to show that this constraint is satisfied by the pair $(R_1,R_2) = (1 - H_2(\delta_1), 1 - H_2(\delta_2))$.

Fixing any $\Ptilde_{X_1X_2Y}$ satisfying the constraints in \eqref{eq:MAC_R12_LM}, we have
\begin{align}
    &D(\Ptilde_{X_{1}X_{2}Y}\|Q_1 \times Q_2 \times P_{Y}) \nonumber \\
        &= \EE_{\Ptilde}\bigg[\log \frac{\Ptilde_{X_1X_2Y}(X_1,X_2,Y)}{ \Ptilde_{X_1}(X_1)\Ptilde_{X_2}(X_2)\Ptilde_{Y}(Y)} \bigg] \label{eq:div12_1} \\
        &= I_{\Ptilde}(X_1,X_2;Y) + I_{\Ptilde}(X_1;X_2) \label{eq:div12_2} \\
        &\ge I_{\Ptilde}(X_1,X_2;Y),
\end{align}
where \eqref{eq:div12_1} makes use of the marginal constraints, and \eqref{eq:div12_2} follows by multiplying and dividing by $\Ptilde_{X_1X_2}$ in \eqref{eq:div12_1}.  The mutual information $I_{\Ptilde}(X_1,X_2;Y)$ was already analyzed in Section \ref{sec:parallel}, where it was shown that
\begin{equation}
     I_{\Ptilde}(X_1,X_2;Y) \ge \big(1 - H_2(\deltatilde_1)\big) +  \big(1 - H_2(\deltatilde_2)\big), \label{eq:parallel_mac_I1}
\end{equation}
where $\deltatilde_1$ and $\deltatilde_2$ are crossover probabilities corresponding to $\Ptilde_{Y_1|X_1}$ and $\Ptilde_{Y_2|X_2}$, both of which must take the form of a BSC.

The analysis now departs from that of Section \ref{sec:parallel} by incorporating the constraints $I_{\Ptilde}(X_{1};Y)\le R_{1}$ and $I_{\Ptilde}(X_{2};Y)\le R_{2}$.  Since we are considering the rate pair with $R_{\nu} = 1 - H_2(\delta_{\nu})$ for $\nu = 1,2$, and since $I_{\Ptilde}(X_{\nu};Y_{\nu}) = 1 - H_2(\deltatilde_{\nu})$, the constraints reduce to $H_2(\delta_{\nu}) \le H_2(\deltatilde_{\nu})$, and we deduce from \eqref{eq:parallel_mac_I1} that
\begin{equation}
    I_{\Ptilde}(X_1,X_2;Y) \ge \big(1 - H_2(\delta_1)\big) +  \big(1 - H_2(\delta_2)\big),
\end{equation}
and as a result, the condition \eqref{eq:MAC_R12_LM} is indeed satisfied by $(R_1,R_2) = (1 - H_2(\delta_1), 1 - H_2(\delta_2))$.  This, in turn, implies the achievability of all rate pairs within the rectangle with corners $(0,0)$ and $(1 - H_2(\delta_1), 1 - H_2(\delta_2))$, which is the capacity region of the  matched MAC, i.e., the capacity of two parallel BSCs.

\paragraph{Discussion.} Due to the parallel structure of the codebooks and the channel, if we let $\log q$ be {\em any} positive weighted sum of $\openone\{ x_1 = y_1 \}$ and $\openone\{ x_2 = y_2 \}$ ({\em cf.}, \eqref{eq:parallel_metric_mac}), we obtain a decoding rule that separately minimizes the Hamming distance between $(\Xv_1,\Yv_1)$ and $(\Xv_2,\Yv_2)$, which in turn is equivalent to maximum-likelihood decoding.  Hence, it should not be surprising that we achieve the matched capacity region.

Nevertheless, the above analysis highlights the importance of the constraints $I_{\Ptilde}(X_{\nu};Y_{\nu}) \le R_{\nu}$; the analysis of Section \ref{sec:parallel} shows that in the absence of these constraints, the right-hand side of \eqref{eq:MAC_R12_LM} is equal to the suboptimal value $2\big( 1 - H_2\big( \frac{\delta_1 + \delta_2}{2} \big) \big)$.

This example also highlights an interesting phenomenon that we will return to in detail in Section \ref{ch:multiuser}:  Multi-user coding techniques can provide improved rates even for single-user mismatched channels.  Indeed, we saw that the LM rate fails to achieve the capacity $C = \big(1 - H_2(\delta_1)\big) +  \big(1 - H_2(\delta_2)\big)$ under a uniform input distribution, and if we deviate from a uniform input, then even optimal decoding fails to achieve capacity.  In contrast, we are always free to generate two codebooks in parallel and treat the single-user channel as if it were a MAC, and the above analysis reveals that doing so achieves the matched capacity (and hence also the mismatch capacity). 

\subsection{Non-Convexity of $\RegLM(Q_1,Q_2)$ } \label{sec:MAC_non_convex}

It is well-known that even in the matched case, the union of the region \eqref{eq:MAC_matched} over all pairs $(Q_1,Q_2)$ can be non-convex \cite[Ch.~4]{Elg11}, meaning that the convex hull operation in \eqref{eq:MAC_matched_conv} is not redundant.  However, the rate region \eqref{eq:MAC_matched} for a given pair $(Q_1,Q_2)$ is always convex, and typically pentagonal; in certain special cases, triangular and quadrilateral shapes are also possible.

In contrast, in the mismatched setting, due to the constraints $I_{\Ptilde}(X_{1};Y)\le R_{1}$ and $I_{\Ptilde}(X_{2};Y)\le R_{2}$ in \eqref{eq:MAC_R12_LM}, the achievable rate region $\RegLM(Q_1,Q_2)$ corresponding to fixed $(Q_1,Q_2)$ can be non-convex \cite{ScarlettThesis}, and can have curved parts on the boundary rather than only straight lines.  To demonstrate this, we consider the mismatched MAC with  $\Xc_{1}=\Xc_{2}=\{0,1\}$, $\Yc=\{0,1,2\}$, and
\begin{align}
    W(y|x_{1},x_{2})&=\begin{cases}
    1-2\delta_{x_{2}} & y=x_{1}+x_{2}\\
    \delta_{x_{2}} & \mathrm{otherwise},
    \end{cases}   \label{eq:MAC_Example_W} \\
    q(x_1,x_2,y)&=\begin{cases}
    1-2\delta & y=x_{1}+x_{2}\\
    \delta & \mathrm{otherwise},
    \end{cases}   \label{eq:MAC_Example_q}
\end{align}
where we choose $\delta_0 = 0.25$ and $\delta_1 = 0.01$, and let $\delta$ equal an arbitrary value in $\big(0,\frac{1}{3}\big)$, any of which yields an equivalent metric.  Hence, the mismatched decoder incorrectly assumes both of the transition probabilities $\delta_{x_2}$ are equal.

\begin{figure}
    \begin{centering}
        \includegraphics[width=0.7\columnwidth]{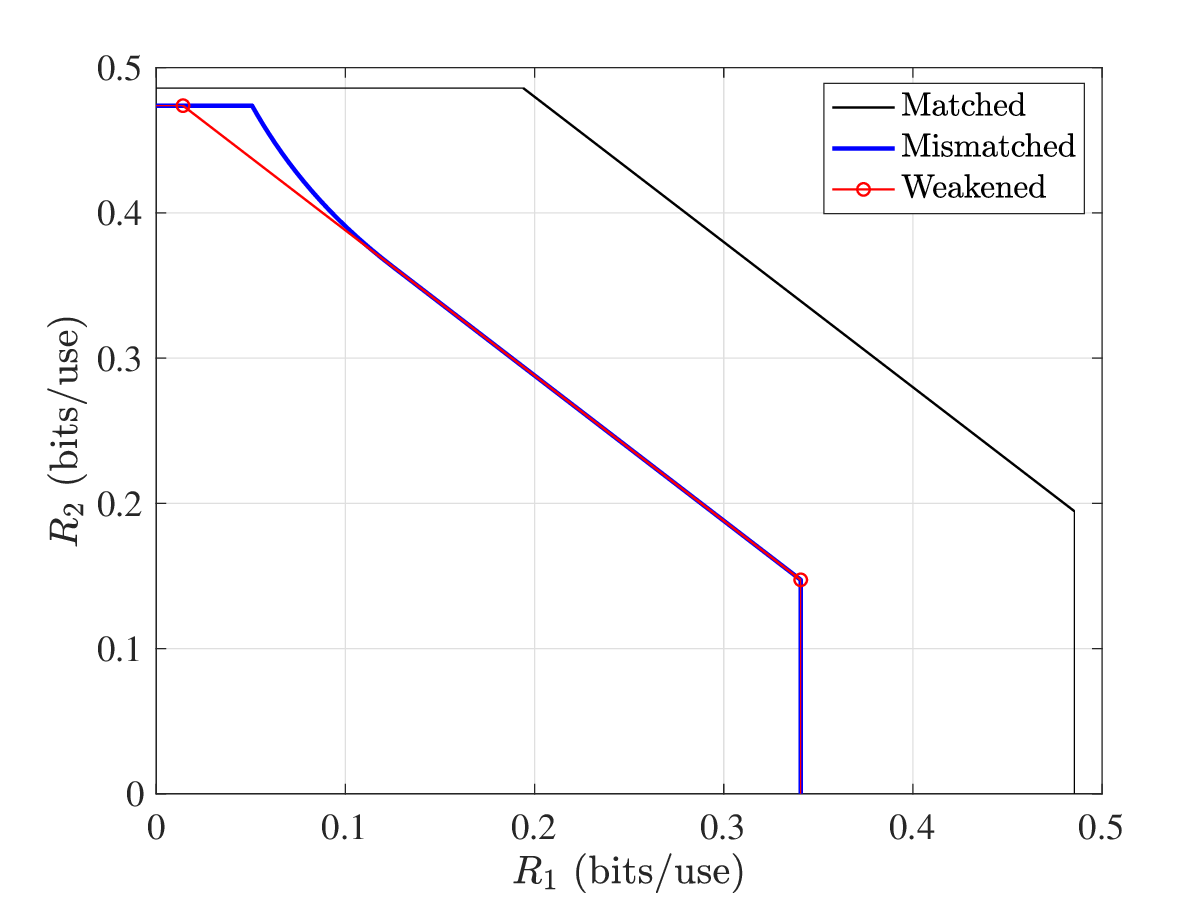}
        \par
    \end{centering}
    
    \caption{Achievable rate regions for the MAC in \eqref{eq:MAC_Example_W}--\eqref{eq:MAC_Example_q} in the matched and mismatched settings, with input distributions $Q_1 = Q_2 = \big(\frac{1}{2},\frac{1}{2}\big)$.  The weakened region removes the two mutual information constraints in \eqref{eq:MAC_R12_LM}.} \label{fig:NonConvexMAC}
\end{figure}

Figure \ref{fig:NonConvexMAC} plots the achievable rate region $\RegLM(Q_1,Q_2)$ corresponding to $Q_1 = Q_2 = \big(\frac{1}{2},\frac{1}{2}\big)$.  We also plot the weakened region obtained when the constraints  $I_{\Ptilde}(X_{1};Y)\le R_{1}$ and $I_{\Ptilde}(X_{2};Y)\le R_{2}$ are removed from \eqref{eq:MAC_R12_LM}, as well as the matched region \eqref{eq:MAC_matched}.

We see that while the weakened and matched regions take the usual pentagonal shape, the ensemble-tight region $\RegLM(Q_1,Q_2)$ curves upward near its upper ``corner point'', and forms a non-convex shape.  This is due to the fact that the constraint $I_{\Ptilde}(X_1;Y) \le R_1$ is active in this part of the curve.  In this example, the constraint $I_{\Ptilde}(X_2;Y) \le R_2$ has no effect on the rate region.


\section{Proof of Achievable Rate Region} \label{sec:mac_proof}

In this subsection, we present the proof of Theorem \ref{thm:MAC_Rate}.  The original proof of Lapidoth \cite{Lap96} and the more recent proof of Scarlett {\em et al.} \cite{Sca16a} share several common ideas.  In the following, we focus on the latter, presenting the analysis in several steps.

\subsection{Separation into Three Error Events} \label{sec:mac_sep_error}

We study the random-coding error probability by considering the following three error types, where we assume without loss of generality that the true messages are $(m_1,m_2) = (1,1)$:
\begin{tabbing}
    ~{\emph{(Type 1)}}~~~ \= $q^n(\Xv_1^{(i)},\Xv_2,\Yv) \ge q^n(\Xv_1,\Xv_2,\Yv)$ for some $i \ne 1$; \\
    ~{\emph{(Type 2)}}~~~ \> $q^n(\Xv_1,\Xv_2^{(j)},\Yv) \ge q^n(\Xv_1,\Xv_2,\Yv)$ for some $j \ne 1$; \\
    ~{\emph{(Type 12)}}~~~ \> $q^n(\Xv_1^{(i)},\Xv_2^{(j)},\Yv) \ge q^n(\Xv_1,\Xv_2,\Yv)$ for some $i \ne 1$, $j \ne 1$.
\end{tabbing}
The corresponding probabilities of the error events are denoted by  $\peibar(n,M_{1})$, $\peiibar(n,M_{2})$, and $\peiiibar(n,M_{1},M_{2})$, respectively.  Writing the overall random-coding error probability as $\pebar(n,M_{1},M_{2})$, we immediately obtain from the union bound that
\begin{equation}
    \pebar(n,M_1,M_2) \le \peibar(n,M_1)+\peiibar(n,M_2)+\peiiibar(n,M_1,M_2). \label{eq:MAC_ErrorProbs}
\end{equation}
Thus, we have $\pebar \to 0$ as long as all three of $\peibar$, $\peiibar$ and $\peiiibar$ vanish, and we can consider the three error types separately.

The analysis of all three error events is similar, but $\peiiibar$ is the only one that requires new tools compared to the single-user setting, studied in Section \ref{sec:su_proofs}.  We therefore focus only on $\peiiibar$ in the remainder of the proof.

\subsection{Separation Into Joint Types}

The type-12 error probability $\peiiibar(n,M_1,M_2)$ admits the following exact expression:
\begin{equation}
    \peiiibar=\EE\Bigg[\PP\Bigg[\bigcup_{i\ne1,j\ne1}\bigg\{\frac{q^{n}(\Xv_{1}^{(i)},\Xv_{2}^{(j)},\Yv)}{q^{n}(\Xv_{1},\Xv_{2},\Yv)}\ge1\bigg\}\bigg|\Xv_{1},\Xv_{2},\Yv\Bigg]\Bigg]. \label{eq:MAC_pe12_ExactIsh}
\end{equation}
In analogy with the primal-domain analysis of the single-user setting (Section \ref{sec:pf_LM_primal}), we will rewrite \eqref{eq:MAC_pe12_ExactIsh} in terms of joint types.   
To lighten the notion, we introduce the sets
\begin{gather}
    \SetSncc(\Qv_{n}) \defeq\Big\{ P_{X_{1}X_{2}Y}\in\Pc_{n}(\Xc_{1}\times\Xc_{2}\times\Yc)\,:\, P_{X_{1}}=Q_{1,n},\, P_{X_{2}}=Q_{2,n}\Big\}, \label{eq:MAC_SetSn} \\
    \SetTiiincc(P_{X_{1}X_{2}Y}) \triangleq \Big\{ \Ptilde_{X_{1}X_{2}Y}\in\Pc_n(\Xc_{1}\times\Xc_{2}\times\Yc)\,:\,
    \Ptilde_{X_{1}}=P_{X_{1}}, \nonumber \\ \Ptilde_{X_{2}}=P_{X_{2}},\Ptilde_{Y}=P_{Y},\EE_{\Ptilde}[\log q(X_{1},X_{2},Y)]\ge\EE_{P}[\log q(X_{1},X_{2},Y)] \Big\},\label{eq:MAC_SetT12n}                    
\end{gather}
where $\SetSncc(\Qv_{n})$ can be thought of as the set of possible joint types of $(\Xv_{1},\Xv_{2},\Yv)$ when each user has a constant-composition codebook, and $\SetTiiincc(P_{X_{1}X_{2}Y})$ can be thought of as the set of joint types of $(\Xvbar_{1},\Xvbar_{2},\Yv)$ that lead to a decoding error when $(\Xv_{1},\Xv_{2},\Yv) \in \Tclass(P_{X_{1}X_{2}Y})$.

Fixing $P_{X_{1}X_{2}Y}\in\SetSncc(\Qv_{n})$ and letting $(\xv_{1},\xv_{2},\yv)$ be an arbitrary triplet such that $(\xv_{1},\xv_{2},\yv)\in \Tclass(P_{X_{1}X_{2}Y})$, we find that the event in \eqref{eq:MAC_pe12_ExactIsh} can be written as 
\begin{equation}
    \bigcup_{i\ne1,j\ne1}\bigcup_{\Ptilde_{X_{1}X_{2}Y}\in\SetTiiincc(P_{X_{1}X_{2}Y})}\Big\{(\Xv_{1}^{(i)},\Xv_{2}^{(j)},\Yv)\in \Tclass(\Ptilde_{X_{1}X_{2}Y})\Big\}.\label{eq:MAC_Event2}
\end{equation}
Writing the probability and expectation in \eqref{eq:MAC_pe12_ExactIsh} as summations over joint types, substituting \eqref{eq:MAC_Event2}, and interchanging  the order of the unions, we obtain the following:
\begin{align}
    \peiiibar &= \sum_{P_{X_{1}X_{2}Y}\in\SetSncc(\Qv_{n})}\PP\big[(\Xv_{1},\Xv_{2},\Yv)\in \Tclass(P_{X_{1}X_{2}Y})\big] \nonumber \\ 
        & \hspace*{-1.5ex}\times \PP\Bigg[\bigcup_{\Ptilde_{X_{1}X_{2}Y}\in\SetTiiincc(P_{X_{1}X_{2}Y})}\bigcup_{i\ne1,j\ne1}\Big\{(\Xv_{1}^{(i)},\Xv_{2}^{(j)},\yv)\in \Tclass(\Ptilde_{X_{1}X_{2}Y})\Big\}\Bigg] \label{eq:MAC_ExpProof1} \\
        & \hspace*{-3ex}\le \sum_{P_{X_{1}X_{2}Y}\in\SetSncc(\Qv_{n})}\PP\big[(\Xv_{1},\Xv_{2},\Yv)\in \Tclass(P_{X_{1}X_{2}Y})\big] \min\Bigg\{1, \nonumber \\ 
        &  \hspace*{-1.5ex}  \sum_{\Ptilde_{X_{1}X_{2}Y}\in\SetTiiincc(P_{X_{1}X_{2}Y})}\PP\Bigg[\bigcup_{i\ne1,j\ne1}\Big\{(\Xv_{1}^{(i)},\Xv_{2}^{(j)},\yv)\in \Tclass(\Ptilde_{X_{1}X_{2}Y})\Big\}\Bigg] \Bigg\}, \label{eq:MAC_ExpProof2}
\end{align}
where $\yv$ is an arbitrary element of $\Tclass(P_{Y})$ (hence depending implicitly on $P_{X_1X_2Y}$), and \eqref{eq:MAC_ExpProof2} follows from the truncated union bound (i.e., the minimum of one and the union bound). 

\subsection{A Refined Union Bound}

It is tempting, particularly based on the primal single-user analysis in Section \ref{sec:su_proofs}, to apply the truncated union bound to the inner probability in \eqref{eq:MAC_ExpProof2} to obtain 
\begin{align}
    & \PP\Bigg[\bigcup_{i\ne1,j\ne1}\bigg\{(\Xv_{1}^{(i)},\Xv_{2}^{(j)},\yv)\in \Tclass(\Ptilde_{X_{1}X_{2}Y})\bigg\}\Bigg] \nonumber \\
    &\quad \le \min\Big\{1, (M_1-1)(M_2-1) \PP\big[(\Xvbar_{1},\Xvbar_{2},\yv)\in \Tclass(\Ptilde_{X_{1}X_{2}Y})\big] \Big\}, \label{eq:ub_standard}
\end{align}
where $(\Xvbar_1,\Xvbar_2) \sim P_{\Xv_1} \times P_{\Xv_2}$.
However, proceeding with this standard union bound would lead to a condition of the form \eqref{eq:MAC_R12_LM} without the constraints $I_{\Ptilde}(X_{1};Y)\le R_{1}$ and $I_{\Ptilde}(X_{2};Y)\le R_{2}$, and we know that these constraints can strictly improve the achievable rate region.

Fortunately, a straightforward solution is available.  Observe that in order to have $(\Xv_{1}^{(i)},\Xv_{2}^{(j)},\yv)\in \Tclass(\Ptilde_{X_{1}X_{2}Y})$, it must be the case that $(\Xv_{1}^{(i)},\yv)\in \Tclass(\Ptilde_{X_{1}Y})$.  This means that the left-hand side of \eqref{eq:ub_standard} is upper bounded by
\begin{equation}
    \PP\Bigg[\bigcup_{i\ne1}\bigg\{(\Xv_{1}^{(i)},\yv)\in \Tclass(\Ptilde_{X_{1}Y})\bigg\}\Bigg] \le (M_1 -1) \PP\big[(\Xvbar_{1},\yv)\in \Tclass(\Ptilde_{X_{1}Y})\big].
\end{equation}
A similar bound follows by noting that $(\Xv_{1}^{(i)},\Xv_{2}^{(j)},\yv)\in \Tclass(\Ptilde_{X_{1}X_{2}Y})$ implies $(\Xv_{2}^{(j)},\yv)\in \Tclass(\Ptilde_{X_{2}Y})$, and putting these bounds together with \eqref{eq:ub_standard} yields
\begin{align}
& \PP\Bigg[\bigcup_{i\ne1,j\ne1}\bigg\{(\Xv_{1}^{(i)},\Xv_{2}^{(j)},\yv)\in \Tclass(\Ptilde_{X_{1}X_{2}Y})\bigg\}\Bigg] \nonumber \\
&\quad \le \min\Big\{1, (M_1 -1) \PP\big[(\Xvbar_{1},\yv)\in \Tclass(\Ptilde_{X_{1}Y})\big], \nonumber \\
    &\hspace*{1.8cm} (M_2 -1) \PP\big[(\Xvbar_{2},\yv)\in \Tclass(\Ptilde_{X_{2}Y})\big], \nonumber \\
    &\hspace*{1.8cm} (M_1-1)(M_2-1) \PP\big[(\Xvbar_{1},\Xvbar_{2},\yv)\in \Tclass(\Ptilde_{X_{1}X_{2}Y})\big] \Big\}. \label{eq:ub_refined}
\end{align}
This refined bound turns out to be sufficient, and in fact, we will see in Section \ref{sec:mac_ens_tight} that it is tight to within a factor of four.

\subsection{Deducing the Rate Condition} \label{sec:mac_deducing}

The remainder of the analysis is analogous to the single-user setting, so we provide a relatively concise treatment.  In the following, we let $\delta > 0$ be an arbitrarily small positive constant.

By standard properties of types similar to Section \ref{sec:prop_types}, the three probabilities in \eqref{eq:ub_refined} respectively behave as $e^{-nI_{\Ptilde}(X_1;Y)}$, $e^{-nI_{\Ptilde}(X_2;Y)}$, and $e^{-nD(\Ptilde_{X_{1}X_{2}Y}\|Q_1 \times Q_2 \times P_{Y})}$ times sub-exponential factors.  In addition, the number of terms in the inner summation of \eqref{eq:MAC_ExpProof2} is polynomial in $n$, so the summation can be upper bounded by a polynomial times the corresponding maximum.   It follows that for a given joint type $P_{X_1X_2Y} \in \Sc_n(\Qv_n)$ corresponding to the outer summation of \eqref{eq:MAC_ExpProof2}, the $\min\{1, \cdot\}$ term vanishes as $n \to \infty$ as long as
\begin{align}
    &\min_{\Ptilde_{X_{1}X_{2}Y} \in \SetTiiincc(P_{X_{1}X_{2}Y}) } 
        \max\bigg\{ I_{\Ptilde}(X_1;Y) - R_1, I_{\Ptilde}(X_2;Y) - R_2, \nonumber \\ 
    & \qquad\qquad\quad D(\Ptilde_{X_{1}X_{2}Y}\|Q_1 \times Q_2 \times P_{Y}) - (R_1 + R_2) \bigg\} \ge \delta. \label{eq:MAC_Final0}
\end{align}
Using the definition of $\SetTiiincc$ in \eqref{eq:MAC_SetT12n} and lower bounding the minimum over joint types by a minimum over all joint distributions, we find that \eqref{eq:MAC_Final0} holds as long as
\begin{align}
 &\min_{\substack{\Ptilde_{X_{1}X_{2}Y} \,:\, \Ptilde_{X_{1}}=P_{X_{1}},\Ptilde_{X_{2}}=P_{X_{2}},\Ptilde_{Y}=P_{Y}\\
        \EE_{\Ptilde}[\log q(X_{1},X_{2},Y)]\ge\EE_{P}[\log q(X_{1},X_{2},Y)]}} 
\max\bigg\{ I_{\Ptilde}(X_1;Y) - R_1, I_{\Ptilde}(X_2;Y) - R_2,\nonumber \\ 
    &\qquad\qquad\qquad\quad D(\Ptilde_{X_{1}X_{2}Y}\|Q_1 \times Q_2 \times P_{Y}) - (R_1 + R_2) \bigg\} \ge \delta. \label{eq:MAC_Final1}
\end{align}
Next, recall that the outer summation of \eqref{eq:MAC_ExpProof2} sums over the possible joint types $P_{X_1X_2Y}$ of $(\Xv_1,\Xv_2,\Yv)$.   Again using standard properties of types similar to Section \ref{sec:prop_types}, we have
\begin{equation}
    \PP\big[ (\Xv_1,\Xv_2,\Yv) \in \Tclass(P_{X_1X_2Y}) \big] \le e^{-n(D(P_{X_1X_2Y} \| Q_1 \times Q_2 \times W) - \delta')} \label{eq:MAC_TypeX1X2Y}
\end{equation}
for arbitrarily small $\delta' > 0$ and sufficiently large $n$.
In particular, since the divergence is only zero when the arguments are identical, the contribution of joint types with $\|P_{X_1X_2Y} - Q_1 \times Q_2 \times W\|_{\infty} > \delta$ is asymptotically negligible.  This means that the error probability  vanishes as long as \eqref{eq:MAC_Final1} holds for all $P_{X_1X_2Y}$ such that $\|P_{X_1X_2Y} - Q_1 \times Q_2 \times W\|_{\infty} \le \delta$.  

Following a similar continuity argument to the single-user setting ({\em cf.}, Appendix \ref{sec:continuity_app}), one finds that the left-hand side of \eqref{eq:MAC_Final1} is continuous as a function of $P_{X_1X_2Y}$.  Hence, and since $\delta$ can be arbitrarily small, it is sufficient for \eqref{eq:MAC_Final1} to hold under the specific choice $P_{X_1X_2Y} = Q_1 \times Q_2 \times W$.  Finally, since the $\max\{\cdot\}$ in \eqref{eq:MAC_Final1} trivially exceeds $\delta$ when $I_{\Ptilde}(X_1;Y) > R_1 + \delta$ or $I_{\Ptilde}(X_2;Y) > R_2 + \delta$, we can limit the minimum to distributions satisfying $I_{\Ptilde}(X_1;Y) \le R_1 + \delta$ and $I_{\Ptilde}(X_2;Y) \le R_2 + \delta$, yielding
\begin{align}
	 &R_1 + R_2 + \delta  \nonumber \\ &\le \min_{\substack{\Ptilde_{X_{1}X_{2}Y} \,:\, \Ptilde_{X_{1}}=P_{X_{1}},\Ptilde_{X_{2}}=P_{X_{2}},\Ptilde_{Y}=P_{Y}\\
			\EE_{\Ptilde}[\log q(X_{1},X_{2},Y)]\ge\EE_{P}[\log q(X_{1},X_{2},Y)] \\ I_{\Ptilde}(X_1;Y) \le R_1 + \delta,~I_{\Ptilde}(X_2;Y) \le R_2 + \delta} }
	D(\Ptilde_{X_{1}X_{2}Y}\|Q_1 \times Q_2 \times P_{Y}). \label{eq:MAC_Final2}
\end{align}
Note that the first two terms of the $\max\{\cdot\}$ in \eqref{eq:MAC_Final1} become redundant under the new constraints, and are therefore not included here.  Finally, a continuity argument reveals that the right-hand side of \eqref{eq:MAC_Final2} is continuous with respect to $\delta \ge 0$ \cite[p.~1447]{Lap96}, and taking $\delta \to 0$, we obtain the final rate condition \eqref{eq:MAC_R12_LM}.

\section{Ensemble Tightness} \label{sec:mac_ens_tight}

In Section \ref{sec:ens_tight}, we introduced the notion of {\em ensemble tightness} of the single-user GMI and LM rate, showing that although these do not equal the mismatch capacity in general, no higher rates can be attained using the respective random coding ensembles. The ensemble tightness proofs were based on the fact that the properties of types used in deriving the primal expressions are exponentially tight.

For the achievable rate region of the MAC in Theorem \ref{thm:MAC_Rate}, establishing ensemble tightness requires additional effort.  As with the achievability part, the difficulty lies in the rate condition \eqref{eq:MAC_R12_LM}, and more specifically, establishing the tightness of the refined union bound \eqref{eq:ub_refined}.  To do this, we follow the approach of Scarlett {\em et al.} \cite{Sca16a} and make use of a general lower bound on the probability of a union due to de Caen \cite{Dec97}.  An alternative proof based on the {second moment method} 
was given in the earlier work of Lapidoth \cite{Lap96}.

Before proceeding, we formally state the ensemble tightness result.

\begin{theorem}
    {\em (Ensemble tightness for the MAC)}
    For any DM-MAC $(W,q$), under constant-composition random coding ({\em cf.}, \eqref{eq:MAC_Q_X1}) with input distributions $(Q_1,Q_2)$, for any rate pair $(R_1,R_2)$ failing to satisfy \eqref{eq:MAC_R1_LM}--\eqref{eq:MAC_R12_LM}, we have
    \begin{equation}
        \pebar(n,\lfloor e^{nR_1} \rfloor, \lfloor e^{nR_2} \rfloor) \to 1
    \end{equation}
    as $n \to \infty$.
\end{theorem} 

The remainder of the subsection is devoted to the proof.  

\subsection{Separation Into Three Error Events}

As with the achievability part, we consider the separation into three error types in Section \ref{sec:mac_sep_error}, and note that
\begin{equation}
    \pebar(n,M_1,M_2) \ge \max\{\peibar(n,M_1),\peiibar(n,M_2),\peiiibar(n,M_1,M_2)\},
\end{equation}
where $M_{\nu} = \lfloor e^{nR_{\nu}} \rfloor$ for $\nu = 1,2$.  The single-user analysis of Section \ref{sec:ens_tight} readily extends to the analysis of $\peibar$ and $\peiibar$, revealing that the respective error probabilities tend to one when \eqref{eq:MAC_R1_LM} or \eqref{eq:MAC_R2_LM} fail to hold.  We therefore focus our attention on the most difficult error event; we will show that $\peiiibar \to 1$ whenever \eqref{eq:MAC_R12_LM} fails to hold.

\subsection{Separation Into Joint Types}

Recall that the exact type-12 error probability can be written as in \eqref{eq:MAC_ExpProof1}. Lower bounding the union over $\Ptilde_{X_{1}X_{2}Y}$ therein by a maximum, we obtain
\begin{align}
    \peiiibar \ge &\sum_{P_{X_{1}X_{2}Y}\in\SetSncc(\Qv_{n})}\PP\big[(\Xv_{1},\Xv_{2},\Yv)\in \Tclass(P_{X_{1}X_{2}Y})\big] \nonumber \\ &\qquad\qquad\quad\times \max_{\Ptilde_{X_{1}X_{2}Y}\in\SetTiiincc(P_{X_1X_2Y})} \peiiibar(\Ptilde_{X_1X_2Y}), \label{eq:mac_lb_init}
\end{align}
where, with a slight abuse of notation, we define
\begin{equation}
    \peiiibar(\Ptilde_{X_1X_2Y}) \triangleq  \PP\bigg[\bigcup_{i\ne1,j\ne1}\big\{ \Ec_{ij}(\Ptilde_{X_1X_2Y}) \big\} \bigg] \label{eq:pe_iii_Ptilde}
\end{equation}
in terms of the event
\begin{equation}
    \Ec_{ij}(\Ptilde_{X_1X_2Y}) \triangleq \Big\{(\Xv_{1}^{(i)},\Xv_{2}^{(j)},\yv)\in \Tclass(\Ptilde_{X_{1}X_{2}Y})\Big\}.
\end{equation}

\subsection{Lower Bound on the Doubly-Indexed Union}

The key step in the analysis is to lower bound \eqref{eq:pe_iii_Ptilde} using the following result of de Caen \cite{Dec97}.

\begin{lemma} \label{lem:deCaen}
    {\em (de Caen's bound \cite{Dec97})}
    For any finite sequence of events $\Ac_1,\dotsc,\Ac_N$ on a probability space, we have
    \begin{equation}
        \PP\Big[\bigcup_{l=1}^N \Ac_{l}\Big] \ge \sum_{l=1}^N \frac{\PP[\Ac_{l}]^{2}}{\sum_{l'=1}^N \PP[\Ac_{l} \cap \Ac_{l'}]}. \label{eq:deCaen}
    \end{equation}
\end{lemma}

To apply this result to \eqref{eq:pe_iii_Ptilde}, we identify $l$ with $(i,j)$ and $N$ with $(M_1-1)(M_2-1)$, and distinguish between four cases for the denominator term $\PP\big[ \Ec_{ij}(\Ptilde_{X_1X_2Y}) \,\cap\, \Ec_{i'j'}(\Ptilde_{X_1X_2Y}) \big]$:
\begin{itemize}
    \item If $i = i'$ and $j=j'$, then the probability of the intersection is simply
    \begin{align}
        \Psi_{00} &\triangleq \PP\big[ (\Xvbar_{1},\Xvbar_2,\yv)\in \Tclass(\Ptilde_{X_{1}X_{2}Y}) \big], \label{eq:Psi00}
    \end{align}
    where here and subsequently, $(\Xvbar_1,\Xvbar_2) \sim P_{\Xv_1}(\xv_1)P_{\Xv_2}(\xv_2)$ denote generic random codewords drawn independently from $(\Xv_1,\Xv_2,\Yv)$.
    \item If $i \ne i'$ and $j = j'$, then the probability of the intersection simplifies to 
    \begin{align}
        \Psi_{10} &\triangleq \PP\big[ (\Xvbar_{2},\yv)\in \Tclass(\Ptilde_{X_{2}Y}) \big] \nonumber \\ 
            &\quad\times \PP\big[ (\Xvbar_{1},\xv_2,\yv)\in \Tclass(\Ptilde_{X_{1}X_{2}Y}) \big]^2. \label{eq:Psi10}
    \end{align}
    where in the second probability, $\xv_2$ is an arbitrary sequence such that $(\xv_{2},\yv) \in \Tclass(\Ptilde_{X_{2}Y})$.  This is because the two relevant $(\Xv_1^{(i)},\Xv_2^{(j)},\yv)$ triplets share the same $\Xv_2$ sequence, which must have joint type $\Ptilde_{X_2Y}$, whereas the two different $\Xv_1$ sequences (which are independent of each other) both need to induce the joint type $\Ptilde_{X_1X_2Y}$.
    \item Similarly, if $i = i'$ and $j \ne j'$, then the probability of the intersection simplifies to 
    \begin{align}
        \Psi_{01} & \triangleq \PP\big[ (\Xvbar_{1},\yv)\in \Tclass(\Ptilde_{X_{1}Y}) \big] \nonumber \\ 
            &\quad \times \PP\big[ (\xv_1,\Xvbar_{2},\yv)\in \Tclass(\Ptilde_{X_{1}X_{2}Y}) \big]^2. \label{eq:Psi01}
    \end{align}
    \item If $i \ne i'$ and $j \ne j'$, then the two events in the intersection are independent, and hence their joint probability is $\Psi_{11} \triangleq \Psi_{00}^2$.
\end{itemize}
For any given pair $(i,j)$, the number of $(i',j')$ pairs falling into these four categories is given by $1$, $M_1-2$, $M_2-2$ and $(M_1-2)(M_2-2)$, respectively.  As a result, Lemma \ref{lem:deCaen} yields the following:
\begin{align}
    &\peiiibar(\Ptilde_{X_1X_2Y}) \nonumber \\
    & \ge \frac{(M_1-1)(M_2-1)\Psi_{00}^2 }{ \Psi_{00} + (M_1-2)\Psi_{10} + (M_2-2)\Psi_{01} + (M_1-2)(M_2-2)\Psi_{00}^2 } \\
    &\ge \frac{(M_1-1)(M_2-1)\Psi_{00}^2 }{ \Psi_{00} + (M_1-1)\Psi_{10} + (M_2-1)\Psi_{01} + (M_1-1)(M_2-1)\Psi_{00}^2 } \\
    &= \frac{1}{\frac{1}{(M_1-1)(M_2-1)\Psi_{00}} + \frac{\Psi_{10}}{(M_2-1)\Psi_{00}^2} + \frac{\Psi_{01}}{(M_1-1)\Psi_{00}^2} + 1}  \label{eq:lower12_3}\\
    &\ge \frac{1}{4} \min\bigg\{ (M_1-1)(M_2-1)\Psi_{00}, (M_2-1)\frac{\Psi_{00}^2}{\Psi_{10}}, (M_1-1)\frac{\Psi_{00}^2}{\Psi_{01}}, 1 \bigg\}. \label{eq:lower12_4}
\end{align}
Next, observe that $\Psi_{00}$ in \eqref{eq:Psi00} can be rewritten as 
\begin{align}
 \Psi_{00}& =  \PP\big[ (\Xvbar_{2},\yv)\in \Tclass(\Ptilde_{X_{2}Y}) \big]  \PP\big[ (\Xvbar_{1},\xv_2,\yv)\in \Tclass(\Ptilde_{X_{1}X_{2}Y}) \big]\\
 &=         \PP\big[ (\Xvbar_{1},\yv)\in \Tclass(\Ptilde_{X_{1}Y}) \big]  \PP\big[ (\xv_1,\Xvbar_{2},\yv)\in \Tclass(\Ptilde_{X_{1}X_{2}Y}) \big],
\end{align}
which implies that
\begin{gather}
    \frac{\Psi_{00}^2}{\Psi_{10}} = \PP\big[ (\Xvbar_{2},\yv)\in \Tclass(\Ptilde_{X_{2}Y}) \big], \\
    \frac{\Psi_{00}^2}{\Psi_{01}} = \PP\big[ (\Xvbar_{1},\yv)\in \Tclass(\Ptilde_{X_{1}Y}) \big].
\end{gather}
Substituting into \eqref{eq:lower12_4}, we deduce that the upper bound in \eqref{eq:ub_refined} is tight to within a factor of four, though we need the stronger bound \eqref{eq:lower12_3} to deduce conditions under which $\peiiibar$ approaches 1 (rather than only $\frac{1}{4}$).  Specifically, we will use the fact that \eqref{eq:lower12_3} approaches 1 whenever the first three terms in the denominator approach zero.

\subsection{Deducing Ensemble Tightness}

In the following, let $\delta > 0$ be arbitrarily small.  As we mentioned in Section \ref{sec:mac_deducing}, the three probabilities in \eqref{eq:ub_refined} behave as $e^{-nI_{\Ptilde}(X_1;Y)}$, $e^{-nI_{\Ptilde}(X_2;Y)}$, and $e^{-nD(\Ptilde_{X_{1}X_{2}Y}\|Q_1 \times Q_2 \times P_{Y})}$ times sub-exponential factors, and the joint type $P_{X_1X_2Y}$ in \eqref{eq:mac_lb_init} satisfies $\| P_{X_1X_2Y} - Q_1 \times Q_2 \times W\|_{\infty} \le \delta$ with probability approaching one.  Substituting these bounds into \eqref{eq:lower12_3}, we deduce that $\peiii \to 1$ whenever the following condition holds for all such $P_{X_1X_2Y}$:
\begin{align}
    \min_{\Ptilde_{X_{1}X_{2}Y} \in \SetTiiincc(P_{X_{1}X_{2}Y}) } &\max\bigg\{ I_{\Ptilde}(X_1;Y) - R_1, I_{\Ptilde}(X_2;Y) - R_2, \nonumber \\ & D(\Ptilde_{X_{1}X_{2}Y}\|Q_1 \times Q_2 \times P_{Y}) - (R_1 + R_2) \bigg\} < -\delta. \label{eq:MAC_Final0_LB}
\end{align}
The main challenge now is to upper bound the minimum over joint types (see \eqref{eq:MAC_SetT12n}) in terms of the corresponding minimum over all joint distributions, with some care in considering the possibility that $I_{\Ptilde}(X_1;Y) = R_1$ or $I_{\Ptilde}(X_2;Y) = R_2$ (which may hold even when \eqref{eq:MAC_R12_LM} is violated, but makes the $\max\{\cdot\}$ equal zero in \eqref{eq:MAC_Final0_LB}).  This is done in a similar way to the single-user setting, which itself was rather technical; further details are given in Appendix \ref{sec:continuity_app}.  Once this is done, the necessity of the condition \eqref{eq:MAC_R12_LM} follows. 

\section{Dual Expressions and Continuous Alphabets} \label{sec:MAC_Dual}

The rate region of Theorem \ref{thm:MAC_Rate} is analogous to the primal expression for the single-user LM rate.  It is therefore natural to ask whether there also exists a dual expression, written as a maximization over real-valued parameters rather than a minimization over joint distributions.  The answer is affirmative, as the following theorem shows \cite{Sca16a}.  Here and subsequently, we write $\EE_{Z}[\cdot]$ to denote averaging over some random variable $Z$ while all other random variables are held fixed.

\begin{theorem} \label{thm:MAC_DualRate}
    {\em (Dual expression for the achievable rate region)}
    The region $\RegLM(Q_1,Q_2)$ described by \eqref{eq:MAC_R1_LM}--\eqref{eq:MAC_R12_LM} can be expressed as the set of rate pairs $(R_{1},R_{2})$ satisfying
    \begin{align}
    R_{1} &\le \sup_{\sgz,a_{1}(\cdot)}\EE\left[\log\frac{q(X_{1},X_{2},Y)^{s}e^{a_{1}(X_{1})}}{\EE\big[q(\Xbar_{1},X_{2},Y)^{s}e^{a_{1}(\Xbar_{1})}\,|\, X_{2},Y\big]}\right], \label{eq:MAC_R1_DualLM} \\
    R_{2} &\le \sup_{\sgz,a_{2}(\cdot)}\EE\left[\log\frac{q(X_{1},X_{2},Y)^{s}e^{a_{2}(X_{2})}}{\EE\big[q(X_{1},\Xbar_{2},Y)^{s}e^{a_{2}(\Xbar_{2})}\,|\, X_{1},Y\big]}\right], \label{eq:MAC_R2_DualLM}
    \end{align}
    and at least one of the following: 
    \begin{align}
    &R_{1} \le \sup_{\rho_{2}\in[0,1],\sgz,a_{1}(\cdot),a_{2}(\cdot)} - \rho_{2}R_{2} \nonumber \\
        &~~+\EE\left[\log\frac{\big(q(X_{1},X_{2},Y)^{s}e^{a_{2}(X_{2})}\big)^{\rho_{2}}e^{a_{1}(X_{1})}}{\EE_{\Xbar_1}\Big[\Big(\EE_{\Xbar_2}\big[q(\Xbar_{1},\Xbar_{2},Y)^{s}e^{a_{2}(\Xbar_{2})}\big]\Big)^{\rho_{2}}e^{a_{1}(\Xbar_{1})}\Big]}\right], \label{eq:MAC_R12_1_DualLM} \\
    &R_{2} \le \sup_{\rho_{1}\in[0,1],\sgz,a_{1}(\cdot),a_{2}(\cdot)} - \rho_{1}R_{1} \nonumber \\
        &~~+ \EE\left[\log\frac{\big(q(X_{1},X_{2},Y)^{s}e^{a_{1}(X_{1})}\big)^{\rho_{1}}e^{a_{2}(X_{2})}}{\EE_{\Xbar_{2}}\Big[\Big(\EE_{\Xbar_{1}}\big[q(\Xbar_{1},\Xbar_{2},Y)^{s}e^{a_{1}(\Xbar_{1})}\big]\Big)^{\rho_{1}}e^{a_{2}(\Xbar_{2})}\Big]}\right] . \label{eq:MAC_R12_2_DualLM}
    \end{align}
    where $(X_{1},X_{2},Y,\Xbar_{1},\Xbar_{2})\sim Q_{1}(x_{1})Q_{2}(x_{2})W(y|x_{1},x_{2})Q_{1}(\xbar_{1})Q_{2}(\xbar_{2})$.
\end{theorem}

The most interesting feature of this result is the presence of the additional optimization variables $\rho_1$ and $\rho_2$ in \eqref{eq:MAC_R12_1_DualLM}--\eqref{eq:MAC_R12_2_DualLM}.  The presence of these variables is closely related to the presence of the mutual information constraints in \eqref{eq:MAC_R12_LM}: If we were to remove these primal-domain constraints, then the equivalent dual expression would only allow $\rho_1 = \rho_2 = 1$ in \eqref{eq:MAC_R12_1_DualLM}--\eqref{eq:MAC_R12_2_DualLM}, in which case the two conditions would be identical.

\subsection{Overview of the Analysis} \label{sec:mac_overview}

The equivalence to the primal expression is proved using Lagrange duality, but with a non-trivial additional step to handle the constraints $I_{\Ptilde}(X_{1};Y)\le R_{1}$ and $I_{\Ptilde}(X_{2};Y)\le R_{2}$ in \eqref{eq:MAC_R12_LM}.  Indeed, forming a simple dual expression directly from \eqref{eq:MAC_R12_LM} via Lagrange duality appears to be difficult.  However, the following result from \cite{Sca16a} establishes that at most one of the two mutual information constraints is active at a time, leading to a formulation more amenable to forming a Lagrange dual.

\begin{lemma} \label{lem:MAC_split}
    {\em (Equivalent formulation of sum rate condition)}
    For any mismatched DM-MAC $(W,q)$ and input distributions $(Q_1,Q_2)$, the condition \eqref{eq:MAC_R12_LM} holds if and only if at least one of the following holds:
    \begin{gather}
        R_{1}+R_{2} \le\min_{\substack{\Ptilde_{X_{1}X_{2}Y} \,:\, \Ptilde_{X_{1}}=P_{X_{1}},\Ptilde_{X_{2}}=P_{X_{2}},\Ptilde_{Y}=P_{Y}\\
            \EE_{\Ptilde}[\log q(X_{1},X_{2},Y)]\ge\EE_{P}[\log q(X_{1},X_{2},Y)], \\ I_{\Ptilde}(X_{1};Y)\le R_{1}}} D(\Ptilde_{X_{1}X_{2}Y}\|P_{X_1X_2}\times P_{Y}), \label{eq:MAC_R12a} \\
        R_{1}+R_{2} \le\min_{\substack{\Ptilde_{X_{1}X_{2}Y} \,:\, \Ptilde_{X_{1}}=P_{X_{1}},\Ptilde_{X_{2}}=P_{X_{2}},\Ptilde_{Y}=P_{Y}\\
            \EE_{\Ptilde}[\log q(X_{1},X_{2},Y)]\ge\EE_{P}[\log q(X_{1},X_{2},Y)], \\ I_{\Ptilde}(X_{2};Y)\le R_{2}}} D(\Ptilde_{X_{1}X_{2}Y}\|P_{X_1X_2}\times P_{Y}),  \label{eq:MAC_R12b}
    \end{gather}
    where $P_{X_1X_2Y} = Q_1 \times Q_2 \times W$. 
\end{lemma}

The expressions in \eqref{eq:MAC_R12a}--\eqref{eq:MAC_R12b} constitute the primal forms of the dual expressions in \eqref{eq:MAC_R12_1_DualLM}--\eqref{eq:MAC_R12_2_DualLM}, with the equivalence proved using Lagrange duality.
Much like the single-user setting, the dual expressions also permit a direct derivation using cost-constrained random coding, which again extends to continuous-alphabet settings.  The conditions \eqref{eq:MAC_R1_DualLM}--\eqref{eq:MAC_R2_DualLM} are attained using a near-identical analysis to the single-user setting (Section \ref{sec:su_proofs}).  While deriving the conditions  \eqref{eq:MAC_R12_1_DualLM}--\eqref{eq:MAC_R12_2_DualLM} is less straightforward, the steps are still standard {once the right counterpart of the non-asymptotic bound \eqref{eq:rcu} is in place}. For a fixed triplet $(\xv_1,\xv_2,\yv)$, we are interested in the probability
\begin{equation}
    \PP\left[\bigcup_{i\ne1,j\ne1}\bigg\{\frac{q^{n}(\Xv_{1}^{(i)},\Xv_{2}^{(j)},\yv)}{q^{n}(\xv_{1},\xv_{2},\yv)}\ge1\bigg\} \right]. \label{eq:mac_direct1}
\end{equation}
The key idea is to avoid applying the truncated union bound $\PP\big[ \cup_{ij} \Ec_{ij}\big] \le \min\big\{ 1, \sum_{i,j} \PP[ \Ec_{ij} ] \big\}$ over all $(M_1-1)(M_2-1)$ incorrect message pairs simultaneously, and to instead apply it to {one user's indices at a time}.  There are two possible orders in which this can be done; one leads to an upper bound on \eqref{eq:mac_direct1} of the form
\begin{equation}
    \min\left\{ 1,(M_{1}-1)\EE_{\Xvbar_1}\Bigg[\min\bigg\{1,(M_{2}-1)\PP_{\Xvbar_2}\bigg[\frac{q^{n}(\Xvbar_{1},\Xvbar_{2},\yv)}{q^{n}(\xv_{1},\xv_{2},\yv)}\ge 1\bigg]\bigg\} \Bigg]\right\} \label{eq:mac_direct2}
\end{equation}
and the other leads to a similar bound with the roles of $\Xvbar_1$ and $\Xvbar_2$ reversed.  Here the subscript of $\EE$ indicates which random variable the averaging is with respect to, and similarly for $\PP$.  Once \eqref{eq:mac_direct2} has been obtained, the direct derivation uses the same tools as the single-user setting (e.g., Markov's inequality), along with the standard inequality $\min\{1,\alpha\} \le \alpha^\rho$ for $\rho \in [0,1]$.

Another feature that distinguishes the analysis from the single-user setting is that we use cost-constrained coding with {multiple auxiliary costs}, as introduced in Section \ref{sec:cost_multi}.  Specifically, each user's codeword distribution takes the form \eqref{eq:PX_cost}, with {two auxiliary costs per user}.  By doing so, we can introduce two different functions $a_1(x_1)$ in \eqref{eq:MAC_R1_DualLM} and \eqref{eq:MAC_R12_1_DualLM}--\eqref{eq:MAC_R12_2_DualLM}, and similarly for $a_2(x_2)$ in \eqref{eq:MAC_R2_DualLM} and \eqref{eq:MAC_R12_1_DualLM}--\eqref{eq:MAC_R12_2_DualLM}.  It suffices to let the choices of $a_1$ and $a_2$ in \eqref{eq:MAC_R12_1_DualLM}--\eqref{eq:MAC_R12_2_DualLM} be identical, since the theorem statement only requires one of the two to hold.

The interested reader is referred to \cite{Sca16a} for the details of the above-outlined techniques.



\chapter[Multi-User Coding Techniques for Single-User Channels]{Multi-User Coding Techniques \\ for Single-User Channels} \label{ch:multiuser}

\section{Introduction}

The parallel channel example of Section \ref{sec:mu_parallel} reveals the somewhat surprising fact that multiple-access coding techniques can lead to improved achievable rates when applied to single-user channels.  This suggests that other multi-user coding techniques may also be beneficial for the single-user setting, i.e., that certain types of structure in the codebook can be preferable to letting all of the codewords be independent.  In this section, we explore this phenomenon in detail, with a particular emphasis on superposition coding techniques.  We consider the point-to-point setup described in Section \ref{sec:setup_general}, assuming finite alphabets except where stated otherwise.

The improvements of constant-composition and cost-constrained random coding over i.i.d.~random coding ({\em cf.}, Sections \ref{ch:single_user}--\ref{ch:single_user_cont}) can also be interpreted as being due to a form of structure in the code.  The difference is that such improvements are a result of introducing dependencies between the symbols of individual codewords, whereas the techniques of this section introduce dependencies between the different codewords of the codebook.
 
The random coding constructions of this section, as well as those of Section \ref{ch:single_user}, can be summarized as follows, in non-increasing order of achievable rate:
\begin{enumerate}
    \item Refined superposition coding (Theorem \ref{thm:RSC_Main});
    \item Standard superposition coding (Theorem \ref{thm:SC_Rate_SC});
    \item Expurgated parallel coding (Theorem \ref{thm:expurg_mac});
    \item Constant-composition or cost-constrained coding with independent codewords (Theorem \ref{thm:LM});
    \item i.i.d.~coding with independent codewords (Theorem \ref{thm:GMI}).
\end{enumerate}
We will see that the gap between 1) and 2) can be strict for a given input distribution; no examples are known where the gap between 2) and 3) is strict; and the gaps between the remaining three can be strict even for an optimized input distribution.

Throughout the section, we will focus primarily on proving the achievability of primal-domain expressions.  Similarly to the previous sections, the rates will exhibit ensemble tightness and permit equivalent dual expressions and continuous-alphabet extensions.  However, many of these results become increasingly cumbersome to state and prove, so for the most part, we will refer the interested reader to \cite{Sca16a,ScarlettThesis} for the details.

This section is predominantly based on the works of Lapidoth \cite{Lap96}, Scarlett {\em et al.} \cite{Sca16a,ScarlettThesis}, and Somekh-Baruch \cite{Som15}.

\section{Expurgated Parallel Coding}

We saw in Section \ref{ch:mac} that by applying the achievable rate of the mismatched MAC to the single-user setting, it is possible to improve on the LM rate.  In particular, we saw an example where such an approach proves that the mismatch capacity equals the matched capacity, but the LM rate is strictly smaller.

While we only demonstrated this phenomenon in a single example, one can use the same idea to establish a general achievable rate: Fix any finite alphabets $\Xc_1,\Xc_2$, input distributions $Q_1,Q_2$, and a function $\psi \,:\, \Xc_1 \times \Xc_2 \to \Xc$, and apply Theorem \ref{thm:MAC_Rate} to the mismatched DM-MAC $W(y|\psi(x_1,x_2))$ with metric $q(\psi(x_1,x_2),y)$.  It follows that any rate $R = R_1 + R_2$ satisfying the conditions therein is achievable for the mismatched single-user channel $W(y|x)$ with metric $q(x,y)$.

It was shown by Lapidoth \cite{Lap96} that we can in fact do better by considering the collection of random codeword pairs $(\Xv_{1}^{(i)}, \Xv_2^{(j)})$, and keeping only the pairs whose empirical distribution is close to $Q_1 \times Q_2$.  This {\em expurgation of non-typical codeword pairs} has a negligible impact on the coding rates, but can lead to improved conditions for vanishing error probability.  One cannot apply this argument to the mismatched MAC itself, as it breaks the requirement of having two distinct codebooks (one per user).  In the single-user setting, there is no such requirement, so we are free to pick subsets of the pairs $(\Xv_{1}^{(i)}, \Xv_2^{(j)})$ to our liking.

The resulting single-user achievable rate is stated as follows; we omit the proof, since in Section \ref{sec:sc} we will use a simpler superposition coding technique to derive an achievable rate that can be weakened to this one.

\begin{theorem} \label{thm:expurg_mac}
    {\em (Expurgated parallel coding rate)}
    Consider a mismatched DMC $(W,q)$, and fix the finite alphabets $(\Xc_{1},\Xc_{2})$, input distributions $(Q_1,Q_2)$, and function $\psi\,:\,\Xc_{1}\times\Xc_{2}\to\Xc$.  Then, the rate
    \begin{equation}
        R = R_1 + R_2
    \end{equation}
    is achievable for $(W,q)$ provided that the pair $(R_1,R_2)$ satisfies
    \begin{align}
    R_{1}       & \le\min_{\substack{\Ptilde_{X_{1}X_{2}Y} \,:\, \Ptilde_{X_1X_2}=P_{X_1X_2}, \, \Ptilde_{X_{2}Y}=P_{X_{2}Y} \\
            \EE_{\Ptilde}[\log q(\psi(X_{1},X_{2}),Y)]\ge\EE_{P}[\log q(\psi(X_{1},X_{2}),Y)]}} I_{\Ptilde}(X_{1};Y|X_{2}), \label{eq:Ex_R1} \\
    R_{2}       & \le\min_{\substack{\Ptilde_{X_{1}X_{2}Y} \,:\,       \Ptilde_{X_1X_2}=P_{X_1X_2}, \, \Ptilde_{X_{1}Y}=P_{X_{1}Y} \\
            \EE_{\Ptilde}[\log q(\psi(X_{1},X_{2}),Y)]\ge\EE_{P}[\log q(\psi(X_{1},X_{2}),Y)]}}I_{\Ptilde}(X_{2};Y|X_{1}), \label{eq:Ex_R2} \\
    R_{1}+R_{2} & \le\min_{\substack{\Ptilde_{X_{1}X_{2}Y} \,:\, \Ptilde_{X_1X_2}=P_{X_1X_2}, \, \Ptilde_{Y}=P_{Y}\\
            \EE_{\Ptilde}[\log q(\psi(X_{1},X_{2}),Y)]\ge\EE_{P}[\log q(\psi(X_{1},X_{2}),Y)], \\ I_{\Ptilde}(X_{1};Y)\le R_{1},\, I_{\Ptilde}(X_{2};Y)\le R_{2}}} I_{\Ptilde}(X_1,X_2;Y), \label{eq:Ex_R12}
    \end{align}
    where $P_{X_1X_2Y}(x_1,x_2,y) = Q_1(x_1)Q_2(x_2)W(y|\psi(x_1,x_2))$.
\end{theorem}

The above rate conditions are identical to those of the achievable rate region for the MAC in Theorem \ref{thm:MAC_Rate}, except that each minimization is further subject to $\Ptilde_{X_1X_2}= P_{X_1X_2}$.  This results from the fact that each codeword pair is constructed to have empirical distribution close to $P_{X_1X_2} = Q_1 \times Q_2$.  Since the minimization problems in the rate conditions are more constrained, the resulting rate is at least as high as that attained directly from Theorem \ref{thm:MAC_Rate}.  We also recall that the constraints $I_{\Ptilde}(X_{1};Y) \le R_1$ and $I_{\Ptilde}(X_{2};Y) \le R_2$ in \eqref{eq:Ex_R12} are non-standard; their main implications discussed following Theorem \ref{thm:MAC_Rate} also apply here.

A disadvantage of the achievable rate in Theorem \ref{thm:expurg_mac} is that it has numerous auxiliary parameters, namely, $\Xc_1$, $\Xc_2$, $Q_1$, $Q_2$, and $\psi$.  While natural choices arose in the parallel channel example of Section \ref{sec:mac_numerical}, it is generally difficult to jointly optimize them.  This will be another advantage of superposition coding in Section \ref{sec:sc}, which will have fewer auxiliary parameters.

We briefly mention that equivalent dual expressions for \eqref{eq:Ex_R1}--\eqref{eq:Ex_R12} can be found in \cite{ScarlettThesis}; these are nearly identical to those of Theorem \ref{thm:MAC_DualRate}, except that the separate functions $a_1(x_1)$ and $a_2(x_2)$ are replaced by {\em joint} functions $a(x_1,x_2)$. 

\subsection{Comparison to the LM Rate}

The easiest way to recover the LM rate (in its primal form \eqref{eq:INTR_PrimalLM}) from Theorem \ref{thm:expurg_mac} is to trivially choose $|\Xc_2| = 1$, $R_2 = 0$, and $\psi(x_1,x_2) = x_1$.  By doing so, conditions \eqref{eq:Ex_R1} and \eqref{eq:Ex_R12} both weaken to $R_1 \le \LM(Q_{X_1})$.  In fact, under the preceding choices, we not only recover the LM rate, but the random-coding ensemble itself reduces to constant-composition random coding with independent codewords.

It is also worth noting the following general statement: Upon removing the two mutual information constraints from \eqref{eq:Ex_R12}, the right-hand side reduces to the LM rate with input distribution $Q_X$ equal to the marginal distribution of $(X_1,X_2,X) \sim Q_1(x_1)Q_2(x_2)\openone\{ x = \psi(x_1,x_2) \}$.  This is formalized as follows.

\begin{lemma} \label{lem:weaken_expurg}
    {\em (Expurgated parallel coding and the LM rate)}
    Under the setup of Theorem \ref{thm:expurg_mac}, we have
    \begin{align}
        &\min_{\substack{\Ptilde_{X_{1}X_{2}Y} \,:\, \Ptilde_{X_1X_2}=P_{X_1X_2}, \, \Ptilde_{Y}=P_{Y} \\ \EE_{\Ptilde}[\log q(\psi(X_{1},X_{2}),Y)]\ge\EE_{P}[\log q(\psi(X_{1},X_{2}),Y)]}} I_{\Ptilde}(X_1,X_2;Y) \nonumber \\
        &\qquad\qquad\qquad\qquad=\min_{\substack{\Ptilde_{XY} \,:\, \Ptilde_X= P_X, \Ptilde_Y= P_Y, \\ \EE_{\Ptilde}[\log q(X,Y)]\ge\EE_{P}[\log q(X,Y)]}} I_{\Ptilde}(X;Y), \label{eq:ex_equiv}
    \end{align}
    where $P_{X_1X_2XY}(x_1,x_2,x,y)$ is the joint distribution $Q_1(x_1)Q_2(x_2)\openone\{x=\psi(x_1,x_2)\} W(y|\psi(x_1,x_2))$. 
\end{lemma}
\begin{proof}
    Fixing any feasible $\Ptilde_{X_1X_2Y}$ on the left-hand side of \eqref{eq:ex_equiv}, and letting $\Ptilde_{X_1X_2XY}$ be an arbitrary joint distribution consistent with $\Ptilde_{X_1X_2Y}$ such that $X = \psi(X_1,X_2)$, we have
    \begin{align}
        I_{\Ptilde}(X_1,X_2;Y) 
            &=  I_{\Ptilde}(X_1,X_2,X;Y) \\
            &= I_{\Ptilde}(X;Y) + I_{\Ptilde}(X_1,X_2;Y|X) \\
            &\ge I_{\Ptilde}(X;Y). \label{eq:ex_mi_step3}
    \end{align}
    We now minimize both sides of this inequality over all $\Ptilde_{X_1X_2XY}$ such that $\Ptilde_{X_1X_2Y}$ is feasible on the left-hand side of \eqref{eq:ex_equiv}, and $X = \psi(X_1,X_2)$.  The left-hand side is as in \eqref{eq:ex_equiv}, and the right-hand side yields the minimization problem
    \begin{equation}
        \min_{\substack{\Ptilde_{X_{1}X_{2}XY} \,:\, \Ptilde_{X_1X_2}=P_{X_1X_2}, \, X = \psi(X_1,X_2), \, \Ptilde_{Y}=P_{Y}, \\ \EE_{\Ptilde}[\log q(X,Y)]\ge\EE_{P}[\log q(X,Y)]}} I_{\Ptilde}(X;Y).
    \end{equation}
    Since the inequality constraint and objective depend on $\Ptilde_{X_1X_2XY}$ only through $\Ptilde_{XY}$, the first two constraints only amount to constraining $\Ptilde_X = P_X$, and we therefore recover the right-hand side of \eqref{eq:ex_equiv}.
    
    We have established that the left-hand side of \eqref{eq:ex_equiv} is lower-bounded by the right-hand side.  However, the only step containing an inequality is \eqref{eq:ex_mi_step3}, and this holds with equality when $\Ptilde_{X_1X_2XY}$ is chosen such that $(X_1,X_2) \to X \to Y$ forms a Markov chain.  Hence, the minimizer must have this property, implying \eqref{eq:ex_equiv}.
\end{proof}

Lemma \ref{lem:weaken_expurg} further highlights the importance of the mutual information constraints in \eqref{eq:Ex_R12}: For all mismatched DMCs, if these constraints are removed, then we cannot hope to improve on the LM rate.

\section{Superposition Coding} \label{sec:sc}

One of the most fundamental multi-user coding techniques in network information theory is superposition coding, which has been applied to broadcast channels, interference channels, and more \cite{Elg11}.  The idea is to generate a number of independent ``cloud centers'' $\{\Uv^{(i)}\}$, and then for each corresponding index $i$, generate a number of codewords $\{\Xv^{(i,j)}\}$ that are {\em conditionally independent} given $\Uv^{(i)}$.  In the degraded broadcast channel model \cite[Ch.~5]{Elg11}, this is done in a manner such that one user can identify the cloud center, while the other user can recover the precise codeword within the cloud.

In this subsection, we derive an achievable rate based on superposition coding with constant-composition codewords \cite{Som15,Sca16a}.  We will show in Section \ref{sec:sc_cmp} that this rate is at least as high as that of expurgated parallel coding (Theorem \ref{thm:expurg_mac}).  We will also see that superposition coding requires fewer auxiliary parameters, and hence, even when a strict improvement over Theorem \ref{thm:expurg_mac} is not possible, the difficult task of finding a good set of parameters may be simplified.

\subsection{Codebook Construction} \label{sec:sc_code}

The random-coding ensemble depends on an auxiliary alphabet $\Uc$, an auxiliary  codeword distribution $P_{\Uv} \in \Pc(\Uc^n)$, and a conditional codeword distribution $P_{\Xv|\Uv} \in \Pc(\Xc^n|\Uc^n)$.  We fix a pair $(R_{0},R_{1})$ and generate the codewords in two steps:
\begin{enumerate}
    \item An auxiliary codebook $\{\Uv^{(i)}\}_{i=1}^{M_{0}}$  with $M_{0}\defeq\lfloor e^{nR_{0}} \rfloor$ codewords of length $n$ is generated, with each auxiliary codeword independently drawn from $P_{\Uv}$. 
    \item For each $i=1,\dotsc,M_{0}$, a codebook $\{\Xv^{(i,j)}\}_{j=1}^{M_{1}}$ with $M_{1} \defeq \lfloor e^{nR_{1}} \rfloor$ codewords of length $n$ is generated, with each codeword conditionally independently drawn from $P_{\Xv|\Uv}(\,\cdot\, | \Uv^{(i)})$.  
\end{enumerate}
The message $m$ at the input to the encoder is indexed as $(m_{0},m_{1})$, and for any such pair, the corresponding transmitted codeword is $\Xv^{(m_{0},m_{1})}$.  Stated concisely, we have
\begin{equation}
    \Big\{\Big( \Uv^{(i)}, \big\{\Xv^{(i,j)}\big \}_{j=1}^{M_1}\Big) \Big\}_{i=1}^{M_0}\sim \prod_{i=1}^{M_0} P_{\Uv}(\uv^{(i)}) \prod_{j=1}^{M_1} P_{\Xv|\Uv}(\xv^{(i,j)} | \uv^{(i)}).
\end{equation}
We consider constant-composition coding with an input distribution $Q_{UX} \in \Pc(\Uc \times \Xc)$.  Specifically, letting $Q_{UX,n} \in \Pc_n(\Uc \times \Xc)$ be a joint type with the same support as $Q_{UX}$ such that $\| Q_{UX,n} - Q_{UX} \|_{\infty} \le \frac{1}{n}$, we set
\begin{gather}
    P_{\Uv}(\uv) = \frac{1}{|\Tclass(Q_{U,n})|} \openone\{ \uv \in \Tclass(Q_{U,n}) \}, \\
    P_{\Xv|\Uv}(\xv|\uv) = \frac{1}{|\Tclass_{\uv}(Q_{UX,n})|} \openone\{ (\uv,\xv) \in \Tclass(Q_{UX,n}) \},
\end{gather}
where $\Tclass_{\uv}(Q_{UX,n})$ is a {\em conditional type class}, i.e., the set of all $\xv$ such that $(\uv,\xv)$ has the specified joint type.  This construction is referred to as {\em constant-composition superposition coding}.

One could also consider i.i.d.~superposition coding, which is the standard choice in matched multi-user settings (e.g., the broadcast channel) \cite{Elg11}.  Similarly to Section \ref{ch:single_user}, this choice leads to achievable rates that are similar to the constant-composition ensemble, but generally weaker.  For brevity, we focus our attention on the constant-composition case.  We briefly discuss a cost-constrained ensemble for continuous-alphabet channels in Section \ref{sec:sc_proof}.

\subsection{Statement of Achievable Rate}

We proceed by stating the achievable rate, first in the primal form and then in the dual form.  The proofs are discussed in Section \ref{sec:sc_proof}.

\begin{theorem} \label{thm:SC_Rate_SC}
    {\em (Superposition coding rate)}
    Consider a DMC $(W,q)$.  For any finite auxiliary alphabet $\Uc$ and input distribution
    $Q_{UX} \in \Pc(\Uc \times \Xc)$, the rate
    \begin{equation}
        R = R_0 + R_1
    \end{equation} 
    is achievable provided that $(R_0,R_1)$ satisfy
    \begin{align}
        R_{1}       & \le\min_{\substack{\Ptilde_{UXY} \,:\, \Ptilde_{UX}=P_{UX},\,\Ptilde_{UY}=P_{UY}, \\ \EE_{\Ptilde}[\log q(X,Y)]\ge\EE_{P}[\log q(X,Y)]}}I_{\Ptilde}(X;Y|U) \label{eq:SC_R1_CC} \\
        R_{0}+R_{1} & \le \min_{\substack{\Ptilde_{UXY} \,:\, \Ptilde_{UX}=P_{UX},\,\Ptilde_{Y}=P_{Y}, \\  \EE_{\Ptilde}[\log q(X,Y)]\ge\EE_{P}[\log q(X,Y)], \\ I_{\Ptilde}(U;Y)\le R_{0}}} I_{\Ptilde}(U,X;Y). \label{eq:SC_Rsum_CC}
    \end{align}
    Moreover, this rate is ensemble-tight for constant-composition superposition coding.
\end{theorem}

We notice that these conditions bear a strong resemblance to those of Theorem \ref{thm:expurg_mac}.  In fact, the two constructions are not as different as they may seem; both can be viewed as constructing a 2D grid of codewords, one indexed by $(\Xv_1^{(i)}, \Xv_2^{(j)})$, and the other by $(\Uv^{(i)},\Xv^{(i,j)})$.  The main difference is that under superposition coding, the rows are independent; this simplifies the construction and analysis, as it allows us to let all $(\Uv,\Xv)$ pairs have a fixed {\em joint} composition without any need for expurgation.

It may appear that this extra independence of superposition coding makes it ``less structured'' than expurgated parallel coding, or that it weakens the rate due to a single mutual information constraint in \eqref{eq:SC_Rsum_CC} in contrast with the two in \eqref{eq:Ex_R12}.  However, we will see in Section \ref{sec:sc_cmp} that Theorem \ref{thm:SC_Rate_SC} can be weakened to Theorem \ref{thm:expurg_mac}, and this will crucially rely on the fact that only one of the two mutual information constraints in \eqref{eq:Ex_R12} can be active at a time (similarly to Lemma \ref{lem:MAC_split}).

The preceding rate conditions admit a similar dual formulation to Theorem \ref{thm:MAC_DualRate}, as stated in the following result.  Recall that $\EE_{Z}[\cdot]$ denotes averaging over some random variable $Z$ with all other random variables held fixed.

\begin{theorem} \label{thm:sc_dual}
    {\em (Dual form of superposition coding rate)} For any mismatched DMC $(W,q)$ and input distribution $Q_{UX}$, the rate conditions in \eqref{eq:SC_R1_CC}--\eqref{eq:SC_Rsum_CC} can be written as
    \begin{gather}
        R_{1}\le\sup_{\sgz,a(\cdot,\cdot)}\EE\left[\log\frac{q(X,Y)^{s}e^{a(U,X)}}{\EE[q(\Xtilde,Y)^{s}e^{a(U,\Xtilde)}\,|\, U,Y]}\right], \label{eq:SC_R1_Dual} \\
        R_{0} \le \sup_{\rho\in[0,1],\sgz,a(\cdot,\cdot)} \EE\left[\log\frac{\big(q(X,Y)^{s}e^{a(U,X)}\big)^{\rho}}{\EE_{\Ubar}\Big[\Big(\EE_{\Xbar}\big[q(\Xbar,Y)^{s}e^{a(\Ubar,\Xbar)} \big]\Big)^{\rho} \Big]}\right] - \rho R_{1}, \label{eq:SC_Rsum_Dual}
    \end{gather}
    where $(U,X,Y,\Xtilde,\Ubar,\Xbar)\sim Q_{UX}(u,x)W(y|x)Q_{X|U}(\xtilde|u)Q_{UX}(\ubar,\xbar)$.
\end{theorem}

We will briefly discuss a direct derivation of this dual form in Section \ref{sec:sc_proof}, as well as its extension to continuous-alphabet channels.

\subsection{Example 1: Zero-Undetected Error Capacity}  \label{sec:mu_zuec}

We return to the example of Section \ref{sec:su_zuec}, where $\Xc=\Yc=\{0,1,2\}$, and the channel and metric are described by the entries of the matrices
\begin{align} 
    \Wv & = \left[\begin{array}{ccc}
    0.75 & 0.25 & 0 \\
    0 & 0.75 & 0.25 \\
    0.25 & 0 & 0.75 \\
    \end{array}\right], \qquad
    \qv = \left[\begin{array}{ccc}
    ~~1~\, & ~~1\,~ & ~~0~~ \\
    ~~0~\, & ~~1\,~ & ~~1~~ \\
    ~~1~\, & ~~0\,~ & ~~1~~ \\
    \end{array}\right], \label{eq:Example_q1_v2}
\end{align}
where $x$ indexes the rows, and $y$ indexes the columns.  We saw in Section \ref{sec:su_zuec} that the optimized LM rate is $\CLM=0.599$ bits/use, and is attained using the input distribution $Q_X=(0.449,0.551,0)$.  

It was stated in \cite{Ahl96} that the rate $\CLM^{(2)}$ obtained by considering the second-order product of the channel and metric (see Section \ref{sec:multi_LM}) is equal to $\CLM^{(2)}=0.616$ bits/use.  Using local optimization techniques, we verified that this rate is achieved by the input distribution $(0,0.250,0,0.319,0,0,0,0.181,0.250)$, where the order of the inputs is $(0,0),(0,1),(0,2),(1,0),\dotsc,(2,2)$.  

The global optimization of \eqref{eq:SC_R1_CC}--\eqref{eq:SC_Rsum_CC}
over $\Uc$ and $Q_{UX}$ appears to be difficult. However, setting $|\Uc|=2$
and applying local optimization techniques using a number of starting
points, we obtained an achievable rate of $R_{\textsc{sc}}^{*}=0.695$ bits/use, with $Q_{U}=(0.645,0.355)$, $Q_{X|U}(\cdot|1)=(0.3,0.7,0)$, and $Q_{X|U}(\cdot|2)=(0,0,1)$. Hence, even if this is not the globally optimal choice of parameters, it reveals the interesting fact that superposition coding not only yields a strict improvement over the single-letter  LM rate, but also over the two-letter version.

\subsection{Example 2: The LM rate is Not Tight for Binary-Input DMCs} \label{sec:binary_input}

It was claimed in an early work of Balakirsky \cite{Bal95} that for {binary-input} DMCs, the LM rate with an optimized input distribution is tight, i.e., $\CM = \CLM$.  However, a recent work of the present authors provided a counter-example to this claim \cite{Sca15a}, which we describe in the following.  We let $\Xc=\{0,1\}$ and $\Yc=\{0,1,2\}$, and consider the channel and metric described by the entries of the $|\Xc|\times|\Yc|$ matrices
\begin{align}
    \Wv & =\left[\begin{array}{ccc}
    0.97 & 0.03 & 0\\
    0.1 & 0.1 & 0.8
    \end{array}\right],~~\qv=\left[\begin{array}{ccc}
    1 & 1 & 1\\
    1 & 0.5 & 1.36
    \end{array}\right].\label{eq:CNV_Channel}
\end{align}
\paragraph{LM rate.} Using numerical evaluations and a brute force search over $Q_X$, along with theoretical results quantifying the approximation errors incurred, it was shown in \cite{Sca15a} that the LM rate with an optimized input distribution, $\CLM = \max_{Q_X} \LM(Q_X)$, satisfies
\begin{equation}
0.19746 \le \CLM \le 0.19751 \quad\mathrm{bits/use}.\label{eq:CNV_BoundLM}
\end{equation}
The optimal input distribution, rounded to five decimal places, is given by $Q_X = (0.75597, 0.24403)$.

\paragraph{Superposition coding rate.} When we applied the superposition coding rate of Theorem \ref{thm:SC_Rate_SC} directly to $(W,q)$, we only managed to obtain a rate coinciding with $\CLM$.  However, the technique of passing to the $k$-th order product alphabet, introduced in Section \ref{sec:multi_LM}, is equally valid for other random-coding rates, including the superposition coding rate of Theorem \ref{thm:SC_Rate_SC}.  We consider the case $k=2$, and denote the resulting rate with an optimized input distribution $Q_{UX^2}$ as $\CSC^{(2)}$.  The counter-example to \cite{Bal95} follows by combining \eqref{eq:CNV_BoundLM} with the following:
\begin{equation}
    \CSC^{(2)} \ge 0.19908\quad\mathrm{bits/use},\label{eq:CNV_BoundSC2}
\end{equation}
which clearly implies $\CM>\CLM$.

\begin{figure}
    \begin{centering}
        \includegraphics[width=0.6\columnwidth]{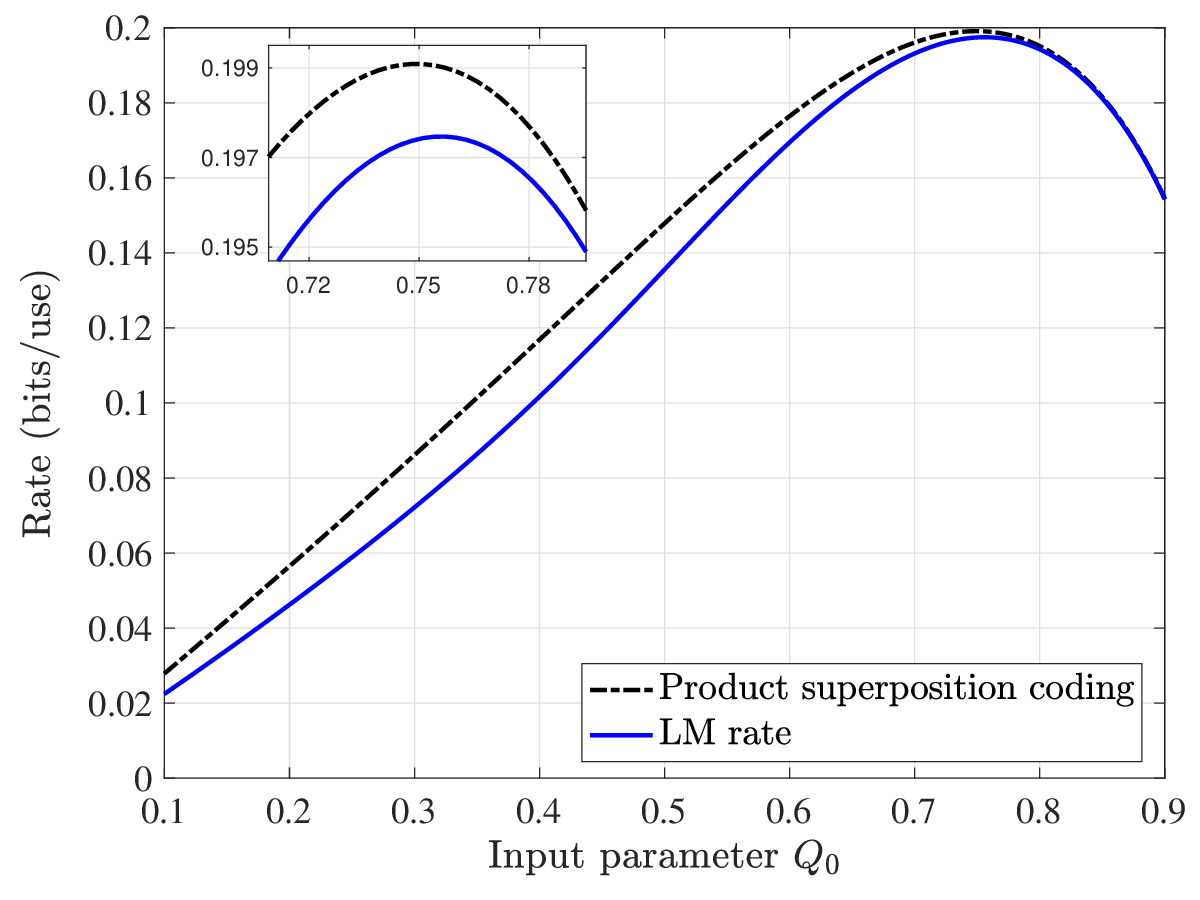}
        \par
    \end{centering}
    
    \caption{Binary-input example: LM rate and second-order superposition coding rate as a function of $Q_0$ for the mismatched channel described by \eqref{eq:CNV_Channel}.} \label{fig:BI_CounterExample}
\end{figure}

To establish \eqref{eq:CNV_BoundSC2}, we set $|\Uc| = 2$, and choose the input distribution $Q_{UX^2}$ on $\Uc \times \Xc^2$ as follows:
\begin{align}
    Q_{U} & =\left[\begin{array}{cc}
    1-Q_{1}^{2} & Q_{1}^{2}
    \end{array}\right]\label{eq:CNV_ChoiceQU}\\
    Q_{X^2|U=0} & =\frac{1}{1-Q_{1}^{2}}\left[\begin{array}{cccc}
    Q_{0}^{2} & Q_{0}Q_{1} & Q_{0}Q_{1} & 0
    \end{array}\right] \label{eq:CNV_ChoiceQX1} \\
    Q_{X^2|U=1} & =\left[\begin{array}{cccc} 
    0 & 0 & 0 & 1
    \end{array}\right]\label{eq:CNV_ChoiceQX2}
\end{align}
with $Q_0 = 0.749$ and $Q_1 = 0.251$. Here, the alphabet $\Xc^2$ is listed in the order $(0,0)$, $(0,1)$, $(1,0)$, $(1,1)$.   Note that this choice of $Q_{UX^2}$ ensures that the marginal distribution $Q_{X^2}$ is the two-fold product of $Q_X = (Q_0,Q_1)$.

\paragraph{Other input distributions.} The construction in \eqref{eq:CNV_ChoiceQX2} is valid for any pair $(Q_0,Q_1)$ summing to one, and it is interesting to compare the resulting rate with the LM rate applied to $Q_X = (Q_0,Q_1)$.  The rates are shown in Figure \ref{fig:BI_CounterExample}, where we observe that superposition coding consistently improves on the LM rate for the values of $Q_0$ shown.

\subsection{Comparison to Expurgated Parallel Coding} \label{sec:sc_cmp}

The following lemma formally states that Theorem \ref{thm:SC_Rate_SC} provides an achievable rate at least as high as that of Theorem \ref{thm:expurg_mac}.  No examples are known where a strict improvement is attained, suggesting that the two rates could be equivalent.  Superposition coding, however, has the advantage of having fewer parameters (namely, $(\Uc,Q_{UX})$ instead of $(\Xc_1,\Xc_2,\psi,Q_1,Q_2)$), and in the authors' experience, this usually makes it easier to find good choices (e.g., based on intuition, or local optimization techniques).

\begin{lemma}
    {\em (Comparison of superposition coding and expurgated parallel coding)} 
    For any mismatched DMC $(W,q)$, we have
    \begin{equation}
         \sup_{\Xc_1,\Xc_2,\psi,Q_1,Q_2} \, \max_{(R_1,R_2) \in \RegMAC} R_1 + R_2 \le \sup_{\Uc,Q_{UX}} \, \max_{(R_0,R_1) \in \RegSC} R_0 + R_1,
    \end{equation}
    where $\RegMAC = \RegMAC(\Xc_1,\Xc_2,\psi,Q_1,Q_2)$ is the set of rate pairs satisfying \eqref{eq:Ex_R1}--\eqref{eq:Ex_R12}, and $\RegSC = \RegSC(\Uc,Q_{UX})$ is the set of rate pairs satisfying \eqref{eq:SC_R1_CC}--\eqref{eq:SC_Rsum_CC}.
\end{lemma}

We only outline the proof here, and refer the reader to \cite{ScarlettThesis} for the full details.  First, analogously to Lemma \ref{lem:MAC_split}, the condition in \eqref{eq:Ex_R12} holds if and only if at least one of the following hold:
\begin{align}
R_{1}+R_{2} & \le\min_{\substack{\Ptilde_{X_{1}X_{2}Y} \,:\, \Ptilde_{X_1X_2}=P_{X_1X_2}, \, \Ptilde_{Y}=P_{Y}\\
        \EE_{\Ptilde}[\log q(\psi(X_{1},X_{2}),Y)]\ge\EE_{P}[\log q(\psi(X_{1},X_{2}),Y)], \\ I_{\Ptilde}(X_{1};Y)\le R_{1}}} I_{\Ptilde}(X_1,X_2;Y) \label{eq:Ex_R12a} \\
R_{1}+R_{2} & \le\min_{\substack{\Ptilde_{X_{1}X_{2}Y} \,:\, \Ptilde_{X_1X_2}=P_{X_1X_2}, \, \Ptilde_{Y}=P_{Y}\\
        \EE_{\Ptilde}[\log q(\psi(X_{1},X_{2}),Y)]\ge\EE_{P}[\log q(\psi(X_{1},X_{2}),Y)], \\ I_{\Ptilde}(X_{2};Y)\le R_{2}}} I_{\Ptilde}(X_1,X_2;Y). \label{eq:Ex_R12b}
\end{align}
In other words, the two mutual information constraints in \eqref{eq:Ex_R12} can be considered one at a time without affecting the rates. 

Once this is established, the proof proceeds in two parts: 
\begin{enumerate}
    \item Show that by identifying $\Usc=\Xiimac$, $\Risc=\Rimac$ and $\Rosc=\Riimac$, we can weaken \eqref{eq:SC_R1_CC} to \eqref{eq:Ex_R1}, and \eqref{eq:SC_Rsum_CC} to \eqref{eq:Ex_R12b}. 
    \item Show that by identifying $\Usc=\Ximac$, $\Risc=\Riimac$ and $\Rosc=\Rimac$, we can weaken \eqref{eq:SC_R1_CC} 
    to \eqref{eq:Ex_R2}, and \eqref{eq:SC_Rsum_CC} to \eqref{eq:Ex_R12a}. 
\end{enumerate}
Here, the superscript explicitly denotes which coding technique the parameters correspond to.  The above two claims are proved using identical steps, so we consider the former.  We choose the input distribution in terms of the expurgated parallel coding parameters as
\begin{equation}
Q_{UX}(u,x)=\sum_{x_{1}}Q_{1}(x_{1})Q_{2}(u)\openone\{x=\psi(x_{1},u)\}. \label{eq:SC_Comparison4}
\end{equation}
By doing so, simple manipulations to the objective functions and constraints in \eqref{eq:SC_R1_CC} and \eqref{eq:SC_Rsum_CC} allow us to recover the conditions \eqref{eq:Ex_R1} and \eqref{eq:Ex_R12b}.  For example, if we substitute this choice of $Q_{UX}$ into the superposition coding metric constraint
\begin{equation}
    \sum_{u,x,y}Q_{UX}(u,x)\Ptilde_{Y|UX}(y|u,x)\log q(x,y) \ge \sum_{x,y}Q_{X}(x)W(y|x)\log q(x,y),
\end{equation}
then we get the following upon renaming $u$ as $x_2$:
\begin{align}
    &\sum_{x_{1},x_{2},y}Q_{1}(x_{1})Q_{2}(x_{2})\Ptilde_{Y|X_{1}X_{2}}(y|x_{1},x_{2})\log q(\psi(x_{1},x_{2}),y) \nonumber \\
    &\qquad\qquad \ge\sum_{x_{1},x_{2},y}Q_{1}(x_{1})Q_{2}(x_{2})W(y|\psi(x_{1},x_{2}))\log q(\psi(x_{1},x_{2}),y),
\end{align}
where $\Ptilde_{Y|X_{1}X_{2}}(y|x_1,x_2) = \Ptilde_{Y|X_{1}X}(y|x_{1},\psi(x_{1},x_{2}))$.  This is precisely the metric constraint for expurgated parallel coding.  The desired claim follows by treating the other constraints and objective functions similarly, and then minimizing over $\Ptilde_{Y|X_{1}X_{2}}$.

We note that a similar argument is also given \cite[Prop.~1]{Som15}, in which a form of superposition coding is considered with MAC-like parameters $\Xc_1$, $\Xc_2$, and $\psi$, as opposed to the parameters $\Uc$ and $Q_{UX}$ considered here.  The parameters $(\Xc_1,\Xc_2, \psi)$ are viewed as forming a {\em cognitive} MAC in which one user knows both messages, meaning that superposition coding can be used to induce general joint input distributions $Q_{X_1X_2}$, rather than only product distributions of the form $Q_{X_1} \times Q_{X_2}$.  An advantage of this perspective is that it permits the direct comparison of superposition coding and expurgated parallel coding for {\em fixed} parameters $(\Xc_1,\Xc_2, \psi)$.  

In particular, the following argument from \cite{Som15} highlights that the superposition coding rates can indeed be strictly higher with fixed parameters: Fix $(\Xc_1,\Xc_2)$ and $\Xc$ with $|\Xc| = |\Xc_1| \cdot |\Xc_2|$, and let $\psi$ be any one-to-one mapping from $\Xc_1 \times \Xc_2$ to $\Xc$.  Then, varying $Q_{X_1X_2}$ can induce any distribution on $X$, but varying $Q_{X_1}$ and $Q_{X_2}$ can only induce a narrower range of distributions (e.g., any symbol having zero probability forces others to also be zero, due to the product structure).  This restriction lowers the achievable rate, for example, when matched decoding ($q=W$) is used and the capacity-achieving input distribution does not conform to the product structure.

\subsection{Proof techniques for the achievable rate} \label{sec:sc_proof}

The proof of Theorem \ref{thm:SC_Rate_SC} is comparatively simpler than that of the multiple-access channel (Theorem \ref{thm:MAC_Rate}), so we focus mostly on the ideas and intuition rather than the full details.

\paragraph{Separation into two error events.} Recalling the codebook construction in Section \ref{sec:sc_code}, we assume without loss of generality that the transmitted codeword corresponds to $(m_0,m_1) = (1,1)$, and we define the following two error events:
\begin{tabbing}
    ~~~{\emph{(Type 0)}}~~~ \= $q^n(\Xv^{(i,j)},\Yv) \ge q^n(\Xv,\Yv)$ for some $i \ne 1$, $j$; \\
    ~~~{\emph{(Type 1)}}~~~ \> $q^n(\Xv^{(1,j)},\Yv) \ge q^n(\Xv,\Yv)$ for some $j \ne 1$.
\end{tabbing}
Denoting the probabilities of these events by $\peobar(n,M_{0},M_{1})$ and $\peibar(n,M_{1})$ respectively, it follows that the overall random-coding error probability $\pebar(n,M_0,M_1)$ satisfies
\begin{equation}
    \max\{\peobar,\peibar\} \le \pebar \le \peobar + \peibar. \label{eq:SC_OverallProb}
\end{equation}

\paragraph{Non-asymptotic bounds.} The first step of the analysis is to establish the following non-asymptotic bounds:
\begin{align}
    \peobar \le \EE_{\Uv,\Xv,\Yv}\Big[\min\Big\{ 1,&(M_{0}-1)\EE_{\Uvbar}\Big[\min\big\{1, \nonumber \\ &M_{1}\PP_{\Xvbar}\big[q^{n}(\Xvbar,\Yv)\ge q^{n}(\Xv,\Yv)\big]\big\}\Big]\Big\} \Big], \label{eq:sc_rcu0}
\end{align}
\begin{gather}
    \peibar \le \EE\Big[\min\Big\{1,(M_1-1)\PP\big[q^{n}(\Xvtilde,\Yv) \ge q^{n}(\Xv,\Yv) \,\big|\, \Uv,\Xv,\Yv\big]\Big\}\Big], \label{eq:sc_rcu1}
\end{gather}
where 
\begin{align}
    (\Uv,\Xv,\Yv,\Xvtilde,\Uvbar,\Xvbar)  \sim &P_{\Uv}(\uv)P_{\Xv|\Uv}(\xv|\uv)W^{n}(\yv|\xv) \nonumber \\ & \times P_{\Xv|\Uv}(\xvtilde|\uv)P_{\Uv}(\uvbar)P_{\Xv|\Uv}(\xvbar|\uvbar). \label{eq:SC_VecDistr}
\end{align}
For the type-1 error probability, \eqref{eq:sc_rcu1} follows by conditioning on $(\Uv,\Xv,\Yv)$ and applying the truncated union bound (i.e., the minimum of one and the standard union bound) over the error events.  On the other hand, more care is needed for the type-0 error probability.  We first write
\begin{equation}
    \peobar = \EE\Bigg[ \PP\bigg[\bigcup_{i\ne1,j}\Big\{q^{n}(\Xv^{(i,j)},\Yv) \ge q^{n}(\Xv,\Yv)\Big\} \,\Big|\, \Uv,\Xv,\Yv \bigg] \Bigg], \label{eq:sc_p0_exact}
\end{equation}
where we assume without loss of generality that $(m_0,m_1) = (1,1)$.  The upper bound \eqref{eq:sc_rcu0} then follows by applying the truncated union bound {first to the union over $i$, and then to the union over $j$}, similarly to the analysis of the DM-MAC in Section \ref{sec:MAC_Dual}.

\paragraph{Primal analysis techniques.} Once the above bounds are in place, we can  apply similar techniques to the single-user setting ({\em cf.}, Section \ref{ch:single_user}).  Focusing on the more interesting type-0 error event, we write \eqref{eq:sc_rcu0} in terms of joint types as follows:
\begin{align}
    &\peobar \le \sum_{\substack{P_{UXY} \in \Pc_n(\Uc\times\Xc\times\Yc) \,:\, \\ P_{UX} = Q_{UX,n}}}\PP\Big[\big(\Uv,\Xv,\Yv\big)\in \Tclass(P_{UXY})\Big] \nonumber \\ 
        &~~\times\min\Bigg\{1,
    (M_{0}-1)\sum_{\substack{\Ptilde_{UXY} \in \Pc_n(\Uc\times\Xc\times\Yc) \,:\, \\ P_{UX} = Q_{UX,n}, \, \Ptilde_Y = P_Y \\ \EE_{\Ptilde}[\log q(X,Y)]\ge\EE_{P}[\log q(X,Y)] }}\PP\Big[\big(\Uvbar,\yv\big)\in \Tclass(\Ptilde_{UY})\Big] \nonumber \\
        &~~\qquad\qquad\quad\times \min\bigg\{1, M_1 \PP\Big[\big(\uvbar,\Xvbar,\yv\big)\in \Tclass(\Ptilde_{UXY}) \,\big|\, \Uvbar = \uvbar\Big] \bigg\}\bigg]\Bigg\},\label{eq:SC_rewritten} 
\end{align}
where $(\uvbar,\yv)$ denotes an arbitrary pair such that 
$\yv\in \Tclass(P_{Y})$ and $(\uvbar,\yv)\in \Tclass(\Ptilde_{UY})$.  Here the constraints on $(U,X)$ arise due the constant-composition construction, and the constraint $\Ptilde_{Y}=P_{Y}$ arises since $(\Uv,\Xv,\Yv)$ and $(\Uvbar,\Xvbar^{(j)},\Yv)$ share the same $\Yv$ sequence.

The proof now proceeds as in the previous sections by applying standard properties of types, using the law of large numbers to establish $\|P_{UXY} - Q_{UX} \times W\|_{\infty} \le \delta$, and applying a continuity argument similar to Appendix \ref{sec:continuity_app}.  The probabilities in the second and third lines of \eqref{eq:SC_rewritten} give rise to the mutual information terms $I_{\Ptilde}(U;Y)$ and $I_{\Ptilde}(X;Y|U)$ respectively, and the resulting rate condition is
\begin{equation}
    R_0 \le \min_{\substack{\Ptilde_{UXY} \,:\, \Ptilde_{UX}=P_{UX},\,\Ptilde_{Y}=P_{Y}, \\  \EE_{\Ptilde}[\log q(X,Y)]\ge\EE_{P}[\log q(X,Y)]}} I_{\Ptilde}(U;Y) + [I_{\Ptilde}(X;Y|U) - R_1]_+.
\end{equation}
This condition is trivially satisfied when $I_{\Ptilde}(U;Y) > R_0$, so we can further constrain $I_{\Ptilde}(U;Y) \le R_0$ in the minimization, and then lower bound the function $[\cdot]^+$ by its argument to obtain \eqref{eq:SC_Rsum_CC}.

\paragraph{Ensemble tightness.} The ensemble tightness proof for the type-1 error probability (i.e., the condition \eqref{eq:SC_R1_CC}) directly follows that of the single-user setting.  For the type-0 error probability (i.e., the condition \eqref{eq:SC_Rsum_CC}), we need to avoid the two union bounds used in weakening \eqref{eq:sc_p0_exact} to \eqref{eq:sc_rcu0}.  However, this is also straightforward: For the union over $i$ we instead resort to the exact expression similarly to the single-user setting ({\em cf.}, \eqref{eq:pe_exact}), and for the union over $j$, it suffices to note that for independent events the truncated union bound is tight to within a factor of $\frac{1}{2}$ (e.g., see \cite[Lemma A.2]{Shu03}).

\paragraph{Dual analysis techniques.} To obtain the dual expression of Theorem \ref{thm:sc_dual}, we can weaken \eqref{eq:sc_rcu0} and \eqref{eq:sc_rcu1} using standard tools such as Markov's inequality and $\min\{1,z\} \le z^{\rho}$, as well as the fact that for any function $a(u,x)$, the quantity $\sum_{i=1}^n a(u_i,x_i)$ takes the same value for any $(\Uv,\Xv)$ occurring with positive probability.  Beyond handling the unions one at a time to obtain \eqref{eq:sc_rcu0}, no new tools are needed compared to the single-user dual analysis for constant-composition random coding ({\em cf.}, Section \ref{ch:single_user}).

Similarly to Section \ref{ch:single_user_cont}, it is of interest to extend the dual analysis to channels with continuous alphabets.  To do so, we need to define $P_{\Uv}$ and $P_{\Xv|\Uv}$ in terms of an input distribution $Q_{UX}$ and auxiliary cost $a(u,x)$ in such a way that (i) $\sum_{i=1}^n a(U_i,X_i)$ is almost surely close to its mean, and (ii) $P_{\Uv}$ and $P_{\Xv|\Uv}$ can be upper bounded by i.i.d.~distributions on $Q_U$ and $Q_{X|U}$ times sub-exponential pre-factors.  It was shown in \cite{ScarlettThesis} that such distributions exist, requiring only simple tricks following the cost-constrained codeword distributions used in Section \ref{ch:single_user_cont}.

\section{Refined Superposition Coding} \label{sec:rsc}

Superposition coding leads to a structured codebook in which the codewords $\{ \Xv^{(i,j)} \}_{j=1}^{M_1}$ are correlated through their dependence on $\Uv^{(i)}$, and we have seen that this can lead to strict improvements over independent codewords.  In this subsection, we present a variation of this idea that adds another level of structure, namely, the codewords $\{ \Xv^{(i,j)} \}_{j=1}^{M_1}$ are dependent even conditioned on $\Uv^{(i)}$.  We call the technique {\em refined superposition coding} \cite{Sca16a}, and we will see that it yields a rate at least as good as the standard version, with strict improvements being possible (at least for fixed $(\Uc,Q_{UX})$).

\subsection{Codebook Construction} \label{sec:rsc_code}

As with superposition coding, the main parameters to the ensemble are the finite alphabet $\Uc$ and input distribution $Q_{UX}\in\Pc(\Uc\times\Xc)$.  In addition, we fix the rates $R_{0}$ and $\{R_{1u}\}_{u\in\Uc}$; the corresponding codebook sizes $M_0$ and $M_{1u}$ will be defined shortly.

The codewords are generated in three steps:
\begin{itemize}
    \item Let $M_{0} \defeq \lfloor e^{nR_{0}} \rfloor$, and generate the length-$n$ auxiliary  
    codewords $\{\Uv^{(i)}\}_{i=1}^{M_{0}}$ independently according to the following constant-composition distribution $P_{\Uv}$:
    \begin{equation} 
        P_{\Uv}(\uv) =\frac{1}{|\Tclass(Q_{U,n})|}\openone\Big\{\uv\in \Tclass(Q_{U,n})\Big\}, \label{eq:RSC_PU}
    \end{equation}
    where $Q_{U,n} \in \Pc_n(\Uc)$ is a type with the same support as $Q_U$ such that
    $\|Q_{U,n} - Q_U\|_{\infty} \le \frac{1}{n}$. 
    \item For each $u\in\Uc$, we define
    \begin{equation}
        n_{u} \defeq Q_{U,n}(u)n\label{eq:RSC_BC_n1}
    \end{equation}
    and let $M_{1u} \defeq \lfloor e^{n_u R_{1u}} \rfloor$.  We fix a partial codeword distribution $P_{\Xv_{u}}\in\Pc(\Xc^{n_{u}})$, and
    for each $i=1,\dotsc,M_{0}$ and $u\in\Uc$, we draw the length-$n_{u}$ 
    partial codewords $\{\Xv_{u}^{(i,j_{u})}\}_{j_{u}=1}^{M_{1u}}$
    independently from $P_{\Xv_{u}}$. 
    \item A given message $m$ is indexed as $(m_{0},m_{11},\dotsc,m_{1|\Uc|})$, and the corresponding codeword $\Xv_{(m)}$ is constructed by treating $\Uv^{(m_{0})}$ as a time-sharing sequence: At the indices where $\Uv^{(m_{0})}$
    equals $u$, we transmit the symbols of $\Xv_{u}^{(m_{0},m_{1u})}$.
\end{itemize}

\begin{figure}
    \begin{centering}
        \includegraphics[width=0.85\columnwidth]{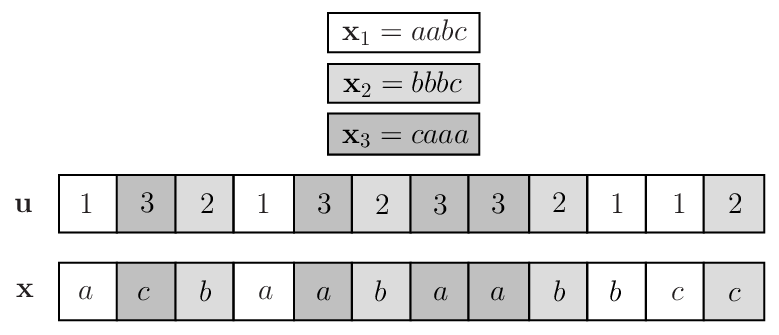}
        \par
    \end{centering}
    
    \caption{The construction of the codeword from the auxiliary sequence $\uv$ and the partial codewords $\xv_{1}$, $\xv_{2}$ and $\xv_{3}$ for refined SC.  Here we have $\Uc = \{1,2,3\}$, $\Xc = \{a,b,c\}$, $n_1 = n_2 = n_3 = 4$, and $n=12$. }
    \label{fig:RSC_Codewords}
\end{figure}

The first two steps are summarized as follows:
\begin{align}
    &\bigg\{\Big(\Uv^{(i)},\big\{\Xv_{1}^{(i,j_{1})}\big\}_{j_{1}=1}^{M_{11}},\dotsc,\big\{\Xv_{|\Uc|}^{(i,j_{|\Uc|})}\big\}_{j_{|\Uc|}=1}^{M_{1|\Uc|}}\Big)\bigg\}_{i=1}^{M_{0}} \nonumber \\ 
    &\qquad\qquad\qquad \sim\prod_{i=1}^{M_{0}}\bigg(P_{\Uv}(\uv^{(i)})\prod_{u \in \Uc}\bigg(\prod_{j_{u}=1}^{M_{1u}}P_{\Xv_{u}}(\xv_{u}^{(i,j_{u})})\bigg)\bigg). \label{eq:RSC_Distr}
\end{align}
From the third step, there are $M=M_{0}\times\prod_{u=1}^{|\Uc|}M_{1u}$ codewords, and hence the overall rate is
\begin{equation}
    R=R_{0}+\sum_{u=1}^{|\Uc|}Q_{U,n}(u)R_{1u}.
\end{equation}
An example of the construction of the codeword $\xv$ from the auxiliary sequence $\uv$ and partial codewords $\xv_{1}$, $\xv_{2}$ and $\xv_{3}$ is shown in Figure \ref{fig:RSC_Codewords}, where we have $\Uc=\{1,2,3\}$ and $\Xc=\{a,b,c\}$.

It remains to specify the partial codeword distributions $P_{\Xv_{u}}$.  Letting $Q_{UX,n}$ be a joint type with the same support as $Q_{UX}$ satisfying $\| Q_{UX,n} - Q_{UX} \|_{\infty} \le \frac{1}{n}$, we choose $P_{\Xv_{u}}$ to be the uniform distribution on the type class $\Tc^{n_{u}}\big(Q_{X|U,n}(\cdot|u)\big)$:
\begin{equation}
    P_{\Xv_{u}}(\xv_{u}) =\frac{1}{\big|\Tc^{n_{u}}\big(Q_{X|U,n}(\cdot|u)\big)\big|}\openone\Big\{\xv_{u}\in \Tc^{n_{u}}\big(Q_{X|U,n}(\cdot|u)\big)\Big\}. \label{eq:RSC_PX}
\end{equation}
Combining this with \eqref{eq:RSC_PU}, we have by symmetry
that each pair $(\Uv^{(i)},\Xv^{(i,j_1,\dotsc,j_{|\Uc|})})$ is uniformly distributed on $\Tclass(Q_{UX})$.  Hence, the {marginal} distribution of each $(\Uv,\Xv)$ pair is the same as superposition coding, but the {joint} codebook distribution has some additional structure.  We refer to the overall construction as {\em constant-composition refined superposition coding}.

Considering the case $|\Uc| = 2$, one can picture the structure of the codewords $\{ \Xv^{(i,j_1,j_2)} \}_{j_1,j_2}$ by imagining them arranged in an $M_{11} \times M_{12}$ grid.  Given $\{\Uv^{(i)}\}_{i=1}^{M_0}$, refined superposition coding generates two separate codebooks of partial codewords $\{\Xv_1\}_{j_1=1}^{M_{11}}$ and $\{\Xv_2\}_{j_2=1}^{M_{12}}$ and fills in the grid accordingly, whereas standard superposition coding fills every entry of the grid in a conditionally independent manner given  $\{\Uv^{(i)}\}_{i=1}^{M_0}$.  As a result, the added structure is analogous to that of using multiple-access coding (Section \ref{ch:mac}) in place of mutually independent codewords (Section \ref{ch:single_user}).

\subsection{Statement of Achievable Rate}

The achievable rate of constant-composition refined superposition coding is stated as follows \cite{Sca16a}.  We make use of the definition of the LM rate in \eqref{eq:INTR_PrimalLM}.

\begin{theorem} \label{thm:RSC_Main}
    {\em (Refined superposition coding rate)}
    Consider a DMC $(W,q)$.  For any finite set $\Uc$ and input distribution $Q_{UX} \in \Pc(\Uc\times\Xc)$, the rate
    \begin{equation}
    R=R_{0}+\sum_{u=1}^{|\Uc|}Q_{U}(u)R_{1u} \label{eq:RSC_Main_R}
    \end{equation}
    is achievable provided that $R_{0}$ and $\{R_{1u}\}_{u=1}^{|\Uc|}$ satisfy
    \begin{equation}
    R_{1u} \le \LM\big(Q_{X|U}(\cdot|u)\big), \quad \forall u\in\Uc, \label{eq:RSC_R1u}
    \end{equation}
	\vspace*{-3ex}
    \begin{align}
        &R_{0}\le\min_{\substack{\Ptilde_{UXY} \,:\, \Ptilde_{UX}=P_{UX},\,\Ptilde_{UY}=P_{UY}, \\  \EE_{\Ptilde}[\log q(X,Y)]\ge\EE_{P}[\log q(X,Y)]}} I_{\Ptilde}(U;Y) \nonumber \\ &\qquad\quad+ \bigg[\max_{\Kc\subseteq\Uc,\Kc\ne\emptyset}\sum_{u\in\Kc}Q_{U}(u)\Big(I_{\Ptilde}(X;Y|U=u)-R_{1u}\Big)\bigg]^{+}.\label{eq:RSC_R0} 
    \end{align}
    Moreover, this rate is ensemble-tight for constant-composition refined superposition coding.
\end{theorem}

We will show in Section \ref{sec:rsc_cmp} that this rate is at least high as that of standard superposition coding (Theorem \ref{thm:SC_Rate_SC}) for any pair $(\Uc,Q_{UX})$ of auxiliary parameters.

A dual formulation of Theorem \ref{thm:RSC_Main} was given in \cite{Sca16a} for the special case $|\Uc| = 2$, along with a direct derivation using cost-constrained coding that generalizes to continuous-alphabet channels.  The statement and proof of the dual equivalence are rather cumbersome even in the simplest case $|\Uc| = 2$, and they are therefore omitted here.

Before comparing to standard superposition and discussing the proof of Theorem \ref{thm:RSC_Main}, we present some examples from \cite{Sca16a}.

\subsection{Example 1: Sum Channel}

The notions of {\em product channels} and {\em sum channels} were introduced in Shannon's original paper \cite{Sha48}.  We have already studied product channels (i.e., parallel channels) in detail in Sections \ref{sec:parallel} and \ref{sec:mac_numerical}; here we provide an analogous example regarding the sum channel.

Given two channels $(W_{1},W_{2})$ respectively defined on the alphabets $(\Xc_{1},\Yc_{1})$ and $(\Xc_{2},\Yc_{2})$, the {\em sum channel} is defined to be the channel $W(y|x)$ with $|\Xc|=|\Xc_{1}|+|\Xc_{2}|$ and $|\Yc|=|\Yc_{1}|+|\Yc_{2}|$ such that one of the two subchannels is used on each transmission. One can similarly combine two metrics $q_{1}(x_{1},y_{1})$ and $q_{2}(x_{2},y_{2})$ to form a sum metric $q(x,y)$:  Assuming without loss of generality that $\Xc_{1}$ and $\Xc_{2}$ are disjoint and $\Yc_{1}$  and $\Yc_{2}$ are disjoint, we have
\begin{equation}
    q(x,y) =
    \begin{cases}
    q_1(x_{1},y_{1}) & x_{1}\in\Xc_{1}\text{ and }y_{1}\in\Yc_{1} \\
    q_2(x_{2},y_{2}) & x_{2}\in\Xc_{2}\text{ and }y_{2}\in\Yc_{2} \\
    0 & \text{otherwise},
    \end{cases}
\end{equation}
and similarly for $W(y|x)$. 

\begin{figure}
    \begin{centering}
        \includegraphics[width=0.5\columnwidth]{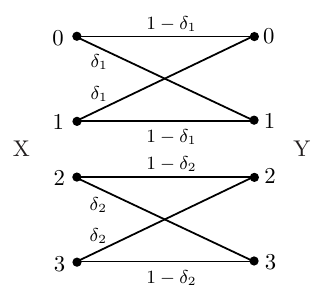}
        \par
    \end{centering}
    
    \caption{Sum of two binary symmetric channels with crossover probabilities $(\delta_1,\delta_2)$. } \label{fig:SumBSC}
\end{figure}

An example of a sum channel is shown in Figure \ref{fig:SumBSC}, where both subchannels are binary symmetric channels (BSCs).   The corresponding channel matrix is given by
\begin{align}
    \Wv & = \left[\begin{array}{cccc}
    1-\delta_1 & \delta_1 & 0 & 0\\
    \delta_1 & 1-\delta_1 & 0 & 0\\
    0 & 0 & 1-\delta_2 & \delta_2\\
    0 & 0 & \delta_2 & 1-\delta_2~
    \end{array}\right]\label{eq:Example_sumBSC}
\end{align}
While the channel diagram is similar to that of the product channel of Figure \ref{fig:ParallelBSC}, the two are fundamentally different: Here we are not transmitting over both subchannels simultaneously, but rather, choosing one to transmit over at each time instant. To appreciate the difference, observe that we can transmit one bit across the sum channel with {\em zero error} by using only the symbols $\{0,2\}$, whereas we can never achieve an error probability of zero in the product channel.

We use Theorem \ref{thm:RSC_Main} to provide an achievable rate for a general sum channel, and then specialize it to the BSC example.  Let $Q_{1}$ and $Q_{2}$ be the distributions that maximize the LM rate in \eqref{eq:INTR_PrimalLM} on the two subchannels.  We set $\Uc=\{1,2\}$, $Q_{X|U}(\cdot|1)=(Q_{1},\bzero)$ and $Q_{X|U}(\cdot|2)=(\bzero,Q_{2})$, where $\bzero$ denotes the zero vector of length two. We leave $Q_{U}$ to be specified.

Combining the constraints $\Ptilde_{UX}=Q_{UX}$ and $\EE_{\Ptilde}[\log q(X,Y)]\ge\EE_{P}[\log q(X,Y)]$ in \eqref{eq:RSC_R0}, it can be shown that the minimizing $\Ptilde_{UXY}$ only takes non-zero values for $(u,x,y)$ such that (i) $u=1$, $x\in\Xc_{1}$ and $y\in\Yc_{1}$, or (ii) $u=2$, $x\in\Xc_{2}$ and $y\in\Yc_{2}$.  It follows that $U$ is a deterministic function of $Y$ under the minimizing $\Ptilde_{UXY}$, and hence
\begin{equation}
    I_{\Ptilde}(U;Y)=H(Q_{U})-H_{\Ptilde}(U|Y)=H(Q_{U}).
\end{equation}
Therefore, the right-hand side of \eqref{eq:RSC_R0} is lower bounded by $H(Q_{U})$. Using \eqref{eq:RSC_Main_R}, it follows that we can achieve the rate
\begin{align}
    R^* &= H(Q_U) + Q_U(1)\LM(Q_{1},W_1) + Q_U(2)\LM(Q_{2},W_2) \\
    &= \log\big(e^{\LM(Q_{1},W_{1})}+e^{\LM(Q_{2},W_{2})}\big), \label{eq:SC_SumChRate}
\end{align}
where $\LM(Q_{u},W_{u})$ is the LM rate for subchannel $u \in \{1,2\}$,  and \eqref{eq:SC_SumChRate} holds under the optimal auxiliary distribution $Q_U(1) = 1 - Q_U(2) = \frac{e^{\LM(Q_{1},W_{1})}}{e^{\LM(Q_{1},W_{1})}+e^{\LM(Q_{2},W_{2})}}$, analogously to the matched case \cite[Sec. 16]{Sha48}.   

\begin{figure}
    \begin{centering}
        \includegraphics[width=0.7\columnwidth]{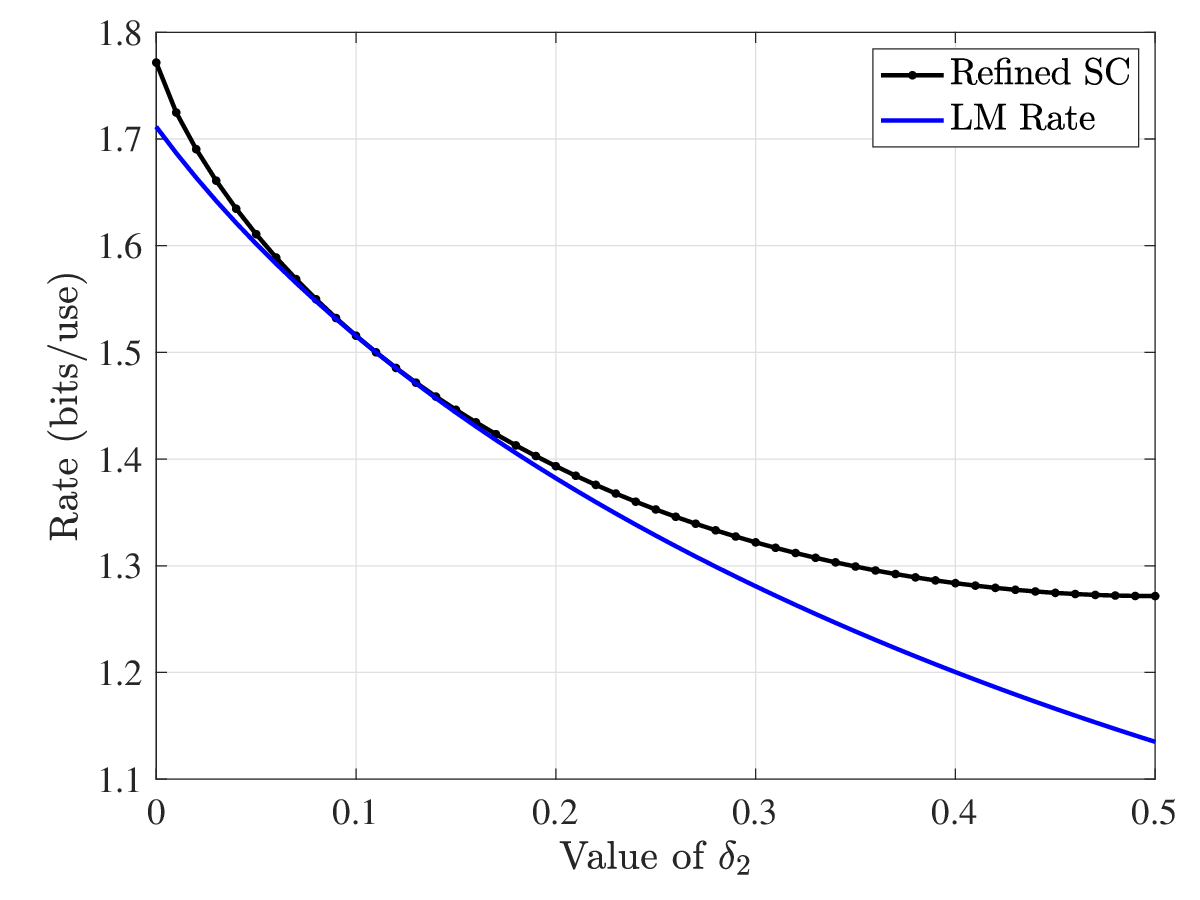}
        \par
    \end{centering}
    
    \caption{Sum channel example: Achievable rates with $\delta_1 = 0.11$, as a function of $\delta_2$.  The refined SC rate and the LM rate only coincide when $\delta_1 = \delta_2$.} \label{fig:SumBSC_Rates}
\end{figure}

Now, suppose that the two subchannels are BSCs with different crossover probabilities $\delta_1,\delta_2 \in \big(0,\frac{1}{2}\big)$, whereas the mismatched decoder assumes a common crossover probability $\delta \in \big(0,\frac{1}{2}\big)$.  By the binary example in Section \ref{sec:binary}, we have for $u = 1,2$ that the choice $Q_{u} = \big(\frac{1}{2},\frac{1}{2}\big)$ yields $\LM(Q_{u},W_{u}) = 1 - H_2(\delta_u)$ bits/use, which equals the matched capacity $C_u$ of the subchannel.  As a result, from \eqref{eq:SC_SumChRate} and the sum channel capacity formula $C^* = \log\big( e^{C_1} + e^{C_2} \big)$ \cite[Sec. 16]{Sha48}, the mismatch capacity is identical to the matched capacity.  In contrast, it can be shown that the LM rate is strictly less than the matched capacity whenever $\delta_1 \ne \delta_2$. We illustrate this numerically in Figure \ref{fig:SumBSC_Rates}, where we plot the matched rate (achieved by refined SC), as well as the LM rate under the (matched) capacity-achieving input distribution.

\subsection{Example 2: Comparison of Superposition Coding Methods} \label{sec:rsc_vs_sc}

We now turn to an example comparing the standard superposition coding rate of Theorem \ref{thm:SC_Rate_SC} with the refined superposition coding rate of Theorem \ref{thm:RSC_Main}.  We consider the channel and decoding metric described by the entries of the matrices
\begin{align}
    \Wv & = \left[\begin{array}{cccc}
    0.99 & 0.01 & 0 & 0\\
    0.01 & 0.99 & 0 & 0\\
    0.1~ & 0.1~ & 0.7 & \,0.1~\\
    0.1 & 0.1 & 0.1 & \,0.7~
    \end{array}\right], \qquad
    \qv = \left[\begin{array}{cccc}
    1 & 0.5 & 0 & 0\\
    0.5 & 1 & 0 & 0\\
    0.05 & 0.15 & 1 & 0.05\\
    0.15 & 0.05 & 0.5 & 1
    \end{array}\right].\label{eq:Example_q}
\end{align}
This is an admittedly artificial example, but we will see that it constitutes an interesting example for the purpose of comparison.  

\begin{center}
    \begin{table}[t]   
        \begin{center}  
            \caption{Achievable rates (bits/use) for the mismatched channel in \eqref{eq:Example_q}.  In addition, $\CLM = 1.111$ bits/use, and $\CGMI = 0.954$ bits/use.}   
            \label{tab:Rates}   
            \begin{tabular}{@{}ccc@{}}     
                \toprule  Input Distribution & Refined SC & Standard SC \\  
                \midrule         
                $Q_{UX}^{(1)}$               & 1.313      & 1.060       
                \vspace{0.5mm}\\     
                $Q_{UX}^{(2)}$               & 1.236      & 1.236       \\
                \bottomrule   
            \end{tabular}     
        \end{center} 
    \end{table}
    \par
\end{center}

Using an exhaustive search to three decimal places, we obtained the following: (i) The optimized LM rate is $\CLM=1.111$ bits/use, and is achieved by the input distribution $Q_{X}=(0.403,0.418,0,0.179)$; (ii) The optimized GMI is $\CGMI=0.954$ bits/use, and is achieved by the input distribution $Q_{X}=(0.330, 0.331, 0.155, 0.184)$.

For refined superposition coding, setting $|\Uc|=2$ and applying local optimization techniques using a number of starting points, we obtained an achievable rate of $R_{\textsc{rsc}}^{*}=1.313$ bits/use, with $Q_{U}=(0.698,0.302)$, $Q_{X|U}(\cdot|1)=(0.5,0.5,0,0)$ and $Q_{X|U}(\cdot|2)=(0,0,0.528,0.472)$. We denote the corresponding input distribution by $Q_{UX}^{(1)}$.

Applying similar local optimization techniques for standard superposition coding, we obtained an achievable rate of $R_{\textsc{sc}}^{*}=1.236$ bits/use, with $Q_{U}=(0.830,0.170)$, $Q_{X|U}(\cdot|1)=(0.435,0.450,0.115,0)$ and $Q_{X|U}(\cdot|2)=(0,0,0,1)$. We denote the corresponding input distribution by $Q_{UX}^{(2)}$.

The superposition coding rates are summarized in Table \ref{tab:Rates}. 
While the achievable rate of Theorem \ref{thm:RSC_Main} coincides with that of Theorem \ref{thm:SC_Rate_SC} under $Q_{UX}^{(2)}$, the former is significantly higher under $Q_{UX}^{(1)}$. Both types of superposition coding yield a strict improvement over the LM rate, which in turn provides a strict improvement over the GMI.

Since we did not perform a global optimization of $(\Uc,Q_{UX})$, we cannot necessarily conclude that refined superposition coding beats the standard version for the best possible choices of $\Uc$ and $Q_{UX}$.  Nevertheless, at the very least, this example highlights that refined superposition coding can facilitate the search for a good set of parameters, and can provide significant improvements for a fixed set of parameters.

\subsection{Comparison to Standard Superposition Coding} \label{sec:rsc_cmp}

As mentioned previously, the refined superposition coding rate is always at least as high as that of standard superposition coding.  This is formally stated in the following.

\begin{lemma} 
    {\em (Comparison of superposition coding rates)}
    For any mismatched DMC $(W,q)$, any pair of parameters $(\Uc,Q_{UX})$ such that $|\Uc|$ is finite, we have
    \begin{equation}
        \max_{(R_0,R_1) \in \RegSC(Q_{UX})} R_0 + R_1 \le \max_{(R_0,R_1,\dotsc,R_{|\Uc|}) \in \RegRSC(Q_{UX})} R_0 + \sum_{u} Q_U(u) R_{1u},
    \end{equation}
    where $\RegSC(Q_{UX})$ is the set of rate pairs satisfying \eqref{eq:SC_R1_CC}--\eqref{eq:SC_Rsum_CC}, and $\RegRSC(Q_{UX})$ is the set of rate tuples satisfying \eqref{eq:RSC_R1u}--\eqref{eq:RSC_R0}.
\end{lemma}
\begin{proof}
    We will show that for the tuple $(R_0,R_1,\dotsc,R_{|\Uc|})$ achieving equality in the refined superposition coding conditions \eqref{eq:RSC_R1u}--\eqref{eq:RSC_R0}, we can lower bound $R_0$ to recover \eqref{eq:SC_Rsum_CC}, and we can lower bound
    \begin{equation}
        R_{1}\triangleq\sum_{u}Q_{U}(u)R_{1u} \label{eq:RSC_Weaken0}
    \end{equation}
    to recover \eqref{eq:SC_R1_CC}.
    
    For the condition on $R_0$, we lower bound the right-hand side of \eqref{eq:RSC_R0} by replacing the maximum over $\Kc$ by the  
    particular choice $\Kc=\Uc$, implying the following sufficient condition for \eqref{eq:RSC_R0} to hold:
    \begin{equation}
    R_0 \le \min_{\substack{\Ptilde_{UXY} \,:\, \Ptilde_{UX}=P_{UX},\,\Ptilde_{UY}=P_{UY}, \\  \EE_{\Ptilde}[\log q(X,Y)]\ge\EE_{P}[\log q(X,Y)]}}  I_{\Ptilde}(U;Y) + \Big[I_{\Ptilde}(X;Y|U)-R_{1}\Big]^{+}, \label{eq:RSC_Weaken1}
    \end{equation}
    where we have used \eqref{eq:RSC_Weaken0} and the definition of conditional mutual information.  Since the objective \eqref{eq:RSC_Weaken1} trivially exceeds $R_0$ when $I_{\Ptilde}(U;Y) > R_0$, we can safely constrain the minimum to satisfy $I_{\Ptilde}(U;Y) \le R_0$.  By doing so, and lower bounding the $[\cdot]^+$ function by its argument, we recover \eqref{eq:SC_Rsum_CC}.
    
    Next, defining $\Ptilde_{XY|U}(\cdot,\cdot|u)$ to be the minimizer in \eqref{eq:INTR_PrimalLM} corresponding to a given $u \in \Uc$ in \eqref{eq:RSC_R1u}, and recalling \eqref{eq:RSC_Weaken0}, we have
    \begin{align}
        R_{1} &= \sum_{u}Q_{U}(u)I_{\Ptilde}(X;Y|U=u) \label{eq:RSC_Weaken2} \\
            &= I_{\Ptilde}(X;Y|U), \label{eq:RSC_Weaken3} 
    \end{align}
    where in \eqref{eq:RSC_Weaken3}, we define $\Ptilde_{UXY}$ according to the above marginals along with $\Ptilde_{U}=Q_{U}$.  The proof now simply amounts to showing that the constraints on the LM rate in \eqref{eq:INTR_PrimalLM} for each $u \in \Uc$ imply the superposition coding constraints in \eqref{eq:SC_R1_CC}.  This is a straightforward exercise; for instance, we have $\EE_{\Ptilde}[\log q(X,Y)\,|\,U=u]\ge\EE_{P}[\log q(X,Y)\,|\,U=u]$ for all $u\in\Uc$, and averaging over $U$ gives the desired constraint $\EE_{\Ptilde}[\log q(X,Y)]\ge\EE_{P}[\log q(X,Y)]$.
\end{proof} 

\subsection{Proof Techniques} \label{sec:rsc_proof}

Here we describe the key idea behind the proof of Theorem \ref{thm:RSC_Main}; we omit the full details, which can be found in \cite{Sca16a}.  We focus on the case $\Uc=\{1,2\}$ for clarity, but the same arguments apply more generally.

\paragraph{Separation into three error events.} Recalling the codebook construction described in Section \ref{sec:rsc_code}, we rewrite the maximum-metric decoding rule as follows:
\begin{align}
    (\hat{m}_{0},\hat{m}_{11},\hat{m}_{12}) &= \argmax_{(i,j_{1},j_{2})}q^{n}(\xv^{(i,j_{1},j_{2})},\yv) \\
    &= \argmax_{(i,j_{1},j_{2})}q^{n_{1}}\big(\xv_{1}^{(i,j_{1})},\yv_{1}(\uv^{(i)})\big)q^{n_{2}}\big(\xv_{2}^{(i,j_{2})},\yv_{2}(\uv^{(i)})\big),\label{eq:RSC_SplitMetric}
\end{align}
where $\yv_{u}(\uv)$ denotes the subsequence of $\yv$ corresponding to the indices
where $\uv$ equals $u$.  The objective in \eqref{eq:RSC_SplitMetric} follows by separating the indices where $u=1$ from those where $u=2$.  

From \eqref{eq:RSC_SplitMetric}, we see that for any given $i$, the pair $(j_{1},j_{2})$ with the highest metric is the one for which $j_{1}$ maximizes $q^{n_{1}}(\xv_{1}^{(i,j_{1})},\yv_{1}(\uv^{(i)}))$ and $j_{2}$ maximizes $q^{n_{2}}(\xv_{2}^{(i,j_{2})},\yv_{2}(\uv^{(i)}))$.  As a result, we can split the error event into the following three types, assuming without loss of generality that the transmitted message corresponds to $(m_0,m_{11},m_{12}) = (1,1,1)$:
\begin{tabbing}
    ~~~{\emph{(Type 0)}}~~~ \= $q^n(\Xv^{(i,j_1,j_2)},\Yv) \ge q^n(\Xv,\Yv)$ for some $i \ne 1$, $j_1$, $j_2$; \\
    ~~~{\emph{(Type 1)}}~~~ \> $q^{n_1}(\Xvbar_1^{(1,j_1)},\Yvi(\Uv)) \ge q^{n_1}(\Xv_1,\Yvi(\Uv))$ for some $j_1 \ne 1$; \\
    ~~~{\emph{(Type 2)}}~~~ \> $q^{n_2}(\Xvbar_2^{(1,j_2)},\Yvii(\Uv)) \ge q^{n_2}(\Xv_2,\Yvii(\Uv))$ for some $j_2 \ne 1$.
\end{tabbing} 
By the union bound, the overall error probability is upper bounded by the sum over these three error types.

Observe that the type-1 error event corresponds to the error event for the standard constant-composition ensemble ({\em cf.}, Section \ref{sec:cc_LM}) with rate $R_{11}$, length $n_{1}=nQ_{U}(1)$, and input distribution $Q_{X|U}(\cdot|1)$. A similar statement holds for the type-2 error probability $\peiibar$.  As a result, we do not need to re-analyze these events; we immediately obtain the condition \eqref{eq:RSC_R1u} from Theorem \ref{thm:LM}.  In the following, we focus on the more difficult type-0 error event.

\paragraph{Separation into joint types.} The error probability for the type-0 error event is given by
\begin{equation}
    \peobar = \PP\bigg[\bigcup_{i\ne1}\bigcup_{j_{1},j_{2}}\big\{q^{n}(\Xv^{(i,j_{1},j_{2})},\Yv)\ge q^{n}(\Xv,\Yv)\big\}\bigg],\label{eq:RSC_RefProof1}
\end{equation}
where $(\Yv|\Xv=\xv)\sim W^{n}(\cdot|\xv)$.  Writing the probability as an expectation given $(\Uv,\Xv,\Yv)$ and applying the truncated union bound, we obtain
\begin{align}
    &\peobar \le \EE_{\Uv,\Xv,\Yv}\Bigg[\min\Bigg\{1,(M_{0}-1) \nonumber \\ & \times \EE_{\Uvbar}\Bigg[\PP_{\{\Xvbar^{(j_1,j_2)}\}}\bigg[\bigcup_{j_{1},j_{2}}\big\{ q^{n}(\Xvbar^{(j_{1},j_{2})},\Yv)\ge q^{n}(\Xv,\Yv)\big\}\bigg] \Bigg]\Bigg\}\Bigg], \label{eq:RSC_RefProof2}
\end{align}
where the codewords $\Xvbar^{(j_1,j_2)}$ correspond to the auxiliary codeword $\Uvbar$ (e.g., $\Uvbar = \Uv^{(2)}$ and $\Xvbar^{(j_1,j_2)} = \Xv^{(2,j_1,j_2)}$).

As with the other achievable rates in the monograph, one can write \eqref{eq:RSC_RefProof2} in terms of joint types.  The resulting counterpart to \eqref{eq:SC_rewritten} is given by
\begin{align}
    &\peobar \le \sum_{\substack{P_{UXY} \in \Pc_n(\Uc\times\Xc\times\Yc) \,:\, \\ P_{UX} = Q_{UX,n}}} \PP\Big[\big(\Uv,\Xv,\Yv\big)\in \Tclass(P_{UXY})\Big] \nonumber \\ 
        &~~\times\min\Bigg\{1,
    (M_{0}-1)\sum_{\substack{\Ptilde_{UXY} \in \Pc_n(\Uc\times\Xc\times\Yc) \,:\, \\ P_{UX} = Q_{UX,n}, \, \Ptilde_Y = P_Y \\ \EE_{\Ptilde}[\log q(X,Y)]\ge\EE_{P}[\log q(X,Y)] }}\PP\Big[\big(\Uvbar,\yv\big)\in \Tclass(\Ptilde_{UY})\Big] \nonumber \\
        &~~\qquad\qquad\qquad\times \PP\bigg[\bigcup_{j_{1},j_{2}}\Big\{\big(\uvbar,\Xvbar^{(j_{1},j_{2})},\yv\big)\in \Tclass(\Ptilde_{UXY})\Big\}\bigg]\Bigg\},\label{eq:RSC_RefProof5} 
\end{align}
where $(\uvbar,\yv)$ denotes an arbitrary pair such that 
$\yv\in \Tclass(P_{Y})$ and $(\uvbar,\yv)\in \Tclass(\Ptilde_{UY})$.

\paragraph{Bounding the double union.}  We bound the doubly-indexed union in \eqref{eq:RSC_RefProof5} using a similar approach to the multiple-access channel in Section \ref{sec:mac_proof}.  Recalling the definition of $\yv_u$ following \eqref{eq:RSC_SplitMetric}, we have
\begin{align}
    &\big(\uvbar,\Xvbar^{(j_{1},j_{2})},\yv\big)\in \Tclass(\Ptilde_{UXY}) \nonumber \\ 
        &\iff \big(\Xvbar_{u}^{(j_{u})},\yv_{u}(\uvbar)\big)\in \Tc^{n_{u}}(\Ptilde_{XY|U}(\cdot,\cdot|u)), \quad u=1,2,
\end{align}
where $\{\Xvbar_u\}_{u=1,2}$ are the partial codewords such that $\Xvbar^{(j_{1},j_{2})}$ is constructed from $(\uvbar,\Xvbar_1,\Xvbar_2)$.  We then claim that
\begin{align}
    &\PP\bigg[\bigcup_{j_{1},j_{2}}\Big\{\big(\uvbar,\Xvbar^{(j_{1},j_{2})},\yv\big)\in \Tclass(\Ptilde_{UXY})\Big\}\bigg] \nonumber \\ 
    & \quad \le \min\bigg\{1,\min_{u=1,2}M_{1u}\PP\Big[\big(\Xvbar_{u},\yv_{u}(\uvbar)\big)\in \Tc^{n_{u}}\big(\Ptilde_{XY|U}(\cdot,\cdot|u)\big)\Big], \nonumber \\
    & \quad\quad~ M_{11}M_{12}\PP\Big[\bigcap_{u=1,2}\Big\{\big(\Xvbar_{u},\yv_{u}(\uvbar)\big)\in \Tc^{n_{u}}\big(\Ptilde_{XY|U}(\cdot,\cdot|u)\big)\Big\}\Big]\bigg\}, \label{eq:RSC_RefProof6} 
\end{align}
The first term is trivial, the last term is the standard union bound, and the middle term corresponds to restricting the above condition for $u=1,2$ to only hold for {\em either} $u=1$ or $u=2$, rather than both.  One can view \eqref{eq:RSC_RefProof6} as taking the minimum over the four subsets $\Kc \subseteq \Uc$, including the empty set.

\paragraph{Wrapping up.} The remainder of the proof amounts to the same techniques as the previous sections, namely, applying standard properties of types, using the law of large numbers to establish $\| P_{UXY} - Q_{UX} \times W \|_{\infty} \le \delta$, and applying a continuity argument similarly to Appendix \ref{sec:continuity_app}.  Regarding the first of these, we note that up to sub-exponential pre-factors, $\PP\big[\big(\Uvbar,\yv\big)\in \Tclass(\Ptilde_{UY})\big]$ behaves as $e^{-nI_{\Ptilde}(U;Y)}$, and $\PP\big[\big(\Xvbar_{u},\yv_{u}(\uvbar)\big)\in \Tc^{n_{u}}\big(\Ptilde_{XY|U}(\cdot,\cdot|u)\big)\big]$ behaves as $e^{-n_u I_{\Ptilde}(X;Y|U=u)}$; these mutual information quantities arise accordingly in the rate condition \eqref{eq:RSC_R0}.

\paragraph{Further analysis techniques.} The ensemble tightness of the rate in Theorem \ref{thm:RSC_Main} can be established using similar techniques to those of the multiple-access channel ({\em cf.}, Section \ref{sec:mac_ens_tight}).  The details can be found in \cite{Sca16a}, along with the corresponding dual expressions and continuous-alphabet extensions.
 

\chapter{Error Exponents} \label{ch:exponents}

\section{Introduction}

Throughout the monograph, our focus has been on achievable rates via random coding. 
A notable limitation of studying achievable rates is that they only establish the first-order conditions for attaining asymptotically vanishing error probability, meaning that in principle the block length may need to be very large to achieve a given target.

In this section, we survey developments in {random-coding error exponents}, which provide refined asymptotic achievability bounds characterizing the speed of convergence of the error probability to zero at fixed rates.  While error exponents are a classical topic in Shannon theory \cite{Gal68,Csi11}, their study in the presence of mismatched decoding comes with interesting new challenges and insights, e.g., regarding the number of auxiliary costs ({\em cf.}, Section \ref{sec:cost_multi}) required to maximize the exponent.
While other refined asymptotics (e.g., speed of convergence to the first-order rate for a fixed error probability) have also been considered under mismatched decoding  \cite{Sca14c}, we focus on error exponents, which are generally  better understood.


Before proceeding, we formalize the notion of an achievable error exponent and ensemble tightness, focusing on the single-user mismatched decoding setting of Sections \ref{ch:single_user} and \ref{ch:single_user_cont}.

\begin{definition} \label{def:error_exp}
    {\em (Error exponents)}
    For a mismatched memoryless channel described by $(W,q)$, we say that $E(R)$ is an {\em achievable error exponent} at rate $R$ if, for any $\delta > 0$, there exists a sequence of codebooks $\Cc_n$ with $M \ge e^{n(R-\delta)}$ codewords such that 
    \begin{equation}
        \liminf_{n \to \infty} - \frac{1}{n} \log \pe(\Cc_n) \ge E(R).
    \end{equation}
    In addition, for a given random coding ensemble with random-coding error probability $\pebar(n,M)$, we say that $E(R)$ is the {\em ensemble-tight} error exponent at rate $R$ if 
    \begin{equation}
        \lim_{n \to \infty} - \frac{1}{n} \log \pebar(n,\lfloor e^{nR} \rfloor) = E(R).
    \end{equation}
\end{definition}

As with the achievable rates, we will focus on i.i.d.~random coding (Definition \ref{def:iid}), constant-composition random coding (Definition \ref{def:cc}), and various forms of cost-constrained random coding (Sections \ref{sec:cost} and \ref{sec:cost_multi}), and we will first consider independently-generated codewords before turning to multi-user coding techniques (Sections \ref{ch:mac} and \ref{ch:multiuser}).

The results of this section are due to Csisz\'ar and K\"orner \cite{Csi81}, Kaplan and Shamai \cite{Kap93}, Scarlett {\em et al.}~\cite{Sca14c,Sca14f,Sca16a}, and Somekh-Baruch {\em et al.}~\cite{Som19}.

\section{Random-Coding Exponents for DMCs}  \label{sec:exponents}

We first consider discrete memoryless channels (DMCs), in which the input and output alphabets are finite.  As is the case for the GMI and the LM rate ({\em cf.}, Section \ref{sec:su_rates}), the random-coding error exponents of i.i.d.~random coding and constant-composition random coding can be written in both primal and dual forms.  In fact, analogous primal and dual forms have long been known in the matched case (e.g., see \cite{GallagerCC}), with Csisz\'ar and K\"orner's type-based analysis giving the primal form \cite{Csi11}, and Gallager's analysis giving the dual form \cite{Gal68}.

We proceed by defining the primal and dual forms of the relevant error exponents under mismatched decoding, and then formally state their achievability and ensemble tightness.  For i.i.d.~random coding, the primal expression is given by
\begin{align}
    &\Eriid(Q_X,R) =  \min_{P_{XY}}\min_{ \substack{\Ptilde_{XY} \,:\, \Ptilde_Y = P_Y, \\ \EE_{\Ptilde}[\log q(X,Y)] \ge \EE_{P}[\log q(X,Y)] } } \nonumber \\ &\qquad D(P_{XY}\|Q_X\times W)  +\big[D(\Ptilde_{XY}\|Q_X\times\Ptilde_{Y})-R\big]^{+}, \label{eq:SU_Er_iid} 
\end{align}
and the dual expression is given by \cite{Kap93}
\begin{equation}
    \Eriid(Q_X,R) \triangleq \max_{\rho\in[0,1]}\Eziid(Q_X,\rho)-\rho R,\label{eq:SU_Er_IID_Dual} 
\end{equation}
where
\begin{equation} 
    \Eziid(Q_X,\rho) \triangleq \sup_{\sgz}\,-\log\sum_{x}Q_X(x)\sum_{y}W(y|x)\bigg(\frac{\sum_{\xbar}Q_X(\xbar)q(\xbar,y)^{s}}{q(x,y)^{s}}\bigg)^{\rho}. \label{eq:SU_E0_IID}    
\end{equation}
For constant-composition random coding, the primal expression is given by \cite{Csi81}
\begin{align}
    \Ercc(Q_X,R) = &\min_{P_{XY} \,:\, P_X = Q_X}\min_{ \substack{\Ptilde_{XY}\,:\, \Ptilde_X = Q_X, \Ptilde_Y = P_Y, \\ \EE_{\Ptilde}[\log q(X,Y)] \ge \EE_{P}[\log q(X,Y)] } } \\ & \qquad D(P_{XY}\|Q_X\times W)+\big[I_{\Ptilde}(X;Y)-R\big]^{+}, \label{eq:SU_Er_cc} 
\end{align}
and the dual expression is given by \cite{Sca14c}
\begin{equation}
    \Ercc(Q_X,R) = \max_{\rho\in[0,1]}\Ezcc(Q_X,\rho)-\rho R\label{eq:SU_Er_LM_Dual},
\end{equation}
where
\begin{align}
    &\Ezcc(Q_X,\rho) \nonumber \\ & = \sup_{\sgz,a(\cdot)}\,-\sum_{x}Q_X(x)\log\sum_{y}W(y|x)\bigg(\frac{\sum_{\xbar}Q_X(\xbar)q(\xbar,y)^{s}e^{a(\xbar)}}{q(x,y)^{s}e^{a(x)}}\bigg)^{\rho}. \label{eq:SU_E0_LM}
\end{align} 
Similarly to the primal-dual equivalences for achievable rates in Section \ref{subsec:primaldual}, the proofs are based on Lagrange duality.  The details can be found in \cite{Sca14c}.

\begin{theorem} \label{thm:exponents}
    {\em (Ensemble-tight error exponents for DMCs)}
    For any mismatched DMC $(W,q)$, input distribution $Q_X \in \Pc(\Xc)$, and rate $R > 0$, we have under i.i.d.~random coding that 
    \begin{equation}
    \lim_{n \to \infty} -\frac{1}{n} \log \pebar(n,\lfloor e^{nR} \rfloor) = \Eriid(Q_X,R),
    \end{equation}
    and under constant-composition random coding that 
    \begin{equation}
    \lim_{n \to \infty} -\frac{1}{n} \log \pebar(n,\lfloor e^{nR} \rfloor) = \Ercc(Q_X,R).
    \end{equation}
\end{theorem}

It is straightforward to show that the i.i.d.~and constant-composition exponents are positive for 
$R < \GMI(Q_X)$ and $R < \LM(Q_X)$, respectively.  We outline the proof of this claim for the constant-composition case \cite{Csi95}, and note that similar arguments hold for the i.i.d.~case:
\begin{itemize}
    \item In the primal expression \eqref{eq:SU_Er_cc}, the term $D(P_{XY}\|Q_X \times W)$ is always positive when $P_{XY}$ is not equal to $Q_X \times W$.  Hence, we can take $P_{XY} \to Q_X \times W$, and upon doing so, it only remains to apply the continuity of the LM rate ({\em cf.}, Lemma \ref{lem:continuity}).
    \item In the dual expression, we can lower bound $\Ezcc(Q_X,\rho)$ in \eqref{eq:SU_E0_LM} by the value $\Ezcc(Q_X,\rho,s,a)$ corresponding to any fixed choices of $\sgz$ and $a(\cdot)$.  Following the analysis of Gallager \cite{Gal68}, one finds that the exponent is positive for all rates up to $\frac{\partial \Ezcc(Q_X,\rho,s,a)}{ \partial \rho }\big|_{\rho = 0}$, and a direct calculation reveals that this derivative equals the objective function in \eqref{eq:INTR_RateLM}.  Optimizing $\sgz$ and $a(\cdot)$ recovers the LM rate.
\end{itemize}
The mismatched decoding error exponents also bear a strong resemblance to the above-mentioned error exponents for maximum-likelihood decoding, and in fact recover them as a special case \cite{Csi95,Sca14c} by setting $q(x,y)=W(y|x)$.  In this case, the minimizations in the primal expressions are achieved with $\Ptilde_{XY} = P_{XY}$, and the maximizations in the dual expressions are achieved with $s = \frac{1}{1+\rho}$.

In analogy with the fact that $\GMI(Q_X) \le \LM(Q_X)$, we observe that
\begin{equation}
   \Eriid(Q_X,R)\leq  \Ercc(Q_X,R) 
\end{equation}
for any channel $(W,q)$, input distribution $Q_X$, and rate $R$.  This can be seen from the primal expressions by noting that the minimizations in \eqref{eq:SU_Er_cc} are more constrained than those in \eqref{eq:SU_Er_iid} (note also that $D(\Ptilde_{XY} \| Q_X \times P_Y)$ and $I_{\Ptilde}(X;Y)$ are identical under the constraints $\Ptilde_X = Q_X$ and $\Ptilde_Y = P_Y$). 
Alternatively, using the dual expressions, we see that setting $a(\cdot) = 0$ in \eqref{eq:SU_E0_LM} and applying Jensen's inequality recovers \eqref{eq:SU_E0_IID}.

\subsection{Proof Outline} 

\paragraph{Initial non-asymptotic bounds.} The starting point of the analysis is the following upper bound on the random-coding error probability when the $M$ codewords are independently drawn from a common distribution $P_{\Xv}$ (see \eqref{eq:rcu}):
\begin{equation}
    \pebar \le \EE\Big[ \min\big\{ 1, (M-1) \PP\big[ q^n(\Xvbar,\Yv) \ge q^n(\Xv,\Yv) \,|\, \Xv,\Yv \big] \big\} \Big], \label{eq:rcu_rep}
\end{equation}
where $(\Xv,\Yv,\Xvbar) \sim P_{\Xv}(\xv)W^n(\yv|\xv)P_{\Xv}(\xvbar)$.  We will also use the fact that
\begin{equation}
    \pebar \ge \frac{1}{2}\, \EE\Big[ \min\big\{ 1, (M-1) \PP\big[ q^n(\Xvbar,\Yv) \ge q^n(\Xv,\Yv) \,|\, \Xv,\Yv \big] \big\} \Big], \label{eq:pe_lower}
\end{equation}
which follows by recalling the step from \eqref{eq:pebar_exact} to \eqref{eq:rcu}, and noting that the truncated union bound is tight to within a factor of $\frac{1}{2}$ for independent events \cite[Lemma A.2]{Shu03}.

With \eqref{eq:rcu_rep}--\eqref{eq:pe_lower} in place, Theorem \ref{thm:exponents} is attained via suitable modifications of the achievable rate analysis of Section \ref{sec:su_proofs}.  We first outline the primal analysis, and then the dual analysis.

\paragraph{Achievability and ensemble tightness -- primal expressions.}  For the primal expressions, the idea is to avoid using the law of large numbers to establish $\|P_{XY} - Q \times W\|_{\infty} \le \delta$ (as was done in Section \ref{sec:su_proofs}), and to instead use the following exponential characterization of the probability of a joint type $P_{XY}$ occurring:
\begin{equation}
    e^{-n(D(P_{XY} \| Q_X \times W) + \delta)} \le \PP\big[ (\Xv,\Yv) \in \Tclass(P_{XY}) \big] \le e^{-n(D(P_{XY} \| Q_X \times W) - \delta)} \label{eq:div_bound}
\end{equation}
for arbitrarily small $\delta > 0$ and sufficiently large $n$.  This bound holds under both i.i.d.~and constant-composition coding, with the latter requiring $P_{X}$ to coincide with the input distribution $Q_X$ (otherwise the probability in \eqref{eq:div_bound} is zero).

In more detail, we first write the expectation in \eqref{eq:rcu_rep} in terms of the joint type $P_{XY}$ of $(\Xv,\Yv)$ as
\begin{align}
    &\pebar \le \sum_{P_{XY} \in \Pc_n(\Xc \times \Yc)} \PP[(\Xv,\Yv) \in \Tclass(P_{XY})] \nonumber \\ 
        &\qquad\qquad\times \min\Big\{1, (M-1)\PP\big[ q^n(\Xvbar,\yv) \ge q^n(\xv,\yv) \big]  \Big\},
\end{align}
where in the inner probability, $(\xv,\yv)$ is an arbitrary pair with joint type $P_{XY}$.  Upon substituting \eqref{eq:div_bound} and the previously-obtained bounds on $\PP[q^n(\Xvbar,\yv) \ge q^n(\xv,\yv)]$ (see \eqref{eq:pf_GMI_primal4} for the i.i.d.~ensemble and \eqref{eq:pf_LM_primal4} for the constant-composition ensemble) into \eqref{eq:rcu_rep}, and upper bounding the outer summation by $(n+1)^{|\Xc|\cdot|\Yc|-1}$ times the maximum, we readily obtain the exponents \eqref{eq:SU_Er_iid} and \eqref{eq:SU_Er_cc}.   The ensemble tightness is proved similarly by starting from \eqref{eq:pe_lower} and using the lower bound counterparts when applying each property of types, along with a continuity argument ({\em cf.}, Appendix \ref{sec:continuity_app}).

\paragraph{Achievability -- dual expressions.} The dual analysis is similarly straightforward given the corresponding rate derivations. Applying Markov's inequality and the inequality $\min\{1,z\} \le z^{\rho}$ (for $\rho \in [0,1]$) in \eqref{eq:rcu_rep}, we obtain
\begin{equation}
    \pebar \le \EE\Bigg[ \bigg( M \EE\bigg[ \bigg(\frac{q^n(\Xvbar,\Yv)}{q^n(\Xv,\Yv)}\bigg)^s \,\bigg|\, \Xv,\Yv \bigg] \Bigg)^{\rho} \bigg].
\end{equation}
This step strengthens the step \eqref{eq:rcu_s_weakened} that we used for the derivation of the rate, and leads to the presence of the free parameter $\rho \in [0,1]$ in both \eqref{eq:SU_Er_IID_Dual} and \eqref{eq:SU_Er_LM_Dual}.  Since the subsequent analysis under i.i.d.~or constant-composition random coding is similar to that of cost-constrained coding in the following subsection, and also straightforward given the dual analysis in Section \ref{sec:su_proofs}, we omit the details.


\section{Random-Coding Exponents with Continuous Alphabets} \label{sec:exponents_cont}

In this subsection, we consider the setup of Section \ref{ch:single_user_cont}, in which the channel alphabets may be continuous, and the transmission is subject to a system cost $\frac{1}{n}\sum_{i=1}^n c(x_i) \le \Gamma$.

In this setting, we introduced the cost-constrained random coding ensemble in Section \ref{sec:cost} with a single {auxiliary cost}, which is used to recover the optimization variable $a(\cdot)$ in the dual form of the LM rate.  In Section \ref{sec:cost_multi}, we generalized this ensemble to contain multiple auxiliary costs $a_1(\cdot),\dotsc,a_L(\cdot)$, with the motivation being that one may not always have knowledge of the optimal (single) choice of $a(x)$.  We also briefly mentioned in Section \ref{sec:MAC_Dual} that the use of multiple auxiliary costs is beneficial in the multiple-access setting.  In this subsection and the subsequent subsections, we will see that yet another benefit of using multiple auxiliary costs is in attaining improved error exponents \cite{Sca14c}. 

For convenience, we repeat the definition of the codeword distribution: With input distribution $Q_X$, system cost $c(\cdot)$, auxiliary costs $\{a_l\}_{l=1}^L$, and parameter $\delta > 0$, we have
\begin{equation}
    P_{\Xv}(\xv) = \frac{1}{\Omega_n} \prod_{i=1}^{n}Q_X(x_{i})\openone\big\{\xv\in\Dc_{n}\big\},\label{eq:PX_cost_multi2}
\end{equation}
where
\begin{align}
    \Dc_{n}\triangleq\bigg\{\xv\,:\,&\bigg|\frac{1}{n}\sum_{i=1}^{n}c(x_{i})-\phi_{c}\bigg| \le \delta, \nonumber \\
    &\bigg|\frac{1}{n}\sum_{i=1}^{n}a_l(x_{i})-\phi_l\bigg| \le \delta, \, \forall l=1,\dotsc,L \bigg\}, \label{eq:SU_SetDn_multi2}
\end{align}
and where $\phi_c = \EE_Q[c(X)]$, $\phi_l = \EE_Q[a_l(X)]$, and $\Omega_n$ is a normalizing constant.

\paragraph{Two auxiliary costs.}
Here we show that two suitably optimized auxiliary costs suffice to obtain an error exponent matching that of the constant-composition ensemble.  Specifically, we generalize the achievability of $\Ercc$ ({\em cf.}, Theorem \ref{thm:exponents}) to continuous alphabets. The analogous generalization of the achievability of $\Eriid$ is similar, but is omitted here.
The generalized exponent is given by 
\begin{equation}
    \Ercc(Q_X,R) = \max_{\rho\in[0,1]}\Ezcc(Q_X,\rho)-\rho R\label{eq:SU_Er_LM_Dual_cont},
\end{equation}
where
\begin{equation}
    \Ezcc(Q_X,\rho) = \sup_{\sgz,a(\cdot)}\EE\Bigg[-\log\EE\bigg[\bigg(\frac{\EE\big[q(\Xbar,Y)^{s}e^{a(\Xbar)}\,|\, Y\big]}{q(X,Y)^{s}e^{a(X)}}\bigg)^{\rho}\,\bigg|\, X\bigg]\Bigg] \label{eq:SU_E0_LM_cont}
\end{equation}
and where $(X,Y,\Xbar) \sim Q_X(x)W(y|x)Q_X(\xbar)$, and the supremum is over all $a(\cdot)$ such that $\EE_Q[a(X)] < \infty$.  

While the expression \eqref{eq:SU_E0_LM_cont} only contains a single auxiliary function $a(\cdot)$ (corresponding to one of the auxiliary costs), a second auxiliary cost is used to ensure that the average over $X$ in \eqref{eq:SU_E0_LM_cont} is {\em outside} the logarithm, which is preferable due to Jensen's inequality.  To derive an exponent with the expectation inside the logarithm, a single auxiliary cost suffices \cite{Sha12}; such an exponent is positive for all $R < \LM(Q_X)$, but can be smaller than $\Ercc$ in general \cite{Sca14c}.


\begin{theorem} \label{thm:exponent_cont}   
    {\em (Error exponent for cost-constrained coding)}
    For any mismatched memoryless channel $(W,q)$, we have the following under cost-constrained random coding with $L = 2$ and suitably-chosen $a_1(\cdot)$, $a_2(\cdot)$, and $\delta>0$:
    \begin{equation}
        \pebar(n,\lfloor e^{nR} \rfloor) \le e^{-n (\Ercc(Q_X,R) - \delta')}
    \end{equation}
    for an arbitrarily small constant $\delta'>0$ and sufficiently large $n$.  Hence, under an input constraint $(c,\Gamma)$, if $\EE_Q[c(X)] < \Gamma$, then $\Ercc(Q_X,R)$ is an achievable error exponent at rate $R$.
\end{theorem}

We provide an outline of the proof (see \cite{Sca14c} for the details), highlighting the fact that the two auxiliary costs $a_1(\cdot)$ and $a_2(\cdot)$ play different roles.  As in Section \ref{sec:exponents}, the analysis is based on the bound
\begin{equation}
    \pebar \le  \EE\Bigg[ \bigg(  M \EE\bigg[ \bigg(\frac{q^n(\Xvbar,\Yv)}{q^n(\Xv,\Yv)}\bigg)^s \,\bigg|\, \Xv,\Yv \bigg] \Bigg)^{\rho} \bigg].
\end{equation}
We fix $a_1(\cdot)$ to be specified later, and analyze the inner expectation in the same way as Section \ref{sec:su_cont_proofs}.  After doing so, we find that the exponential decay of the error probability given $\Xv = \xv$ is dictated by the following quantity:
\begin{equation}
    -\frac{1}{n} \sum_{i=1}^n \log\EE\bigg[\bigg(\frac{\EE\big[q(\Xbar,Y)^{s}e^{a_1(\Xbar)}\,|\, Y\big]}{q(X,Y)^{s}e^{a_1(X)}}\bigg)^{\rho}\,\bigg|\, X = x_i\bigg]. \label{eq:cost_empirical_sum}
\end{equation}
In order to replace this empirical average by an average over $Q_X$, we choose the second auxiliary cost as
\begin{equation}
    a_2(x) = -\log\EE\bigg[\bigg(\frac{\EE\big[q(\Xbar,Y)^{s}e^{a_1(\Xbar)}\,|\, Y\big]}{q(X,Y)^{s}e^{a_1(X)}}\bigg)^{\rho}\,\bigg|\, X = x\bigg]. \label{eq:a2_choice}
\end{equation}  
Since $\big|\frac{1}{n} \sum_{i=1}^n a_2(x_i) - \EE_{Q}[ a_2(X) ]\big| \le \delta$ by construction in \eqref{eq:SU_SetDn_multi2}, this means that \eqref{eq:cost_empirical_sum} can be lower bounded by the argument to the supremum in \eqref{eq:SU_E0_LM_cont}, up to an arbitrarily small loss of $\delta$.  The proof is completed by noting that $\delta$ can be arbitrarily small, optimizing the free parameters $\sgz$ and $a_1(\cdot)$, and renaming $a_1(\cdot)$ to $a(\cdot)$.

\paragraph{Additional auxiliary costs.} 
In analogy with Theorem \ref{thm:LM_fixed}, one can use similar analysis techniques to deduce an achievable error exponent for a fixed and possibly suboptimal set of auxiliary costs.  We state the resulting error exponent without proof, and refer the reader to \cite{Sca14c} for the details:
\begin{equation}
    \Ercost(Q_X,R,\{a_{l}\}) =\max_{\rho\in[0,1]}\Ezcost(Q_X,\rho,\{a_{l}\})-\rho R,\label{eq:SU_Er_Cost_Dual} 
\end{equation}
where
\begin{align}
    &\Ezcost(Q_X,\rho,\{a_{l}\}) \triangleq \sup_{s\ge0,\{r_{l}\},\{\rbar_{l}\}} \nonumber \\ &\qquad\qquad-\log\EE\left[\bigg(\frac{\EE\big[q(\Xbar,Y)^{s}e^{\sum_{l=1}^{L}\rbar_{l}(a_{l}(\Xbar)-\phi_{l})}\,|\, Y\big]}{q(X,Y)^{s}e^{\sum_{l=1}^{L}r_{l}(a_{l}(X)-\phi_{l})}}\bigg)^{\rho}\right], \label{eq:SU_E0_Cost} 
\end{align}
and where $\{a_l\}_{l=1}^L$ are the auxiliary costs, and $\phi_{\ell} = \EE_Q[a_l(X)]$.  

Similarly to the achievable rates, the use of $L > 2$ auxiliary costs may be useful when one does not have knowledge of the optimal $a_1(\cdot)$ and $a_2(\cdot)$ in the proof of Theorem \ref{thm:exponent_cont}.  It is tempting to ask whether using $L > 2$ {\em optimized} auxiliary costs can lead to a better exponent than that of Theorem \ref{thm:exponent_cont}; however, the answer turns out to be negative, since
\begin{equation}
    \Eriid(Q_X,R)\le \Ercost(Q_X,R,\{a_{l}\})\le \Ercc(Q_X,R).\label{eq:SU_Connection1}
\end{equation}
The left inequality follows by setting each $r_l = \rbar_l = 0$ in \eqref{eq:SU_E0_Cost}, and the right inequality follows from two steps: (i) use Jensen's inequality to take the outer expectation over $X$ outside the logarithm in \eqref{eq:SU_E0_Cost}; (ii) define $a(x) = \sum_{l=1}^{L}\rbar_{l}(a_{l}(\Xbar)-\phi_{l})$ to recover the objective function in \eqref{eq:SU_E0_LM_cont}.

\section{Expurgated Exponents} \label{sec:expurg}

Expurgation is a classical technique for attaining improved error exponents at low rates \cite{Gal65}. The technique consists of removing, from a randomly generated codebook, a given fraction of codewords whose conditional error probability is the highest.  In this subsection, we overview an extension of Gallager's expurgation technique \cite[Sec.~5.7]{Gal68} to the mismatched setting \cite{Sca14f}.  In Section \ref{sec:best_of_both}, we will discuss alternative techniques that also provide these low-rate improvements.

The derivation of primal and dual expurgated exponents follows similar steps to those of the random coding exponent once the appropriate initial non-asymptotic bound (i.e., a counterpart to \eqref{eq:rcu_rep}) is obtained.  Such a bound is given in the following theorem.

\begin{theorem} \label{thm:rcux}
    {\em (Non-asymptotic expurgated bound)}
    For any mismatched memoryless channel $(W,q)$ and codeword distribution $P_{\Xv}$, there exists a codebook $\Cc_n$ with $M$ codewords of length $n$ whose error probability satisfies
    \begin{equation}
        \pe(\Cc_n) \le \inf_{\rho \ge 1} \Big(4(M-1)\EE\Big[\PP\big[q^{n}(\Xvbar,\Yv)\ge q^{n}(\Xv,\Yv)\,\big|\,\Xv,\Xvbar\big]^{1/\rho}\Big]\Big)^{\rho}, \label{eq:EXP_RCX}
    \end{equation}
    where $(\Xv,\Yv,\Xvbar) \sim P_{\Xv}(\xv)W^n(\yv|\xv)P_{\Xv}(\xvbar)$.
\end{theorem}

The proof closely follows the analysis of Gallager \cite[Sec.~5.7]{Gal68}, and is outlined as follows:
\begin{itemize}
    \item Let $\Csf_n$ be a codebook with $M' = 2M-1$ codewords drawn independently from $P_{\Xv}$, and let $\pem(\Csf_n)$ be the conditional error probability associated with the $m$-th message.  We have from Markov's inequality that at least half of the codewords satisfy $\pem(\Csf_n)^{1/\rho} \le 2\EE\big[\pem(\Csf_n)^{1/\rho}\big]$; hence, there exists a codebook $\Cc_n$ with $M$ codewords such that
    \begin{equation}
        p_{e}(\Cc_n) \le \Big(2\EE\big[\pem(\Csf_n)^{1/\rho}\big]\Big)^{\rho}. \label{eq:EXP_GalBound}
    \end{equation}
    \item We obtain \eqref{eq:EXP_RCX} by using the union bound to upper bound $\EE\big[\pem(\Csf_n)^{1/\rho}\big]$ in terms of $M' - 1 = 2(M-1)$ error events $\{ q^{n}(\Xv^{(\mbar)},\Yv)\ge q^{n}(\Xv^{(m)},\Yv) \}_{\mbar \ne m}$, and then applying the inequality $\big(\sum_{i} \alpha_{i}\big)^{1/\rho} \le \sum_{i}\alpha_{i}^{1/\rho}$ for $\rho \ge 1$.
\end{itemize}

With Theorem \ref{thm:rcux} in place, we can consider the same choices of $P_{\Xv}$ as those of Sections \ref{sec:exponents} and \ref{sec:exponents_cont}: i.i.d., constant-composition, and cost-constrained.  We focus on the latter two, as they attain higher exponents.
%
Starting with the discrete memoryless setting, the primal form of the constant-composition exponent is given by
\begin{equation}
    \Eexcc(Q_X,R) \defeq \min_{\substack{P_{X\Xtilde Y} \,:\, P_X = P_{\Xtilde} = Q_X, \\ \EE_{P}[\log q(\Xtilde,Y)]\ge\EE_{P}[\log q(X,Y)] \\ I_{P}(X;\Xtilde) \le R}} D(P_{X \Xtilde Y}\|Q_X \times Q_X \times W)-R, \label{eq:EXP_PrimalAlt_CC}
\end{equation}
and the dual expression is given by
\begin{equation}
    \Eexcc(Q_X,R) = \sup_{\rho\ge1}\Excc(Q_X,\rho)-\rho R, \label{eq:EXP_Eex_CC}
\end{equation}
where
\begin{align}
    &\Excc(Q_X,\rho) = \sup_{s \ge 0,a(\cdot)} -\rho \sum_{x} Q_X(x) \nonumber \\ 
    &\qquad\qquad\times \log \sum_{\xbar} Q_X(\xbar) \Bigg(\sum_{y}W(y|x)\bigg(\frac{q(\xbar,y)}{q(x,y)}\bigg)^{s}\frac{e^{a(\xbar)}}{e^{a(x)}} \Bigg)^{1/\rho}. \label{eq:EXP_Ex_CC}
\end{align} 
Once again, the equivalence between the two is proved via Lagrange duality \cite{Sca14f}.

\begin{theorem} \label{thm:expurg}
    {\em (Expurgated error exponent for DMCs)}
    For any mismatched DMC $(W,q)$, input distribution $Q_X \in \Pc(\Xc)$, and rate $R > 0$, there exists a sequence of codebooks $\Cc_n$ with $M = \lfloor e^{nR} \rfloor$ messages of length $n$ such that
    \begin{equation}
    \liminf_{n \to \infty} -\frac{1}{n} \log \pe(\Cc_n) \ge \Eexcc(Q_X,R).
    \end{equation}
\end{theorem}

The primal expression is derived from \eqref{eq:EXP_RCX} using the method of types. The dual expression is derived by fixing $\sgz$ and weakening \eqref{eq:EXP_RCX} as follows via Markov's inequality:
\begin{equation}
    \pe(\Cc_n) \le  \Bigg(4(M-1)\EE\Bigg[\EE\Bigg[\bigg(\frac{q^{n}(\Xvbar,\Yv)}{q^{n}(\Xv,\Yv)}\bigg)^{s}\,\Bigg|\,\Xv,\Xvbar\Bigg]^{1/\rho}\Bigg]\Bigg)^{\rho}. \label{eq:EXP_RCX_s}
\end{equation}
With this bound in place, the desired error exponent is obtained via further bounding techniques similar to the dual derivation of the random-coding exponent.  

The extension of Theorem \ref{thm:expurg} to continuous alphabets follows similar steps to Section \ref{sec:exponents_cont} via cost-constrained random coding with $L = 2$:  We let one auxiliary cost $a_1(\cdot)$ correspond to $a(\cdot)$ in \eqref{eq:EXP_Ex_CC}, and we choose the other auxiliary cost as
\begin{equation}
    a_2(x) = -\log\EE\Bigg[\Bigg(\EE\bigg[\bigg(\frac{q(\Xbar,Y)}{q(x,Y)}\bigg)^{s}\frac{e^{a_1(\Xbar)}}{e^{a_1(x)}} \,\Big|\, \Xbar, X=x\bigg] \Bigg)^{1/\rho} \Bigg],
\end{equation}
in analogy with \eqref{eq:a2_choice}.  The details can be found in \cite{Sca14f}.

\section{Attaining the Best of Both Exponents} \label{sec:best_of_both}

In addition to the constant-composition expurgated exponent $\Eexcc$ in \eqref{eq:EXP_PrimalAlt_CC}, recall from \eqref{eq:SU_Er_cc} that the primal form of the constant-composition random-coding exponent $\Ercc$ is given by 
\begin{align}
    &\Ercc(Q_X,R) =  \min_{P_{XY} \,:\, P_X = Q_X}\min_{ \substack{\Ptilde_{XY}\,:\, \Ptilde_X = Q_X, \Ptilde_Y = P_Y, \\ \EE_{\Ptilde}[\log q(X,Y)] \ge \EE_{P}[\log q(X,Y)] } }  \nonumber \\ &\qquad\qquad\qquad D(P_{XY}\|Q_X\times W)+\big[I_{\Ptilde}(X;Y)-R\big]^{+}. \label{eq:SU_Er_cc2} 
\end{align}
With these two achievable error exponents established, we highlight the following limitations of Theorem \ref{thm:expurg}:
\begin{itemize}
    \item Despite the low-rate improvement, the exponent $\Eexcc$ is loose at high rates, in particular falling to zero at a rate strictly below $\LM(Q_X)$, in contrast with $\Ercc$ (see Section \ref{sec:exp_numerical} for an example).  While one could overcome this by simply using the fact that $\max\{ \Ercc, \Eexcc  \}$ is clearly an achievable exponent, it is also of interest to find a construction that directly attains the best of both exponents simultaneously.
    \item While a random codebook $\Csf_n$ is considered as a step towards constructing the codebook $\Cc_n$, the structure of $\Cc_n$ is less clear (i.e., it is difficult to characterize properties of the removed vs.~retained codewords).  Hence, the standard expurgation argument is, in a sense, less constructive than standard random-coding methods.
    \item In contrast with the second part of Theorem \ref{thm:exponents} concerning $\Ercc$, Theorem \ref{thm:expurg} does not make any claim of ensemble tightness, so it is unclear to what extent the exponent could be improved via refined analysis techniques.
\end{itemize}
The first of these limitations was addressed in an early work of Csisz\'ar and K\"orner \cite{Csi81}, who derived the following  error exponent: 
    \begin{align}
        &E_{\textsc{ck}}(Q_X,R) \nonumber \\ &= \min_{\substack{P_{X\Xtilde Y} \,:\, P_X=P_{\Xtilde}=Q_X, \\ \EE_P[\log q(\Xtilde,Y)] \geq \EE_P[\log q(X,Y)], \\ I_P(X;\Xtilde)\leq R }}D(P_{Y|X}\|W|Q_X)+\big[I(\Xtilde;Y,X)-R\big]_+. \label{eq:E_CK}
    \end{align}
This can be weakened to $\Eexcc$ by lower bounding $[\cdot]_+$ by its argument, or can be weakened to $\Ercc$ by dropping the constraint $I_P(X;\Xtilde) \le R$ and lower bounding $I_P(\Xtilde;Y,X) \ge I_P(\Xtilde;Y)$. In the latter case, the resulting optimization problem in \eqref{eq:E_CK} depends on $P_{X\Xtilde Y}$ only through $(P_{XY},\Ptilde_{XY})$, and the minimization can be split into two minimizations of the form \eqref{eq:SU_Er_cc2} accordingly.

The derivation of $E_{\textsc{ck}}$ in \cite{Csi81} uses combinatorial arguments to establish the existence of a constant-composition code with certain properties, but the proof is highly non-constructive in the sense discussed above.  In addition, the analysis in \cite{Csi81} appears to be unsuitable for channels with continuous alphabets, and leaves open the question of ensemble tightness.  In the following, we outline a recent alternative derivation of \eqref{eq:E_CK} that addresses both of these limitations \cite{Som19}.


The analysis in \cite{Som19} is based on a sequential random coding construction resembling the Gilbert-Varshamov construction for binary codes \cite{Gil52,Var57}.  Fixing an input distribution $Q_X$, a generic ``distance'' function $d(\xv,\xv')$, and a constant $\Delta$, the codebook is generated as follows:
\begin{enumerate}
\item The first codeword, $\Xv_1$, is drawn uniformly from the type class $\Tclass(Q_{X,n})$, where $Q_{X,n}$ is a type approximating $Q_X$;
\item Given $\Xv_1 = \xv_1$, the second codeword $\Xv_2$ is drawn uniformly from the set
\begin{align}
        \Tc^n(Q_{X,n},\xv_1)&\triangleq\left\{\xvbar\in \Tclass(Q_{X,n}) \,:\, d(\xvbar,\xv_1)> \Delta\right\}\\
         &=\Tclass(Q_{X,n})\backslash \left\{\xvbar\in \Tclass(Q_{X,n}) \,:\, d(\xvbar,\xv_1)\leq \Delta\right\},
     \end{align}
i.e., the set of sequences with composition $Q_{X,n}$ whose distance to $\xv_1$ exceeds $\Delta$;
\item Continuing recursively, if the first $i-1$ codewords are $\xv_1^{i-1} \triangleq (\xv_1,\dotsc,\xv_{i-1})$, then the $i$-th codeword $\Xv_i$ is drawn uniformly from the set
 \begin{align}
         &\Tc^n(Q_{X,n},\xv_1^{i-1}) \nonumber \\
         &\triangleq\left\{\xvbar\in \Tc^n(Q_{X,n})  \,:\,  d(\xvbar,\xv_j)> \Delta, \, \forall j=1\dotsc, i-1\right\}\\
         &=\Tc^n(Q_{X,n},\xv_1^{i-2})
         \backslash \left\{\xvbar\in \Tc^n(Q_{X,n},\xv_1^{i-2}) \,:\, d(\xvbar,\xv_{i-1})\leq \Delta\right\}.
     \end{align}
\end{enumerate}
Essentially, we sequentially perform random coding while avoiding any codewords that are too close to those already selected, according to a generic distance $d(\cdot,\cdot)$.  This procedure was termed the {\em (generalized) random Gilbert-Varshamov} (RGV) construction in \cite{Som19}.

While we use the terminology ``distance'', $d(\cdot,\cdot)$ need not be a distance function in the topological sense.  However, for the analysis, it is required to be symmetric, bounded, and dependent only on the joint empirical distribution of its arguments.  To highlight the latter condition, we henceforth write $d(P_{X\Xtilde})$ as a function on the probability simplex.  For instance, the choice $d(P_{X\Xtilde}) = \PP[X \ne \Xtilde]$ corresponds to a normalized Hamming distance, and the negative mutual information $d(P_{X\Xtilde}) = -I_{P}(X; \Xtilde)$ will be highlighted below.

If the recursive procedure is continued for too many iterations, the entire space of codewords may be exhausted.  To ensure that this does not occur, it suffices to enforce that $M$ times the volume of the appropriate $d$-ball is smaller than the size of the type class $|\Tc(Q_{X,n})|$. Asymptotically, this translates into the following rate condition \cite{Som19}:
\begin{align} \label{eq:RGV_rate}
    R\leq \min_{\substack{P_{X\Xtilde} \,:\, d(P_{X\Xtilde}) \leq\Delta,\,\\ P_X=P_{\Xtilde}=Q_X}} I_P(X;\Xtilde)-2\delta
\end{align}
for arbitrarily small $\delta > 0$.  Under this condition, it was shown in \cite{Som19} that the above sequential random coding scheme attains the following strengthened (albeit asymptotic) variant of \eqref{eq:rcu_rep}:
\begin{flalign}
    &\hspace{-2mm} \pebar \doteq   \sum_{\xv \in \Tclass(Q_{X,n}),\yv}\frac{1}{|\Tclass(Q_{X,n})|}W^n(\yv|\xv)\nonumber\\
    &\times\min\Biggl\{1,(M-1) \sum_{\substack{\xv' \in \Tclass(Q_{X,n}) \,:\, q^{n}(\xv',\yv)\geq q^{n}(\xv,\yv)\\ d(\xv,\xv')\geq \Delta }} \frac{1}{|\Tclass(Q_{X,n})|} \Biggr\}, \label{eq:rcu-like}
\end{flalign}
where $\doteq$ denotes asymptotic equality up to a sub-exponential pre-factor.  Note that the key difference compared to \eqref{eq:rcu_rep} is the constraint $d(\xv,\xv')\geq \Delta$. With this expression in place, an analysis based on the method of types yields the following theorem \cite{Som19}.

\begin{theorem} \label{thm:RGV}
    {\em (Ensemble-tight exponent for the RGV construction)}
    For any mismatched DMC $(W,q)$, input distribution $Q_X \in \Pc(\Xc)$, bounded symmetric type-dependent distance function $d(P_{X\Xtilde})$, parameters $\Delta$ and $\delta > 0$, and rate $R$ satisfying \eqref{eq:RGV_rate}, we have under the RGV construction that
    \begin{align}
        \liminf_{n \to \infty} -\frac{1}{n} \log \pebar(n,\lfloor e^{nR} \rfloor) &\ge E_{\textsc{rgv}}(Q_X,R,\Delta), \label{eq:RGV_tight1} \\
        \limsup_{n \to \infty} -\frac{1}{n} \log \pebar(n,\lfloor e^{nR} \rfloor) &\le E_{\textsc{rgv}}(Q_X,R,\Delta+\epsilon), \label{eq:RGV_tight2}
    \end{align}
    where $\epsilon > 0$ is arbitrarily small, and
    \begin{align}\label{eq: E_ex dfn}
        &E_{\textsc{rgv}}(Q_X,R,\Delta) \nonumber \\
        &= \min_{\substack{P_{X\Xtilde Y} \,:\, P_X=P_{\Xtilde}=Q_X,\, \\ \EE_P[\log q(\Xtilde,Y)] \geq \EE_P[\log q(X,Y)], \,\\ d(P_{X\Xtilde})\geq \Delta}}D(P_{Y|X}\|W|Q_X)+\big[I_P(\Xtilde;Y,X)-R\big]_+.
    \end{align}
    In addition, \eqref{eq:RGV_tight1} holds even in the case that $d(\cdot)$ is non-symmetric.
\end{theorem}

Observe that \eqref{eq:RGV_tight1} and \eqref{eq:RGV_tight2} establish ensemble tightness under the technical condition that $E_{\textsc{rgv}}$ is continuous in $\Delta$ at the specified input.  The final claim of the theorem concerning non-symmetric $d(\xv,\xv')$ is proved by considering a symmetrized distance of the form $\min\{ d(\xv,\xv'), d(\xv',\xv) \}$ and applying the first part of the theorem \cite{Som19}.

Setting $d(P_{X\Xtilde}) = -I_P(X;\Xtilde)$ in \eqref{eq: E_ex dfn} readily recovers $E_{\textsc{ck}}$ in \eqref{eq:E_CK}, but the more general form here also provides insight into other distance functions (though it can be shown that no choice provides a better exponent than $E_{\textsc{ck}}$), e.g., see \eqref{eq:ds} below.  

In the case of additive distance functions, i.e., $d(\xv,\xv') = \frac{1}{n} \sum_{i=1}^n d(x_i,x'_i)$,\footnote{Here we normalize by $\frac{1}{n}$ in contrast with the previous sections; this is for consistency with the assumption that the distance can be expressed in terms of the joint type: $d(P_{X\Xtilde}) = \EE_{P}[d(X,\Xtilde)]$ is equivalent to $d(\xv,\xv') = \frac{1}{n} \sum_{i=1}^n d(x_i,x'_i)$.} one can use Lagrange duality to establish an equivalent dual form of \eqref{eq: E_ex dfn}, given by \cite{Som19} 
\begin{equation}
    E_{\textsc{rgv}}(Q_X, R, \Delta)  = \sup_{\rho\in[0,1]} E_{0,\textsc{rgv}}(Q_X,\rho,\Delta)-\rho R, \label{eq:RGV_dual}
\end{equation}
where
\begin{align}
&E_{0,\textsc{rgv}}(Q_X,\rho,\Delta) = \sup_{r\ge0,s\ge0,a(\cdot)} -\sum_{x}Q_X(x) \nonumber \\ 
    &\quad \times \log\sum_{y}W(y|x)\bigg(\frac{\sum_{\xbar}Q_X(\xbar)q(\xbar,y)^s e^{a(\xbar)}e^{r(d(x,\xbar)-\Delta)}}{q(x,y)^se^{a(x)}}\bigg)^{\rho}. \label{eq:E0_rgv}
\end{align}
Similarly, one can obtain the following dual expression for the rate condition \eqref{eq:RGV_rate}:
\begin{equation}
    R \le \sup_{r\ge0,a(\cdot)} - \sum_{x}Q_X(x)\log\sum_{\xbar}Q_X(\xbar)e^{a(\xbar)-\phi_a}e^{-r(d(x,\xbar)-\Delta)} - 2\delta,
\label{eq:dual_r_cond}
\end{equation}
where $\phi_a = \EE_Q[a(X)]$.
Setting $r=0$ in \eqref{eq:E0_rgv} readily recovers $\Ercc$ ({\em cf.}, \eqref{eq:SU_Er_cc2}). With some additional effort, we can also show that \eqref{eq:RGV_dual} recovers $\Eexcc$ ({\em cf.}, \eqref{eq:EXP_Eex_CC}).  The idea is to use the final claim of Theorem \ref{thm:RGV} concerning non-symmetric distance functions, setting $\rho=1$ in \eqref{eq:E0_rgv} and choosing
\begin{equation}
    d(x,\xbar) = -\log\sum_y W(y|x) \Big( \frac{q(\xbar,y)}{q(x,y)} \Big)^s \label{eq:ds}
\end{equation}
for some $\sgz$.  With a suitable choice of $\Delta$, it can be shown that \eqref{eq:E0_rgv} recovers $\Excc$ in \eqref{eq:EXP_Ex_CC}, while also maintaining the rate condition \eqref{eq:dual_r_cond} \cite{Som19}.

Finally, we briefly discuss the generalization of \eqref{eq:RGV_dual}--\eqref{eq:dual_r_cond} to continuous alphabets, which can be achieved (for additive distance functions) by suitably modifying the recursive random coding construction.  The idea is to recursively draw from a cost-constrained codeword distribution {\em conditioned} on the distance to each previously-selected codeword being at least $\Delta$, and to derive an upper bound analogous to \eqref{eq:rcu-like}, with the cost-constrained codeword distribution in place of $\frac{1}{|\Tclass(Q_{X,n})|}$.  The interested reader is referred to \cite{Som19} for the details.

\section{Numerical Example} \label{sec:exp_numerical}

\begin{figure}
    \begin{centering}
        \includegraphics[width=0.7\columnwidth]{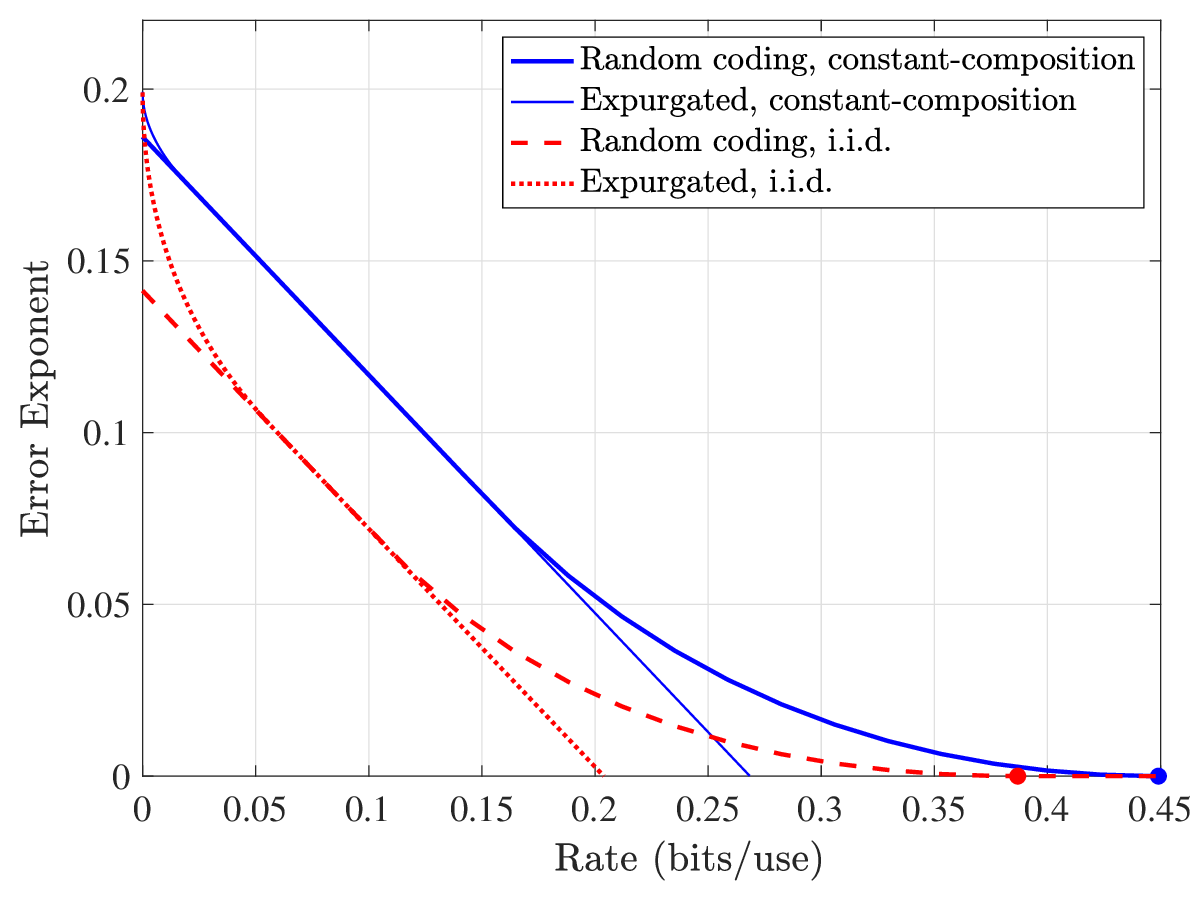}
        \par
    \end{centering}
    
    \caption{Mismatched random-coding and expurgated exponents under constant-composition and i.i.d.~random coding.  The GMI and LM rate are marked on the horizontal axis.} \label{fig:ExponentExample}
\end{figure}

Here we present a numerical example from \cite{Sca14c} that serves to compare the exponents attained by i.i.d.~vs.~constant-composition coding, as well as random coding vs.~expurgated exponents.  We consider an asymmetric channel, as we found that this better highlights these differences.  Specifically, the channel and metric are defined by the entries of the $|\Xc|\times|\Yc|$ matrices
\begin{equation}
    \Wv =\left[\begin{array}{ccc}
        1-2\delta_{0} & \delta_{0} & \delta_{0}\\
        \delta_{1} & 1-2\delta_{1} & \delta_{1}\\
        \delta_{2} & \delta_{2} & 1-2\delta_{2}
    \end{array}\right],~\qv=\left[\begin{array}{ccc}
        1-2\delta & \delta & \delta \\
        \delta & 1-2\delta & \delta \\
        \delta & \delta & 1-2\delta
    \end{array}\right] \label{eq:SU_MatrixW}
\end{equation}
with $\Xc=\Yc=\{0,1,2\}$, and with $\delta_0$, $\delta_1$, $\delta_2$, and $\delta$ taking values in $\big(0,\frac{1}{3}\big)$.  We set $\delta_{0}=0.01$, $\delta_{1}=0.05$, $\delta_{2}=0.25$, and $Q_X=(0.1,0.3,0.6)$, and note that any choice of $\delta \in \big(0,\frac{1}{3}\big)$ yields an equivalent decoder (namely, minimum-distance decoding).
Under these parameters, we have $\GMI(Q_X)=0.387$, $\LM(Q_X)=0.449$, and $I(X;Y)=0.471$ bits/use.  The resulting error exponents are shown in Figure \ref{fig:ExponentExample}.

As expected, constant-composition random coding consistently yields a higher exponent than i.i.d.~random coding.  For both ensembles, the random-coding exponent is positive for all rates below the corresponding achievable rate, whereas this fails to hold for the expurgated exponents.  On the other hand, the expurgated exponents are better at low rates.  

In the limit of zero rate, the i.i.d.~and constant-composition expurgated exponents approach the same value, i.e., $\Eexcc(0) = \Eexiid(0)$.  In fact, this behavior is not specific to this example, but rather a general phenomenon for any channel, decoding metric, and input distribution \cite{Sca14f}.

\section{Error Exponents for Multi-User Coding Techniques}


The analysis leading to the achievable rates for the multiple-access channel (Section \ref{ch:mac}) and superposition coding (Section \ref{ch:multiuser}) can also be adapted to obtain achievable error exponents.  The modifications are similar to those above for the single-user channel, but somewhat more technical \cite{ScarlettThesis}.

Starting with the multiple-access channel, the main modification required is to use the exponential upper bound \eqref{eq:MAC_TypeX1X2Y} directly (as well as a matching lower bound for the ensemble tightness part), rather than only using it to infer that $\|P_{X_1X_2Y} - Q_1 \times Q_2 \times W\|_{\infty} \le \delta$.  By doing so, we find that the three rate conditions in \eqref{eq:MAC_R1_LM}--\eqref{eq:MAC_R12_LM} each have a corresponding error exponent containing the divergence term $D(P_{X_{1}X_{2}Y}\|Q_{1}\times Q_{2}\times W)$ from \eqref{eq:MAC_TypeX1X2Y}, and the overall error exponent is the minimum of the three. For instance, the error exponent associated with \eqref{eq:MAC_R12_LM} is given by
\begin{align}
&\Eiiircc(\Qv,R_{1},R_{2}) \nonumber \\
    &= \min_{P_{X_{1}X_{2}Y} \,:\, P_{X_1} = Q_1, P_{X_2} = Q_2} \min_{\substack{\Ptilde_{X_{1}X_{2}Y} \,:\, \Ptilde_{X_{1}}=P_{X_{1}},\Ptilde_{X_{2}}=P_{X_{2}},\Ptilde_{Y}=P_{Y}\\
            \EE_{\Ptilde}[\log q(X_{1},X_{2},Y)]\ge\EE_{P}[\log q(X_{1},X_{2},Y)]}}
      \nonumber \\  
    &\qquad D(P_{X_{1}X_{2}Y}\|Q_{1}\times Q_{2}\times W) + \Big[\max\Big\{I_{\Ptilde}(X_{1};Y)-R_{1}, \nonumber \\ 
    &\qquad I_{\Ptilde}(X_{2};Y)-R_{2}, D\big(\Ptilde_{X_{1}X_{2}Y}\|Q_{1}\times Q_{2}\times P_{Y}\big)-R_{1}-R_{2}\Big\}\Big]^{+}.
\end{align}
%
Dual expressions for the error exponents, along with direct derivations, can be obtained using analogous (albeit more cumbersome) techniques to those for the achievable rate region.  As was the case in the single-user setting, the number of auxiliary costs used is higher when it comes to attaining the error exponents; it was shown in \cite{Sca16a} that five per user is sufficient.

Similar ideas apply in the context of superposition coding ({\em cf.}, Section \ref{ch:multiuser}) \cite{ScarlettThesis}.  
While we saw that refined superposition coding ({\em cf.}, Section \ref{sec:rsc}) provides an achievable rate at least as high as all other known achievable rates, it is worth noting that the error exponent may be small due to the use of shorter ``sub-codes'' of length $n_u = nQ_{U}(u)$ for each auxiliary symbol $u \in \Uc$.  If such a sub-code has error probability decaying as $e^{-n_u E_u}$ for some $E_u > 0$, then its error exponent with respect to the full block length $n$ is $Q_U(u) E_u$, which may be low if $Q_U(u)$ is small.  In contrast, for standard superposition coding ({\em cf.}, Section \ref{sec:sc}), all of the error events directly correspond to a block length of $n$. 

\begin{figure}
    \begin{centering}
        \includegraphics[width=0.7\columnwidth]{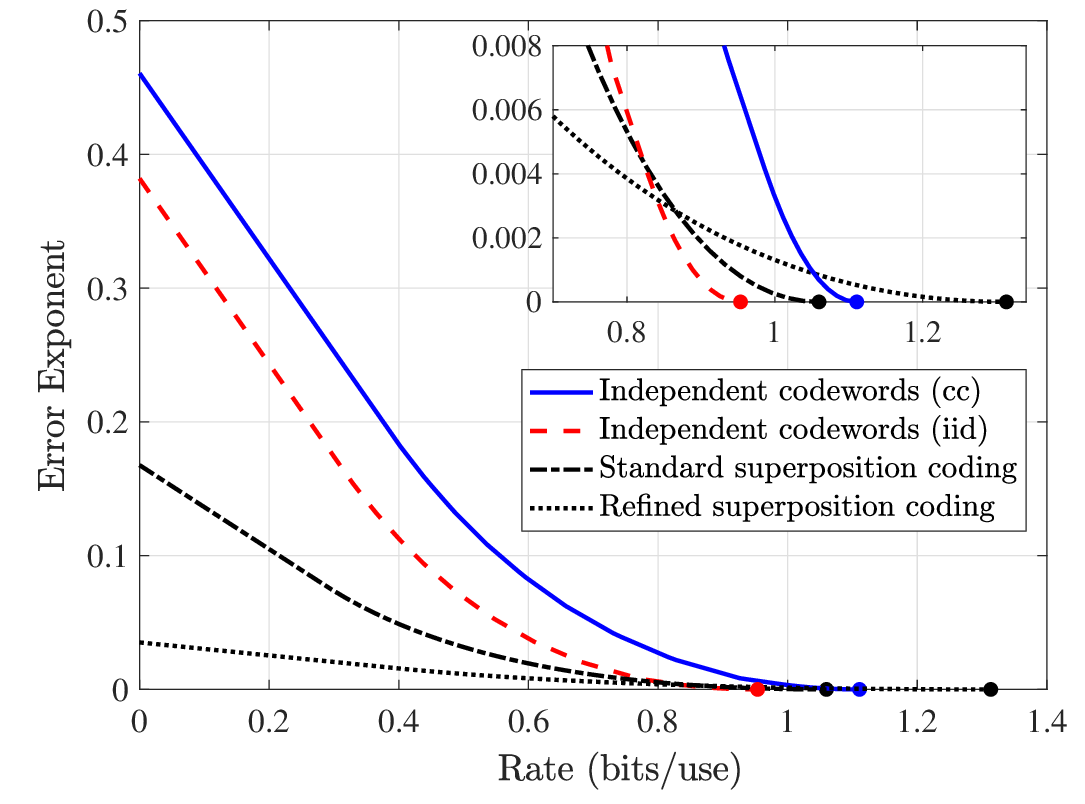}
        \par
    \end{centering}
    
    \caption{Error exponents of superposition coding for the example in Section \ref{sec:rsc_vs_sc}, using the input distribution optimized for the refined superposition coding rate.  The exponents under independent codewords are also shown, in which the input distribution is chosen to optimize the corresponding rate (GMI or LM rate).  The achievable rates are marked on the horizontal axis.} \label{fig:RSC_vs_SC}
\end{figure}

To demonstrate this difference, and also compare against the exponents of standard random coding (i.e., independent codewords), we revisit the example from Section \ref{sec:rsc_vs_sc}, in which the refined superposition coding rate was strictly higher than the standard version.  We compare the exponents as follows:
\begin{itemize}
    \item We let $Q_{UX}^*$ be the optimized input distribution attaining the highest refined superposition coding rate, and let $(R_0^*,R_{11}^*,R_{12}^*)$ be the corresponding triplet attaining equality in each rate condition of Theorem \ref{thm:RSC_Main}. For a range of $\alpha \in [0,1]$, we compute the error exponent for the rate triplets $(R_0,R_{11},R_{12}) = (\alpha R_0^*,\alpha R_{11}^*,\alpha R_{12}^*)$, with rate $R = R_0 + Q_U^*(1)R_{11} + Q_U^*(2)R_{12}$.
    \item Similarly, for the same $Q_{UX}^*$, letting $(R_0^*,R_1^*)$ be the pair attaining equality in the rate conditions for standard superposition coding (Theorem \ref{thm:SC_Rate_SC}), we plot the exponent for various pairs $(R_0,R_1) = (\alpha R_0^*, \alpha R_1^*)$, with rate $R = R_0 + R_1$.
\end{itemize}
The resulting error exponents are shown in Figure \ref{fig:RSC_vs_SC}, along with the error exponents for i.i.d.~and constant-composition random coding with independent codewords and an input distribution chosen to maximize the respective rate (see Section \ref{sec:rsc_vs_sc}).

Although refined superposition coding has the highest achievable rate, we see that the corresponding error exponent is much smaller than the standard version at lower rates, likely as a result of using shorter sub-codes.  Interestingly, a similar phenomenon also occurs when moving from standard superposition coding to independent codewords: The superposition coding rate is comparable to the LM rate and higher than the GMI, yet its exponent is smaller at lower rates.   A possible reason for this is that superposition coding yields the minimum of two error exponents -- one for each error type -- whereas with independent codewords we only have a single error type.

This example indicates a trend of ensembles with higher rates tending to yield lower exponents at smaller rates (with the exception of i.i.d.~vs.~constant-composition random coding), and suggests an interesting open problem of finding an ensemble that naturally attains the best of both worlds.

We conclude this discussion by mentioning another example highlighting that the error exponents attained by multi-user coding techniques may have no general ordering (i.e., neither exponent can be weakened to one another).  In \cite{Som15}, a {\em cognitive} variant of the multiple-access channel was considered, in which one user has access to both messages.  Under this setting, the standard superposition coding technique was compared to a random binning technique, and it was shown that neither of the two dominate each other in general.  This suggests that unlike most achievable rates, one may not typically expect to be able to weaken one mismatched decoding error exponent to another without imposing additional assumptions.

 



%

\chapter{Upper Bounds on the Mismatch Capacity}  \label{ch:converse}



\section{Introduction}

In contrast with the extensive results on achievable rates that we surveyed in the preceding sections, studies of mismatch capacity upper bounds (i.e., converse results) have produced relatively fewer results. In this section, we survey the main upper bounds that are known to date. We focus on discrete memoryless channels (DMCs), as considered in Section \ref{ch:single_user}.

In Section \ref{sec:conv_elem}, we overview some early observations that were made having a converse flavor.  We then return to the Csisz\'ar-Narayan conjecture on the multi-letter extension of the LM rate ({\em cf.}, Section \ref{sec:multi_LM}) in Section \ref{sec:cn}, and give some partial results suggesting its validity.  In Section \ref{sec:partial results CN conjecture}, we provide related results for other decoders beyond the standard maximum-metric decoder.  Additional multi-letter upper bounds are presented in Section \ref{sec:conv_multi}, and we conclude by presenting a recent single-letter upper bound in Section \ref{sc:single letter upper bound}.

Before proceeding, we recall some of the notation from Sections \ref{ch:intro} and \ref{ch:single_user}.  The mismatch capacity (Definition \ref{def:CM}) is denoted by $\CM$.  We will frequently make use of the LM rate $\LM(Q_X)$, whose primal expression \eqref{eq:INTR_PrimalLM} is repeated as follows:
\begin{equation}
    \LM(Q_X)=\min_{\substack{\Ptilde_{XY} \in \Pc(\Xc \times \Yc) \,:\, \Ptilde_{X}=Q_X, \Ptilde_{Y}=P_Y \\ \EE_{\Ptilde}[\log q(X,Y)] \ge \EE_{P}[\log q(X,Y)]}}I_{\Ptilde}(X;Y), \label{eq:CNV_PrimalLM}
\end{equation} 
where $P_{XY} = Q_X \times W$.  The input-optimized LM rate ({\em cf.}, \eqref{eq:CLM}) is defined as 
\begin{equation}
    \CLM = \max_{Q_X \in \Pc(\Xc)} \LM(Q_X), \label{eq:CNV_CLM}
\end{equation}
and its $k$-letter extension ({\em cf.}, \eqref{eq:CLMk})  is given by 
\begin{equation}
    \CLM^{(k)} = \frac{1}{k} \max_{ Q_{X^k} } \LM( Q_{X^k}, W^k, q^k ). \label{eq:CLMk_Ch8}
\end{equation}
We will often make the dependence of the preceding quantities on the channel $W$ and decoding metric $q$ explicit by writing $\LM(Q_X,W,q)$, $\CLM(W,q)$, $\CM(W,q)$, $\CLM^{(k)}(W,q)$, and so on.

This section is predominantly based on the works of Csisz\'ar and Narayan \cite{Csi95}, Somekh-Baruch \cite{Som14,Som15b}, and Asadi Kangarshahi and Guill\'en i F\`abregas \cite{KangarshahiGuilleniFabregasFull2020}.

\section{Initial Results}
\label{sec:conv_elem}


Perhaps the most elementary known result with a converse flavor is that of \cite[Thm.~2]{Csi95}, which gives a necessary and sufficient condition for the positivity of the mismatch capacity.  This result was stated in Lemma \ref{lem:pos_conds}, and is repeated as follows in a slightly different form.

\begin{lemma}\label{lem:positivity_condition_CN}
    {\em (Condition for positive mismatch capacity)} 
    For any mismatched DMC $(W,q)$, we have
    \begin{equation}
        \CM > 0 \iff \CLM > 0,
    \end{equation}
    with these equivalent conditions holding if and only if there exists $Q_X \in \Pc(\Xc)$ such that
    \begin{equation}
        \EE_{Q_X \times W}[\log q(X,Y)] > \EE_{Q_X \times P_Y}[\log q(X,Y)], \label{eq:pos_cond_Ch8}
    \end{equation}
    where $P_Y$ is the $Y$-marginal of $P_{XY} = Q_X \times W$. 
\end{lemma}


While this result precisely establishes the conditions under which $\CM = 0$ vs.~$\CM > 0$, it does not provide any further information about how high $\CM$ may be when it is positive; for a given value $C > 0$ of the matched capacity, all we can immediately deduce is that $\CM \in (0,C]$.


As we saw in Section \ref{sec:binary}, the mismatch capacity is completely characterized in the special case of binary channels, i.e., $|\Xc| = |\Yc| = 2$ \cite{Csi95}.  We formalize this statement in the following lemma, in which we let $C$ denote the matched capacity and write the sign function as
\begin{equation}
    \sign(z) = \begin{cases}
        1 & z > 0 \\
        0 & z = 0 \\
        -1 & z < 0,
    \end{cases}
\end{equation}
where we also allow $\sign(\infty) = 1$ and $\sign(-\infty) = -1$.

\begin{lemma}\label{lem:binary}
    {\em (Mismatch capacity of binary channels)} 
    For any mismatched DMC $(W,q)$ with $|\Xc| = |\Yc| = 2$, we have
    \begin{equation}
        \CM = \CLM =
            \begin{cases}
                C & \sign\big( \log\frac{q(1,1)q(2,2)}{q(1,2)q(2,1)} \big) = \sign\big( \log\frac{W(1|1)W(2|2)}{W(2|1)W(1|2)} \big) \\
                0 & {\rm otherwise},
            \end{cases}
    \end{equation}
    where $C$ is the matched capacity of $W$.
\end{lemma}



Throughout the monograph, we have presented several ensemble tightness results \cite{Mer95}, upper bounding the maximal rate for which the random-coding error probability $\pebar$ tends to zero for a given random-coding ensemble.  While such results do not provide upper bounds on $\CM$, they are helpful in ruling out the possibility of weaknesses in the proofs of the achievable rates.  For convenience, Lemma \ref{lem:ens_tight} is re-stated as follows (see also Section \ref{sec:cont_tightness} regarding the GMI for continuous-alphabet channels).

\begin{lemma} \label{lem:ens_tight2}
    {\em (Ensemble tightness)} 
     For any mismatched DMC $(W,q)$ and any given input distribution $Q_X \in \Pc(\Xc)$, we have the following:
     \begin{itemize}
         \item Under i.i.d.~random coding, $\pebar(n,\lfloor e^{nR} \rfloor) \to 1$ as $n \to \infty$ for any $R > \GMI(Q_X)$;
         \item Under constant-composition random coding, $\pebar(n,\lfloor e^{nR} \rfloor) \to 1$ as $n \to \infty$ for any $R > \LM(Q_X)$.
     \end{itemize}
\end{lemma}


Another result having a converse flavor is that of \cite[Lemma 1]{Mer95}, which we briefly outline as follows.  Consider the special case that $q(x,y)$ represents a conditional distribution, written as $V(y|x)$ to highlight this distinction. Consider a constant-composition codebook whose codewords have type $Q_X$,\footnote{This is without loss of generality, since any code has a constant-composition sub-code with the same asymptotic rate, as outlined in Section \ref{sec:su_properties}.} let $P_Y$ be the induced output distribution, and for a small constant $\epsilon > 0$, let $\Tc_{\epsilon}^n(P_Y)$ be the set of sequences whose empirical distribution is $\epsilon$-close to $P_Y$ in the $\ell_{\infty}$ sense.  That is, $\Tc_{\epsilon}^n(P_Y)$ represents a typical (high-probability) set of output sequences.  For a fixed codeword $\xv$, the set
\begin{equation}    
    S_d(\xv) = \big\{ \yv \in \Tc_{\epsilon}^n(P_Y) \,:\, -\log V^n(\yv|\xv) \le n d \big\}
\end{equation}
can be interpreted as a decoding sphere with normalized radius $d$.  By a concentration argument, the transmitted codeword will, with high probability, lie near the surface of this sphere when $d =  \EE_{Q_X \times W}[-\log V(Y|X)]$.  In \cite[Lemma 1]{Mer95}, it is shown that if there exist $e^{nR}$ {\em disjoint} decoding spheres $\{S_d(\xv_j)\}_{j=1}^{e^{nR}}$ in $\Tc_{\epsilon}^n(P_Y)$ for some $R$ and the preceding choice of $d$, then it must be the case that $R < \CLM$.  A strengthened statement is also given in \cite[Lemma 2]{Mer95}, characterizing the amount of overlap in the case that $R > \CLM$.  In any case, overlapping decoding spheres does not imply an upper bound on $\CM$, since it may be the case that the overlapping parts have low probability with respect to $W^n$.

Finally, we recall that an early work of Balakirsky \cite{Bal95} reported that $\CLM = \CM$ for general binary-input channels, which would considerably generalize the fact that this holds for $|\Xc| = |\Yc| = 2$ ({\em cf.},  Lemma \ref{lem:binary}).  However, as we saw in Section \ref{sec:binary_input}, this claim was refuted via a counter-example in which the $2$-letter extension of superposition coding attains a rate exceeding $\CLM$ for a binary-input ternary-output channel \cite{Sca15a}.

\section{The Csisz\'ar-Narayan Conjecture} \label{sec:cn}

As discussed in Section \ref{sec:multi_LM}, applying the LM rate to the product channel $W^k$ from ${\cal X}^k$ to ${\cal Y}^k$ yields an achievable rate $\CLM^{(k)}$ (see \eqref{eq:CLMk_Ch8}) for the original channel.  It is known that even when $k=2$, this rate can exceed the LM rate $\CLM = \CLM^{(1)}$ (e.g., see Section \ref{sec:mu_zuec}). Additionally, since $\CLM^{(k)}$ is achievable for any positive integer $k$, we deduce that 
\begin{equation}\label{eq:C_infty_def}
    \CLM^{(\infty)} \triangleq \sup_{k \in \ZZ_{>0}} \CLM^{(k)}
\end{equation}
is also an achievable rate.  In other words,
\begin{equation}\label{eq:achievable_side_of_CN_Conjecture}
    \CM \geq \CLM^{(\infty)}.
\end{equation}
In addition, the supremum over $k$ in \eqref{eq:achievable_side_of_CN_Conjecture} can be replaced by the limit as $k$ tends to infinity, yielding
\begin{equation}
    \CLM^{(\infty)}
    = \lim_{k \rightarrow \infty} \CLM^{(k)}.
\end{equation}
This is seen by noting that $\CLM^{(k)}$ is a uniformly bounded sequence, and for any fixed positive integer $c$ we have $\CLM^{(ck)} \ge \CLM^{(k)}$ by choosing the $ck$-fold input distribution to be $c$ independent copies of the optimal $k$-fold input distribution.  For non-integer multiples of $k$, a similar argument applies with the ``remainder'' padded by a deterministic input distribution placing probability 1 on a fixed symbol (making the decoder effectively ignore that part of the codeword), which has a vanishing impact on the rate as $c$ increases.

Csisz\'ar and Narayan \cite{Csi95} conjectured that the mismatch capacity is equal to the multi-letter extension of the LM rate, in the sense that inequality \eqref{eq:achievable_side_of_CN_Conjecture} holds with equality:
\begin{equation}\label{eq:CN_Conjecture}
    \CM \stackrel{?}{=} \CLM^{(\infty)}.
\end{equation}
Even if this conjecture were proved to be true, this would not directly provide a {\em computable} expression for the mismatch capacity, since (i) in itself, \eqref{eq:CN_Conjecture} does not indicate how close $\CLM^{(k)}$ is to $\CLM^{(\infty)}$ for any finite value of $k$, and (ii) the optimization over $Q_{X^k}$ in the definition $\CLM^{(k)}(W,q)= \frac{1}{k} \max_{ Q_{X^k} } \LM( Q_{X^k}, W^k, q^k)$ becomes computationally intractable as $k$ increases, particularly since $\LM$ is non-concave with respect to the input distribution in general (see Lemma \ref{lem:non_concave}).

Nevertheless, the conjecture would yield some interesting observations; for instance, if \eqref{eq:CN_Conjecture} is true, then all of the following statements also hold:
\begin{itemize}
    \item[(i)] We can attain a rate arbitrarily close to $\CM$ by employing random coding over the $k$-letter extension $W^k$ (with metric $q^k$) for sufficiently large $k$;
    \item[(ii)] At any rate below $\CM$, there exist sequences of codes whose error probability decays exponentially in the block length;
    \item[(iii)] Mismatched maximum-metric decoding and mismatched threshold-based decoding lead to the same mismatch capacity (see Section \ref{subsec:CN_threshold}).
\end{itemize}
Given the (hypothetical) validity of \eqref{eq:CN_Conjecture}, the first two implications follow immediately from the fact that constant-composition random coding achieves $\LM(Q_X)$ (see Section \ref{ch:single_user}) and yields a positive error exponent (see Section \ref{ch:exponents}), and the third follows since derivations of the LM rate already exist that upper bound the maximum-metric decoding error probability by that of threshold decoding (see Section \ref{subsec:CN_threshold}).

At the time of writing, the Csisz\'ar-Narayan conjecture remains unresolved in general.  It was proved in \cite{Csi95} that the conjecture is true in the case of the erasures-only metrics, and further partial results suggesting its validity were given by Somekh-Baruch \cite{Somekh-Baruch_IT_18}.  These findings are surveyed in more detail below.


\subsection{Erasures-Only Metrics}\label{subsec:CN_erasures_only}

Recall from Section \ref{sec:app_zuec} that the mismatched decoding problem with metric $q(x,y)=\openone\{W(y|x)>0\}$ yields the zero-undetected error capacity, which is a problem of independent interest.  It was shown in \cite[Thm.~3]{Csi95} that \eqref{eq:CN_Conjecture} is true not only in this special case, but also for any decoding metric $q(x,y)$ satisfying the condition
\begin{equation}\label{eq: eo_condition}
    q(x,y)=q_{\max},\; \forall (x,y)~{\rm such~that }~W(y|x)>0, 
\end{equation}
where $q_{\max}=\max_{x\in{\cal X},y\in{\cal Y}}q(x,y)$.   Metrics satisfying \eqref{eq: eo_condition} are referred to as {\em erasures-only metrics}, because they lead to an error whenever there exist two or more codewords consistent with the output, which implies that all errors are detectable.  Not all metrics satisfying \eqref{eq: eo_condition} yield the zero-undetected error capacity, since there could also exist $(x,y)$ pairs such that $q(x,y)=q_{\max}$ and $W(y|x) = 0$; the metric used to recover the zero-error capacity (see Section \ref{sec:app_zero}) is one such example.

The validity of the Csisz\'ar-Narayan conjecture for erasures-only metrics is formally stated as follows.

\begin{theorem}\label{th:zero undetected error theorem}
    {\em (Mismatch capacity for erasures-only metrics)}
     For any mismatched DMC $(W,q)$, if $q$ is an erasures-only metric (i.e., satisfies \eqref{eq: eo_condition}), then we have
    \begin{equation}
    \CM = \CLM^{(\infty)}.
    \end{equation}
\end{theorem} 
%
\begin{proof}
    We know that $\CM \ge \CLM^{(\infty)}$, so it suffices to establish that $\CM \le \CLM^{(\infty)}$. 
    We consider a sequence of codebooks $\Cc_k$ with block length $k$ such that the error probability satisfies $\lim_{k \to \infty} \pe(\Cc_k) \to 0$.  For brevity, we denote $\epsilon_k = \pe(\Cc_k)$.  Here we use $k$ instead of $n$ for the block length to highlight the fact that we consider $\CLM^{(k)}$ in \eqref{eq:CLMk_Ch8} and choose $Q_{X^k}$ depending on $\Cc_k$.

    We first consider the case that all codewords of $\Cc_k$ are distinct, and then discuss the general case at the end of the proof.  Let $P_{X^k}$ be the equiprobable distribution over the codebook $\Cc_k$, i.e., for $\xv\in\Xc^k$,
    \beq
    P_{X^k}(\xv) = \begin{cases}
    \frac{1}{M} & \xv \in \Cc_k\\
    0 & \text{otherwise,}
    \end{cases}
    \eeq
     and let $P_{X^k Y^k} = P_{X^k}\times W^k$ be the corresponding joint distribution under the channel $W^k$. For a given input distribution $Q_{X^k}$, we define the joint distribution $Q_{X^k Y^k} = Q_{X^k}\times W^k$, yielding
    \begin{align}\label{eq:rederivation CN}
        \CLM^{(k)}
        &= \frac{1}{k} \max_{ Q_{X^k} } \min_{\substack{\Ptilde_{{X}^k Y^k}  \,:\, \Ptilde_{{X}^k}=Q_{X^k}, \Ptilde_{Y^k}=Q_{Y^k} \\ \EE_{\Ptilde}[\log q^k({X}^k,Y^k)] \ge \EE_{Q}[\log q^k(X^k,Y^k)] }}I_{\Ptilde}({X}^k;{Y}^k)  \\
        &\geq \frac{1}{k}  \min_{\substack{\Ptilde_{{X}^k Y^k}  \,:\,  \Ptilde_{{X}^k}=P_{X^k} , \Ptilde_{Y^k}=P_{Y^k} \\ \EE_{\Ptilde}[\log q^k({X}^k,Y^k)] \ge \EE_P[\log q^k(X^k,Y^k)]}}I_{\Ptilde}({X}^k;Y^k) \label{eq:cond_ent_b} \\
        &= \frac{1}{k}H_P(X^k)- \frac{1}{k} \max_{\substack{\Ptilde_{{X}^k Y^k}  \,:\,  \Ptilde_{{X}^k}=P_{X^k}, \Ptilde_{Y^k} = P_{Y^k}, \\ \EE_{\Ptilde}[\log q^k({X}^k,Y^k)] \ge \EE_P[\log q^k(X^k,Y^k)]}}H_{\Ptilde}({X}^k|Y^k), \label{eq:cond_ent}
    \end{align}
    where \eqref{eq:cond_ent_b} follows by bounding $\max_{Q_{X^k}}$ by the specific choice $Q_{X^k} = P_{X^k}$, and \eqref{eq:cond_ent} follows by expanding the mutual information.

   Since we are considering the case that all codewords in $\Cc_k$ are distinct, and $P_{X^k}$ is uniform over the codebook, we obtain
    \begin{equation}\label{eq: normalized Hk bound}
        \frac{1}{k}H_P(X^k) = \frac{1}{k}\log|\Cc_k|. 
    \end{equation}
    The second term in \eqref{eq:cond_ent} requires more effort.  Since the metric satisfies \eqref{eq: eo_condition}, we have $ \EE_P[\log q^k(X^k,Y^k)] = k \log q_{\max}$ and $q^k({x}^k,y^k) \leq q_{\max}^k$ for all $(x^k,y^k)$, so the constraint $\EE_{\Ptilde}[\log q^k({X}^k,{Y}^k)] \ge \EE_P[\log q^k(X^k,Y^k)]$ is equivalent to $\PP_{\Ptilde}[ q^k({X}^k,Y^k)= q_{\max}^k ] = 1$.

    Let $\Ac_k$ denote the set of all $y^k \in \Yc^k$ for which there is a unique maximizer of $\max_{x^k\in\Cc_k} q^k(x^k,y^k)$. 
    If $q^k({X}^k,Y^k)= q_{\max}^k$ with probability one, then given $Y^k = y^k$ with $y^k \in \Ac_k$, the sequence $X^k$ can only equal this unique maximizer, implying that $H_{\Ptilde}({X}^k|Y^k=y^k)=0$.  This observation allows us to simplify the second term in \eqref{eq:cond_ent} to the following:
    \begin{align}
        &\frac{1}{k} \max_{\substack{\Ptilde_{{X}^k Y^k}  \,:\,  \Ptilde_{{X}^k}=P_{X^k}, \Ptilde_{Y^k}=P_{Y^k}, \\ \PP_{\Ptilde}[ q^k({X}^k,Y^k)=  q_{\max}^k]=1}}H_{\Ptilde}(X^k|Y^k)\\
            &\quad= \frac{1}{k} \max_{\substack{\Ptilde_{{X}^k Y^k}  \,:\,  \Ptilde_{{X}^k}=P_{X^k}, \Ptilde_{Y^k}=P_{Y^k}, \\ \PP_{\Ptilde}[ q^k({X}^k,Y^k)=  q_{\max}^k]=1}}\sum_{y^k\in\Ac_k^c}P_{Y^k}(y^k)H_{\Ptilde}({X}^k|Y^k=y^k)\label{eq:step a}\\
            &\quad\leq   \frac{1}{k}\sum_{y^k\in\Ac_k^c}P_{Y^k}(y^k)\cdot \log|\Cc_k|,\label{eq:step b}
    \end{align}
    since $X^k$ takes one of at most $|\Cc_k|$ values under $P_{X^k}$.
    
    Substituting \eqref{eq: normalized Hk bound} and \eqref{eq:step b} into \eqref{eq:cond_ent}, we obtain
    \begin{align}\label{eq:rederivation CN 2}
        \CLM^{(k)} \geq \frac{1}{k} \log|\Cc_k|\cdot\bigg(1- \sum_{y^k\in\Ac_k^c}P_{Y^k}(y^k)\bigg). 
    \end{align} 
    Now, since we assume that ties are broken as errors ({\em cf.}, Section \ref{sec:setup_general}), any $y^k\in\Ac_k^c$ produces an error, and hence  $\sum_{y^k\in\Ac_k^c}P_{Y^k}(y^k) \to 0$ due to our assumption $\epsilon_k \to 0$.  Taking $k \to \infty$ on both sides of \eqref{eq:rederivation CN 2}, we deduce that
    \begin{equation}
        \limsup_{k\rightarrow \infty} \frac{1}{k} \log|\Cc_k| 
        \le \CLM^{(\infty)}. \label{eq:er_final} 
    \end{equation}
    Having shown this to be true for codebooks with distinct codewords, it is straightforward to generalize to arbitrary codebooks.  Specifically, since we assume that $\lim_{k \to \infty} \epsilon_k \to 0$, we have for $k$ sufficiently large that at least half of the codewords of $\Cc_k$ must be unique.  By applying the above argument to a smaller codebook with only these unique codewords, we deduce that \eqref{eq:er_final} still holds, since removing half the codewords amounts to a negligible loss in the rate.
    
    Since \eqref{eq:er_final} holds for any sequence of codebooks $\Cc_k$ with vanishing error probability, it follows that $\CM \le \CLM^{(\infty)}$, as desired.
\end{proof}
 
\subsection{A Soft Converse for Rational Metrics}\label{subsec:CN_soft_converse}

A partial result was given in \cite[Thm.~3]{Somekh-Baruch_IT_18} stating conditions under which any length-$n$ code attaining error probability $\pe \ll \frac{1}{n}$ (i.e., $n\pe \to 0$) must have a rate no higher than  $\CLM^{(\infty)}$.  We refer to this as a {\em soft converse}, since it does not directly preclude the possibility of higher-rate codes whose error probability tends to zero at a slower speed, such as $\frac{1}{\sqrt n}$ or $\frac{1}{\log n}$.

We proceed by stating the result formally.  We say that a decoding metric is {\em rational} if its (log-)values can be written in the following form:
\begin{equation}
    \log q(x,y)= \frac{j(x,y)}{j_{\rm denom}} \quad \forall x,y \text{ such that } q(x,y) \ne 0,  \label{eq:rational}
\end{equation}
for some integers $j(x,y)$ and $j_{\rm denom}$.\footnote{The denominator $j_{\rm denom}$ being independent of $(x,y)$ is without loss of generality, since one could always take the lowest common denominator.}  In addition, we assume that the metric can only equal zero when the channel transition probability is also zero:
\begin{equation}
    \quad q(x,y) = 0 \implies W(y|x) = 0. \label{eq:tech_cond}
\end{equation}
This is a mild assumption, since, assuming without loss of generality that all inputs are used, the mismatch capacity is zero when \eqref{eq:tech_cond} fails (see the discussion following Lemma \ref{lem:continuity}).

%

We again let $\Cc_k$ be a codebook with block length $k$, and let $\epsilon_k$ be its associated average error probability when applied to the mismatched DMC $(W,q)$.  The following result is stated in \cite[Thm.~3]{Somekh-Baruch_IT_18}.

\begin{theorem}\label{th:soft converse theorem}
    {\em (Soft converse for rational metrics)} 
    For any mismatched DMC $(W,q)$, if the decoding metric is rational ({\em cf.}, \eqref{eq:rational}) and satisfies \eqref{eq:tech_cond}, and $\Cc_k$ ($k=1,2,...$) is a sequence of length-$k$ codebooks whose average error probability $\epsilon_k = \pe(\Cc_k)$ satisfies $\lim_{k\rightarrow\infty} k\cdot \epsilon_k= 0$, then it holds that 
    \begin{eqnarray}
        \limsup_{k\rightarrow\infty} \frac{1}{k}\log|\Cc_k|\leq \CLM^{(\infty)}. \label{eq:soft_claim}
    \end{eqnarray}
\end{theorem}
%
\begin{proof}
    The proof in \cite{Somekh-Baruch_IT_18} was based on the dual expression \eqref{eq:INTR_RateLM} for the LM rate; here we provide an alternative proof based on the primal expression \eqref{eq:INTR_PrimalLM}, also re-using some of the general steps given in the proof of Theorem \ref{th:zero undetected error theorem}.  By the same argument used therein, it suffices to prove the result for codebooks $\Cc_k$ whose codewords are all distinct; we henceforth assume that this is the case.

    Let $\eta_k > 0$ denote the minimal difference between distinct $k$-letter log-metric values with respect to the same length-$k$ output sequence:
    \begin{equation}\label{eq:defin gammak}
        \eta_k=\min_{\substack{(x^k,\bar{x}^k,y^k)\in{\cal X}^k\times{\cal X}^k\times  {\cal Y}^k \,:\, \\ q^k(x^k,y^k)> q^k(\bar{x}^k,y^k)}} \Big(\log q^k(x^k,y^k)-\log q^k(\bar{x}^k,y^k)\Big).
    \end{equation}
    Note that if the minimization in \eqref{eq:defin gammak} is empty, then we must have $q(x,y) = q(\bar{x},y)$ for all $(x,\bar{x},y)$.  This case can be excluded, as it trivially gives a mismatch capacity of zero.

    We observe that the assumptions of the theorem imply the following:
    \begin{enumerate}
        \item By \eqref{eq:tech_cond}, we can find a number $B \in (0,\infty)$ such that the metric is bounded as follows for all $(x,y)$:
        \begin{flalign}
            \log q(x,y)&\leq B, \quad~ \forall x,y, \nonumber \\
            \log q(x,y)&\geq -B \quad \forall x,y ~\text{ s.t. }~ W(y|x)>0. \label{eq:bounded metric}
        \end{flalign}
        Indeed, it suffices to set $B$ to the higher of the two quantities $\max_{(x,y) \,:\, q(x,y) > 0} \log q(x,y)$ and $\max_{(x,y) \,:\, W(y|x) > 0} -\log q(x,y)$, with \eqref{eq:tech_cond} ensuring that $B$ remains finite.
        
        \item By the assumption of a rational metric, it is straightforward to show that $\eta_k$ is bounded away from zero uniformly in $k$; see \cite[Lemma 4]{Somekh-Baruch_IT_18} for a formal argument.\footnote{In \cite{Somekh-Baruch_IT_18} the log-metric is normalized by $\frac{1}{k}$, so in the notation therein the statement is that the decay rate of $\eta_k$ to zero is no faster than $\frac{1}{k}$ (rather than being bounded away from zero). }
    \end{enumerate}
    Hence, to prove the theorem, it suffices to establish the following more general claim:  For any mismatched DMC $(W,q)$, if the metric is bounded according to $\eqref{eq:bounded metric}$,  and $\lim_{k\rightarrow\infty} \frac{k \epsilon_k}{\eta_k} = 0$, then
    \begin{eqnarray}
        \limsup_{k\rightarrow\infty} \frac{1}{k}\log|\Cc_k|\leq \CLM^{(\infty)}. \label{eq:new_claim}
    \end{eqnarray}
    We proceed with the proof of this claim.

    As was done in the proof of Theorem \ref{th:zero undetected error theorem}, we let $P_{X^k}$ be uniformly distributed over the codebook $\Cc_k$, and define $P_{X^k Y^k} = P_{X^k} \times W^k$.
    %
    Recall the chain of inequalities leading to \eqref{eq:cond_ent}, which is re-stated here for convenience:
    \begin{equation}
        \CLM^{(k)} \ge \frac{1}{k}H_P(X^k)- \frac{1}{k} \max_{\substack{\Ptilde_{{X}^k Y^k}  \,:\,  \Ptilde_{{X}^k}=P_{X^k}, \Ptilde_{Y^k} = P_{Y^k}, \\ \EE_{\Ptilde}[\log q^k({X}^k,Y^k)] \ge \EE_P[\log q^k(X^k,Y^k)]}}H_{\Ptilde}({X}^k|Y^k). \label{eq:cond_ent2}
    \end{equation}
    We fix $\Ptilde_{X^k Y^k}$ satisfying the constraints in \eqref{eq:cond_ent2}, and seek to simplify the condition $\EE_{\Ptilde}[\log q^k({X}^k,Y^k)] \ge \EE_P[\log q^k(X^k,Y^k)]$, which is equivalent to 
    \begin{equation}
        \EE[ \log q^k(\Xbar^k,Y^k)-\log q^k(X^k,Y^k) ] \ge 0 \label{eq:metric_cond}
    \end{equation}
    for any triplet $(X^k,Y^k,\Xbar^k)$ with marginals $(X^k,Y^k) \sim P_{X^k Y^k}$ and $(\Xbar^k,Y^k) \sim \Ptilde_{X^k Y^k}$.  In addition to these random variables, we define 
    \begin{equation}
        q_{\max}^k(Y^k)=\max_{x^k\in\Cc_k}q^k(x^k,Y^k), \label{eq:qmaxk}
    \end{equation}
    and we let $\Xhat^k$ be the estimate of $X^k$ produced when the mismatched decoder is applied to $Y^k$ (i.e., $\Xhat^k$ is the codeword in $\Cc_k$ corresponding to the message estimate $\hat{m}$).  Then, we simplify \eqref{eq:metric_cond} by considering the following:
    %
    \begin{align}
        &\EE[ \log q^k(\Xbar^k,Y^k)-\log q^k(X^k,Y^k) ] \nonumber\\
        &= \EE\Big[\openone\{\Xhat^k\neq X^k\}\big( \log q^k(\Xbar^k,Y^k)-\log q^k(X^k,Y^k) \big) \Big] \nonumber \\
            &\quad + \EE\Big[\openone\{\Xhat^k= X^k\}\big(\log q^k(\Xbar^k,Y^k)-\log q^k(X^k,Y^k)\big)\Big]\\
        &\leq 2kB\cdot \epsilon_k +\EE\Big[\openone\{\Xhat^k= X^k\}\big(\log q^k(\Xbar^k,Y^k)-\log q^k(X^k,Y^k)\big)\Big]\label{eq: q bound 1} \\
        &= 2kB\cdot \epsilon_k +\EE\Big[\openone\{\Xhat^k= X^k\}\big(\log q^k(\Xbar^k,Y^k)-\log q_{\max}^k(Y^k)\big)\Big]\label{eq: q bound 2} \\
        &= 2kB\cdot \epsilon_k +\EE\Big[\openone\{\Xhat^k= X^k, q^k(\Xbar^k,Y^k)< q_{\max}^k(Y^k)\} \nonumber \\ 
            &\quad \times \big(\log q^k(\Xbar^k,Y^k)-\log q_{\max}^k(Y^k)\big) \Big]\label{eq: delta exchange} \\
        &\leq 2kB\cdot \epsilon_k-\eta_k \PP[\Xhat^k= X^k,q^k(\Xbar^k,Y^k)<q_{\max}^k(Y^k)]\label{eq: q bound 3}\\
        &\leq 2kB\cdot \epsilon_k-\eta_k\big((1-\epsilon_k)+ \PP[q^k(\Xbar^k,Y^k)<q_{\max}^k(Y^k)]-1\big)\label{eq: q bound 4} \\
        &= 2kB\cdot \epsilon_k+\eta_k\big(\epsilon_k-\PP[q^k(\Xbar^k,Y^k)<q_{\max}^k(Y^k)] \big), \label{eq:deduced}
    \end{align}
    where:
    \begin{itemize}
        \item \eqref{eq: q bound 1} follows from the upper bound $\log q^k(\Xbar^k,Y^k)-\log q^k(X^k,Y^k)\leq 2 k B$, which holds due to \eqref{eq:bounded metric};
        \item \eqref{eq: q bound 2} holds since whenever $\Xhat^k= X^k$, one has $q^k(X^k,Y^k)=q_{\max}^k(Y^k)$ by the definition of $\Xhat^k$ as a codeword attaining the highest metric;
        \item \eqref{eq: delta exchange} follows since the difference in log-metric values is trivially zero when $q^k(\Xbar^k,Y^k) = q_{\max}^k(Y^k)$;
        \item \eqref{eq: q bound 3} holds since the condition $q^k(\Xbar^k,Y^k)< q_{\max}^k(Y^k)$ combined with the definition of $\eta_k$ yields $\log q^k(\Xbar^k,Y^k)-\log q^k(X^k,Y^k)\leq-\eta_k$;
        \item \eqref{eq: q bound 4} follows from the fact that $\PP[\Ec_1\cap \Ec_2]\geq \PP[\Ec_1]+\PP[\Ec_2]-1$ for any two events $\Ec_1$ and $\Ec_2$, and since $\PP[\Xhat^k= X^k] = 1-\epsilon_k$ under our assumption that the codewords of $\Cc_k$ are distinct. 
    \end{itemize}
    Re-arranging \eqref{eq:deduced}, we deduce that if \eqref{eq:metric_cond} holds, then
    \begin{equation}
        \PP[q^k(\Xbar^k,Y^k)<q_{\max}^k(Y^k)]\leq 2kB\cdot \frac{\epsilon_k}{\eta_k}+\epsilon_k,
    \end{equation}
    or equivalently, taking one minus both sides,
    \begin{equation}
        \PP[q^k(\Xbar^k,Y^k) = q_{\max}^k(Y^k)] \geq 1 - \left(\frac{2kB}{\eta_k}+1\right)\epsilon_k \label{eq:p_max_ge}
    \end{equation}
    Letting $\Ac_k$ denote the set of $y^k$ sequences for which the maximizer of $\max_{x^k\in\Cc_k}q^k(x^k,y^k)$ is unique, we can upper bound the left-hand side of \eqref{eq:p_max_ge} as follows:
    \begin{align}
        &\PP\big[ q^k(\Xbar^k,Y^k)= q_{\max}^k(Y^k)\big] \\
        &~~\leq \PP\big[ q^k(\Xbar^k,Y^k)= q_{\max}^k(Y^k), Y^k\in\Ac_k \big]
        + \PP[ Y^k\in\Ac_k^c ]\\
        &~~\leq \PP\big[ q^k(\Xbar^k,Y^k)= q_{\max}^k(Y^k), Y^k\in\Ac_k \big] +\epsilon_k, \label{eq:last_step0}
    \end{align}
    where \eqref{eq:last_step0} follows since the event $\Ac_k^c$ implies a decoding error, due to the fact that ties are decoded as errors (see Section \ref{sec:setup_general}).  Hence, combining \eqref{eq:p_max_ge} and \eqref{eq:last_step0}, we have
    \begin{flalign}
        \PP\big[ q^k(\Xbar^k,Y^k)= q_{\max}^k(Y^k), Y^k\in\Ac_k \big] \geq 1-2 \left(\frac{kB}{\eta_k}+1\right)\epsilon_k. \label{eq:last_step}
    \end{flalign}

    Recall that the preceding analysis (and hence \eqref{eq:last_step}) holds whenever $(X^k,Y^k) \sim P_{X^k Y^k}$ and $(\Xbar^k,Y^k) \sim \Ptilde_{X^k Y^k}$ with $\Ptilde_{X^k Y^k}$ satisfying the constraints in \eqref{eq:cond_ent2}.  Hence, reverting back from the notation $(X^k,\Xbar^k,Y^k)$ to the notation $\PP_{\Ptilde}[\cdot]$, we can write \eqref{eq:last_step} as
    \begin{flalign}\label{eq: enlarge max set dgjh}
        \PP_{\Ptilde}\big[ q^k({X}^k,Y^k)= q_{\max}^k(Y^k), Y^k\in\Ac_k \big] \geq 1-2 \left(\frac{kB}{\eta_k}+1\right)\epsilon_k.
    \end{flalign}
    Now, similarly to the derivation of Fano's inequality, we introduce the binary random variable
    \begin{equation}
        S = \openone\{ q^k({X}^k,Y^k)= q_{\max}^k(Y^k), Y^k\in\Ac_k\}.
    \end{equation} 
    By the definition of $\Ac_k$ following \eqref{eq:p_max_ge}, and the definition of $q_{\max}^k$ in \eqref{eq:qmaxk}, we see that when $S=1$ it must be the case that ${X}^k$ is the unique maximizer of the $k$-letter decoding metric.  This means that $H_{\Ptilde}({X}^k|Y^k,S=1)=0$, and hence
    \begin{align}
        H_{\Ptilde}({X}^k|Y^k)
        &\leq H_{\Ptilde}({X}^k,S|Y^k) \label{eq:FanoStep1} \\
        &= H_{\Ptilde}(S|Y^k)+
        H_{\Ptilde}({X}^k|Y^k,S \label{eq:FanoStep2} )\\
        &\leq 1+
        \PP_{\Ptilde}[S=0]H_{\Ptilde}({X}^k|Y^k,S=0) \label{eq:FanoStep3} \\
        &\leq 1+2\left(\frac{kB}{\eta_k}+1\right)\epsilon_k\log|\Cc_k|, \label{eq:this_means}
    \end{align}
    where \eqref{eq:FanoStep2} follows from the chain rule, \eqref{eq:FanoStep3} uses the fact that $H_{\Ptilde}({X}^k|Y^k,S=1)=0$, and \eqref{eq:this_means} follows by upper bounding $\PP_{\Ptilde}[S=0] = 1-\PP_{\Ptilde}[S=1]$ using \eqref{eq: enlarge max set dgjh}, and by trivially upper bounding the conditional entropy by $\log|\Cc_k|$.

    We are now ready to simplify the expression given in \eqref{eq:cond_ent2}.  As we already established in \eqref{eq: normalized Hk bound}, the first term equals $\frac{1}{k}\log|\Cc_k|$ under our assumption that the codewords of $\Cc_k$ are distinct.  As for the second term, we have
    \begin{align}
        & \frac{1}{k} \max_{\substack{\Ptilde_{X^k, Y^k}  \,:\,  \Ptilde_{X^k}=P_{X^k}, \Ptilde_{Y^k} = P_{Y^k}, \\ \EE_{\Ptilde}[\log q^k(X^k,Y^k)] \ge \EE_{P}[\log q^k(X^k,Y^k)] }}H_{\Ptilde}(X^k|Y^k)\\
        &~~~\leq \frac{1}{k} \max_{\substack{\Ptilde_{X^k,Y^k}  \,:\,  \Ptilde_{X^k}=P_{X^k}, \Ptilde_{Y^k} = P_{Y^k}, \\ \PP_{\Ptilde}[q^k(X^k,Y^k)=q_{\max}^k(Y^k),Y^k\in\Ac_k]\geq 1-2\left(\frac{kB}{\eta_k}+1\right)\epsilon_k}}H_{\Ptilde}(X^k|Y^k)\label{eq: enlarge max set}\\
        &~~~\leq \frac{1}{k}\left(1+ 2\left(\frac{kB}{\eta_k}+1\right)\epsilon_k\cdot \log|\Cc_k|\right), \label{eq: frac H_k eq}
    \end{align}
    where \eqref{eq: enlarge max set} follows from the fact established above that the constraint $\EE_{\Ptilde}[\log q^k(X^k,Y^k)] \ge \EE_{\Ptilde}[\log q^k(X^k,Y^k)]$ implies \eqref{eq: enlarge max set dgjh}, and \eqref{eq: frac H_k eq} follows from \eqref{eq:this_means}.

    Substituting \eqref{eq: normalized Hk bound} and \eqref{eq: frac H_k eq} into \eqref{eq:cond_ent2}, we find that
    \begin{align}\label{eq:rederivation CN re}
        \CLM^{(k)}
        &\geq \left(1-2\left(\frac{kB}{\eta_k}+1\right)\epsilon_k
        \right) \cdot \frac{1}{k}\log|\Cc_k|-\frac{1}{k}.
    \end{align}
    Taking the limit as $k$ tends to infinity, and recalling the assumption $\lim_{k\rightarrow\infty} \frac{k \epsilon_k}{\eta_k} = 0$ stated above \eqref{eq:new_claim}, we obtain \eqref{eq:soft_claim} as desired.
\end{proof}

By the results of Section \ref{ch:exponents} (applied to the product channel), we know that at any rate below $\CLM^{(k)}$, there exist sequences of codes whose error probability vanishes exponentially fast.  Combined with Theorem \ref{th:soft converse theorem}, we deduce that for rational metrics, $\CLM^{(\infty)}$ is the supremum of all rates for which such exponential decay can be attained.


\section{Other Decoding Rules} \label{sec:partial results CN conjecture}

Throughout the monograph, we have focused on the standard maximum-metric decoder defined in \eqref{eq:decoder}.  In this subsection, we consider two variations of this decoder in which $\CLM^{(\infty)}$ can be deduced as an upper bound on the mismatch capacity: maximum-metric decoding with a margin, and threshold decoding.

\subsection{Maximum-Metric Decoding with a Margin}\label{subsec:CN_margin}

We consider mismatched decoding with a {\em $\delta$-margin}, meaning that the decoder selects a message whose codeword yields a normalized log-metric exceeding that of all other codewords by at least $\delta$, i.e., the decoder chooses message $\hat m$ if and only if the codeword $\xv^{(\hat m)}$ is the only codeword such that
\begin{flalign}
\frac{1}{n}\log q^n(\xv^{(\hat m)},\yv) \geq \frac{1}{n}\log q^n(\xv^{(j)},\yv)+\delta , ~~\forall j \neq \hat m.\label{eq: margin decision rule 1}
\end{flalign}
If no such codeword exists, then an error is declared.  This decoder is clearly no better than the standard maximum-metric decoder, since whenever $\hat{m}$  is selected it must be the case that $\xv^{(\hat m)}$ is the maximum-metric codeword.


For this setup, the following was proved in \cite{Somekh-Baruch_IT_18}.

\begin{theorem} \label{thm:margin}
    %
    {\em (Mismatch capacity for maximum-metric decoding with a margin)}
    For any mismatched DMC $(W,q)$, under maximum-metric decoding with a margin according to \eqref{eq: margin decision rule 1}, we have the following:
    \begin{enumerate}
        \item[(i)] For any given $\epsilon>0$, there exists $\delta>0$ such that the rate $\CLM^{(\infty)}-\epsilon$ is achievable when the margin parameter is $\delta$.
        \item[(ii)] For any given $\delta>0$ and $\epsilon > 0$, there does not exist any sequence of codes having rate $R>\CLM^{(\infty)}+\epsilon$ such that the error probability vanishes when the margin parameter is $\delta$.
    \end{enumerate}
\end{theorem}

The first of these claims (i.e., achievability) is proved via a straightforward extension of that of the analogous claim for maximum-metric decoding, with the additional step of showing that the function
\begin{equation}
    \CLM^{\rm marg}(\delta)=\max_{ Q_{X} } \min_{\substack{\Ptilde_{{X} Y}  \,:\, \Ptilde_{{X}}=Q_{X}, \Ptilde_{Y}=P_{Y} \\ \EE_{\Ptilde}[\log q({X},Y)] \ge \EE_{P}[\log q(X,Y)]-\delta }}I_{\Ptilde}({X};Y)
\end{equation}
(which equals the LM rate $\CLM$ when $\delta=0$) is continuous as $\delta\rightarrow 0$ from above.  
The second claim above (i.e., the converse result) is proved using similar ideas to the proof of Theorem \ref{th:soft converse theorem}; the details can be found in \cite[Thm.~1]{Somekh-Baruch_IT_18}.

\subsection{Mismatched Threshold Decoding} \label{subsec:CN_threshold}

Another important decoding rule that is related to the mismatched maximum-metric decoder is the mismatched threshold decoder, defined as follows: For a fixed threshold $\tau \in \RR$, the decoded message $\hat{m}$ is the unique message whose decoding metric is above the threshold, i.e., the decoder outputs $\hat m$ if and only if
\begin{equation}
    \frac{1}{n}\log q^n(\xv^{(\hat m)},\yv) \geq  \tau ~~\text{and}~~ \frac{1}{n}\log q^n(\xv^{(j)},\yv)<\tau,\;\forall j \neq \hat m.\label{eq: threshold decision rule 2}
\end{equation}
If no such codeword exists, then an error is declared.  Similarly to the margin decoder introduced above, this decoder is no better than the standard maximum-metric decoder, since whenever the decoder declares $\hat{m}$ it must be the case that $\hat m$ yields the highest metric.  In principle, the threshold $\tau$ could vary with $n$, and with the channel output $\yv$, but the result given below has only been proved in the fixed-threshold setting.


For this setup, we have the following result \cite{Somekh-Baruch_IT_18}, which is analogous to Theorem \ref{thm:margin}, but has additional technical assumptions in the converse part.

\begin{theorem} \label{thm:threshold}
    %
    {\em (Mismatch capacity for threshold decoding)}
    For any mismatched DMC $(W,q)$, under threshold decoding according to \eqref{eq: threshold decision rule 2}, we have the following:
    \begin{enumerate} 
        \item[(i)] For any given $\epsilon>0$, there exists $\tau \in \RR$ such that the rate $\CLM^{(\infty)}-\epsilon$ is achievable when the threshold parameter is $\tau$.
        \item[(ii)] Fix $Q_X \in \Pc(\Xc)$, and suppose that 
        there exists some $x^* \in \Xc$ and $y',y'' \in \Yc$ such that
        \begin{equation}    \label{eq: xstar yytag condition}
            Q_X(x^*) > 0, \quad W(y'|x^*) > 0, \quad W(y''|x^*) > 0, \quad \text{ and } \quad q(x^*,y') \ne q(x^*,y'').
        \end{equation}
        Then, for any fixed threshold $\tau \in \RR$ and constant $\epsilon > 0$, and any sequence of codebooks (indexed by the block length $n$) of rate $R$ having codewords of the same type $Q_{X,n}$ converging to $Q_X$ sufficiently fast so that $\lim_{n\to\infty} \sqrt{n} \max_{x} |Q_{X,n}(x) - Q_X(x)| = 0$, if $R > \CLM^{(\infty)} + \epsilon$, then the error probability is bounded away from zero as $n \to \infty$.
    \end{enumerate}
\end{theorem}



The first claim (i.e., achievability) is implicit in the derivation of the LM rate in \cite{Gan00} that is based on upper bounding the maximum-metric decoding achievable rate in terms of that of threshold decoding.  The idea is that by employing constant-composition random coding and setting $\tau=\EE_{P}[\log q(X,Y)]-\epsilon$ for some $\epsilon>0$, the resulting achievable rate is
\begin{equation}
    \LM^{\rm thresh}(Q_X,\tau)= \min_{\substack{\Ptilde_{{X} Y}  \,:\, \Ptilde_{{X}}=Q_{X}, \Ptilde_{Y}=P_{Y} \\ \EE_{\Ptilde}[\log q({X},Y)] \ge \tau }}I_{\Ptilde}({X};Y),
\end{equation}
which can be made arbitrarily close to the LM rate $\LM(Q_X)$ for small enough $\epsilon>0$.

For the second claim (i.e., the converse) we outline the main ideas of the proof as follows. First, it is shown that for successful threshold decoding, the threshold must be strictly less than $\mathbb{E}_{Q_X \times W}[\log q(X,Y)]$, where $Q_X$ is the limiting distribution in the theorem statement.  Intuitively, this is because the normalized log-metric $\frac{1}{n}\log q^n(\Xv,\Yv)$ for the transmitted codeword $\Xv$ concentrates around this value.  This is formalized using the central limit theorem (CLT), and making use of the assumptions in the theorem: (i) the assumption \eqref{eq: xstar yytag condition} ensures that $\frac{1}{n}\log q^n(\Xv,\Yv)$ has positive variance;
(ii) the assumption $\lim_{n\to\infty} \sqrt{n} \max_{x} |Q_{X,n}(x) - Q_X(x)| = 0$ is utilized since the CLT concerns deviations on the order of $\frac{1}{\sqrt n}$ from the mean.  Since we consider the case that $\tau$ does not depend on $n$, we deduce that $\tau=\mathbb{E}_{Q\times W}[\log q(X,Y)]-\epsilon'$ for some $\epsilon'>0$.

The second idea is to relate the performance of threshold decoding to that of maximum-metric decoding with a margin ({\em cf.}, Section \ref{subsec:CN_margin}).  Specifically, letting $p_{{\rm e}, {\rm thresh}}^{(\tau)}$ and $p_{{\rm e}, {\rm marg}}^{(\delta')}$ be the corresponding error probabilities, it can be shown that $p_{{\rm e}, {\rm marg}}^{(\delta')} \leq p_{{\rm e}, {\rm thresh}}^{(\tau)} + \PP\big[ \frac{1}{n} \log q^n(\Xv,\Yv) \leq \tau+\delta'\big]$.  By the above choice of $\tau$ and the law of large numbers, the second term on the right-hand side tends to zero under the choice $\delta'=\epsilon'/2$, and the second claim of Theorem \ref{thm:threshold} then follows from the analogous claim for maximum-metric decoding with a margin.  The details can be found in \cite[Thm.~2]{Somekh-Baruch_IT_18}.

\section{Further Multi-Letter Upper Bounds} \label{sec:conv_multi}


The quantity $\CLM^{(\infty)}$ ({\em cf.}, \eqref{eq:C_infty_def}) is multi-letter, in the sense that it is not expressed as the optimization of information measures on $\Xc$ and $\Yc$, but rather, on $\Xc^k$ and $\Yc^k$ (with $k \to \infty$).  Further multi-letter upper bounds on $\CM$ were derived in \cite{Somekh-Baruch_IT_18,Som14}, some in the spirit of $\CLM^{(\infty)}$, and others building on information-spectrum methods \cite{Ver94,Han02}.  In this subsection, we overview some of the results attained.

Before proceeding, we briefly discuss the merit in deriving multi-letter upper bounds on $\CM$, which may appear to be of limited value due to being non-computable.  Despite this limitation, such bounds can serve to deduce certain properties of the mismatch capacity $\CM$ and related quantities.  For instance, by a suitable comparison of multi-letter expressions, a precise characterization was given in \cite{Som15a} on how much the mismatch capacity increases under certain list decoding recovery criteria.  Multi-letter bounds could also potentially serve as starting points towards further results, such as settling the Csisz\'ar and Narayan  conjecture, attaining new single-letter bounds, and so on.  In fact, such benefits have already been observed in other topics in information theory, e.g., a multi-letter characterization of the interference channel capacity region \cite{Ahl73} was used to derive single-letter outer bounds, and to characterize certain extreme points \cite{Mot09}.


We also briefly mention that multi-letter {\em achievability} bounds have been established, including a multi-letter counterpart to the LM rate \cite{Gan00} and lower bounds on $\CM$ with a similar flavor to the upper bounds summarized below \cite{Somekh-Baruch_IT_18}.  These achievability results are omitted from our discussion, as the focus of this section is on converse results.

\subsection{A Max-Min Multi-Letter Upper Bound} \label{sc: Max-Min Upper Bounds}

Substituting the LM rate expression \eqref{eq:CNV_PrimalLM} into \eqref{eq:C_infty_def}, we find that $\CLM^{(\infty)}$ can be written as
\begin{equation}
     \CLM^{(\infty)} = \lim_{k \to \infty} \max_{Q_{X^k}} \min_{\substack{\Ptilde_{X^k {Y}^k} \,:\, \Ptilde_{X^k} = Q_{X^k}, \Ptilde_{Y^k} = P_{Y^k}, \\ \EE_{\Ptilde}[\log q^k(X^k,Y^k)] \ge \EE_{P}[\log q^k(X^k, Y^k)] } }
        \frac{1}{k}  I_{\Ptilde}(X^k;{Y}^k), \label{eq:substituted}
\end{equation}
where $P_{X^k Y^k} = Q_{X^k} \times W^k$.
The following upper bound on $\CM$ has the same form as \eqref{eq:substituted}, but differs in the set over which the inner minimization is performed.  

\begin{theorem}\label{th: second theorem}
    {\em (Multi-letter upper bound)}
    For any mismatched DMC $(W,q)$, we have
    \begin{flalign}\label{eq: bound of Theorem 1 a}
        &\CM \leq  \liminf_{k\rightarrow\infty} \max_{Q_{X^k}} \min_{\substack{\Ptilde_{X^k{Y}^k} \,:\, \Ptilde_{X^k} = Q_{X^k}, \\ P_{{Y}^k,q^k(X^k,{Y}^k) }=\widetilde{P}_{Y^k,q^k(X^k,Y^k) } } }
        \frac{1}{k}  I_{\Ptilde}(X^k;{Y}^k),
    \end{flalign}
    where $\Ptilde_{{Y}^k,q^k(X^k,{Y}^k) }$ is the joint distribution of ${Y}^k$ and the metric value $q^k(X^k,{Y}^k)$ at the output of the auxiliary channel $\Ptilde_{{Y}^k|X^k}$, and $P_{Y^k,q^k(X^k,Y^k) }$ is the corresponding joint distribution for the true channel $W^k$. 
\end{theorem}

The idea of the proof is the following \cite{Somekh-Baruch_IT_18}: If two (multi-letter) channels lead to the same joint distribution of $Y^k$ and $q^k(X^k,Y^k)$, then they have the same average probability of error under metric $q$.  For any such channel, the amount of information transmitted is at most $I_{\Ptilde}(X^k;{Y}^k)$, and the result follows by minimizing over all such channels.

\subsection{Information-Spectrum Formula for the Mismatch Capacity}\label{sc: Information Spectrum Upper Bounds}

Next, we turn to an {\em exact} formula for the mismatch capacity derived in \cite{Som14} using the information-spectrum method.  While the result in \cite{Som14} applies to general channels and metrics with memory, we restrict our attention to the case of mismatched DMCs here. For a sequence of real-valued random variables $\{X_i\}_{i=1}^{\infty}$, the limit inferior in probability is defined as follows
\begin{flalign}
    \text{p}\,\mhyphen\liminf_{n\to\infty} X_n&= \sup\Big\{t :\; \lim_{n\rightarrow\infty} \PP[X_n < t]=0\Big\}.
\end{flalign}

In the following, we consider an $n$-letter input distribution $Q_{\Xm}$, and define the pairwise error probability random variable
\begin{align}\label{eq: Phin dfn}
    \Phi_n(\Xm,\Ym) 
    &=\PP\left[ q^n(\Xvbar,\Ym)\geq q^n(\Xm,\Ym) \,\big|\, \Xm,\Ym\right], 
\end{align}
where $(\Xm,\Ym,\Xvbar)\sim Q_{\Xm}(\xv) W^n(\yv|\xv) Q_{\Xm}(\bar\xv)$.
This is the conditional probability of a codeword $\Xvbar$ independently drawn from $Q_{\Xm}$ yielding a higher decoding metric than $q^n(\Xm,\Ym)$, and this also appeared in the derivation of the GMI and LM rate (see Section \ref{sec:su_proofs}).
%
\begin{theorem} \label{th: General formula expression}
    {\em (Multi-letter expression for the mismatch capacity)}
    For any mismatched DMC $(W,q)$, the mismatch capacity is given by
    \begin{align}\label{eq: the General formula ksdjhfkjhsk}
        \CM = \sup_{\boldsymbol{Q}} ~ \text{p}\,\mhyphen\liminf_{n\to\infty} -\frac{1}{n}\log \Phi_n(\Xm, \Ym),
    \end{align}
    where $(\Xm,\Ym) \sim Q_{\Xm} \times W^n$, $\Phi_n$ is defined in \eqref{eq: Phin dfn}, and $\boldsymbol{Q}$ denotes the infinite sequence of multi-letter input distributions $(Q_{X^1}, Q_{X^2}, Q_{X^3}, \dotsc)$.
\end{theorem}

The proof is based on the following observation, providing an exact expression for the error probability for a given length-$n$ codebook $\Cc_n$ in terms of $\Phi_n$. Let $\Xm$ be a random variable equiprobable over $\Cc_n$, and let $\Ym$ be the output of  $W^n$ with $\Xm$ as the input. Then, we have
\begin{equation}\label{eq: VerduHan UB}
    \pe(\Cc_n) = \PP\left[-\frac{1}{n}\log \Phi_n(\Xm,\Ym) < \frac{1}{n}\log |\Cc_n| \right],
\end{equation}
where $\Phi_n$ is computed with respect to $Q_{\Xm}$, the equiprobable distribution on $\Cc_n$.  Theorem \ref{th: General formula expression} can then be deduced using the definition of $\liminf$ in probability.  The details can be found in \cite{Som14}.

The exact expression of Theorem \ref{th: General formula expression} can be upper bounded by the same expression with the limit inferior in probability replaced by the limit inferior of the series of expectations (see \cite[Cor.~2]{Som14}), yielding
\begin{equation}\label{eq: the General formula 2}
    \CM \leq \sup_{{\boldsymbol{Q}}} ~ \liminf_{n\rightarrow \infty}-\frac{1}{n}\EE\big[ \log \Phi_n(\Xm,\Ym) \big].
\end{equation}
Moreover, this bound was shown to be tight for finite-input channels whenever the strong converse property holds for mismatched decoding (see \cite[Lemma 7]{Som14}), though understanding when the latter property holds remains an open problem.

The bound \eqref{eq: the General formula 2} was also shown to be tight in the case of the erasures-only metric $q(x,y)=\openone\{W(y|x)>0\}$, in which case we have
\begin{equation}\label{eq: Phin dfn for EO}
    \Phi_n(\Xm,\Ym)
    = \PP\big[ W^n(\Ym|\Xvbar) > 0 \,|\,\Ym \big]. 
\end{equation}
Hence, in this case, the multi-letter capacity formula becomes \cite[Proposition 1]{Som14}
\begin{equation}\label{eq: the General formula for eo with expectation}
    \CM =\sup_{{\boldsymbol{Q}}} ~\liminf_{n\rightarrow \infty} ~ -\frac{1}{n}\EE\Big[\log \PP\big[ W^n(\Ym|\Xvbar) > 0 \,|\,\Ym \big] \Big].
\end{equation}


\section{A Single-Letter Upper Bound}\label{sc:single letter upper bound}

The task of attaining single-letter upper bounds on the mismatch capacity for general mismatched DMCs has proved to be notoriously difficult since the introduction of the mismatched decoding problem.  In this subsection, we present a recent result providing a single-letter upper bound on the mismatch capacity that can be strictly smaller than the matched capacity \cite{KangarshahiGuilleniFabregasFull2020} (see also \cite{KangarshahiGuilleniFabregasIZS2020,KangarshahiGuilleniFabregasISIT2019}).

The high-level idea of the upper bound is to relate the error probability of the mismatched DMC $(W,q)$ to the error probability associated with an auxiliary channel $P_{\Ybar|X}$ under optimal maximum-likelihood (ML) decoding.  Then, a standard mutual information based upper bound on the capacity of $P_{\Ybar|X}$ translates to a non-trivial upper bound on the mismatch capacity of $(W,q)$.

In order to state the upper bound, we introduce some definitions.  For any two outputs $y,\ybar\in{\cal Y}$, we define the following set of inputs maximizing the difference of log-metric values:\footnote{Some additional care is needed when some $q(x,y)$ values are zero, to avoid expressions such as $(-\infty) - (-\infty)$; we omit these technical details here and refer the interested reader to \cite{KangarshahiGuilleniFabregasFull2020}.}
\begin{equation}
    \Xc_q^*(y,\ybar)=\Big\{x\in{\cal X}:\; x \in \argmax_{x'\in\Xc}\big( \log q(x',\ybar)- \log q(x',y)\big)\Big\}. \label{eq:star}
\end{equation} 
In addition, we define
\begin{equation}
    \Mc_{\max}(q)  \triangleq \Big\{P_{Y\Ybar|X}\in{\cal P}({\cal Y}^2|{\cal X}):\; P_{Y\Ybar|X}(y,\ybar|x)=0,\; \forall x\notin \Xc_q^*(y,\ybar)\Big\}, \label{eq:Mmax}
\end{equation}
which is the set of conditional joint distributions such that $x$ maximizes the log-metric difference in \eqref{eq:star} with probability one.  We henceforth refer to any conditional joint distribution in $\Mc_{\max}(q)$ as being {\em maximal}.

\begin{theorem}\label{th:SL upper bound}
    {\em (Single letter upper bound on the mismatch capacity)}
    For any mismatched DMC $(W,q)$, we have
    \begin{equation}
        \CM(W,q) \leq \Rbar(W,q),
    \end{equation}
    where
    \begin{equation}
        \Rbar(W,q)\triangleq \max_{Q_X \in \Pc(\Xc)} 
        \min_{\substack{P_{Y\Ybar|X}\in{\cal M}_{\max}(q) \,:\, \\ P_{Y|X}=W}}I_{P}(X;\Ybar), \label{eq:SingleLetterUpperBound}
    \end{equation}
    where $I_{P}(X;\Ybar)$ is computed with respect to $Q_X \times P_{\Ybar|X}$.
\end{theorem}

Observe that $\Rbar(W,q)$ has a similar form to the input-optimized LM rate $\CLM$ in its primal form ({\em cf.}, \eqref{eq:CNV_PrimalLM}--\eqref{eq:CNV_CLM}), with a maximization over the input distribution and a minimization over the (joint) auxiliary channel.  The defining feature here is the constraint that $P_{Y\Ybar|X}$ is maximal.


Before outlining the proof of Theorem \ref{th:SL upper bound}, we present some examples and properties of $\Rbar(W,q)$.

\paragraph{Examples.}
We return to the binary-input ternary-output example of Section \ref{sec:binary_input}, in which $\Xc = \{0,1\}$, $\Yc = \{0,1,2\}$, and the channel and metric are described by the entries of the $|\Xc|\times|\Yc|$ matrices
\begin{align}
    \Wv & =\left[\begin{array}{ccc}
    0.97 & 0.03 & 0\\
    0.1 & 0.1 & 0.8
    \end{array}\right],~~~~\qv=\left[\begin{array}{ccc}
    1 & 1 & 1\\
    1 & 0.5 & 1.36
    \end{array}\right].\label{eq:CNV_Channel_rep}
\end{align}
We saw in Section \ref{sec:binary_input} that the $2$-letter superposition coding rate $ \CSC^{(2)} = 0.19908$ bits/use is strictly higher than the LM rate $0.19746 \le \CLM \le 0.19751$, implying that $\CM>\CLM$ \cite{Sca15a}.  

Since \eqref{eq:SingleLetterUpperBound} minimizes over $P_{Y\Ybar|X}$, we can substitute any specific choice therein and still have a valid upper bound on $\CM$.  We consider the choice in Table \ref{tbl:maximal_example2} (with all triplets not shown therein having probability zero), which is easily verified to be maximal and to satisfy $P_{{Y}|X}=W$.  Marginalizing out the $Y$ variable, we find that 
	 \begin{align}
	 \big[P_{\Ybar|X}(\ybar|x)\big]_{x \in \Xc, \ybar \in \Yc} = \begin{bmatrix}
	0.5 &0.5  &0\\
	0.1  &0.1   &0.8
	\end{bmatrix}.
\end{align}
A numerical calculation with an efficient iterative algorithm (discussed below) establishes that the joint conditional distribution in Table \ref{tbl:maximal_example2} attains the minimum in \eqref{eq:SingleLetterUpperBound}, and gives the following (to five decimal places):
\begin{equation}
    \Rbar(W,q) = 0.61823 ~~~ \text{bits/use}.
\end{equation}
On the other hand, a basic numerical calculation establishes that $W$ has matched capacity 
\begin{equation}
    C(W) = 0.71329 ~~~ \text{bits/use}.
\end{equation}
Hence, the upper bound $\Rbar(W,q)$ is strictly smaller than the matched capacity, albeit still far from the best known achievable rate (i.e., lower bound on the mismatch capacity) in this example.

\begin{table}[h]
    \centering
    \caption{Non-zero entries of $P_{Y\Ybar|X}(y,\ybar|x)$ for the example in \eqref{eq:CNV_Channel_rep}}
    \label{tbl:maximal_example2}
    \begin{tabular}{lclclcl}
        \toprule
        $(x,y,\ybar)$ & $P_{Y\Ybar|X}(y,\ybar|x)$ & $(x,y,\ybar)$ & $P_{Y\Ybar|X}(y,\ybar|x)$ \\
        \midrule
        $(0,0,0)$ & $0.5$ & $(1,0,0)$ & $0.1$ \\
        $(0,0,1)$ & $0.47$ & $(1,1,1)$ & $0.1$ \\
        $(0,1,1)$ & $0.03$ & $(1,2,2)$ & $0.8$ \\
        \bottomrule
    \end{tabular}
\end{table}

In addition to the above example, the following observations were made in \cite{KangarshahiGuilleniFabregasFull2020}:
\begin{itemize}
    \item If we change $q(1,1)$ from $0.5$ to $1$ in \eqref{eq:CNV_Channel_rep}, then the choice of $P_{Y\Ybar|X}$ in Table \ref{tbl:maximal_example2} remains the optimal solution of the iterative algorithm, meaning that $\Rbar = 0.61823$ bits/use.  In this case, however, a numerical evaluation of the LM rate also gives $\CLM = 0.61823$ bits/use, and hence, the upper and lower bounds coincide (at least numerically). 
    \item In the binary-input binary-output case $|\Xc|=|\Yc|=2$, the upper bound $\Rbar(W,q)$ always equals the mismatch capacity, as characterized in Lemma \ref{lem:binary}.
    \item A sufficient condition can be given for a binary-input mismatched DMC to have $\CM < C$, where $C$ is the matched capacity. To state this condition, we start with the following definition: Two sequences $\{\alpha_k\}_{k=1}^{K}$ and $\{\beta_k\}_{k=1}^{K}$ are said to have the same order if, for all $k_1,k_2 \in \{1,\dotsc,K\}$, it holds that
    \begin{align}
        \alpha_{k_1} \geq \alpha_{k_2} \Rightarrow \beta_{k_1} \geq \beta_{k_2}.
    \end{align}
   Then, if $\min_{x,y} W(y|x) > 0$ and the sequences $\big\{\log W(y_k|1) - \log W(y_k|2)\big\}_{k=1}^{|\Yc|}$ and $\big\{\log q(1,y_k) - \log q(2,y_k)\big\}_{k=1}^{|\Yc|}$ do not have the same order, then $\CM < C$.
    \item As for a non-binary example, consider the following ternary-input quaternary-output channel and the corresponding erasures-only metric (Section \ref{sec:app_zuec}):
    \begin{align}
    		\Wv = \begin{bmatrix}
    		0.25 \ &0 \ &0.05 \ &0.7 \\
    		0.3  \ &0.55 \ &0  \  &0.15 \\
    		0.05  \  &0.5  \  &0.45 \ &0
    		\end{bmatrix},~~
    		\qv = \begin{bmatrix}
    		1 \ &0 \ &1 \ &1 \\
    		1  \ &1 \ &0  \  &1 \\
    		1  \  &1  \  &1 \ &0
    		\end{bmatrix}.
		\end{align}
    The matched capacity of $W$ is $C = 0.78537$ bits/use, and the above-mentioned iterative algorithm gives $\Rbar = 0.62318$ bits/use. The LM rate computed by an exhaustive search over the input distributions is approximately $\CLM=0.42922$ bits/use.
\end{itemize}


\paragraph{Properties of $\Rbar(W,q)$.}  
Here we discuss some convexity properties of $\Rbar(W,q)$ and their implications for its computation.  We first present a lemma stating that the minimization in \eqref{eq:SingleLetterUpperBound} is performed over a convex set.  The proof is straightforward, and can be found in \cite{KangarshahiGuilleniFabregasFull2020}.

\begin{lemma} \label{lem:Rbar_convexity}
    {\em (Convexity of the constraint set)} For any mismatched DMC $(W,q)$, the set of conditional joint distributions $P_{Y\Ybar|X} \in \Mc_{\text{max}}(q)$ such that $P_{Y|X} = W$ is convex.
\end{lemma}
Since the mutual information is concave in $Q_X$ and convex in $P_{\Ybar|X}
$, Lemma \ref{lem:Rbar_convexity} implies that the optimization problem in \eqref{eq:SingleLetterUpperBound} is a convex-concave saddlepoint problem, and can be simplified as follows:
\begin{align} 
    \Rbar(W,q) &= \max_{Q_X}\min_{\substack{P_{Y\Ybar|X} \in \Mc_{\text{max}}(q)\,:\,\\ P_{Y|X} = W}} I_{P}(X;\Ybar)\\
    &= \min_{\substack{P_{Y\Ybar|X} \in \Mc_{\text{max}}(q)\,:\,\\ P_{Y|X} = W}} \max_{Q_X}I_{P}(X;\Ybar)\\
    &= \min_{\substack{P_{Y\Ybar|X} \in \Mc_{\text{max}}(q)\,:\,\\ P_{Y|X} = W}} C(P_{\Ybar|X}).
\end{align}
Two further implications of the convex-concave structure of \eqref{eq:SingleLetterUpperBound} are as follows (see \cite{KangarshahiGuilleniFabregasFull2020} for details): (i) One can establish an efficient iterative algorithm to compute $\Rbar(W,q)$; (ii) 
Multi-letter extensions of the bound do not yield any improvement, i.e., we have for any positive integer $k$ that
\beq
\frac{1}{k} \Rbar^{(k)}(W,q) = \Rbar(W,q),
\eeq
where $\Rbar^{(k)}(W,q)$ denotes the multi-letter extension of $\Rbar(W,q)$. This is shown by first deriving the KKT conditions for both the single-letter and multi-letter problems, and then showing that the $k$-fold product of any optimal distribution for the single-letter problem also satisfies the KKT conditions for the multi-letter problem.

\paragraph{Outline of the derivation of $\Rbar(W,q)$.} 
We provide a brief overview of the proof of Theorem \ref{th:SL upper bound}, and refer the reader to \cite{KangarshahiGuilleniFabregasFull2020} for the details.
As we mentioned previously, the key idea is to lower bound the error probability of a codebook $\Cc_n$ applied to the mismatched DMC $(W,q)$ by that of the same codebook over a different channel $P_{\Ybar|X}$ with optimal (maximum-likelihood) decoding. 

We first state some standard assumptions on the codebook $\Cc_n = \{ \xv^{(1)}, \dotsc, \xv^{(M)} \}$ that incur no loss of generality.  Since any code contains a constant-composition sub-code with the same asymptotic rate (see Section \ref{sec:su_properties}), we may assume that $\Cc_n$ is a constant-composition codebook, say with type $Q_{X,n}$.  In addition, it suffices to consider the maximal error probability $\pemax$ (with respect to $m=1,\dotsc,M$) instead of the average error probability $\pe$, since these two notions lead to the same mismatch capacity (see Remark \ref{rem:max_avg}).


We introduce the following counterpart of \eqref{eq:Mmax}:
\begin{align}
    \Mc_{\max,n}(q) \triangleq&\Big\{P_{Y\Ybar|X,n}\in\Pc_n(\Yc^2|\Xc):\;\nonumber \\ &P_{Y\Ybar|X,n}(y,y'|x)=0,\; \forall x\notin \Xc_q^*(y,y')\Big\},
\end{align}
which has the same constraints as $\Mc_{\max}(q)$, but has the further constraint of being a conditional joint type.  Then, for a given codeword $\xv \in \Tc^n(Q_{X,n})$ and a maximal conditional joint type $P_{Y\Ybar|X,n} \in \Mc_{\max,n}(q)$,  consider the bipartite graph  
\beq
    \Gc_\xv(P_{Y\Ybar|X,n}) = \big\{\Tc_{\xv}^n(P_{Y|X,n}),\Tc_{\xv}^n(P_{\Ybar|X,n}), \Ec_{\xv}\big\}
    \label{eq:graph}
\eeq 
with left-node set $\Tc_{\xv}^n(P_{Y|X,n})$, right-node set $\Tc_{\xv}^n(P_{\Ybar|X,n})$, and edge set
\beq
    \Ec_{\xv} = \big\{(\yv,\yvbar) \,|\,\widehat{P}_{\yv\yvbar|\xv} = P_{Y\Ybar|X,n}\big\},
    \label{eq:edge}
\eeq	
with $\widehat{P}_{\yv\yvbar|\xv}$ denoting the joint conditional type induced by $(\xv,\yv,\yvbar)$.

Thus, the graph connects $\yv$, the output of the channel $P_{Y|X,n}$, to $\yvbar$, the output of the channel  $P_{\Ybar|X,n}$, whenever their joint conditional type $\widehat{P}_{\yv\yvbar|\xv}$ is equal to $P_{Y\Ybar|X,n}$. It can be shown that the graph $\Gc_\xv(P_{Y\Ybar|X,n})$ is regular, i.e., all left-nodes have the same degree, and all right-nodes have the same degree. 

With the preceding definitions in place, the key step in the analysis is as follows: Show that for two constant-composition codewords $\xv, \xvbar\in\Cc_n$ and an output sequence $\yv$, if $P_{Y\Ybar|X,n}$ is a maximal distribution whose marginal distribution satisfies $P_{{Y}|X,n} = \Phat_{\yv|\xv}$, then there exists some $\yvbar$ connected to $\yv$ in $\Gc_\xv(P_{Y\Ybar|X,n})$ for which we have
\begin{align}
\Phat_{\yvbar|\xv} = \Phat_{\yvbar|\xvbar}. \label{eq:type_error}
\end{align}
Here, as usual, $\xv$ represents the transmitted codeword, and $\xvbar$ represents some non-transmitted codeword.    Then, the following crucial lemma shows that \eqref{eq:type_error} implies $q^n(\xv,\yv) \leq q^n(\xvbar,\yv)$.

\begin{lemma}\label{lemmaximal}
    {\em (Connection between conditional type error and decoding error)}
	Consider a type $Q_{X,n} \in \Pc_n(\Xc)$ and sequences $\xv,\xvbar \in \Tc^n(Q_{X,n})$, and let $P_{Y\Ybar|X,n}\in \Mc_{\max,n}(q)$ be a maximal  joint conditional type satisfying $P_{{Y}|X,n} = \Phat_{\yv|\xv}$. If there exists some $\yvbar \in \Tc_{\xv}^n(P_{\Ybar|X,n}) \cap \Tc_{\xvbar}^n(P_{\Ybar|X,n})$ connected to $\yv \in \Tc_{\xv}^n(P_{Y|X,n})$ in $\Gc_\xv(P_{Y\Ybar|X,n})$, then
\begin{align} 
     q^n(\xv,\yv) \leq q^n(\xvbar,\yv).\label{geewt4}
\end{align}
\end{lemma}
\begin{proof}
     First consider the following standard marginalization property for any triplet $(\xv', \yv, \yvbar)$:
	 \begin{align}
	 \Phat_{\yv\yvbar}(y,\ybar) = \sum_{x}\Phat_{\xv'\yv\yvbar}(x,y,\ybar).  \label{equivalent}
	 \end{align}
	 Using this property once with $\xv' = \xv$ and again with $\xv' = \xvbar$, it follows that
	 \begin{align}
	 \sum_{x}\Phat_{\xv\yv\yvbar}(x,y,\ybar) = \sum_{x}\Phat_{\xvbar\yv\yvbar}(x,y,\ybar).\label{rtat4}
	 \end{align}
     We now consider the following difference of log-metric values:
	 \begin{align}
	 &\log q^n(\xvbar,\yvbar) - \log q^n(\xvbar,\yv) \nonumber \\
	 &~= n\sum_{x,y,\ybar}\Phat_{\xvbar\yv\yvbar}(x,y,\ybar)\big(\log q(x,\ybar) - \log q(x,y)\big) \label{werqt}\\ \label{ld;ejjreir}
	  &~\leq n\sum_{y,\ybar}\Big(\sum_{x}\Phat_{\xvbar\yv\yvbar}(x,y,\ybar)\Big)\max_{x\in\Xc}\big(\log q(x,\ybar) - \log q(x,y)\big) \\ \label{sa;efq}
	  &~= n\sum_{y,\ybar}\Big(\sum_{x}\Phat_{\xv\yv\yvbar}(x,y,\ybar)\Big)\max_{x \in \Xc}\big(\log q(x,\ybar) - \log q(x,y)\big)  \\ \label{fwefne}
	  &~= n\sum_{y,\ybar}\sum_{x}\Phat_{\xv\yv\yvbar}(x,y,\ybar)\big(\log q(x,\ybar) - \log q(x,y)\big) \\ \label{fr4werq}
	  &~= \log q^n(\xv,\yvbar) - \log q^n(\xv,\yv)
	 \end{align}
    where \eqref{werqt} uses $\log q^n(\xv,\yv)=n\sum_{x,y}\widehat{P}_{\xv\yv}(x,y)\log q(x,y)$, \eqref{ld;ejjreir} follows by upper bounding $\log q(x,\ybar) - \log q(x,y)$ by its maximum over the inputs, \eqref{sa;efq} follows from \eqref{rtat4}, \eqref{fwefne} follows from the maximality of $P_{Y\Ybar|X,n}$ (see \eqref{eq:star}) and the graph construction $\Gc_\xv(P_{Y\Ybar|X,n})$ (see \eqref{eq:graph}--\eqref{eq:edge}), and \eqref{fr4werq} follows similarly to \eqref{werqt}.

    Now,  the assumption $\yvbar \in \Tc_{\xv}^n(P_{\Ybar|X,n}) \cap \Tc_{\xvbar}^n(P_{\Ybar|X,n})$ implies that $\log q^n(\xv,\yvbar) = \log q^n(\xvbar,\yvbar)$, and combining this with \eqref{fr4werq}, we get the desired result $\log q^n(\xv,\yv) \leq \log q^n(\xvbar,\yv)$.
\end{proof}

The event $\Phat_{\yvbar|\xv} = \Phat_{\yvbar|\xvbar}$ is referred to as a {\em conditional type error}, corresponding to the correct codeword and some incorrect codeword inducing the same conditional type with $\yvbar$. 
Lemma \ref{lemmaximal} implies that whenever there is a conditional type error in the channel $P_{\Ybar|X,n}$, there also is a mismatched decoding error in the channel $P_{{Y}|X,n}$, which in turn should be close to the original channel $W$ by a typicality argument (recall that $P_{{Y}|X,n} = \Phat_{\yv|\xv}$).

As hinted above, it can be shown that for rates $R$ above $I_{P}(X; \Ybar)$, there exists a message that yields a conditional type error in the channel $P_{\Ybar|X,n}$, and hence a mismatched decoding error in the channel $P_{{Y}|X,n}$.  Since this holds for every maximal $P_{Y\Ybar|X,n} \in \Mc_{\max,n}(q)$ for which $P_{{Y}|X,n}$ is close to $W$, we can choose such a $P_{Y\Ybar|X,n}$ with the minimal value of $I_{P}(X;\Ybar)$, in accordance with the minimization present in \eqref{eq:SingleLetterUpperBound}.  Then, the remainder of the proof revolves around the technicalities related to passing from types to general distributions.




\chapter{Overview of Other Topics} \label{ch:other}

In this monograph, we have focused our attention on what we view as the most fundamental topics in mismatched decoding.  Naturally, the material that we have presented is far from exhaustive.  In this section, we provide a brief overview of several other topics in the literature that we did not cover, pointing the reader to the relevant references.

\subsection*{Alternative Achievable Rates and Capacity Notions}

\paragraph{Generalized cutoff rate.} For the standard single-user mismatched decoding problem ({\em cf.}, Section \ref{ch:single_user}), several early works focused on an achievable rate known as the {generalized cutoff rate}  \cite{Div78,Kaz81,Omu82,Sal95}, defined as follows in analogy with the GMI in \eqref{eq:INTR_RateGMI}:
\begin{equation}
    \GCR(Q_X) = \sup_{\sgz} \,-\log\sum_{x,y} Q_X(x)W(y|x)\frac{\sum_{\xbar} Q_X(\xbar) q(\xbar,y)^{s}}{q(x,y)^{s}}. \label{eq:GCR}
\end{equation}
This can be thought of as the rate obtained by naively upper bounding the random-coding error probability under the i.i.d.~ensemble by $(M-1)\PP[ q^n(\Xvbar,\Yv) \ge q^n(\Xv,\Yv) ]$, where $(\Xv,\Yv,\Xvbar) \sim Q_X^n(\xv)W^n(\yv|\xv)Q_X^n(\xvbar)$.  By applying Jensen's inequality in \eqref{eq:INTR_RateGMI} we find that $\GCR(Q_X) \le \GMI(Q_X)$, and it was shown in \cite[Sec.~III]{Lap96a} that the gap between the two can be arbitrarily large.

 \paragraph{Alternative notions of mismatch capacity.} In \cite{Mer95}, it was shown that seemingly innocuous adjustments to the definition of the mismatch capacity can drastically affect what rates are achievable.  In particular, variations based on bit-error probability (instead of block error probability), randomized decoding, and encoders with added flexibility were all shown to potentially increase the mismatch capacity.  As an example, we elaborate on the last of these:  If the true channel is binary and noiseless, but the decoder adopts a decoding metric corresponding to the channel that deterministically flips its input, then the classical mismatch capacity is zero.  However, if the encoder is allowed to pre-process the selected codeword before passing it through the channel, it can simply flip each bit prior to transmission, so that a rate of $1$ bit/use is trivially attained.

\subsection*{Continuous and Fading Channels}

\paragraph{Inter-symbol interference.}  The vast majority of the mismatched decoding literature focuses on memoryless channels, as these are the most convenient to analyze mathematically.  However, an ongoing line of works has studied the important class of inter-symbol interference noise channels under mismatched decoding \cite{Abo00,Hul17,Rus12a,Sad09} (see also \cite{Rus12} for more general linear channel models).  In particular,  ensemble-tight achievable rates were given in \cite{Hul17} for an autoregressive random-coding ensemble and a counterpart with auxiliary costs, generalizing the rates obtained in \cite{Abo00} for the i.i.d.~and shell ensembles.

 \paragraph{Block fading channels.} The block fading channel model is a widely-adopted model in theoretical studies of wireless communication, in which a multiplicative noise variable stays constant within fixed-size blocks, but then independently changes between blocks.  In \cite{Asy12,asyhari2014mimo}, various performance measures for block fading channels were studied under mismatched nearest-neighbor decoding resulting from imperfect channel state information.  In particular, a notion of generalized outage probability was studied, characterizing the probability that the GMI is below a certain target rate.

 \paragraph{Output processing.} In \cite{Zha12}, a variation of the mismatched decoding problem with additive noise channels was considered, in which a quantizer is applied to the output $Y$ before being passed to the decoder.  Adopting the GMI as the performance measure, the authors studied the problem of optimizing the quantizer subject to suitable constraints in order to maximize the GMI.  In addition, gains in the GMI resulting from faster-than-Nyquist sampling of the output were characterized.

 \paragraph{Rescaled codewords at the transmitter.} In \cite{Feh16}, it was shown that a simple re-scaling operation at the transmitter can increase the error exponent of i.i.d.~random codes for a Gaussian channel with a given discrete constellation at the input.  Specifically, the transmitter simply scales the codeword so that the power constraint is met with equality, whereas the decoder still uses the original codebook and is therefore mismatched.  This technique provides a simple procedure for reaping some (but not all) of the gains that constant-composition codes provide over i.i.d.~codes.

\subsection*{Refined Asymptotics}

\paragraph{Fixed-error asymptotics and moderate deviations.} In Section \ref{ch:exponents}, we focused on error exponents, which characterize the exponential speed of decay of the error probability at fixed rates.  In \cite{Sca14c}, building on recent studies in the matched setting \cite{Pol10a,Tan14}, two other notions of refined asymptotics were studied: (i) fixed-error asymptotics, in which one fixes a target error probability and studies the speed of convergence to the ``first-order'' achievable rate; (ii) moderate deviations asymptotics, in which the rate approaches its asymptotic limit and the error probability approaches zero simultaneously.  In addition, a saddlepoint approximation was introduced (see also \cite{Mar11a,Mar11}) that unifies the asymptotics for both fixed and varying rates under i.i.d.~random coding.

 \paragraph{Refined asymptotics for Gaussian sources and channels.} In \cite{Sca17a}, the fixed-error asymptotics were studied for additive non-Gaussian channels (see Section \ref{sec:ex_noise}), providing a refined characterization of Lapidoth's study of the achievable rates \cite{Lap96a}.  The rate-distortion counterpart (see Section \ref{sec:gaussian_rd}) was studied in \cite{Zho17}, and \cite{Zho18} unified these two lines of works by considering {joint-source channel coding} with Gaussian coding techniques applied to general sources and channels.

\subsection*{Modified Channel Coding Settings}

\paragraph{Capacity per unit cost.} In the presence of a system cost constraint ({\em cf.}, Section \ref{sec:cont_setup}), as an alternative to considering the capacity with a given cost constraint, one can seek to obtain the highest possible {capacity per unit cost} for a fixed cost function without a hard constraint.  This problem was studied in the mismatched decoding setting in \cite{Gan00}.  It was shown that the mismatch capacity per unit cost is achieved in the low-cost limit when a zero-cost input exists, but somewhat surprisingly, the counterpart to this result no longer holds in general when the mismatch capacity is replaced by the LM rate.

 \paragraph{Mismatched decoding with feedback.} In \cite{Lap16}, the problem of channel coding with feedback and mismatched decoding was introduced.  It was shown that feedback can strictly increase the mismatch capacity, and sufficient conditions were given under which the mismatch capacity equals the matched capacity.   A related study on the zero-undetected error capacity with feedback was given in \cite{Bun12}, where it was shown that the maximal achievable rate is always equal to either zero or the matched capacity.

 \paragraph{Channels with a state.} In \cite{Fel16}, various random-coding based achievable rates were given for channels with a state, taking the form $W(y|x,s)$ when the state takes the value $s$.  In analogy with the matched setting \cite[Ch.~7]{Elg11}, the achievable rates vary according to whether the state is known at the encoder and/or decoder, and whether it is known causally or non-causally.  This setup is also directly related to fading channels, but the focus in \cite{Fel16} is on discrete memoryless channels, leading to different analysis techniques.

 \paragraph{Identification via channels.} The problem of identification via channels seeks to allow a decoder to reliably determine whether or not a {specific} pre-specified message was transmitted, which is a weaker requirement than determining an arbitrary message.  In \cite{Som17}, this problem was studied in the presence of mismatched decoding.  Generalizing a classical result \cite{Ahl99}, it was shown that the identification problem can be performed reliably whenever the number $M$ of messages satisfies $\limsup_{n \to \infty} \frac{1}{n} \log \log M < \LM(Q_X)$, i.e., a double-exponentially large number of messages.

\subsection*{Other Decoders}

\paragraph{Stochastic likelihood decoding.} In the matched setting, an alternative to maximum-likelihood decoding is {stochastic likelihood decoding} \cite{Yas13}, in which each codeword is selected with probability proportional to its likelihood.  
In \cite{Sca15}, ensemble-tight rates and error exponents were studied for a mismatched variant of this decoder in which the likelihood $W^n(\yv|\xv)$ is replaced by a single-letter decoding metric $q^n(\xv,\yv)$.  These results were further generalized to certain multi-letter metrics in \cite{Mer17}, where it was additionally shown that this decoder provides a novel approach to deriving expurgated error exponents.

 \paragraph{Threshold decoding.} Another well-known alternative to the maximum-likelihood decoder is the threshold decoder: Search for a unique codeword whose likelihood exceeds some threshold, or declare an error if no such codeword exists.  This decoding rule was studied in \cite{Csi95,Mar11a,Som14} in distinct directions: \cite{Csi95} posed the question of whether mismatched threshold-decoding and maximum-metric decoding yield the same mismatch capacity (which remains unsolved in general), \cite{Mar11a} developed a saddlepoint approximation of the random-coding error probability, and \cite{Som14} established a multi-letter expression for the mismatch capacity, as we outlined in Section \ref{sc: Information Spectrum Upper Bounds}.

 \paragraph{Successive decoding.} In Section \ref{ch:mac}, we considered maximum-metric decoding for the multiple-access channel (MAC).  In \cite{Sca18}, a {mismatched multi-letter successive decoding} rule was instead adopted, with the two decoding steps taking the form
\begin{align}
    \hat{m}_{1} & =\arg\max_{i}\sum_{j}q^{n}(\xv_{1}^{(i)},\xv_{2}^{(j)},\yv), \\
    \hat{m}_{2} & =\arg\max_{j}q^{n}(\xv_{1}^{(\hat{m}_{1} )},\xv_{2}^{(j)},\yv).
\end{align}
The first step is motivated by the fact that substituting $q = W$ in the first step yields the optimal rule for estimating $m_1$ alone.  It was demonstrated that for a given random-coding ensemble, there exist rate pairs that can be attained via successive decoding but not maximum-metric decoding, {and} vice versa.

 \paragraph{List decoding.} The list decoding criterion allows the decoder to output a list of messages of a certain size, with the decoding considered successful if the true message is included in the list.  A mismatched variant of list decoding was studied in \cite{Som15a}, with an emphasis on multi-letter formulas analogous to those given in Section \ref{ch:converse}.  Among other things, it was shown that when the list size grows exponentially as $e^{n\Theta}$, the resulting increase in the mismatch capacity is exactly $\Theta$.  In particular, if the list size grows sub-exponentially fast, then the mismatch capacity is the same as that of regular decoding.

\paragraph{Universal decoding.} One of the earliest works in the mismatched decoding literature \cite{Sti66} sought to adopt a single decoding metric that simultaneously works well for all channels in some class.  In a more recent work \cite{Mer13a}, a complementary problem is studied, in which some {\em class of decoding metrics} is fixed, and the goal is to construct a channel-independent decoder that is guaranteed to work nearly as well as the best decoding metric in the class.  More specifically, for a given random coding ensemble, it is shown that the proposed universal decoder has an error probability within a sub-exponential factor of the best fixed decoder in the class.

\subsection*{Mismatch in Practical Codes}

\paragraph{Huffman coding.} Perhaps the most well-known example of mismatch in information theory is that for {Huffman coding} in lossless symbol-wise source coding \cite[Sec.~5.6]{Cov06}: If the true source distribution is $P_X$ but Huffman coding is performed according to some $Q_X$, then the resulting average code length $L_{\rm avg}$ satisfies $H(P_X) + D(P_X\|Q_X) \le L_{\rm avg} < H(P_X) + D(P_X\|Q_X) + 1$.  When $P_X = Q_X$, this reduces to the well-known bound $H(P_X) \le L_{\rm avg} < H(P_X) + 1$.

 \paragraph{Viterbi decoding.} The Viterbi algorithm \cite{For73} is a classical technique for the maximum-likelihood decoding of convolution codes, and related decoding tasks involving Markov models.  In \cite{Lap98a}, a mismatched variant of the Viterbi algorithm was analyzed for convolutional coding over a symmetric channel, using a symmetric decoding metric that may differ from the true likelihood function.  The main result is a lower bound on the bit-error probability; in broad scenarios, this bound is asymptotically tight in the limit of a high signal-to-noise ratio.

 \paragraph{Polar codes.} In a recent line of works \cite{Als13a,Als12,Als14}, mismatched decoding is analyzed in the context of polar codes over binary-input discrete memoryless channels.  The achievable rate derived resembles the dual form of the GMI (see \eqref{eq:INTR_RateGMI}), but with two notable modifications: (i) The input distribution is always equiprobable, i.e., $Q_X(0) = Q_X(1) = \frac{1}{2}$; (ii) The optimization over $\sgz$ is replaced by the specific value $s = 1$ coinciding with Fischer's rate \cite{Fis71}.  This expression can match the GMI in certain special cases, but in general may be significantly lower, or even negative (thus providing no achievable rate).

\chapter{Conclusion} \label{ch:conclusion}

In this monograph, we have surveyed a wide range of classical results and recent developments on the problem of point-to-point communication with mismatched decoding, as well as the analogous multiple-access communication problem, and that of mismatched encoding in rate-distortion theory.  In the following, we provide a brief summary of each section, and list several open problems of interest.

In Section \ref{ch:single_user}, we studied the rates achievable by random coding with independent codewords for mismatched DMCs, with an emphasis on the i.i.d.~ensemble and the constant-composition ensemble.  We introduced a number of properties of the resulting achievable rates, referred to as the GMI and the LM rate.  In Section \ref{ch:single_user_cont}, we presented the generalization of these rates to continuous-alphabet channels, and introduced the cost-constrained random-coding ensemble as a means for proving the achievability of the LM rate.  In addition, we evaluated these rates on a variety of additive noise channels, including the prominent examples of (i) Gaussian codes for additive non-Gaussian noise channels, and (ii) fading channels with mismatched channel state information at the decoder.

In the case of discrete alphabets, ensemble tightness results are well-established for both the GMI and the LM rate for their respective ensembles, whereas for continuous-alphabet channels, only the ensemble tightness of the GMI is known (see Section \ref{sec:cont_tightness}). This raises the following open problem, for which we expect the answer to be affirmative.

\begin{openproblem}
    Is the LM rate tight with respect to the ensemble average for the cost-constrained ensemble for continuous-alphabet memoryless channels with mismatched decoding? 
\end{openproblem}

The following question is relatively open-ended, and serves to highlight the fact that the GMI has been the focus of many specific continuous channels, with the potential gains of the LM rate remaining elusive other than the simple improvement given in Theorem \ref{thm:gaussian}.

\begin{openproblem}
    To what extent can the LM rate improve over the GMI for additive noise channels and/or fading channels with mismatched decoding? 
\end{openproblem}

In Section \ref{ch:rate_dist}, we turned to the problem of rate distortion theory with mismatched encoding, and studied an achievable rate based on random coding that can be viewed as a counterpart to the LM rate for channel coding.  Despite this analogy, there are several concepts and techniques that have remained unexplored in this context despite being studied extensively in channel coding.  We summarize three notable examples as follows.

\begin{openproblem} \label{op:rd_dual}
    Does there exist an equivalent ``dual'' expression for the achievable mismatched distortion-rate function derived in Section \ref{sec:rd_ach} based on constant-composition random coding?  If so, can it also be attained via a suitably-designed cost-constrained random-coding ensemble?
\end{openproblem}

\begin{openproblem} \label{op:rd_cont}
    For the rate-distortion problem with mismatched encoding, what rates can be achieved for continuous sources?
\end{openproblem}

\begin{openproblem}
    For the rate-distortion problem with mismatched encoding, what rates can be achieved via multi-user coding techniques, and to what extent do these rates improve over the rates attained using independent codewords?
\end{openproblem}

Regarding Open Problems \ref{op:rd_dual} and \ref{op:rd_cont}, we note that certain dual expressions and continuous-alphabet extensions were indeed stated in Section \ref{sec:rd_mm_random}, but only for the case of a single (known) distortion measure and a mismatched random coding distribution, as opposed to the setup of Section \ref{sec:rd_ach} in which the encoder adopts a mismatched distortion measure.

In Section \ref{ch:mac}, we introduced the mismatched multiple-access channel, and argued that it is not only of independent interest, but provides valuable insight and results concerning point-to-point communication.  In Section \ref{ch:multiuser}, we took this idea further by studying multi-user coding techniques for point-to-point communication in more detail, with an emphasis on superposition coding and its refined variant.

Despite having established several analogies between the achievable rates of these sections and the LM rate, there remain several ways in which our understanding of the former is comparatively lacking.  For instance, while Lemma \ref{lem:opt_conds} provides precise conditions under which the LM rate equals the matched capacity, no analog has been established for the rates of multi-user coding techniques.

\begin{openproblem}
    Under what conditions does the achievable rate region of Section \ref{sec:mac_ach} recover the matched capacity region?  Similarly, under what conditions do the superposition coding rates of Sections \ref{sec:sc} and \ref{sec:rsc} attain the matched capacity?
\end{openproblem}

In addition, while refined superposition coding is known to attain a rate at least as high as standard superposition coding, and the gap can be strict for a given input distribution, little is known regarding how these rates compare when the input distribution is optimized.

\begin{openproblem}
    Do there exist mismatched DMCs for which the refined superposition coding rate exceeds that of standard superposition coding even after the optimization of the auxiliary alphabet and the input distribution?
\end{openproblem}

In addition, there is no apparent reason to believe that the refined superposition coding rate equals the mismatch capacity in general, leading us to ask the following.

\begin{openproblem}
    Do there exist multi-user random coding techniques, or alternative coding techniques, that attain rates higher than that of refined superposition coding?
\end{openproblem}

In Section \ref{ch:exponents}, we surveyed the notion of error exponents for mismatched memoryless channels, and saw that their study can lead to various insights not seen in the study of achievable rates alone.  In fact, it was observed that the random-coding ensembles attaining higher rates are often coupled with smaller exponents at low rates, raising the following open problem.

\begin{openproblem}
    Do there exist random coding techniques that are suited to achieving both high rates and high error exponents?
\end{openproblem}

In addition, the following basic question proposed by Csisz\'ar and Narayan \cite{Csi95} still remains open.

\begin{openproblem} \label{op:exp_below}
    Is it always possible to attain exponentially decaying error probability for all rates below the mismatch capacity?
\end{openproblem}

%

In Section \ref{ch:converse}, we turned to upper bounds on the mismatch capacity, i.e., converse results, for which there have been comparatively few results compared to lower bounds.  We paid particular attention to the multi-letter achievable rate $\CLM^{(\infty)}$, obtained by considering the $k$-letter improvement of the LM rate in the limit as $k \to \infty$.  We extensively discussed the Csisz\'ar-Narayan conjecture, which remains unresolved in general, and is stated in the following.

\begin{openproblem}
    Does the multi-letter achievable rate $\CLM^{(\infty)}$ equal the mismatch capacity $\CM$ for arbitrary mismatched DMCs?
\end{openproblem}

As we mentioned in Section \ref{sec:binary_input}, the early converse result of Balakirsky \cite{Bal95} claiming tightness of the LM rate was refuted via a numerical counter-example in \cite{Sca15a}, raising the following question.

\begin{openproblem}
    Is there a purely analytical proof that the LM rate with an optimized input distribution can be strictly less than the mismatch capacity for binary-input mismatched DMCs?
\end{openproblem}

The following question asks whether the strong converse holds; this was originally posed by Csisz\'ar and Narayan \cite{Csi95}, and still remains unsolved.

\begin{openproblem} \label{op:strong_conv}
    Does the smallest possible (maximal or average) error probability tend to one for all rates strictly above the mismatch capacity?  If so, does it approach one exponentially fast?
\end{openproblem}

The second part of this open problem serves as a counterpart to Open Problem \ref{op:exp_below}, as it concerns rates above the mismatch capacity rather than rates below the mismatch capacity.  In fact, Open Problem \ref{op:exp_below} is also of direct interest in the development of converse results: An affirmative answer would mean that the soft converse of Section \ref{subsec:CN_soft_converse} immediately implies the corresponding regular converse.

Of course, given the apparent considerable difficulty in establishing single-letter upper bounds \cite{Bal95,Sca15a,KangarshahiGuilleniFabregasISIT2019}, the following open-ended question remains particularly important.

\begin{openproblem} 
    What additional single-letter upper bounds on the mismatch capacity can be established, and how do they compare with the best known achievable rates?
\end{openproblem}

In Section \ref{ch:other}, we highlighted several additional topics that we did not survey in detail, and many of these are still in their early stages and warrant further study.  To name just two examples, we expect many further developments on channels and/or decoding metrics with memory to be possible, and we believe that additional developments in the theory of practical codes with mismatched decoding would be of considerable interest, complementing the purely information-theoretic results surveyed in this monograph.

Finally, there remain both classical notions in information theory (e.g., variable-length codes) and modern applications of information theory (e.g., machine learning) for which the mismatched decoding viewpoint appears to have been largely or completely unexplored, thereby posing potentially exciting avenues for future research.

\appendix


\chapter{Appendix}

\section{Continuity Arguments} \label{sec:continuity_app}

In this appendix, we present some of the technical continuity arguments required for the primal-domain analysis of the GMI and LM rate in Section \ref{ch:single_user}.  We also briefly comment on the analogous arguments for the multi-user random coding techniques of Sections \ref{ch:mac} and \ref{ch:multiuser}.

\subsection{Continuity of the GMI and LM Rate (Lemma \ref{lem:continuity})}

Here we prove Lemma \ref{lem:continuity}, which states that the GMI and LM rate are continuous in the pair $(Q_X,W)$ within the space of pairs satisfying $Q_X(x)>0 \cap q(x,y) = 0 \implies W(y|x) = 0$.  This assumption ensures that the distribution $P_{XY} = Q_X \times W$ yields $\EE_{P}[\log q(X,Y)] > -\infty$.

We present the proof for the LM rate (taken from \cite{Csi95}); similar steps hold for the GMI.  Fix a pair $(Q_{X},W)$, and let $(Q_{X,n},W_n)$ be a sequence converging to $(Q_{X},W)$.  To simplify the notation, we subsequently represent these pairs in terms of their joint distributions, denoted by $P_{XY}$ and $P_{XY,n}$ respectively.  The corresponding LM rates are written as $\LM(P_{XY})$ and $\LM(P_{XY,n})$, and we let $\Ptilde^*_{XY}$ and $\Ptilde^*_{XY,n}$ denote corresponding joint distributions achieving the minimum in \eqref{eq:INTR_PrimalLM}.  Moreover, the constraint sets corresponding to \eqref{eq:INTR_PrimalLM} are written as $\Sc'(P_{XY})$ and $\Sc'(P_{XY,n})$.

We first note that since $\EE_{P}[\log q(X,Y)] > -\infty$ (as established above), every feasible distribution in \eqref{eq:INTR_PrimalLM} assigns zero mass to any $(x,y)$ pair for which $q(x,y) = 0$.  Below, any statements regarding ``continuity with respect to $P_{XY}$'' are implicitly restricted to the space of such distributions.

We establish continuity in two steps, first showing the lower bound $\liminf_{n\to\infty} \LM(P_{XY,n}) \ge \LM(P_{XY})$ and then the upper bound $\limsup_{n\to\infty} \LM(P_{XY,n}) \le \LM(P_{XY})$.  For the lower bound, note that since the probability simplex is compact, any infinite subsequence of $\Ptilde^*_{XY,n}$ must have a further subsequence converging to some $\Ptilde^*_{XY,\infty}$.  Since $P_{XY,n} \to P_{XY}$ and the constraint functions defining $\Sc'(P_{XY})$ are continuous in $P_{XY}$, the fact that $\Ptilde^*_{XY,n} \in \Sc'( P_{XY,n} )$ implies that $\Ptilde^*_{XY,\infty} \in \Sc'( P_{XY} )$.  That is, $\Ptilde^*_{XY,\infty}$ satisfies the constraints in the minimization problem of the LM rate, and hence $ \LM( P_{XY} ) \le I_{\Ptilde^*_{XY,\infty}}(X;Y)$.  Since this argument holds with $\Ptilde^*_{XY,\infty}$ corresponding to any arbitrary subsequence of $P_{XY,n}$, it follows that
\begin{equation}
    \liminf_{n\to\infty} \LM(P_{XY,n}) \ge \LM(P_{XY}).
\end{equation}
To establish the matching upper bound, we will show that there exists a sequence $\Ptilde_{XY,n} \in \Sc'(P_{XY,n})$ converging to $\Ptilde^*_{XY}$.  Once this is done, the continuity of mutual information implies
\begin{equation}
    \LM(P_{XY,n}) \le I_{\Ptilde^*_{XY,n}}(X;Y) \to I_{\Ptilde^*_{XY}}(X;Y) = \LM(P_{XY}),
\end{equation}
and taking $\limsup_{n\to\infty}$ on both sides completes the proof.  

The above-mentioned sequence $\Ptilde_{XY,n}$ is selected as follows for some vanishing sequence of values $\epsilon_n > 0$:
\begin{equation}
    \Ptilde_{XY,n}(x,y) = P_{XY,n}(x,y) + (1-\epsilon_n) \big( \Ptilde^*_{XY}(x,y) - P_{XY}(x,y) \big). \label{eq:Ptilde_n_choice}
\end{equation}
By the assumption that $P_{XY,n}$ converges to $P_{XY}$, we have that $P_{XY,n}(x,y) \ge (1-\epsilon_n) P_{XY}(x,y)$ for sufficiently large $n$ as long as $\epsilon_n$ is chosen to decay to zero sufficiently slowly.  Hence, the right-hand side of \eqref{eq:Ptilde_n_choice} is non-negative.  In addition, since $P_{XY,n}$, $\Ptilde^*_{XY}$, and $P_{XY}$ all sum to one, we can sum both sides of \eqref{eq:Ptilde_n_choice} to obtain $\sum_{x,y} \Ptilde_{XY,n}(x,y) = 1$, meaning that $\Ptilde_{XY,n}$ is a valid joint distribution.  

Recall that the two desired properties of $P_{XY,n}$ are that $\Ptilde_{XY,n}(x,y) \to \Ptilde^*_{XY}(x,y)$, and that $\Ptilde_{XY,n} \in \Sc'(P_{XY,n})$.  The former of these follows immediately by substituting the assumptions $P_{XY,n}(x,y) \to P_{XY}(x,y)$ and $\epsilon_n \to 0$ into \eqref{eq:Ptilde_n_choice}, so it only remains to check that  $\Ptilde_{XY,n} \in \Sc'(P_{XY,n})$, i.e., the constraints defining the LM rate are satisfied under the input distribution and channel corresponding to $P_{XY,n}$.

To see that the marginal constraints $\Ptilde_{X,n} = P_{X,n}$ and $\Ptilde_{Y,n} = P_{Y,n}$ ({\em cf.}, \eqref{eq:INTR_PrimalLM}) are satisfied, observe that if we sum both sides of \eqref{eq:Ptilde_n_choice} over either $x$ or $y$, the second term cancels to zero due to the fact that $\Ptilde^*_{XY} \in \Sc'(P_{XY})$ and hence $\Ptilde^*_{X} = P_X$ and $\Ptilde^*_{Y} = P_Y$.  Similarly, for the metric constraint, we multiply both sides of \eqref{eq:Ptilde_n_choice} by $\log q(x,y)$ and sum over $x,y$.  The fact that $\Ptilde^*_{XY} \in \Sc'(P_{XY})$ implies that $\EE_{\Ptilde^*}[\log q(X,Y)] \ge \EE_{P}[\log q(X,Y)]$, meaning that the second term on the right-hand side of \eqref{eq:Ptilde_n_choice} contributes a non-negative amount, and we are left with $\EE_{\Ptilde_{XY,n}}[\log q(X,Y)] \ge \EE_{P_{XY,n}}[\log q(X,Y)]$ as required.

\subsection{Technical Ensemble Tightness Step (Lemma \ref{lem:lb_technical})} \label{sec:technical_tightness}

Here we prove the technical result stated in Lemma \ref{lem:lb_technical}, in which the minimization over joint types is upper bounded in terms of a matching minimization over all joint distributions.  To our knowledge, the proof that we present here is the first one to hold for arbitrary mismatched DMCs.  See \cite{Lap96,Mer95} for simpler proofs under the assumption that there exists some $\Ptilde_{XY} \in \Pc(\Xc \times \Yc)$ such that $\Ptilde_X = P_X$, $\Ptilde_Y=P_Y$, and $\EE_{\Ptilde}[\log q(X,Y)] > \EE_{P}[\log q(X,Y)]$, i.e., strict inequality in the metric constraint.  This is true for most mismatched DMCs, but can sometimes fail, e.g., for the erasures-only metric introduced in Section \ref{sec:app_zuec}.

For convenience, we write the left-hand side of \eqref{eq:lb_technical} as
\begin{equation}
    \LMn(P_{XY}) = \min_{\substack{\Ptilde_{XY}\in \Pc_n(\Xc \times \Yc) \,:\, \Ptilde_X= P_X, \,\Ptilde_Y = P_Y, \\ \EE_{\Ptilde}[\log q(X,Y)] \ge \EE_{P}[\log q(X,Y)]}} I_{\Ptilde}(X;Y). \label{eq:ens_R0n}
\end{equation}
We need to show that, for any joint type $P_{XY} \in \Pc_n(\Xc \times \Yc)$ and arbitrarily small $\delta > 0$, it holds that
\begin{equation}
\LMn(P_{XY}) \le \LM(P_{XY}) + \delta, \label{eq:ens_LM_cmp}
\end{equation}
 where
\begin{equation}
    \LM(P_{XY}) = \min_{\substack{\Ptilde_{XY}\in \Pc(\Xc \times \Yc) \,:\, \Ptilde_X= P_X, \,\Ptilde_Y = P_Y, \\ \EE_{\Ptilde}[\log q(X,Y)] \ge \EE_{P}[\log q(X,Y)]}} I_{\Ptilde}(X;Y). \label{eq:ens_R0}
\end{equation}
Fix $\Ptilde_{XY} \in \Pc(\Xc \times \Yc)$ (not necessarily a joint type) satisfying the constraints in \eqref{eq:ens_R0}.  Since $P_{XY}$ is a joint type by assumption, the marginal constraints in \eqref{eq:ens_R0} ensure that $\Ptilde_X$ and $\Ptilde_Y$ are types.  We use $\Ptilde_{XY}$ to construct a joint type $\Ptilde_{XY,n}$ as follows:
\begin{itemize}
    \item Initialize $\Ptilde'_{Y|X,n}(y|x) = \frac{1}{n P_X(x)}\lfloor n P_X(x)\Ptilde_{Y|X}(y|x)\rfloor$ (any $x$ values with $P_X(x) = 0$ may be removed from consideration); note that because we are rounding down, this in itself may not be a valid conditional distribution.  The constraint $\Ptilde_X= P_X$ leads to $\sum_{x}\Ptilde_{X}(x)\Ptilde'_{Y|X,n}(y|x)\le P_{Y}(y)$ for all $y$, and $\sum_{y}\Ptilde'_{Y|X,n}(y|x)\le1$ for all $x$, but some of these inequalities might be strict.
    \item Since the marginals $\Ptilde_{X}$ and $\Ptilde_{Y}$ are types, we can think of them in terms of length-$n$ sequences $\xv$ and $\yv$, which we seek to suitably arrange to form a joint type $\Ptilde_{XY,n}$.  The above-defined conditional distribution $\Ptilde'_{Y|X,n}$
    specifies most of the pairs $(x_i,y_i)$ (for $i=1,\dotsc,n$) corresponding to a joint type, but we are still left with some entries of $\xv$ and some entries of $\yv$ that have not yet been assigned.  Specifically, for each $(x,y)$ pair, $nP_X(x)\Ptilde'_{Y|X,n}(y|x) = \lfloor n P_X(x)\Ptilde_{Y|X}(y|x)\rfloor$ pairs have been assigned to $(x,y)$, but the rest remain unassigned.
    \item We let $N_{x}$ (respectively, $N_{y}$) be the number of unassigned entries of $\xv$ with value $x$ (respectively, of $\yv$ with value $y$), and match these up in a way that maximizes the metric: 
    \begin{align}
    \mathrm{maximize}_{\{N_{xy}\}_{x \in \Xc,y \in \Yc}}\quad & \sum_{x,y}N_{xy}\log q(x,y)\label{eq:IP}\\
    \mathrm{subject\,to}\quad & \sum_{x}N_{xy}=N_{y},\,\,\,\forall y,\nonumber \\
    & \sum_{y}N_{xy}=N_{x},\,\,\,\forall x,\nonumber \\
    & N_{xy}\in\ZZ_{\ge 0},\,\,\,\forall x,y,\nonumber 
    \end{align}
    where $N_{xy}$ represents the number of additional occurrences of
    the pair $(x,y)$ formed in this matching-up step, and $\ZZ_{\ge 0}$ is the set of non-negative integers.  This is an {\em integer programming} (IP) problem.  
\end{itemize}
Noting that we established that $\lfloor n P_X(x)\Ptilde_{Y|X}(y|x)\rfloor$ pairs are assigned in the first step above for each $(x,y)$, we can add these counts together to get $\sum_{x,y} \lfloor n P_X(x)\Ptilde_{Y|X}(y|x)\rfloor \ge \sum_{x,y}\big( n P_X(x)\Ptilde_{Y|X}(y|x) - 1\big) = n - |\Xc|\cdot|\Yc|$, so the number of unassigned pairs remaining is at most $|\Xc|\cdot|\Yc|$.  This implies that $\Ptilde_{XY,n}$ constructed above is $\frac{2 |\Xc|\cdot|\Yc|}{n}$-close to $\Ptilde_{XY}$ in terms of the $\ell_1$-norm (i.e., sum of absolute differences), and hence
\begin{equation}
    I_{\Ptilde_{XY,n}}(X;Y) \le I_{\Ptilde_{XY}}(X;Y) + \delta \label{eq:ens_mi}
\end{equation}
for any $\delta > 0$ and sufficiently large $n$ by the continuity of mutual information.  Since the above findings hold for an arbitrary choice of $\Ptilde_{XY}$ satisfying the constraints in \eqref{eq:ens_R0}, establishing \eqref{eq:ens_LM_cmp} now only requires showing that our constructed $\Ptilde_{XY,n}$ is feasible in \eqref{eq:ens_R0n}.  The marginal constraints are already satisfied by construction, so it only remains to check the inequality constraint containing the decoding metric.  

If we relax the integer constraint $N_{xy}\in\ZZ_{+}$ to $N_{xy}\in\RR_{+}$ (i.e., the non-negative reals) in \eqref{eq:IP}, we get a {\em linear program} (LP).  With this modification, the assignments $N_{xy}=nP_X(x)\Ptilde_{Y|X}(y|x)-\lfloor nP_X(x)\Ptilde_{Y|X}(y|x)\rfloor$ corresponding to the joint distribution $\Ptilde_{XY}$ become feasible, since we trivially have $\lfloor nP_X(x)\Ptilde_{Y|X}(y|x)\rfloor + N_{xy} = \Ptilde_{XY}(x,y)$, and the assumptions $\Ptilde_X = P_X$ and $\Ptilde_Y = P_Y$ coincide with the required constraints $\sum_{y} N_{xy} = N_x$ and $\sum_{x} N_{xy} = N_y$.

Since $\Ptilde_{XY}$ satisfies the metric constraint in \eqref{eq:ens_R0} by assumption, the joint distribution corresponding to the maximizer of the LP must do so as well.  As a result, the same will follow for $\Ptilde_{XY,n}$ if we can show that the LP and the IP share a common solution.  This turns out to be a special case of the following technical yet standard result from the theory of discrete optimization.

\begin{lemma} \label{lem:TU}
    {\em (Linear programming and totally unimodular matrices)} \cite[Sec.~III.1]{Nem98} 
    Consider a linear program (LP) of the form 
    \begin{equation}
            \mathrm{\maximize}_{\zv} ~\cv^{T}\zv ~~\text{ subject to }~~ \Av\zv=\bv,\zv\succeq 0 \label{eq:LP}
    \end{equation}
    for some matrix $\Av$ and vectors $(\bv,\cv)$, where $\succeq$ denotes element-wise inequality.  We have the following:
    \begin{enumerate}
        \item If the matrix $\Av$ is totally unimodular (i.e., any square sub-matrix has determinant $1$, $-1$, or $0$), and the vector $\bv$ is integer-valued, then there must exist an integer-valued maximizer $\zv^*$ in \eqref{eq:LP};
        \item A sufficient condition for $\Av$ to be totally unimodular is the following: The rows of $\Av$ index the nodes of an undirected bipartite graph, the columns of $\Av$ index the corresponding edges, and the $(i,j)$-th entry of $\Av$ equals $\openone\{ \text{node } i \text{ is one of the two nodes in edge }j \}$.
    \end{enumerate}
\end{lemma}

We now argue that the LP version of \eqref{eq:IP} (i.e., with $\RR_{+}$ in place of $\ZZ_{+}$) satisfies the conditions of Lemma \ref{lem:TU}.  We first assume that $q(x,y) > 0$ for all $(x,y)$, and then turn to the general case.  When $q(x,y) > 0$ for all $(x,y)$, the constraint matrix corresponds to the {\em complete} bipartite graph with left-nodes $\Xc$ and right-nodes $\Yc$.  For instance, in the case $|\Xc|=|\Yc|=3$, we have
\begin{equation}
    \Av=\left[\begin{array}{ccccccccc}
    \bone & \bone & \bone & 0 & 0 & 0 & 0 & 0 & 0\\
    0 & 0 & 0 & \bone & \bone & \bone & 0 & 0 & 0\\
    0 & 0 & 0 & 0 & 0 & 0 & \bone & \bone & \bone\\
    \bone & 0 & 0 & \bone & 0 & 0 & \bone & 0 & 0\\
    0 & \bone & 0 & 0 & \bone & 0 & 0 & \bone & 0\\
    0 & 0 & \bone & 0 & 0 & \bone & 0 & 0 & \bone
    \end{array}\right], \label{eq:A_example}
\end{equation}
where the rows correspond to the 3 elements of $\Xc$ and then the 3 elements of $\Yc$, and the columns correspond to the 9 pairs $(x,y)$.  We also see that \eqref{eq:IP} corresponds to an integer vector $\bv$ in \eqref{eq:IP}, containing the values $N_x$ and $N_y$.

When there exist $(x,y)$ pairs such that $q(x,y) = 0$ (and hence $\log q(x,y) = -\infty$), more care is needed in equating the relaxed variant of \eqref{eq:IP} with \eqref{eq:LP}.  In particular, any positive value of $N_{xy}$ corresponding to $q(x,y) = 0$ gives an objective value of $-\infty$, which cannot be optimal.  Thus, we reformulate \eqref{eq:IP} as optimizing only over $\{N_{xy} \,:\, q(x,y) > 0\}$, and always setting $N_{xy} = 0$ when $q(x,y) = 0$.  In this case, the only difference in the above analysis is that the complete bipartite graph construction exemplified in \eqref{eq:A_example} is replaced by an {\em incomplete} bipartite graph construction; the edges $(x,y)$ corresponding to $q(x,y) = 0$ are removed.

Hence, Lemma \ref{lem:TU} implies that the IP \eqref{eq:IP} and its LP relaxation share a common solution, implying that $\Ptilde_{XY,n}$ is feasible in \eqref{eq:ens_R0n}, and therefore that \eqref{eq:ens_mi} yields the desired claim \eqref{eq:ens_LM_cmp}.

\subsection{Analog of Lemma \ref{lem:lb_technical} for the GMI}

Recall that $P_{XY}$ is assumed to be a joint type.  When proving ensemble tightness for the GMI, the analog of \eqref{eq:lb_technical} in Lemma \ref{lem:lb_technical} is given by
\begin{align}
    &\min_{\substack{\Ptilde_{XY} \in \Pc_n(\Xc\times\Yc) \,:\, \Ptilde_Y = P_Y, \\ \EE_{\Ptilde}[\log q(X,Y)] \ge \EE_{P}[\log q(X,Y)]}} D(\Ptilde_{XY} \| Q_X \times P_Y) \nonumber \\ &\qquad\quad\le \min_{\substack{\Ptilde_{XY} \in \Pc(\Xc\times\Yc) \,:\, \Ptilde_Y = P_Y, \\ \EE_{\Ptilde}[\log q(X,Y)] \ge \EE_{P}[\log q(X,Y)]}} D(\Ptilde_{XY} \| Q_X \times P_Y) + \delta. \label{eq:lb_technical_GMI}
\end{align}
Compared to the above analysis, this is relatively simple to prove.  First note that since $\Ptilde_Y = P_Y$ is fixed on both sides, we can rewrite both sides in terms of $\Ptilde_{X|Y}$, with the left-hand side constrained to choices such that $P_Y \times \Ptilde_{X|Y}$ is a joint type.

Given any $\Ptilde_{X|Y}$ feasible on the right-hand side of \eqref{eq:lb_technical_GMI}, we construct a feasible $\Ptilde_{X|Y}^{(n)}$ on the left-hand side, by performing {\em rounding} to ensure that each value $nP_Y(y)\Ptilde_{X|Y}(x|y)$ is integer-valued.  More specifically, for each $y \in \Yc$, we sort the values $\{q(x,y)\}_{x\in\Xc}$ in increasing order, and round the values $\Ptilde_{X|Y}(x|y)$ up for the $x$ values with the highest metric, while rounding down for the $x$ values with the lowest metric.  This requires choosing the appropriate number of values (between $0$ and $|\Xc|$) to round up vs.~down to ensure that the probabilities sum to one.  A feasible approach is to work iteratively, rounding another value up (respectively, down) whenever the current sum of rounded and non-rounded values is less than or equal to one (respectively, greater than one).

By doing so, we are guaranteed that $\Ptilde_{X|Y}^{(n)}$ yields a higher decoding metric than $\Ptilde_{X|Y}$, and hence the metric constraint remains satisfied.  Moreover, the entries of $P_Y \times \Ptilde_{X|Y}$ are arbitrarily close to $P_Y \times \Ptilde_{X|Y}^{(n)}$ for large $n$, and hence, the change in the objective function is upper bounded by an arbitrarily small constant $\delta > 0$.

\subsection{Extensions to Multi-User Coding Techniques}

For the most part, the preceding analysis readily extends to the achievable rates obtained via multi-user coding techniques ({\em cf.}, Sections \ref{ch:mac} and \ref{ch:multiuser}).  However, the analysis of Section \ref{sec:technical_tightness} requires some non-trivial changes, which we outline here.  We focus on the MAC considered in Section \ref{ch:mac}, since once this is understood, the ensembles of Section \ref{ch:multiuser} can be handled similarly.

Let $g(\Ptilde_{X_1X_2Y})$ denote the objective in the left-hand side of \eqref{eq:MAC_Final0_LB}, and recall that the constraint set $\SetTiiincc$ is defined in \eqref{eq:MAC_SetT12n}.  For the subsequent analysis, the key property is that $g(\cdot)$ is a continuous function.  We wish to show that
\begin{align}
    &\min_{\substack{\Ptilde_{X_{1}X_{2}Y} \in \Pc_n(\Xc_1 \times \Xc_2 \times \Yc) \,:\, \\ \Ptilde_{X_{1}}=P_{X_{1}},\Ptilde_{X_{2}}=P_{X_{2}},\Ptilde_{Y}=P_{Y}\\
            \EE_{\Ptilde}[\log q(X_{1},X_{2},Y)]\ge\EE_{P}[\log q(X_{1},X_{2},Y)]}} g(\Ptilde_{X_1X_2Y}) \nonumber \\ &\qquad\quad
    \le\min_{\substack{\Ptilde_{X_{1}X_{2}Y} \in \Pc(\Xc_1 \times \Xc_2 \times \Yc) \,:\, \\ \Ptilde_{X_{1}}=P_{X_{1}},\Ptilde_{X_{2}}=P_{X_{2}},\Ptilde_{Y}=P_{Y}\\
            \EE_{\Ptilde}[\log q(X_{1},X_{2},Y)]\ge\EE_{P}[\log q(X_{1},X_{2},Y)]}} g(\Ptilde_{X_1X_2Y}) + \delta \label{eq:mac_two_mins}
\end{align}
for arbitrarily small $\delta > 0$ and sufficiently large $n$; observe that the only difference in the two minimizations is the presence of $\Pc$ vs.~$\Pc_n$.  To prove \eqref{eq:mac_two_mins}, we apply the arguments of Section \ref{sec:technical_tightness} in two steps:
\begin{enumerate}
    \item Write the minimization on the left-hand side of \eqref{eq:mac_two_mins} as $\min_{\Ptilde_{X_1X_2Y}\in\Pc_n} = \min_{\Ptilde_{X_1X_2}\in\Pc_n} \min_{\Ptilde_{Y|X_1X_2}\in\Pc_n}$, where the decoding metric constraint appears in the second minimization.  By the analysis of Section \ref{sec:technical_tightness}, the resulting minimum value is within $\frac{\delta}{2}$ of the corresponding minimization over all conditional distributions (not necessarily yielding a joint type) for sufficiently large $n$, where $\delta > 0$ is arbitrarily small.
    \item Swap the order of the two minimizations from the first step, so that we have $\min_{\Ptilde_{Y|X_1X_2}\in\Pc}\min_{\Ptilde_{X_1X_2}\in\Pc_n}$, and now it is the minimization over $\Ptilde_{X_1X_2}$ that contains the decoding metric constraint.  Again applying the analysis of Section \ref{sec:technical_tightness}, the minimum value is within $\frac{\delta}{2}$ of the corresponding minimization over all joint distributions  for sufficiently large $n$.
\end{enumerate}
Some additional care is required for \eqref{eq:MAC_Final0_LB} because the failure of \eqref{eq:MAC_R12_LM} initially only guarantees the existence of a distribution for which the three quantities in \eqref{eq:MAC_Final0_LB} are non-positive, with the divergence term being strictly negative; it could still be the case that $I_{\widetilde{P}}(X_1;Y) = R_1$ or $I_{\widetilde{P}}(X_2;Y) = R_2$.  We can address this via a rate-backoff argument. The region $\mathcal{R}_{\mathrm{LM}}(Q_1,Q_2)$ is closed (e.g., due to the dual characterization in Theorem~\ref{thm:MAC_DualRate}), so if $(R_1,R_2)$ lies outside this region and satisfies the individual conditions \eqref{eq:MAC_R1_LM}--\eqref{eq:MAC_R2_LM}, then for some arbitrarily small $\eta>0$, the pair $(R_1-\eta,R_2-\eta)$ also lies outside the region and must violate \eqref{eq:MAC_R12_LM}.\footnote{We may assume that $R_1 > 0$ and $R_2 > 0$ here, since either rate being zero recovers a single-user problem.} Thus, there exists a feasible distribution $\widetilde{P}_{X_1X_2Y}$ satisfying $I_{\widetilde{P}}(X_1;Y) \le R_1 - \eta$, $I_{\widetilde{P}}(X_2;Y) \le R_2 - \eta$, and $D\big(\widetilde{P}_{X_1X_2Y} \,\|\, Q_1\times Q_2\times P_Y\big) < R_1 + R_2 - 2\eta$.   Consequently, the maximum in \eqref{eq:MAC_Final0_LB}, evaluated using the original rates $(R_1,R_2)$, is at most $-\eta$. The continuity and type-approximation arguments above preserve a strictly negative margin for all sufficiently large $n$, thus yielding \eqref{eq:MAC_Final0_LB}.

\section{Other Omitted Proofs} \label{sec:omitted_proofs}

\subsection{Condition for Zero Mismatch Capacity (Second Part of Lemma \ref{lem:pos_conds})}

The following proof is taken from \cite{Csi95}.  For convenience, the condition \eqref{eq:pos_cond} for the positivity of the GMI and LM rate is repeated here:
    \begin{equation}
        \EE_{Q_X \times W}[\log q(X,Y)] > \EE_{Q_X \times P_Y}[\log q(X,Y)]. \label{eq:pos_cond2}
    \end{equation}
In particular, we will use the fact (from the first part of the lemma) that $\CLM > 0$ (and $\CGMI > 0$) if and only if \eqref{eq:pos_cond2} holds for some input distribution $Q_X$, with $P_Y$ being the resulting output distribution.

We first note that $\CM$ is clearly positive whenever the GMI and/or LM rate are positive.  Hence, with the first part of the lemma already established, we only need to show that $\CM = 0$ whenever \eqref{eq:pos_cond2} fails for all $Q_X$.  We prove the contrapositive statement: If $\CM > 0$, then \eqref{eq:pos_cond2} holds for some $Q_X$.

In fact, we will show a stronger statement, namely, that \eqref{eq:pos_cond2} holds for some $Q_X$ whenever the error probability can be made arbitrarily small for a codebook with just two codewords, say $\xv$ and $\xvbar$.  We write
\begin{equation}
    \pe(\xv) = \PP\bigg[\sum_{i=1}^n \log q(\xbar_i,Y_i) \ge \sum_{i=1}^n \log q(x_i,Y_i) \bigg ] \label{eq:pe_x}
\end{equation}
with $Y_i \sim W(\,\cdot \,|\, x_i)$, and similarly for $\pe(\xvbar)$.  By the assumption that the error probability can be made arbitrarily small, we have for any $\epsilon > 0$ and sufficiently large $n$ that
\begin{equation}
    \pe(\xv) \le \epsilon, \text{  and  } \pe(\xvbar) \le \epsilon, \label{eq:two_cw_error}
\end{equation}
since otherwise, the average error probability would be at least $\frac{\epsilon}{2}$.  

To simplify \eqref{eq:two_cw_error}, we use the following technical result \cite[Lemma 3]{Csi95}: Given a finite set $\Zc \subseteq \RR$, for sufficiently small $\epsilon > 0$, it holds for any $n$ and any independent random variables $\{Z_i\}_{i=1}^n$ only taking values on $\Zc$ that
\begin{equation}
    \PP\bigg[\sum_{i=1}^n Z_i \ge 0 \bigg] \le \epsilon \implies \EE\bigg[\sum_{i=1}^n Z_i\bigg] < 0. \label{eq:sum_n_tech_cond}
\end{equation}
We can assume that $q(x_i,Y_i) > 0$ with probability one in \eqref{eq:pe_x}, since otherwise we would have $\pe(\xv) \ge W(y|x) > 0$ for the corresponding $(x,y)$ pair that gives $q(x,y) = 0$, contradicting the assumption of arbitrarily small error probability.
 If it is also the case that $q(\xbar_i,Y_i) > 0$ with probability one in \eqref{eq:pe_x}, then we deduce from \eqref{eq:sum_n_tech_cond} (with $Z_i = \log q(\xbar_i,Y_i) - \log q(x_i,Y_i)$) that
\begin{gather}
    \EE\bigg[\sum_{i=1}^n \log q(\xbar_i,Y_i) - \sum_{i=1}^n \log q(x_i,Y_i) \bigg ] < 0, \label{eq:fixed_xy_1}
\end{gather}
where $Y_i \sim W(\,\cdot \,|\, x_i)$.  Moreover, if the preceding assumption on $q(\xbar_i,Y_i) > 0$ fails, then \eqref{eq:fixed_xy_1} holds trivially, since $\log q(\xbar,y) = -\infty$ whenever $q(\xbar,y) = 0$.  By an analogous argument with the roles of $\xv$ and $\xvbar$ reversed, we have
\begin{equation}
    \EE\bigg[\sum_{i=1}^n \log q(x_i,\Ybar_i) - \sum_{i=1}^n \log q(\xbar_i,\Ybar_i) \bigg ] < 0,  \label{eq:fixed_xy_2}
\end{equation}
where $\Ybar_i \sim W(\,\cdot \,|\, \xbar_i)$.  Adding \eqref{eq:fixed_xy_1}--\eqref{eq:fixed_xy_2} together, we conclude that the sum of the quantities
\begin{equation}
    \EE\big[ \log q(\xbar_i,Y_i) + \log q(x_i,\Ybar_i) - \log q(x_i,Y_i) - \log q(\xbar_i,\Ybar_i) \big] \label{eq:neg_exp}
\end{equation}
over $i=1,\dotsc,n$ must be negative, and hence one of the individual terms (i.e., \eqref{eq:neg_exp} for a given choice of $i$) must also be negative.  That is, there exists a pair $(x,\xbar)$ such that
\begin{equation}
    \EE\big[\log q(x,Y) + \log q(\xbar,\Ybar) \big] > \EE\big[ \log q(\xbar,Y) + \log q(x,\Ybar)\big] \label{eq:exists_pair}
\end{equation}
with $Y \sim W(\,\cdot \,|\, x)$ and $\Ybar \sim W(\,\cdot \,|\, \xbar)$.  We claim that this condition implies that \eqref{eq:pos_cond2} holds with $Q_X(x) = \frac{1}{2}$ for $x \in \{x,\xbar\}$, and $Q_X(x) = 0$ elsewhere.  To see this, observe that $\EE_{Q_X \times W}[\log q(X,Y)]$ equals $\frac{1}{2}$ times the left-hand side of \eqref{eq:exists_pair}, whereas $\EE_{Q_X \times P_Y}[\log q(X,Y)]$ equals $\frac{1}{4}$ times the sum of both sides of \eqref{eq:exists_pair}.

\subsection{Equivalence of the GMI and LM Rate Under Output-Symmetry and Equiprobable Inputs (Lemma \ref{lem:gmi_lm_symm})}

Recall that we are considering $(W,q)$ exhibiting matching output symmetry according to Definition \ref{def:output_symm}, along with the equiprobable input distribution $Q_X(x) = \frac{1}{|\Xc|}$.  Since the channel is output-symmetric, the resulting output probability $P_Y(y)$ is the same for all $y$ in a given subset $\Yc_j$ from Definition \ref{def:output_symm} (but may vary with $j$).

In order to show that $\GMI(Q_X) = \LM(Q_X)$, we will show that the objective function in \eqref{eq:INTR_RateLM} is maximized by $a(x) = 0$ for any value of $\sgz$.  The optimality of $a(x) = 0$ also implies the optimality of any $a(x)$ equaling a constant value not depending on $x$, but the value of zero is the most convenient to work with.

First note that by expanding the logarithm in \eqref{eq:INTR_RateLM}, substituting the equiprobable input distribution, and omitting the term $\EE[ \log q(X,Y)^s]$ not depending on $a(\cdot)$, we are left with
\begin{equation}
    f(s,a) \triangleq \frac{1}{|\Xc|} \sum_{x} a(x) -  \sum_{y} P_Y(y) \log \bigg( \frac{1}{|\Xc|} \sum_{\xbar} q(\xbar,y)^s e^{a(\xbar)} \bigg).
\end{equation}
The partial derivative with respect to $a(x)$ is given by
\begin{equation}
    \frac{\partial f}{\partial a(x)} = \frac{1}{|\Xc|} - \sum_{y} P_Y(y) \frac{ q(x,y)^s e^{a(x)} }{ \sum_{\xbar} q(\xbar,y)^s e^{a(\xbar)} }. \label{eq:deriv_a_init}
\end{equation}
We will show that the choice $a(x) = 0$ for all $x$ makes all of these partial derivatives equal to zero.  This will establish the optimality of $a(x) = 0$, since the objective function is concave in $a(\cdot)$ for fixed $s$, and thus any stationary point must be a global maximum.  It is clear from \eqref{eq:INTR_RateGMI} and \eqref{eq:INTR_RateLM} that if $a(x) = 0$ is optimal, then the GMI and LM rate are equal.

Defining $\eta_s(y) \triangleq \sum_{\xbar} q(\xbar,y)^s$, we can write \eqref{eq:deriv_a_init} as
\begin{align}
    \frac{\partial f}{\partial a(x)} \bigg|_{a(\cdot) = 0} 
    &= \frac{1}{|\Xc|} - \sum_{y} \frac{ P_Y(y) }{ \eta_s(y) } q(x,y)^s. \label{eq:deriv_a_2}
\end{align}
It will be useful to establish that the second term in \eqref{eq:deriv_a_2} sums to one when summed over $x \in \Xc$:
\begin{align}
    \sum_{x} \sum_{y} \frac{ P_Y(y) }{ \eta_s(y) } q(x,y)^s = \sum_{y} \frac{ P_Y(y) }{ \eta_s(y) } \sum_{x} q(x,y)^s = \sum_y \frac{ P_Y(y) }{ \eta_s(y) } \eta_s(y) = 1, \label{eq:sum_to_one}
\end{align}
since $\sum_y P_Y(y) = 1$.

Since $W$ and $q$ are output-symmetric with the same partitions in Definition \ref{def:output_symm}, both $P_Y(y)$ and $\eta_s(y)$ depend on $y$ only through the subset $\Yc_j$ it lies in.  Hence, we can also rewrite \eqref{eq:deriv_a_2} as
\begin{align}
    \frac{\partial f}{\partial a(x)} \bigg|_{a(\cdot) = 0} &= \frac{1}{|\Xc|} - \sum_{j=1}^k \frac{ P_Y(y_j) }{ \eta_s(y_j) } \sum_{y \in \Yc_j} q(x,y)^s, \label{eq:deriv_a_3}
\end{align}
where $y_j$ is an arbitrary element of $\Yc_j$.  Again exploiting the symmetry structure of $q$, we see that the second term in \eqref{eq:deriv_a_3} is the same for all $x$; in particular, recall that the $|\Xc| \times |\Yc_j|$ sub-matrix corresponding to $\Yc_j$ contains rows that are permutations of each other.  By \eqref{eq:sum_to_one}, we know that the sum of the second term in \eqref{eq:deriv_a_2} over all $x \in \Xc$ equals one.  Combining these two findings, we conclude that each term is equal to $\frac{1}{|\Xc|}$, and we attain $\frac{\partial f}{\partial a(x)} \big|_{a(\cdot) = 0} = 0$, as desired.  As mentioned above, this establishes that $\GMI(Q_X) = \LM(Q_X)$.

\subsection{Non-Concavity of the GMI (Lemma \ref{lem:non_concave})}

Recall the multi-letter extension $\CGMI^{(k)}$ of the GMI introduced in Section \ref{sec:multi_LM}.  It was shown in \cite{Bun14} that, in the special case of the zero-undetected error capacity ({\em cf.}, Section \ref{sec:app_zuec}), $\CGMI^{(k)}$ approaches the mismatch capacity $\CM$ in the limit as $k \to \infty$.  Here we argue that this result implies that $\GMI(Q_X)$ is, in general, non-concave in $Q_X$.  To our knowledge, this argument has not appeared in the existing literature.

For concreteness, let $(W,q)$ be the mismatched DMC of Section \ref{sec:su_zuec},  in which $\Xc = \Yc = \{0,1,2\}$, and the channel and decoding metric are given by
\begin{align} 
    \Wv & = \left[\begin{array}{ccc}
    0.75 & 0.25 & 0 \\
    0 & 0.75 & 0.25 \\
    0.25 & 0 & 0.75 \\
    \end{array}\right], \qquad
    \qv = \left[\begin{array}{ccc}
    ~~1~\, & ~~1\,~ & ~~0~~ \\
    ~~0~\, & ~~1\,~ & ~~1~~ \\
    ~~1~\, & ~~0\,~ & ~~1~~ \\
    \end{array}\right], \label{eq:Example_q1_rep}
\end{align}
where $x$ indexes the rows and $y$ indexes the columns.  Observe that the channel and metric are strongly symmetric in the sense of exhibiting the output-symmetry condition of Definition \ref{def:output_symm} with just a single partition.  We also know from Section \ref{sec:su_zuec} that the GMI with $Q_X = \big(\frac{1}{3},\frac{1}{3},\frac{1}{3}\big)$ is strictly smaller than $\CM$.

Fix $k$ sufficiently large and $Q_{X^k}$ such that $\GMI(Q_{X^k}) \ge k(\CM - \delta)$ for some small $\delta > 0$; this is possible due to the fact that $\CGMI^{(k)} \to \CM$ as $k \to \infty$.  Writing $Q_{X^k} = Q_{X_1,\dotsc,X_k}$ explicitly, we claim that the symmetry of $(W,q)$ implies that
\begin{equation}
    \GMI(Q_{X_1,\dotsc,X_k}) = \GMI(Q_{\pi_1(X_1),\dotsc,\pi_k(X_k)}), \label{eq:GMI_cyclic}
\end{equation}
where $Q_{\pi_1(X_1),\dotsc,\pi_k(X_k)}$ is the joint distribution obtained by applying a cyclic shift ({\em not} an arbitrary permutation) $\pi_j$ to each ternary variable $X_j$.  For instance, if we set $k=1$ and consider the input distribution $Q_X = (0.2,0.3,0.5)$, then the GMI is unchanged when we switch to $Q_{X} = (0.5,0.2,0.3)$ or $Q_X = (0.3,0.5,0.2)$.  This can be seen via the dual expression \eqref{eq:INTR_RateGMI} and the fact that applying the same cyclic shift to $X$ and $Y$ in \eqref{eq:Example_q1_rep} leaves each matrix unchanged.

Now, letting $\Pi_k$ denote the set of all possible cyclic shifts $(\pi_1,\dotsc,\pi_k)$, of which there are $3^k$ combinations, we have
\begin{equation}
    \frac{1}{3^k} \sum_{(\pi_1,\dotsc,\pi_k) \in \Pi_k} Q_{\pi_1(X_1),\dotsc,\pi_k(X_k)} = \Qunifk, \label{eq:k_unif}
\end{equation}
where $\Qunifk$ is the uniform distribution over all $3^k$ possible $X^k$ sequences.  This is because for any given length-$k$ input sequence (e.g., $(0,\dotsc,0)$), when we sum $Q_{\pi_1(X_1),\dotsc,\pi_k(X_k)}(0,\dotsc,0)$ over all $\pi_1,\dotsc,\pi_k$, each value of $Q_{X_1 \dotsc X_k}(x_1,\dotsc,x_k)$ is included exactly once, so the total sum is one.

If the GMI were concave, then we could use \eqref{eq:k_unif} and Jensen's inequality to deduce that the input distribution $\Qunifk$ also attains a rate of $k(\CM - \delta)$.  However, this is impossible, because i.i.d.~random coding over the product channel with input distribution $\Qunifk$ is equivalent to i.i.d.~random coding over the original channel $(W,q)$ with $Q_X = \big(\frac{1}{3},\frac{1}{3},\frac{1}{3}\big)$.  We know from Section \ref{sec:su_zuec} that $\GMI(Q_X)$ is strictly smaller than the mismatch capacity in this case, and we know from Lemma \ref{lem:ens_tight} that the GMI is tight with respect to the ensemble average for i.i.d.~random coding.  Hence, attaining a rate of $k(\CM - \delta)$ via the multi-letter GMI with distribution $\Qunifk$ (for arbitrarily small $\delta$) is not possible, meaning that the GMI cannot be concave in general.


\chapter*{Acknowledgments}
\markboth{\sffamily\slshape Acknowledgments}{\sffamily\slshape Acknowledgments}
\addcontentsline{toc}{chapter}{Acknowledgments}

We are grateful to Ehsan Asadi Kangarshahi for helpful discussions regarding the single-letter converse bound of Section \ref{sc:single letter upper bound}, and to Ioannis Kontoyiannis for helpful pointers to the relevant literature on mismatched random coding in rate-distortion theory surveyed in Section \ref{sec:rd_mm_random}.

We would also like to thank the reviewers for their helpful feedback and suggestions, including Amos Lapidoth for handling our submission and providing detailed comments.  In addition, we are grateful to the editorial board and staff of the {\em Foundations and Trends in Communications and Information Theory} for their support.

J.~Scarlett was supported by the Singapore National Research Foundation (NRF) under grant number R-252-000-A74-281.  A.~Guill\'en i F\`abregas and A.~Martinez were supported by the European Research Council under ERC grant agreement 725411, and by the Spanish Ministry of Economy and
Competitiveness under grant TEC2016-78434-C3-1-R. 
A.~Somekh-Baruch was supported by the Israel Science Foundation (ISF) under grant number 631/17. 

\bigskip
{\bf Note on arXiv v5:} We are grateful to Mohammad Amin Sarzaeem and Mohammadhadi Hadavi for helpful comments on our monograph, which led to several corrections, including item (i).  In addition, we identified and corrected other typos and issues throughout the monograph, including item (ii) below.

We highlight the following two corrections due to them being the most significant:
\begin{itemize}
    \item[(i)] corrected several issues in Section \ref{sec:mac_ens_tight} including confusion between $\min\{\cdot\}$ and $\max\{\cdot\}$ in equations such as \eqref{eq:MAC_Final0_LB}, and the incorrect reliance on \eqref{eq:lower12_4} when instead \eqref{eq:lower12_3} is needed to show that the error probability approaches one;
    \item[(ii)] corrected the fading example in Section \ref{sec:ex_fading} so that it states and proves an upper bound and conditions for equality, rather than a general equality.
\end{itemize}



\bibliographystyle{myIEEEtran}
\bibliography{refsMerged}

@article{Kos12,
  author={V. {Kostina} and S. {Verd\'u}},
  journal={IEEE Trans. Inf. Theory}, 
  title={Fixed-Length Lossy Compression in the Finite Blocklength Regime}, 
  year={2012},
  volume={58},
  number={6},
  pages={3309-3338},}

@article{Yan99,
  title={On the redundancy of lossy source coding with abstract alphabets},
  author={Yang, E. and Zhang, Z.},
  journal={IEEE Trans. Inf. Theory},
  volume={45},
  number={4},
  pages={1092-1110},
  year={1999}
}

@ARTICLE{Rus12a,
  author={F. {Rusek} and D. {Fertonani}},
  journal={IEEE Trans. Inf. Theory}, 
  title={Bounds on the information rate of intersymbol interference channels based on mismatched receivers}, 
  year={2012},
  volume={58},
  number={3},
  pages={1470-1482},}

@INPROCEEDINGS{Bro12,
  author={S. I. {Bross} and A. {Lapidoth}},
  booktitle={IEEE Inf. Theory Workshop}, 
  title={The additive noise channel with a helper}, 
  year={2019},
}

@INPROCEEDINGS{Abo00,
  author={I. {Abou-Faycal} and A. {Lapidoth}},
  booktitle = {IEEE Conv. Elec. Eng. in Israel},
  title={On the capacity of reduced-complexity receivers for intersymbol interference channels}, 
  year={2000},
}

@article{Sak69,
  title={The rate distortion function for a class of sources},
  author={D. Sakrison},
  journal={Information and Control},
  volume={15},
  number={2},
  pages={165-195},
  year={1969}
}

@article{Sak70,
  title={The rate of a class of random processes},
  author={Sakrison, D},
  journal={IEEE Trans. Inf. Theory},
  volume={16},
  number={1},
  pages={10-16},
  year={1970}
}

@article{Ste93,
  title={An algorithm for source coding subject to a fidelity criterion, based on string matching},
  author={Steinberg, Yossef and Gutman, Michael},
  journal={IEEE Trans. Inf. Theory},
  volume={39},
  number={3},
  pages={877--886},
  year={1993}
}

@article{Zha96,
  title={An on-line universal lossy data compression algorithm via continuous codebook refinement. {I}. {B}asic results},
  author={Zhang, Zhen and Wei, Victor K},
  journal={IEEE Trans. Inf. Theory},
  volume={42},
  number={3},
  pages={803-821},
  year={1996}
}

@article{Yan98,
  title={On the performance of data compression algorithms based upon string matching},
  author={Yang, E. and Kieffer, J. C.},
  journal={IEEE Trans. Inf. Theory},
  volume={44},
  number={1},
  pages={47-65},
  year={1998}
}

@article{Kon06,
  title={Mismatched codebooks and the role of entropy coding in lossy data compression},
  author={Kontoyiannis, Ioannis and Zamir, Ram},
  journal={IEEE Trans. Inf. Theory},
  volume={52},
  number={5},
  pages={1922-1938},
  year={2006}
}

@article{Dem02,
  title={Source coding, large deviations, and approximate pattern matching},
  author={Dembo, Amir and Kontoyiannis, Ioannis},
  journal={IEEE Trans. Inf. Theory},
  volume={48},
  number={6},
  pages={1590-1615},
  year={2002}
}

@article{Mot09,
  title={Capacity bounds for the {G}aussian interference channel},
  author={Motahari, Abolfazl Seyed and Khandani, Amir Keyvan},
  journal={IEEE Trans. Inf. Theory},
  volume={55},
  number={2},
  pages={620-643},
  year={2009}
}

@inproceedings{Ahl73,
  title={Multi-way communication channels},
  author={Ahlswede, Rudolf},
  booktitle={IEEE Int. Symp. Inf. Theory},
  year={1973}
}

@article{asyhari2014mimo,
	Author = {A. T. Asyhari and A. {Guill\'{e}n i F\`{a}bregas}},
	Journal = {IEEE Trans. Inf. Theory},
	Number = {11},
	Pages = {7166--7185},
	Title = {{MIMO} block-fading channels with mismatched {CSI}},
	Volume = {60},
    Month = {Nov.},
	Year = {2014}}

@article{Somekh-Baruch_IT_18,
	Author = {A. {Somekh-Baruch}},
	Journal = {IEEE Trans. Inf. Theory},
	Month = {Sept.},
	Number = {9},
	Pages = {6196-6207},
	Title = {Converse theorems for the {DMC} with mismatched decoding},
	Volume = {64},
	Year = {2018}}

@article{ghozlan2017models,
	Author = {H. Ghozlan and G. Kramer},
	Journal = {IEEE Trans. Inf. Theory},
	Number = {4},
	Pages = {2376--2393},
	Title = {Models and information rates for {W}iener phase noise channels},
	Volume = {63},
	Year = {2017}}

@inproceedings{KangarshahiGuilleniFabregasISIT2019,
	Author = {E. {Asadi Kangarshahi} and A. {Guill\'en i F\`abregas}},
	Booktitle = {IEEE Int. Symp. Inf. Theory},
	Pages = {2873-2877},
	Title = {An upper bound to the mismatch capacity},
	Year = {2019}}

@article{KangarshahiGuilleniFabregasFull2020,
 	author = {E. {Asadi Kangarshahi} and A. {Guill\'en i F\`abregas}},
    title={A single-letter upper bound to the mismatch capacity},
    journal={IEEE Trans. Inf. Theory},
    volume={67},
    number={4},
    pages={2013-2033},
    year={2021}
}

@inproceedings{KangarshahiGuilleniFabregasIZS2020,
	Author = {E. {Asadi Kangarshahi} and A. {Guill\'en i F\`abregas}},
	Booktitle = {Int. Z\"urich Sem. Inf. Comm.},
	Pages = {115-119},
	Title = {Properties of a recent upper bound to the mismatch capacity},
	Year = {2020}}

@article{Bal95,
	Author = {Balakirsky, V.B.},
	Journal = {IEEE Trans. Inf. Theory},
	Month = {Nov.},
	Number = {6},
	Pages = {1889-1902},
	Title = {A converse coding theorem for mismatched decoding at the output of binary-input memoryless channels},
	Volume = {41},
	Year = {1995}}

@book{Boy04,
	Author = {Stephen Boyd and Lieven Vandenberghe},
	Publisher = {Cambridge University Press},
	Title = {Convex Optimization},
	Year = {2004}}

@book{Csi11,
	Author = {Imre Csisz\'{a}r and Janos K\"{o}rner},
	Edition = {2},
	Publisher = {Cambridge University Press},
	Title = {Information Theory: Coding Theorems for Discrete Memoryless Systems},
	Year = {2011}}

@article{Csi81,
	Author = {Imre Csisz\'{a}r and Janos K\"{o}rner},
	Journal = {IEEE Trans. Inf. Theory},
	Month = {Jan.},
	Number = {1},
	Pages = {5-12},
	Title = {Graph decomposition: A new key to coding theorems},
	Volume = {27},
	Year = {1981}}

@article{Csi95,
	Author = {Imre Csisz\'{a}r and Prakash Narayan},
	Journal = {IEEE Trans. Inf. Theory},
	Month = {Jan.},
	Number = {1},
	Pages = {35-43},
	Title = {Channel capacity for a given decoding metric},
	Volume = {41},
	Year = {1995}}

@article{Dec97,
	Author = {D {de Caen}},
	Journal = {Discrete Math.},
	Pages = {217-220},
	Title = {A lower bound on the probability of a union},
	Volume = {169},
	Year = {1997}}

@phdthesis{Div78,
	Author = {D. Divsalar},
	School = {Univ. California L.A.},
	Title = {Performance of mismatched receivers on bandlimited channels},
	Year = {1978}}

@conference{Fis71,
	Author = {T. R. M. Fischer},
	Booktitle = {Trans. 8th Prague Conf. Inf. Theory},
	Pages = {211-226},
	Title = {Some remarks on the role of inaccuracy in {S}hannon's theory of information transmission},
	Year = {1978}}

@unpublished{GallagerCC,
	Author = {Robert~Gallager},
	Note = {http://web.mit.edu/gallager/www/notes/notes5.pdf},
	Title = {Fixed composition arguments and lower bounds to error probability}}

@book{Gal68,
	Author = {Robert Gallager},
	Publisher = {John Wiley \& Sons},
	Title = {Information Theory and Reliable Communication},
	Year = {1968}}

@article{Gan00,
	Author = {Anand Ganti and Amos Lapidoth and Emre Telatar},
	Journal = {IEEE Trans. Inf. Theory},
	Month = {Nov.},
	Number = {7},
	Pages = {2315-2328},
	Title = {Mismatched decoding revisited: {G}eneral alphabets, channels with memory, and the wide-band limit},
	Volume = {46},
	Year = {2000}}

@article{Gui08,
	Acmid = {1482623},
	Address = {Hanover, MA, USA},
	Author = {Guill{\'e}n i F\`{a}bregas, Albert and Martinez, Alfonso and Caire, Giuseppe},
	Issue_Date = {January 2008},
	Journal = {Found. Trends Comms. Inf. Theory},
	Month = {Jan.},
	Number = {1-2},
	Numpages = {153},
	Pages = {1-153},
	Publisher = {Now Publishers Inc.},
	Title = {Bit-Interleaved Coded Modulation},
	Volume = {5},
	Year = {2008}}

@book{Han02,
	Author = {Te Sun Han},
	Publisher = {Springer},
	Title = {Information-Spectrum Methods in Information Theory},
	Year = {2002}}

@phdthesis{Hui83,
	Author = {J.Y.N. Hui},
	School = {MIT},
	Title = {Fundamental issues of multiple accessing},
	Year = {1983}}

@article{Kap93,
	Author = {Gideon Kaplan and Shlomo Shamai},
	Journal = {Arch. Elek. \"{U}ber.},
	Number = {4},
	Pages = {228-239},
	Title = {Information rates and error exponents of compound channels with application to antipodal signaling in a fading environment},
	Volume = {47},
	Year = {1993}}

@conference{Kaz81,
	Author = {D. Kazakos},
	Booktitle = {Proc. Conf. Info. Sci. Sys.},
	Pages = {37-42},
	Title = {Upper and lower bounds for noisy channel coding under mismatch},
	Year = {1981}}

@article{Lap96,
	Author = {Amos Lapidoth},
	Journal = {IEEE Trans. Inf. Theory},
	Month = {Sept.},
	Number = {5},
	Pages = {1439-1452},
	Title = {Mismatched decoding and the multiple-access channel},
	Volume = {42},
	Year = {1996}}

@conference{Mar11,
	Author = {Alfonso Martinez and Albert {Guill\'{e}n i F\`{a}bregas}},
	Booktitle = {Inf. Theory Apps. Workshop},
	Title = {Saddlepoint approximation of random-coding bounds},
	Year = {2011}}

@inproceedings{Mar11a,
	Author = {Alfonso Martinez and Albert {Guill\'{e}n i F\`{a}bregas}},
	Booktitle = {IEEE Int. Symp. Inf. Theory},
	Title = {Random-coding bounds for threshold decoders: Error exponent and saddlepoint approximation},
	Year = {2011}}

@article{Mar09,
	Author = {Alfonso Martinez and Albert {Guill\'{e}n\ i\ F\`{a}bregas} and Guiseppe Caire and Frans Willems},
	Journal = {IEEE Trans. Inf. Theory},
	Month = {June},
	Number = {6},
	Pages = {2756-2765},
	Title = {Bit-interleaved coded modulation revisited: A mismatched decoding perspective},
	Volume = {55},
	Year = {2009}}

@article{Mer95,
	Author = {Neri Merhav and Gideon Kaplan and Amos Lapidoth and Shlomo Shamai},
	Journal = {IEEE Trans. Inf. Theory},
	Month = {Nov.},
	Number = {6},
	Pages = {1953-1967},
	Title = {On information rates for mismatched decoders},
	Volume = {40},
	Year = {1994}}

@article{Pol10a,
	Author = {Yury Polyanskiy and H. Vincent Poor and Sergio Verd\'{u}},
	Journal = {IEEE Trans. Inf. Theory},
	Month = {May},
	Number = {5},
	Pages = {2307-2359},
	Title = {Channel coding rate in the finite blocklength regime},
	Volume = {56},
	Year = {2010}}

@phdthesis{ScarlettThesis,
	Author = {Jonathan Scarlett},
	Url = {http://itc.upf.edu/biblio/1061},
	School = {University of Cambridge},
	Title = {Reliable communication under mismatched decoding},
	Year = {2014}}

@article{Sca16a,
	Author = {Jonathan Scarlett and Alfonso Martinez and Albert {Guill{\'e}n i F\`{a}bregas}},
	Journal = {IEEE Trans. Inf. Theory},
	Month = {July},
	Number = {7},
	Pages = {3950-3970},
	Title = {Multiuser random coding techniques for mismatched decoding},
	Volume = {62},
	Year = {2016}}

@article{Sca14c,
	Author = {Jonathan Scarlett and Alfonso Martinez and Albert {Guill{\'e}n i F\`{a}bregas}},
	Journal = {IEEE Trans. Inf. Theory},
	Month = {May},
	Number = {5},
	Pages = {2647-2666},
	Title = {Mismatched decoding: Error exponents, second-order rates and saddlepoint approximations},
	Volume = {60},
	Year = {2014}}

@conference{Sca15,
	Author = {Jonathan Scarlett and Alfonso Martinez and Albert {Guill\'{e}n i F\`{a}bregas}},
	Booktitle = {IEEE Int. Symp. Inf. Theory},
	Title = {The likelihood decoder: Error exponents and mismatch},
	Year = {2015}}

@article{Sca18,
	Author = {Scarlett, Jonathan and Martinez, Alfonso and Guill{\'e}n i F{\`a}bregas, Albert},
	Journal = {IEEE Trans. Inf. Theory},
	Number = {4},
	Pages = {2253-2266},
	Title = {Mismatched multi-letter successive decoding for the multiple-access channel},
    Month = {April},
	Volume = {64},
	Year = {2018}}

@conference{Sca13e,
	Address = {San Diego, CA},
	Author = {Jonathan Scarlett and Alfonso Martinez and Albert {Guill\'{e}n i F\`{a}bregas}},
	Booktitle = {Inf. Theory and Apps. Workshop},
	Month = {Feb.},
	Title = {Cost-constrained random coding and applications},
	Year = {2013}}

@article{Sca14f,
	Author = {Jonathan Scarlett and Li Peng and Neri Merhav and Alfonso Martinez and Albert {Guill\'{e}n i F\`{a}bregas}},
	Journal = {IEEE Trans. Inf. Theory},
	Month = {Aug.},
	Number = {8},
	Pages = {4449-4462},
	Title = {Expurgated random-coding ensembles: Exponents, refinements and connections},
	Volume = {60},
	Year = {2014}}

@article{Sca15a,
	Author = {Scarlett, J. and Somekh-Baruch, A. and Martinez, A. and Guill\'en i F\`abregas, A.},
	Journal = {IEEE Trans. Inf. Theory},
	Month = {Oct.},
	Number = {10},
	Pages = {5387-5395},
	Title = {A counter-example to the mismatched decoding converse for binary-input discrete memoryless channels},
	Volume = {61},
	Year = {2015}}

@article{Sha12,
	Author = {Shlomo Shamai and Igal Sason},
	Journal = {IEEE Trans. Inf. Theory},
	Month = {Dec.},
	Number = {12},
	Pages = {3029-3051},
	Title = {Variations on the {G}allager bounds, connections, and applications},
	Volume = {48},
	Year = {2002}}

@inproceedings{Als12,
	Author = {Alsan, Mine},
	Booktitle = {Int. Symp. Inf. Theory Apps.},
	Title = {Performance of mismatched polar codes over {BSC}s},
	Year = {2012}}

@inproceedings{Als14,
	Author = {Alsan, Mine and Telatar, Emre},
	Booktitle = {IEEE Inf. Theory Workshop},
	Title = {Polarization as a novel architecture to boost the classical mismatched capacity of {B-DMC}s},
	Year = {2014}}

@book{Elg11,
	Author = {Abbas {El Gamal} and Young Han Kim},
	Publisher = {Cambridge University Press},
	Title = {Network Information Theory},
	Year = {2011}}

@inproceedings{Yas13,
	Author = {Mohammad Hossein Yassaee and Mohammad Reza Aref and Amin Gohari},
  booktitle={IEEE Int. Symp. Inf. Theory}, 
    title={A technique for deriving one-shot achievability results in network information theory}, 
    year={2013}}

@article{Ahl96,
	Author = {Ahlswede, R. and Ning Cai and Zhen Zhang},
	Journal = {IEEE Trans. Inf. Theory},
	Month = {Jan.},
	Number = {1},
	Pages = {55-62},
	Title = {Erasure, list, and detection zero-error capacities for low noise and a relation to identification},
	Volume = {42},
	Year = {1996}}

@inproceedings{Als13a,
	Author = {Alsan, M.},
	Booktitle = {IEEE Inf. Theory Workshop},
	Title = {A lower bound on achievable rates by polar codes with mismatch polar decoding},
	Year = {2013}}

@article{Asy12,
	Author = {Asyhari, A.T. and {Guill\'{e}n\ i\ F\`{a}bregas}, A.},
	Journal = {IEEE Trans. Inf. Theory},
    Month = {March},
	Number = {3},
	Pages = {1483-1517},
	Title = {Nearest neighbor decoding in {MIMO} block-fading channels with imperfect {CSIR}},
	Volume = {58},
	Year = {2012}}

@article{Bin99,
	Author = {Binshtok, N. and Shamai, S.},
	Journal = {IEEE Trans. Commun.},
	Number = {11},
    Month = {Nov.},
	Pages = {1636-1645},
	Title = {Integer metrics for binary input symmetric output memoryless channels},
	Volume = {47},
	Year = {1999}}

@book{Cov06,
	Author = {Thomas M Cover and Joy A Thomas},
	Publisher = {John Wiley \& Sons, Inc.},
	Title = {Elements of Information Theory},
	Year = {2006}}

@conference{Eli55,
	Author = {P. Elias},
	Booktitle = {London Symp. Inf. Theory},
	Title = {Coding for two noisy channels},
	Year = {1955}}

@article{For73,
	Author = {Forney, G.D.},
	Journal = {Proc. IEEE},
	Month = {March},
	Number = {3},
	Pages = {268-278},
	Title = {The {V}iterbi algorithm},
	Volume = {61},
	Year = {1973}}

@article{Gal65,
	Author = {Gallager, R.},
	Journal = {IEEE Trans. Inf. Theory},
	Month = {Jan.},
	Number = {1},
	Pages = {3-18},
	Title = {A simple derivation of the coding theorem and some applications},
	Volume = {11},
	Year = {1965}}

@article{Kor98,
	Author = {K\"{o}rner, J. and Orlitsky, A.},
	Journal = {IEEE Trans. Inf. Theory},
	Month = {Oct.},
	Number = {6},
	Pages = {2207 -2229},
	Title = {Zero-error information theory},
	Volume = {44},
	Year = {1998}}

@article{Lap97,
	Author = {Lapidoth, A.},
	Journal = {IEEE Trans. Inf. Theory},
	Month = {Jan.},
	Number = {1},
	Pages = {38-47},
	Title = {On the role of mismatch in rate distortion theory},
	Volume = {43},
	Year = {1997}}

@article{Lap96a,
	Author = {Lapidoth, A.},
	Journal = {IEEE Trans. Inf. Theory},
	Month = {Sept.},
	Number = {5},
	Pages = {1520-1529},
	Title = {Nearest neighbor decoding for additive non-{G}aussian noise channels},
	Volume = {42},
	Year = {1996}}

@article{Lap98,
	Author = {Lapidoth, A. and Narayan, P.},
	Journal = {IEEE Trans. Inf. Theory},
	Month = {Oct.},
	Number = {6},
	Pages = {2148-2177},
	Title = {Reliable communication under channel uncertainty},
	Volume = {44},
	Year = {1998}}

@article{Lap02,
	Author = {Lapidoth, A. and Shamai, S.},
	Journal = {IEEE Trans. Inf. Theory},
	Number = {5},
	Pages = {1118-1134},
	Title = {Fading channels: How perfect need ``perfect side information'' be?},
	Volume = {48},
	Year = {2002},
    Month = {May}}

@article{Lap98a,
	Author = {Lapidoth, Amos and Shamai, Shlomo},
	Journal = {European Trans. Telecom.},
	Number = {6},
	Pages = {473-482},
	Publisher = {Wiley Subscription Services, Inc., A Wiley Company},
	Title = {A lower bound on the bit-error-rate resulting from mismatched {V}iterbi decoding},
	Volume = {9},
	Year = {1998}}

@article{Sal95,
	Author = {Salz, J. and Zehavi, E.},
	Journal = {IEEE Trans. Commun.},
	Month = {Feb./Mar./Apr.},
	Number = {2/3/4},
	Pages = {307-317},
	Title = {Decoding under integer metrics constraints},
	Volume = {43},
	Year = {1995}}

@article{Sha59,
	Author = {C. E. Shannon},
	Journal = {Bell Sys. Tech. Journal},
	Number = {3},
	Pages = {611-656},
	Title = {Probability of error for optimal codes in a {G}aussian channel},
	Volume = {38},
	Year = {1959}}

@article{Sha56,
	Author = {C. E. Shannon},
	Journal = {IRE Trans. Inf. Theory},
	Month = {Sept.},
	Number = {3},
	Pages = {8-19},
	Title = {The zero error capacity of a noisy channel},
	Volume = {2},
	Year = {1956}}

@article{Sha48,
	Author = {C. E. Shannon},
	Journal = {Bell Sys. Tech. Journal},
	Month = {July and Oct.},
	Pages = {379-423 and 623-656},
	Title = {A mathematical theory of communication},
	Volume = {27},
	Year = {1948}}

@phdthesis{Shu03,
	Author = {N. Shulman},
	School = {Tel Aviv University},
	Title = {Communication over an unknown channel via common broadcasting},
	Year = {2003}}

@article{Som15,
	Author = {Somekh-Baruch, A.},
	Journal = {IEEE Trans. Inf. Theory},
	Month = {Feb.},
	Number = {2},
	Pages = {727-740},
	Title = {On achievable rates and error exponents for channels with mismatched decoding},
	Volume = {61},
	Year = {2015}}

@article{Sti66,
	Author = {Stiglitz, I.},
	Journal = {IEEE Trans. Inf. Theory},
	Month = {April},
	Number = {2},
	Pages = {189-195},
	Title = {Coding for a class of unknown channels},
	Volume = {12},
	Year = {1966}}

@article{Ver94,
	Author = {Verd\'{u}, S. and Te Sun Han},
	Journal = {IEEE Trans. Inf. Theory},
	Month = {July},
	Number = {4},
	Pages = {1147-1157},
	Title = {A general formula for channel capacity},
	Volume = {40},
	Year = {1994}}

@article{Wei04,
	Author = {Weingarten, H. and Steinberg, Y. and Shamai, S.},
	Journal = {IEEE Trans. Inf. Theory},
	Month = {Aug.},
	Number = {8},
	Pages = {1665-1686},
	Title = {Gaussian codes and weighted nearest neighbor decoding in fading multiple-antenna channels},
	Volume = {50},
	Year = {2004}}

@article{Mer13a,
	Author = {Merhav, N.},
	Journal = {IEEE Trans. Inf. Theory},
	Month = {Sept.},
	Number = {9},
	Pages = {5566-5576},
	Title = {Universal decoding for arbitrary channels relative to a given class of decoding metrics},
	Volume = {59},
	Year = {2013}}

@inproceedings{Som15a,
	Author = {A. Somekh-Baruch},
	Booktitle = {IEEE Int. Symp. Inf. Theory},
	Title = {On mismatched list decoding},
	Year = {2015}}

@article{Som14,
	Author = {A. Somekh-Baruch},
	Journal = {IEEE Trans. Inf. Theory},
	Month = {Sept.},
	Number = {9},
	Pages = {4554-4568},
	Title = {A general formula for the mismatch capacity},
	Volume = {61},
	Year = {2015}}

@inproceedings{Fel16,
	Author = {Y. Feldman and A. Somekh-Baruch},
	Booktitle = {IEEE Inf. Theory Workshop},
	Title = {Channels with state information and mismatched decoding},
	Year = {2016}}

@inproceedings{Som17,
	Author = {A. Somekh-Baruch},
	Booktitle = {IEEE Int. Symp. Inf. Theory},
	Title = {Mismatched identification via channels},
	Year = {2017}}

@article{Hul17,
	Author = {W. {Huleihel} and S. {Salamatian} and N. {Merhav} and M. {M{\'e}dard}},
	Journal = {IEEE Trans. Inf. Theory},
	Month = {July},
	Number = {7},
	Pages = {4499-4517},
	Title = {Gaussian intersymbol interference channels with mismatch},
	Volume = {65},
	Year = {2019}}

@inproceedings{Som15b,
	Author = {A. Somekh-Baruch},
	Booktitle = {IEEE Int. Symp. Inf. Theory},
	Title = {Multi-letter converse bounds for the mismatched discrete memoryless channel with an additive metric},
	Year = {2015}}

@article{Gra75,
	Author = {R. Gray and L. Davisson},
	Journal = {IEEE Trans. Commun.},
	Month = {April},
	Number = {4},
	Pages = {439-443},
	Title = {Quantizer mismatch},
	Volume = {23},
	Year = {1975}}

@article{Gra03,
	Author = {R. M. Gray and T. Linder},
	Journal = {IEEE Trans. Inf. Theory},
	Month = {May},
	Number = {5},
	Pages = {1204-1217},
	Title = {Mismatch in high-rate entropy-constrained vector quantization},
	Volume = {49},
	Year = {2003}}

@inproceedings{Bun12,
	Author = {Bunte, Christoph and Lapidoth, Amos},
	Booktitle = {Allerton Conf. Comm., Control, and Comp.},
	Pages = {1838--1842},
	Title = {The zero-undetected-error capacity of discrete memoryless channels with feedback},
	Year = {2012}}

@article{Bun14,
	Author = {Bunte, Christoph and Lapidoth, Amos and Samorodnitsky, Alex},
	Journal = {IEEE Trans. Inf. Theory},
	Number = {7},
	Pages = {3825-3833},
	Publisher = {IEEE},
	Title = {The zero-undetected-error capacity approaches the {S}perner capacity},
	Volume = {60},
	Year = {2014}}

@inproceedings{Lap16,
	Author = {A. Lapidoth},
	Booktitle = {IEEE Int. Conf. Sci. Elec. Eng.},
	Pages = {1-5},
	Title = {A note on feedback communication with mismatched decoding},
	Year = {2016}}

@article{Omu82,
	Author = {Omura, J and Levitt, B},
	Journal = {IEEE Trans. Commun.},
	Number = {5},
	Pages = {896-903},
	Publisher = {IEEE},
    Month = {May},
	Title = {Coded error probability evaluation for antijam communication systems},
	Volume = {30},
	Year = {1982}}

@article{Sca17a,
	Author = {Scarlett, Jonathan and Tan, Vincent Y F and Durisi, Giuseppe},
	Journal = {IEEE Trans. Inf. Theory},
	Number = {1},
    Month = {Jan.},
	Pages = {81-92},
	Publisher = {Ieee},
	Title = {The dispersion of nearest-neighbor decoding for additive non-{G}aussian channels},
	Volume = {63},
	Year = {2017}}

@article{Zho17,
	Author = {L. {Zhou} and V. Y. F. {Tan} and M. {Motani}},
	Journal = {IEEE Trans. Inf. Theory},
	Month = {May},
	Number = {5},
	Pages = {3145-3159},
	Title = {Refined asymptotics for rate-distortion using {G}aussian codebooks for arbitrary sources},
	Volume = {65},
	Year = {2019}}

@book{Nem98,
	Author = {Nemhauser, George L and Wolsey, Laurence A},
	Publisher = {Wiley New York},
	Title = {Integer and Combinatorial Optimization},
	Year = {1999}}

@phdthesis{TelatarThesis,
	Author = {Telatar, Emre},
	Institution = {MIT},
	School = {Massachusetts Institute of Technology},
	Title = {Multi-access communications with decision feedback decoding},
	Year = {1992}}

@book{Tse05,
	Author = {Tse, David and Viswanath, Pramod},
	Publisher = {Cambridge University Press},
	Title = {Fundamentals of Wireless Communication},
	Year = {2005}}

@article{Sad09,
	Author = {Sadeghi, Parastoo and Vontobel, Pascal O and Shams, Ramtin},
	Journal = {IEEE Trans. Inf. Theory},
	Number = {2},
	Pages = {663-688},
	Month = {Feb.},
	Title = {Optimization of information rate upper and lower bounds for channels with memory},
	Volume = {55},
	Year = {2009}}

@article{Rus12,
	Author = {Rusek, Fredrik and Fertonani, Dario},
	Journal = {IEEE Trans. Inf. Theory},
	Number = {3},
	Pages = {1470-1482},
	Title = {Bounds on the information rate of intersymbol interference channels based on mismatched receivers},
    Month = {March},
	Volume = {58},
	Year = {2012}}

@inproceedings{Zho18,
	Author = {Zhou, Lin and Tan, Vincent Y F and Motani, Mehul},
	Booktitle = {IEEE Int. Symp. Inf. Theory},
	Title = {Second-order asymptotics of universal {JSCC} for arbitrary sources and additive channels},
	Year = {2018}}

@article{Zha12,
	Author = {Zhang, Wenyi},
	Journal = {IEEE Trans. Commun.},
	Number = {2},
	Pages = {384-399},
	Publisher = {IEEE},
	Title = {A general framework for transmission with transceiver distortion and some applications},
    Month = {Feb.},
	Volume = {60},
	Year = {2012}}

@article{Tan14,
	Author = {Tan, Vincent Y F},
	Journal = {Found. Trends Comms. Inf. Theory},
	Number = {1-2},
	Pages = {1-184},
	Publisher = {Now Publishers, Inc.},
    Month = {Sept.},
	Title = {Asymptotic Estimates in Information Theory with Non-Vanishing Error Probabilities},
	Volume = {11},
	Year = {2014}}

@inproceedings{Feh16,
	Author = {Fehr, Daniel and Scarlett, Jonathan and Martinez, Alfonso},
	Booktitle = {Int. Z\"urich Sem. Comms.},
	Title = {Fixed-energy random coding with rescaled codewords at the transmitter},
	Year = {2016}}

@article{Mer17,
	Author = {Merhav, Neri},
	Journal = {IEEE Trans. Inf. Theory},
	Number = {8},
	Pages = {5039-5051},
	Title = {The generalized stochastic likelihood decoder: Random coding and expurgated bounds},
    Month = {Aug.},
	Volume = {63},
	Year = {2017}}

@article{Ahl99,
	Author = {Ahlswede, Rudolf and Dueck, Gunter},
	Journal = {IEEE Trans. Inf. Theory},
	Number = {1},
	Pages = {15-29},
	Title = {Identification via channels},
    Month = {Jan.},
	Volume = {35},
	Year = {1989}}

@Article{Som19,
  title                    = {Generalized random {G}ilbert-{V}arshamov codes},
  author                   = {Anelia Somekh-Baruch and Jonathan Scarlett and Albert {Guill{\'e}n i F\`{a}bregas}},
  journal                  = {IEEE Trans. Inf. Theory},
  volume={65},
  number={6},
  month = {June},
  pages={3452-3469},
  year={2019}
}

@inproceedings{Var57,
	Author = {R. R. Varshamov},
	Booktitle = {Dokl. Akad. Nauk SSSR},
	Number = {5},
	Pages = {739-741},
	Title = {Estimate of the number of signals in error correcting codes},
	Volume = {117},
	Year = {1957}}

@article{Gil52,
	Author = {E. N. Gilbert},
	Journal = {Bell Labs Tech. J.},
	Number = {3},
	Pages = {504-522},
	Title = {A comparison of signalling alphabets},
	Volume = {31},
	Year = {1952}}

@article{Csi88,
	Author = {Csisz\'{a}r, Imre and Narayan, Prakash},
	Journal = {IEEE Trans. Inf. Theory},
	Number = {2},
	Pages = {181-193},
    Month = {Feb.},
	Title = {The capacity of the arbitrarily varying channel revisited: Positivity, constraints},
	Volume = {34},
	Year = {1988}}

@article{Alv17,
	Author = {A. {Alvarado} and T. {Fehenberger} and B. {Chen} and F. {Willems}},
	Journal = {J. Lightwave Tech.},
	Month = {Jan.},
	Number = {2},
	Pages = {424-439},
	Title = {Achievable information rates for fiber optics: Applications and computations},
	Volume = {36},
	Year = {2018}}

@article{Sec17,
	Author = {M. {Secondini} and E. {Forestieri}},
	Journal = {J. Lightwave Tech.},
	Month = {Feb.},
	Number = {4},
	Pages = {893-902},
	Title = {Scope and limitations of the nonlinear {S}hannon limit},
	Volume = {35},
	Year = {2017}}

@article{Bla60,
	Author = {Blackwell, David and Breiman, Leo and Thomasian, A. J.},
	Journal = {Ann. Math. Stats.},
	Number = {3},
	Pages = {558-567},
	Title = {The capacities of certain channel classes under random coding},
	Volume = {31},
	Year = {1960}}

@article{Bla59,
	Author = {Blackwell, David and Breiman, Leo and Thomasian, A. J.},
	Journal = {Ann. Math. Stats.},
	Number = {4},
	Pages = {1229-1241},
	Title = {The capacity of a class of channels},
	Volume = {30},
	Year = {1959}}

\newpage

\end{document}